\documentclass[sigconf,screen]{acmart}
\usepackage{nicefrac}
\usepackage{scalerel}
\usepackage{soul}
\usepackage{soulpos}
\usepackage{tabularx}
\usepackage{multirow}
\usepackage{arydshln}
\usepackage{threeparttable}
\usepackage{catchfile}
\usepackage{pifont}
\usepackage{framed}
\usepackage{listings}
\usepackage{lstautogobble}
\usepackage{mfirstuc}
\usepackage[textsize=footnotesize]{todonotes}

\usetikzlibrary{arrows.meta,calc}

\makeatletter
\if@ACM@anonymous
  \newcommand{\AnonymousReview}{1}
\fi
\makeatother

\newcommand{\LinksInRowDescription}{1}

\ifdefined\AnonymousReview
  \setuptodonotes{disable}
\fi

\definecolor{Thistle1}{rgb}{1,.884,1}
\definecolor{NavajoWhite1}{rgb}{1,.87,.68}
\definecolor{Azure1}{rgb}{.94,1,1}

\definecolor{BoxColorBlue}{HTML}{1D89E4}
\definecolor{BoxColorRed}{HTML}{FF0D57}
\definecolor{Silver}{HTML}{C0C0C0}
\definecolor{Orange}{HTML}{FFA500}
\definecolor{MediumVioletRed}{HTML}{C71585}
\definecolor{Sienna}{HTML}{A0522D}
\definecolor{DarkOrange}{HTML}{FF8C00}
\definecolor{DimGray}{HTML}{696969}
\definecolor{WebGray}{HTML}{808080}

\AtBeginDocument{%
  }

\setcitestyle{nosort}

\copyrightyear{2024}
\acmYear{2024}
\setcopyright{cc}
\acmConference[RAID 2024]{The 27th International Symposium on Research in Attacks, Intrusions and Defenses}{September 30-October 02, 2024}{Padua, Italy}
\acmBooktitle{The 27th International Symposium on Research in Attacks, Intrusions and Defenses (RAID 2024), September 30-October 02, 2024, Padua, Italy}
\acmDOI{10.1145/3678890.3678897}
\acmISBN{979-8-4007-0959-3/24/09}

\newcommand{\ymark}{\ding{51}}
\newcommand{\nmark}{\ding{55}}

\newcommand{\risk}{\includegraphics[height=1em]{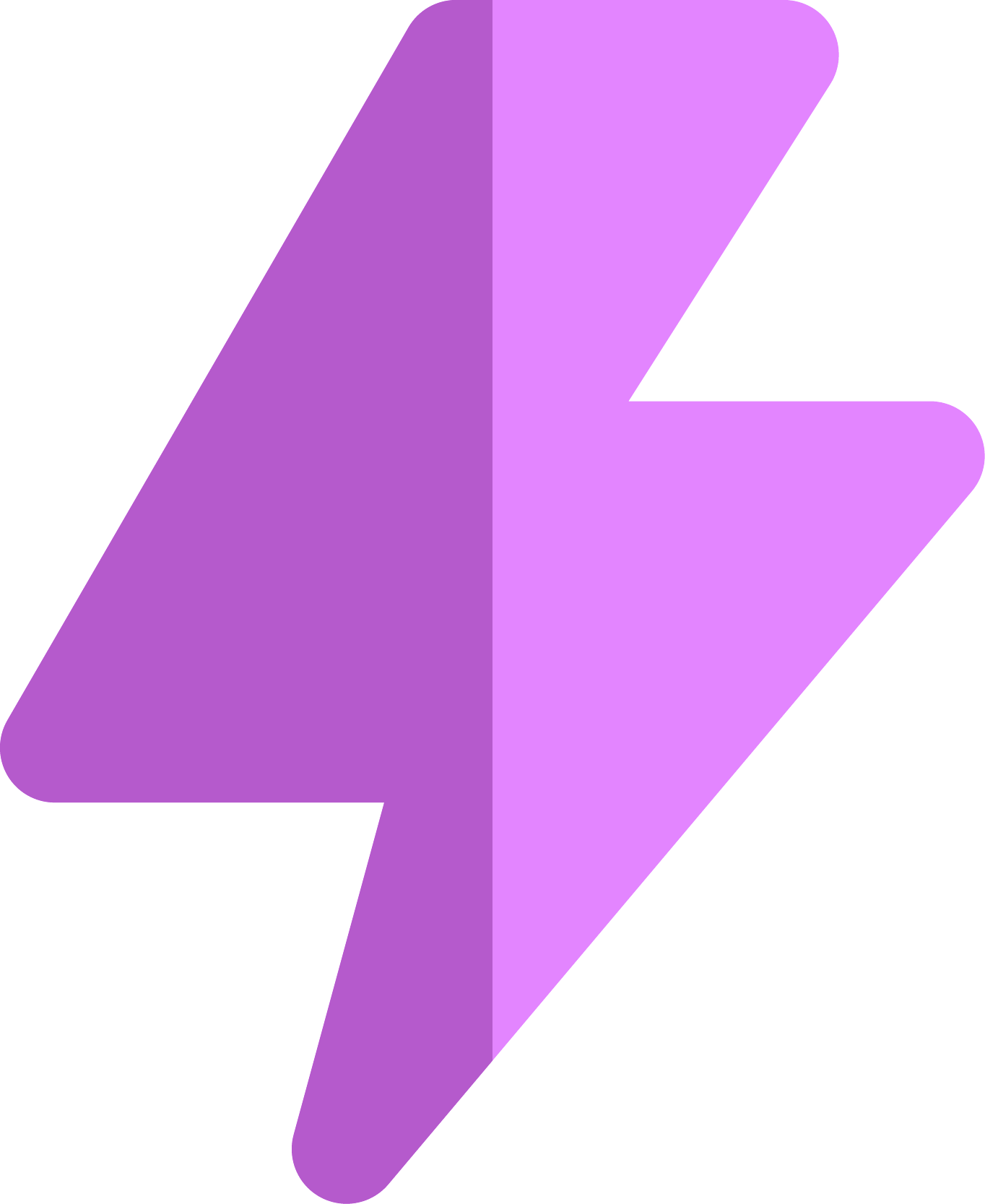}}
\newcommand{\riski}{\scalerel*{\risk}{M}}

\newcommand{\dcpt}{\includegraphics[height=1em]{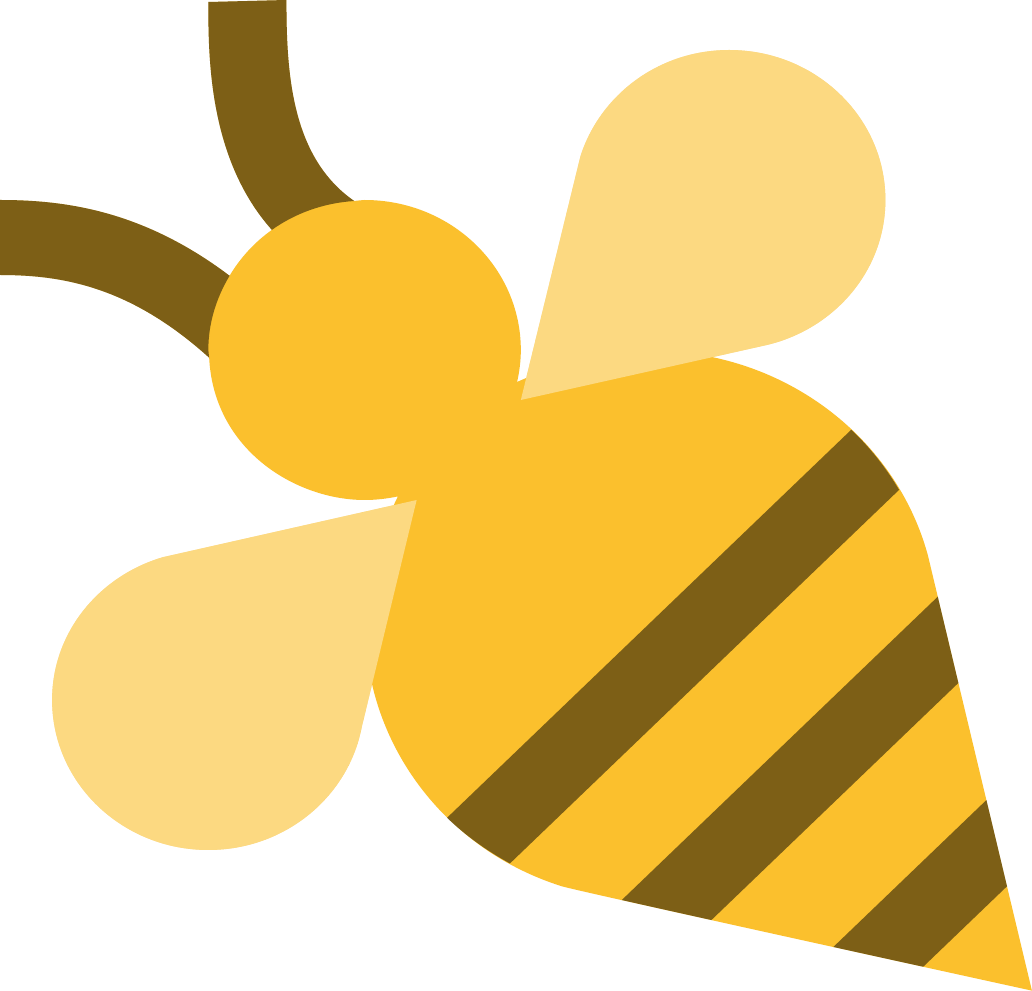}}
\newcommand{\dcpti}{\scalerel*{\dcpt}{M}}

\newcommand{\ntrl}{\includegraphics[height=1em]{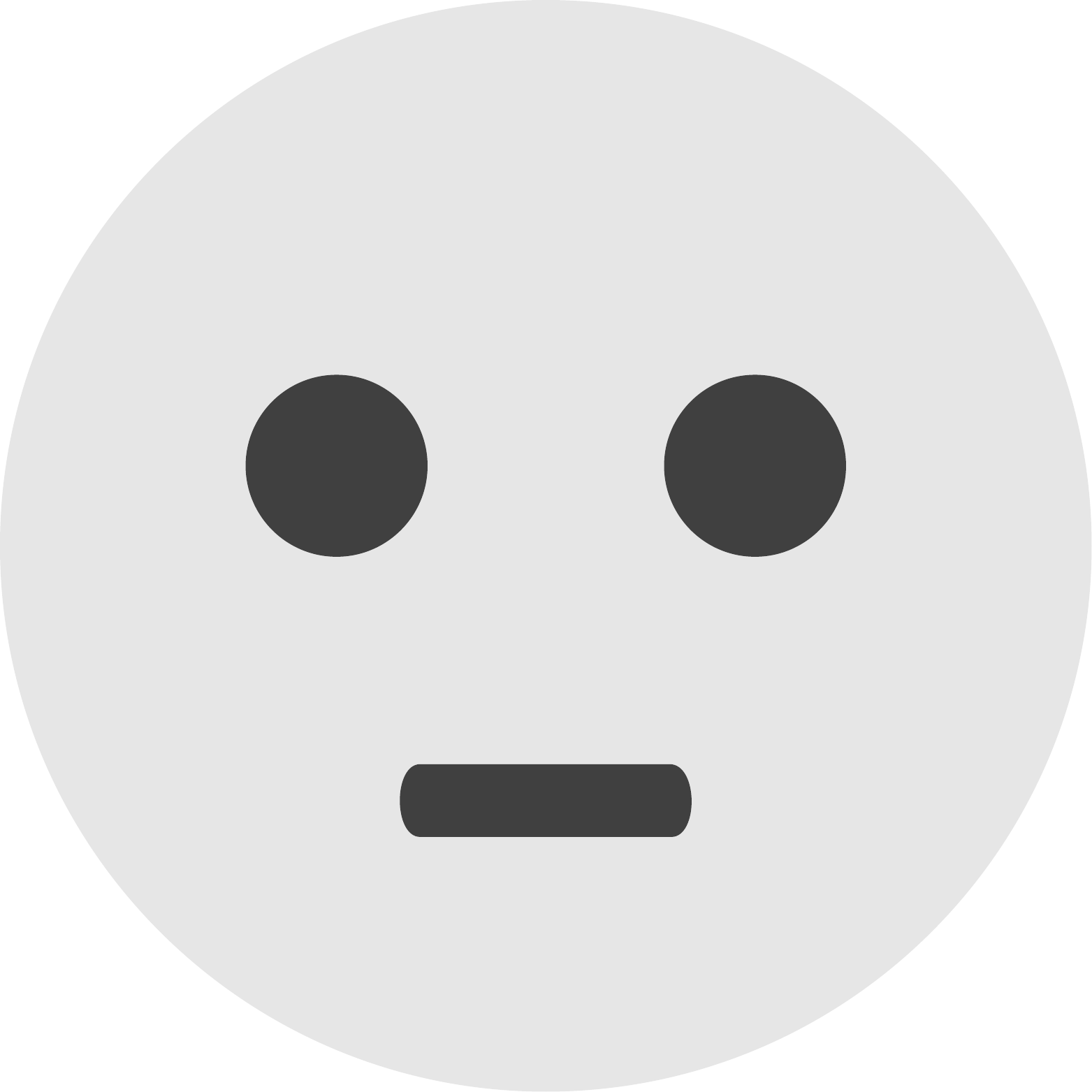}}
\newcommand{\ntrli}{\scalerel*{\ntrl}{M}}

\newcommand{\trap}{\includegraphics[height=1em]{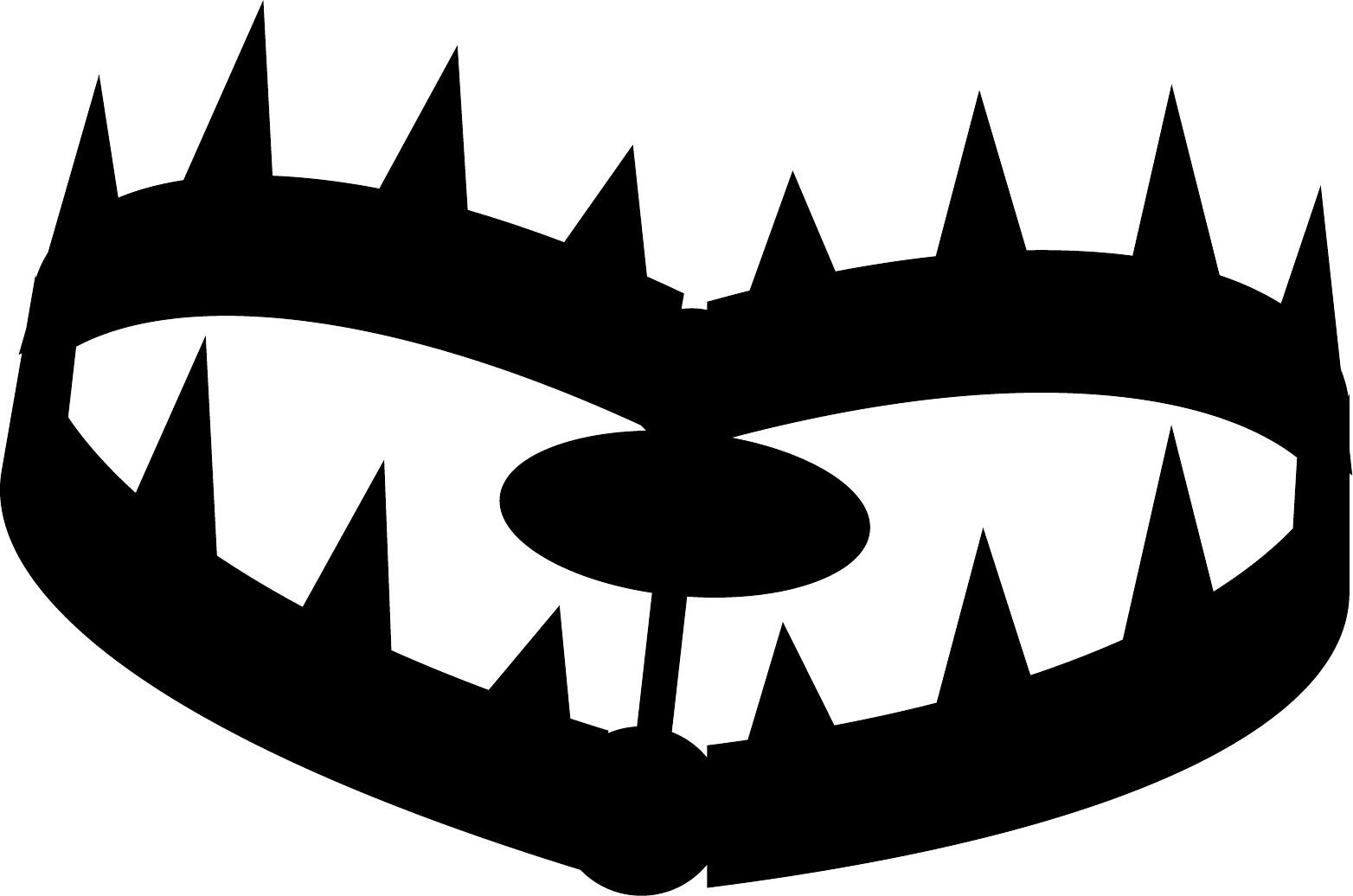}}
\newcommand{\trapi}{\scalerel*{\trap}{M}}

\newcommand{\expl}{\includegraphics[height=1em]{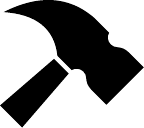}}
\newcommand{\expli}{\scalerel*{\expl}{M}}

\NewDocumentCommand{\rhypertarget}{ m }{%
  \raisebox{\ht\strutbox}{\hypertarget{#1}{}}%
}

\NewDocumentCommand{\RowRefSingle}{ m m }{%
  \hyperlink{row:#1:#2}{\csname VarId#2\endcsname}%
}

\ExplSyntaxOn
\seq_new:N \l_myref_hyperlinks_seq
\NewDocumentCommand{\RowRef}{ m m }{%
  {
  (
  \seq_set_from_clist:Nn \l_myref_hyperlinks_seq {#2}
  \int_zero:N \l_tmpa_int
  \seq_map_inline:Nn \l_myref_hyperlinks_seq
   {
    \int_incr:N \l_tmpa_int
    \hyperlink{row:#1:##1}{\use:c{VarId##1}}
    \int_compare:nNnT { \l_tmpa_int } < { \seq_count:N \l_myref_hyperlinks_seq }
      {,~}
   }
  )
  }
 }
\ExplSyntaxOff

\newcommand{\mfus}{\xmakefirstuc}
\newrobustcmd{\filesystem}{file system}
\newrobustcmd{\header}{HTTP response}
\newrobustcmd{\network}{HTTP request}
\newrobustcmd{\htaccess}{.htaccess file}

\def\secname{\S}

\newcommand{\citereplicatedworks}{\cite{%
  Sahin2022:MeasuringDevelopersWeb,
  Sahin2020:LessonsLearnedSunDEW,
  Han2017:EvaluationDeceptionBasedWeb,
  Petrunic2015:HoneytokensActiveDefense,
  Nikiforakis2011:ExposingLackPrivacy,
  Rowe2006:FakeHoneypotsDefensive,
  Rowe2007:DefendingCyberspaceFake}}

\begin{document}

\setlength{\marginparwidth}{15mm}

%% [usenix] do not print date
% \date{}

%%
%% includes
\newcommand{\RowDeceptiveDeceptive}{}

\newcommand{\RowDeceptiveFilesystemDeceptive}{%
  \itshape CDTs add files with deceptive names to the file system}
\newcommand{\RowDeceptiveHtaccessDeceptive}{%
  \itshape CDTs add directives that seem to leak sensitive paths}
\newcommand{\RowDeceptiveHttpheadersDeceptive}{%
  \itshape CDTs add headers with deceptive tokens, cookies, paths}
\newcommand{\RowDeceptiveNetworkrequestsDeceptive}{%
  \itshape CDTs add parameters or requests that imitate true risks}

\newcommand{\RowDeceptiveFilesystemPrivateKey}{}
\newcommand{\RowDeceptiveFilesystemCardrz}{}
\newcommand{\RowDeceptiveFilesystemPasswords}{}
\newcommand{\RowDeceptiveFilesystemBackup}{}
\newcommand{\RowDeceptiveFilesystemKeys}{}
\newcommand{\RowDeceptiveFilesystemCustomerList}{}
\newcommand{\RowDeceptiveFilesystemConfig}{}
\newcommand{\RowDeceptiveFilesystemSpamList}{}

\newcommand{\RowDeceptiveFilesystemRowe}{%
  e.g., \texttt{examples}, \texttt{gif\_files}, \texttt{idlold}, \texttt{wizard}, ...}

\newcommand{\RowDeceptiveHtaccessAdminRedirect}{%
  \texttt{Redirect 301 "/admin"} line leaks sensitive path}

\newcommand{\RowDeceptiveHttpheadersDevtoken}{%
  \texttt{X-DevToken} header has JWT token with a secret key}

\newcommand{\RowDeceptiveHttpheadersAdminCookie}{%
  \texttt{Set-Cookie} header with \texttt{admin=false} in Base64}

\newcommand{\RowDeceptiveHttpheadersApiserver}{%
  \texttt{X-ApiServer: /hko/api} header leaks API path}

\newcommand{\RowDeceptiveHttpheadersProxyReferer}{%
  \texttt{X-Proxy-Referer} header exposes a fake server's path}

\newcommand{\RowDeceptiveNetworkrequestsCleartextPassword}{%
  Add \texttt{?user=john\&pass=carrot13} to \texttt{/login} request}

\newcommand{\RowDeceptiveNetworkrequestsPathTraversal}{%
  Add \texttt{?file=../dist/Aq.svg} query parameter}

\newcommand{\RowDeceptiveNetworkrequestsIdorReadSecrets}{%
  Extra requests to a few \texttt{/secrets/123} paths\tnote{a}}

\newcommand{\RowDeceptiveNetworkrequestsSessidParameter}{%
  Add \texttt{?SESSID=odq...} query parameter}

\newcommand{\RowDeceptiveNetworkrequestsAdminFalse}{%
  Add \texttt{?admin=false} query parameter}

\newcommand{\RowDeceptiveNetworkrequestsUnescapedJavascript}{%
  Add GET parameter with raw JS}

\newcommand{\RowDeceptiveNetworkrequestsSystemParameter}{%
  Add \texttt{?system=prod} query parameter}

\newcommand{\RowDeceptiveNetworkrequestsDevEndpoint}{%
  Add \texttt{/api.dev} requests}

\newcommand{\RowDeceptiveNetworkrequestsUnescapedJson}{%
  Add GET parameter with raw JSON}

\newcommand{\RowDeceptiveNetworkrequestsMassAssignment}{%
  Extra requests that set fields with GET parameters\tnote{b}}

\newcommand{\RowDeceptiveNetworkrequestsLogEndpoint}{%
  Add \texttt{/log?msg=abc} requests}

\newcommand{\RowRiskyRisky}{}

\newcommand{\RowRiskyFilesystemRisky}{}
\newcommand{\RowRiskyHttpheadersRisky}{}
\newcommand{\RowRiskyNetworkrequestsRisky}{}

\newcommand{\RowRiskyFilesystemPrivateKey}{%
  \raggedright\arraybackslash\texttt{private-key.pem}}

\newcommand{\RowRiskyFilesystemBackup}{%
  \raggedright\arraybackslash\texttt{backup.tar.gz}}

\newcommand{\RowRiskyFilesystemOpenvpnConfig}{%
  \raggedright\arraybackslash\texttt{salphard.ovpn}}

\newcommand{\RowRiskyFilesystemDnsUpdateKey}{%
  \raggedright\arraybackslash\texttt{ddns-update-key}}

\newcommand{\RowRiskyFilesystemKubernetesManifests}{%
  \raggedright\arraybackslash\texttt{k8s-manifests} (directory)}

\newcommand{\RowRiskyHttpheadersOutdatedPhp}{%
  \raggedright\arraybackslash\texttt{X-Powered-By: PHP/5.1.6}}

\newcommand{\RowRiskyHttpheadersProxyAuthLeak}{%
  \raggedright\arraybackslash\texttt{Proxy-Auth.: Basic ...}}

\newcommand{\RowRiskyHttpheadersOutdatedApache}{%
  \raggedright\arraybackslash\texttt{Server: Apache/1.0.3}}

\newcommand{\RowRiskyHttpheadersRequestSmugglingClte}{%
  HTTP Request Smuggling\tnote{a}}

\newcommand{\RowRiskyHttpheadersCrossDomainRefererLeakage}{%
  \texttt{Referer: https://...}}

\newcommand{\RowRiskyNetworkrequestsPasswordHashesInQueryParameters}{%
  Password Hash Parameter, e.g., \raggedright\arraybackslash\texttt{?user=maltier\&hash=...}}

\newcommand{\RowRiskyNetworkrequestsBrokenFunctionLevelAuthorization}{%
  Brk. Fun.-Lvl. Auth.: Unauthenticated user requests privileged data}

\ifdefined\LinksInRowDescription%
  \newcommand{\RowRiskyNetworkrequestsMassAssignment}{%
    Mass Assignment (as in \RowRefSingle{Results}{DcptNetworkrequestsMassAssignment})}
\else%
  \newcommand{\RowRiskyNetworkrequestsMassAssignment}{%
    Mass Assignment (as in \VarIdDcptNetworkrequestsMassAssignment{})}
\fi

\newcommand{\RowRiskyNetworkrequestsBrokenObjectLevelAuthorization}{%
  Brk. Obj.-Lvl. Auth.: User can request sensitive data from other user}

\newcommand{\RowRiskyNetworkrequestsLogSpamEndpoint}{%
  \texttt{/log?msg=...} Endpoint}

\newcommand{\RowRiskyNetworkrequestsInsecureHttp}{%
  Mixing HTTP with HTTPS}

\newcommand{\RowRiskyNetworkrequestsNoRateLimiting}{%
  Huge Payload Sizes}

\newcommand{\RowRiskyNetworkrequestsDevEndpointAccessible}{%
  \texttt{/api.dev} Endpoint}

\newcommand{\RowRiskyNetworkrequestsNosqlInjection}{%
  NoSQL Injection}

% draw a horizontal bar that indicates the various detection scores
\newcommand{\defaultrectwidth}{3cm}
\newcommand{\defaultrectheight}{0.2cm}
\newcommand{\defaultlabelthreshold}{15}
\newcommand{\defaulthackcolor}{MediumVioletRed}
\newcommand{\defaulthackcontrastcolor}{MediumVioletRed}
\newcommand{\defaulthackcolorbrightness}{30}
\newcommand{\defaulthackcolorbrightnesscse}{20}
\newcommand{\defaulthackdrawopacity}{1.0}
\newcommand{\defaulttrapcolor}{DarkOrange}
\newcommand{\defaulttrapcontrastcolor}{DarkOrange}
\newcommand{\defaulttrapcolorbrightness}{30}
\newcommand{\defaulttrapcolorbrightnesscse}{25}
\newcommand{\defaulttrapdrawopacity}{1.0}
\newcommand{\defaultothercolor}{Silver}
\newcommand{\defaultothercontrastcolor}{DimGray}
\newcommand{\defaultothercolorbrightness}{50}
\newcommand{\defaultotherdrawopacity}{1.0}
\newcommand{\defaultnonecolor}{Silver}
\newcommand{\defaultnonecontrastcolor}{WebGray}
\newcommand{\defaultnonecolorbrightness}{10}
\newcommand{\defaultnonedrawopacity}{0.5}

\NewDocumentCommand{\detectionbar}{ m m m m o o }%
{%
  \def\rectwidth{\defaultrectwidth}%
  \pgfmathtruncatemacro\labelthreshold{\defaultlabelthreshold}%
  \def\rectheight{\defaultrectheight}%
  \def\cseoffsetratio{0.2}%
  \def\borderwidth{0mm}%
  \def\borderwidthcse{0mm}%
  \def\tinypercent{\scalebox{1.0}{\%}}%
  \def\ra{#1}%
  \def\racse{\IfNoValueTF{#5}{0}{#5}}%
  \pgfmathtruncatemacro\rat{round(\ra)}%
  \def\rattext{\scalebox{1.0}{\tiny\rat\tinypercent}}%
  \def\rb{#2}%
  \def\rbcse{\IfNoValueTF{#6}{0}{#6}}%
  \pgfmathtruncatemacro\rbt{round(\rb)}%
  \def\rbttext{\scalebox{1.0}{\tiny\rbt\tinypercent}}%
  \def\rc{#3}%
  \pgfmathtruncatemacro\rct{round(\rc)}%
  \def\rcttext{\scalebox{1.0}{\tiny\rct\tinypercent}}%
  \def\rd{#4}%
  \pgfmathtruncatemacro\rdt{round(\rd)}%
  \def\rdttext{\scalebox{1.0}{\tiny\rdt\tinypercent}}%
  \begin{tikzpicture}[baseline=(base)]%
    \coordinate (A0) at (0,0);
    \coordinate (A1) at ({(\ra/100)*\rectwidth}, \rectheight);
    \coordinate (A1LL) at ({((\ra/100)-(\racse/2))*\rectwidth}, \rectheight);
    \coordinate (A1L) at ({((\ra/100)-(\racse/2))*\rectwidth}, {\rectheight*(1+\cseoffsetratio)});
    \coordinate (A1H) at ({((\ra/100)+(\racse/2))*\rectwidth}, {\rectheight*(1+\cseoffsetratio)});
    \coordinate (B0) at ({(\ra/100)*\rectwidth},0);
    \coordinate (B1) at ({((\ra+\rb)/100)*\rectwidth}, {\rectheight});
    \coordinate (B1LL) at ({(((\ra+\rb)/100)-(\rbcse/2))*\rectwidth}, 0);
    \coordinate (B1L) at ({(((\ra+\rb)/100)-(\rbcse/2))*\rectwidth}, {\rectheight*(-\cseoffsetratio)});
    \coordinate (B1H) at ({(((\ra+\rb)/100)+(\rbcse/2))*\rectwidth}, {\rectheight*(-\cseoffsetratio)});
    \coordinate (C0) at ({((\ra+\rb)/100)*\rectwidth}, 0);
    \coordinate (C1) at ({((\ra+\rb+\rc)/100)*\rectwidth}, {\rectheight});
    \coordinate (D0) at ({((\ra+\rb+\rc)/100)*\rectwidth}, 0);
    \coordinate (D1) at (\rectwidth, {\rectheight});
    \coordinate (base) at ($(A0)+(0,0.2mm)$);

    \ifnum\rbt>0
      \IfNoValueTF{#6}{}{%
      \fill[fill=\defaulttrapcolor!\defaulttrapcolorbrightnesscse,draw opacity=\defaulttrapdrawopacity] (B1LL) rectangle (B1H);
      \draw[\defaulttrapcolor,{Bar[scale=0.25]}-{Bar[scale=0.25]},draw opacity=1.0,line width=\borderwidthcse] (B1L) -- (B1H);
      }
    \fi

    \ifnum\rat>0
      \IfNoValueTF{#5}{}{%
      \fill[fill=\defaulthackcolor!\defaulthackcolorbrightnesscse,draw opacity=\defaulthackdrawopacity] (A1LL) rectangle (A1H);
      \draw[\defaulthackcolor,{Bar[scale=0.25]}-{Bar[scale=0.25]},draw opacity=1.0,line width=\borderwidthcse] (A1L) -- (A1H);
      }
    \fi

    \ifnum\rdt>0
      \fill[fill=\defaultnonecolor!\defaultnonecolorbrightness,draw=\defaultnonecolor,draw opacity=\defaultnonedrawopacity,line width=\borderwidth] (D0) rectangle (D1);
      \ifnum\rdt<\labelthreshold\else
        \node[anchor=east,align=center,text=\defaultnonecontrastcolor] at ($(D1)+0.5*(0,-\rectheight)$) {\rdttext};
      \fi
    \fi

    \ifnum\rct>0
      \fill[fill=\defaultothercolor!\defaultothercolorbrightness,draw=\defaultothercolor,draw opacity=\defaultotherdrawopacity,line width=\borderwidth] (C0) rectangle (C1);
      \ifnum\rct<\labelthreshold\else
        \node[anchor=east,align=center,text=\defaultothercontrastcolor] at ($(C1)+0.5*(0,-\rectheight)$) {\rcttext};
      \fi
    \fi

    \ifnum\rbt>0
      \fill[fill=\defaulttrapcolor!\defaulttrapcolorbrightness,draw=\defaulttrapcolor,draw opacity=\defaulttrapdrawopacity,line width=\borderwidth] (B0) rectangle (B1);
      \ifnum\rbt<\labelthreshold\else
        \node[anchor=east,align=center,text=\defaulttrapcontrastcolor] at ($(B1)+0.5*(0,-\rectheight)$) {\rbttext};
      \fi
    \fi

    \ifnum\rat>0
      \fill[fill=\defaulthackcolor!\defaulthackcolorbrightness,draw=\defaulthackcolor,draw opacity=\defaulthackdrawopacity,line width=\borderwidth] (A0) rectangle (A1);
      \ifnum\rat<\labelthreshold\else
        \node[anchor=east,align=center,text=\defaulthackcontrastcolor] at ($(A1)+0.5*(0,-\rectheight)$) {\rattext};
      \fi
    \fi

  \end{tikzpicture}%
}

% just draw one of the small boxes of the detectionbar
\NewDocumentCommand{\detectionbarbox}{ m m m o o o }%
{%
  \def\rectwidth{\IfNoValueTF{#4}{0.22cm}{#4}}%
  \def\rectheight{\IfNoValueTF{#5}{0.22cm}{#5}}%
  \def\borderwidth{\IfNoValueTF{#6}{0mm}{#6}}%
  \begin{tikzpicture}%
    \coordinate (A0) at (0,0);
    \coordinate (A1) at (\rectwidth, \rectheight);
    \fill[fill=#1!#2,draw=#1,draw opacity=#3,line width=\borderwidth] (A0) rectangle (A1);
  \end{tikzpicture}%
}

\newcommand{\dhackbox}{\detectionbarbox{\defaulthackcolor}{\defaulthackcolorbrightness}{1.0}}
\newcommand{\dtrapbox}{\detectionbarbox{\defaulttrapcolor}{\defaulttrapcolorbrightness}{1.0}}
\newcommand{\dotherbox}{\detectionbarbox{\defaultothercolor}{\defaultothercolorbrightness}{1.0}}
\newcommand{\dbothbox}{\detectionbarbox{Sienna}{\defaultothercolorbrightness}{0.5}}
\newcommand{\dnonebox}{\detectionbarbox{\defaultnonecolor}{\defaultnonecolorbrightness}{1.0}}

% draw a number line and boxy arrows that point to the left or right
\NewDocumentCommand{\rewardfigure}{ m }%
{%
  \def\numlinelen{1.1cm}%
  \def\arrowoffset{0.1cm}%
  \def\boxheight{0.2cm}%
  \def\boxwidth{#1}%
  \pgfmathtruncatemacro\boxwidthint{\boxwidth}%
  \ifnum\boxwidthint>0
    \begin{tikzpicture}
      \draw[-] (-\numlinelen/2,0) -- (\numlinelen/2,0);
      \filldraw[fill=BoxColorBlue!30, draw=BoxColorBlue, thick] (0, -\boxheight/2) -- (\boxwidth, -\boxheight/2) -- (\boxwidth+\arrowoffset, 0) -- (\boxwidth, \boxheight/2) -- (0, \boxheight/2) -- (0, -\boxheight/2);
    \end{tikzpicture}
  \else
    \begin{tikzpicture}
      \draw[-] (-\numlinelen/2,0) -- (\numlinelen/2,0);
      \filldraw[fill=BoxColorRed!30, draw=BoxColorRed, thick] (\boxwidth, -\boxheight/2) -- (0, -\boxheight/2) -- (0, \boxheight/2) -- (\boxwidth, \boxheight/2) -- (\boxwidth-\arrowoffset, 0) -- (\boxwidth, -\boxheight/2);
    \end{tikzpicture}
  \fi
}

%%
%% result variables
\newcommand{\VarAbbrvFilesystem}{F}
\newcommand{\VarAbbrvHtaccess}{H}
\newcommand{\VarAbbrvHttpheaders}{P}
\newcommand{\VarAbbrvNetworkrequests}{S}
\newcommand{\VarAspectRankedFirstMinimumPvalue}{0.1719}
\newcommand{\VarAspectThreeDcptOverallSampleSize}{313}
\newcommand{\VarAspectThreeRiskOverallSampleSize}{136}
\newcommand{\VarAspectTwoMinimumSampleSize}{5}
\newcommand{\VarAspectTwoOverallSampleSize}{216}
\newcommand{\VarIdDcptFilesystemBackup}{DF2}
\newcommand{\VarIdDcptFilesystemCardrz}{DF4}
\newcommand{\VarIdDcptFilesystemConfig}{DF7}
\newcommand{\VarIdDcptFilesystemCustomerList}{DF6}
\newcommand{\VarIdDcptFilesystemKeys}{DF3}
\newcommand{\VarIdDcptFilesystemPasswords}{DF5}
\newcommand{\VarIdDcptFilesystemPrivateKey}{DF1}
\newcommand{\VarIdDcptFilesystemRowe}{DF9}
\newcommand{\VarIdDcptFilesystemSpamList}{DF8}
\newcommand{\VarIdDcptHtaccessAdminRedirect}{DH1}
\newcommand{\VarIdDcptHttpheadersAdminCookie}{DP2}
\newcommand{\VarIdDcptHttpheadersApiserver}{DP3}
\newcommand{\VarIdDcptHttpheadersDevtoken}{DP1}
\newcommand{\VarIdDcptHttpheadersProxyReferer}{DP4}
\newcommand{\VarIdDcptNetworkrequestsAdminFalse}{DS5}
\newcommand{\VarIdDcptNetworkrequestsCleartextPassword}{DS2}
\newcommand{\VarIdDcptNetworkrequestsDevEndpoint}{DS8}
\newcommand{\VarIdDcptNetworkrequestsIdorReadSecrets}{DS1}
\newcommand{\VarIdDcptNetworkrequestsLogEndpoint}{DS11}
\newcommand{\VarIdDcptNetworkrequestsMassAssignment}{DS10}
\newcommand{\VarIdDcptNetworkrequestsPathTraversal}{DS4}
\newcommand{\VarIdDcptNetworkrequestsSessidParameter}{DS3}
\newcommand{\VarIdDcptNetworkrequestsSystemParameter}{DS7}
\newcommand{\VarIdDcptNetworkrequestsUnescapedJavascript}{DS6}
\newcommand{\VarIdDcptNetworkrequestsUnescapedJson}{DS9}
\newcommand{\VarIdRiskFilesystemBackup}{RF2}
\newcommand{\VarIdRiskFilesystemDnsUpdateKey}{RF5}
\newcommand{\VarIdRiskFilesystemKubernetesManifests}{RF4}
\newcommand{\VarIdRiskFilesystemOpenvpnConfig}{RF3}
\newcommand{\VarIdRiskFilesystemPrivateKey}{RF1}
\newcommand{\VarIdRiskHttpheadersCrossDomainRefererLeakage}{RP5}
\newcommand{\VarIdRiskHttpheadersOutdatedApache}{RP2}
\newcommand{\VarIdRiskHttpheadersOutdatedPhp}{RP3}
\newcommand{\VarIdRiskHttpheadersProxyAuthLeak}{RP1}
\newcommand{\VarIdRiskHttpheadersRequestSmugglingClte}{RP4}
\newcommand{\VarIdRiskNetworkrequestsBrokenFunctionLevelAuthorization}{RS1}
\newcommand{\VarIdRiskNetworkrequestsBrokenObjectLevelAuthorization}{RS4}
\newcommand{\VarIdRiskNetworkrequestsDevEndpointAccessible}{RS6}
\newcommand{\VarIdRiskNetworkrequestsInsecureHttp}{RS8}
\newcommand{\VarIdRiskNetworkrequestsLogSpamEndpoint}{RS5}
\newcommand{\VarIdRiskNetworkrequestsMassAssignment}{RS3}
\newcommand{\VarIdRiskNetworkrequestsNoRateLimiting}{RS7}
\newcommand{\VarIdRiskNetworkrequestsNosqlInjection}{RS9}
\newcommand{\VarIdRiskNetworkrequestsPasswordHashesInQueryParameters}{RS2}
\newcommand{\VarNumAvgMarksPerResponse}{1.8}
\newcommand{\VarNumCDTs}{25}
\newcommand{\VarNumCDTsWithLiteratureRef}{13}
\newcommand{\VarNumCDTsWithRiskyAndDeceptiveQueries}{5}
\newcommand{\VarNumCDTsWithoutLiteratureRef}{12}
\newcommand{\VarNumDroppedUsers}{30}
\newcommand{\VarNumMarks}{6,659}
\newcommand{\VarNumMinimumResponses}{16}
\newcommand{\VarNumMinimumResponsesWithoutTutorial}{8}
\newcommand{\VarNumPairedQueries}{18}
\newcommand{\VarNumParticipants}{47}
\newcommand{\VarNumParticipantsAndDroppedUsers}{77}
\newcommand{\VarNumParticipantsCtf}{12}
\newcommand{\VarNumParticipantsCtfDevelopers}{3}
\newcommand{\VarNumParticipantsCtfManagers}{0}
\newcommand{\VarNumParticipantsCtfOps}{1}
\newcommand{\VarNumParticipantsCtfPhaseA}{10}
\newcommand{\VarNumParticipantsCtfPhaseB}{2}
\newcommand{\VarNumParticipantsCtfResearcher}{5}
\newcommand{\VarNumParticipantsCtfSecOps}{1}
\newcommand{\VarNumParticipantsCtfStudents}{2}
\newcommand{\VarNumParticipantsCtfYearsMean}{4.4}
\newcommand{\VarNumParticipantsCtfYearsMedian}{5}
\newcommand{\VarNumParticipantsPhaseA}{23}
\newcommand{\VarNumParticipantsPhaseB}{24}
\newcommand{\VarNumParticipantsRes}{35}
\newcommand{\VarNumParticipantsResDevelopers}{19}
\newcommand{\VarNumParticipantsResEligible}{8}
\newcommand{\VarNumParticipantsResIncentivized}{18}
\newcommand{\VarNumParticipantsResLotteryPlayers}{1}
\newcommand{\VarNumParticipantsResManagers}{1}
\newcommand{\VarNumParticipantsResOps}{0}
\newcommand{\VarNumParticipantsResPhaseA}{13}
\newcommand{\VarNumParticipantsResPhaseB}{22}
\newcommand{\VarNumParticipantsResResearcher}{8}
\newcommand{\VarNumParticipantsResSecOps}{2}
\newcommand{\VarNumParticipantsResStudents}{3}
\newcommand{\VarNumParticipantsResYearsMean}{2.9}
\newcommand{\VarNumParticipantsResYearsMedian}{2}
\newcommand{\VarNumParticipantsStudentsRatio}{11}
\newcommand{\VarNumQueries}{174}
\newcommand{\VarNumQueriesDeceptive}{71}
\newcommand{\VarNumQueriesFilesystem}{36}
\newcommand{\VarNumQueriesFilesystemDcpt}{18}
\newcommand{\VarNumQueriesFilesystemNeut}{12}
\newcommand{\VarNumQueriesFilesystemRisk}{6}
\newcommand{\VarNumQueriesHtaccess}{15}
\newcommand{\VarNumQueriesHtaccessDcpt}{5}
\newcommand{\VarNumQueriesHtaccessNeut}{10}
\newcommand{\VarNumQueriesHtaccessRisk}{0}
\newcommand{\VarNumQueriesHttpHeaders}{58}
\newcommand{\VarNumQueriesHttpHeadersDcpt}{23}
\newcommand{\VarNumQueriesHttpHeadersNeut}{27}
\newcommand{\VarNumQueriesHttpHeadersRisk}{8}
\newcommand{\VarNumQueriesNetworkRequests}{65}
\newcommand{\VarNumQueriesNetworkRequestsDcpt}{25}
\newcommand{\VarNumQueriesNetworkRequestsNeut}{31}
\newcommand{\VarNumQueriesNetworkRequestsRisk}{9}
\newcommand{\VarNumQueriesNeutral}{80}
\newcommand{\VarNumQueriesRisky}{23}
\newcommand{\VarNumResponses}{3,669}
\newcommand{\VarNumRisks}{19}
\newcommand{\VarNumTotalDeceptiveLines}{88}
\newcommand{\VarNumTotalDeceptiveLinesPercent}{3.62}
\newcommand{\VarNumTotalQueryLines}{2,433}
\newcommand{\VarNumTotalRiskyLines}{94}
\newcommand{\VarNumTotalRiskyLinesPercent}{3.86}
\newcommand{\VarRatioAnswerAllMarked}{0.19}
\newcommand{\VarRatioQueriesWithGtOneDeceptiveLine}{4.55}
\newcommand{\VarRatioQueriesWithGtOneRiskyLine}{6.82}
\newcommand{\VarRatioVariantExactMax}{66.69}
\newcommand{\VarRatioVariantNoneMax}{48.90}
\newcommand{\VarRatioVariantOtherMax}{32.65}
\newcommand{\VarRatioVariantOverlapMax}{8.53}
\newcommand{\VarRatioVariantSubsetMax}{0.82}
\newcommand{\VarResAnswerQueriesMean}{76}
\newcommand{\VarResAnswerQueriesMedian}{59}
\newcommand{\VarResAnswerTimeMean}{64.3}
\newcommand{\VarResAnswerTimeMedian}{19}
\newcommand{\VarResBeforeAfterHttpheadersApiserverPvalue}{0.0107}
\newcommand{\VarResBeforeAfterHttpheadersApiserverRiskReduction}{32}
\newcommand{\VarResBeforeAfterHttpheadersDevtokenPvalue}{0.0195}
\newcommand{\VarResBeforeAfterHttpheadersDevtokenRiskReduction}{27}
\newcommand{\VarResBeforeAfterHttpheadersProxyRefererPvalue}{0.1719}
\newcommand{\VarResBeforeAfterHttpheadersProxyRefererRiskReduction}{15}
\newcommand{\VarResBeforeAfterNetworkrequestsPathTraversalPvalue}{0.5000}
\newcommand{\VarResBeforeAfterNetworkrequestsPathTraversalRiskReduction}{14}
\newcommand{\VarResBeforeAfterNetworkrequestsSessidParameterPvalue}{0.8750}
\newcommand{\VarResBeforeAfterNetworkrequestsSessidParameterRiskReduction}{-50}
\newcommand{\VarResBeforeAfterNumRejectNull}{2}
\newcommand{\VarResBeforeAfterOverallPvalue}{0.0013}
\newcommand{\VarResBeforeAfterOverallRiskReduction}{22}
\newcommand{\VarResCmDcptAcc}{53}
\newcommand{\VarResCmDcptAccCse}{1.8}
\newcommand{\VarResCmDcptAccRaw}{.53}
\newcommand{\VarResCmDcptFn}{1325}
\newcommand{\VarResCmDcptFone}{24}
\newcommand{\VarResCmDcptFoneRaw}{.24}
\newcommand{\VarResCmDcptFp}{136}
\newcommand{\VarResCmDcptFpr}{9}
\newcommand{\VarResCmDcptFprCse}{1.4}
\newcommand{\VarResCmDcptFprRaw}{.09}
\newcommand{\VarResCmDcptMcc}{10}
\newcommand{\VarResCmDcptMccRaw}{.10}
\newcommand{\VarResCmDcptPpv}{63}
\newcommand{\VarResCmDcptPpvCse}{4.9}
\newcommand{\VarResCmDcptPpvRaw}{.63}
\newcommand{\VarResCmDcptTn}{1403}
\newcommand{\VarResCmDcptTp}{234}
\newcommand{\VarResCmDcptTpr}{15}
\newcommand{\VarResCmDcptTprCse}{1.8}
\newcommand{\VarResCmDcptTprRaw}{.15}
\newcommand{\VarResCmRiskAcc}{53}
\newcommand{\VarResCmRiskAccCse}{2.2}
\newcommand{\VarResCmRiskAccRaw}{.53}
\newcommand{\VarResCmRiskFn}{256}
\newcommand{\VarResCmRiskFone}{30}
\newcommand{\VarResCmRiskFoneRaw}{.30}
\newcommand{\VarResCmRiskFp}{682}
\newcommand{\VarResCmRiskFpr}{44}
\newcommand{\VarResCmRiskFprCse}{2.5}
\newcommand{\VarResCmRiskFprRaw}{.44}
\newcommand{\VarResCmRiskMcc}{0}
\newcommand{\VarResCmRiskMccRaw}{.00}
\newcommand{\VarResCmRiskPpv}{23}
\newcommand{\VarResCmRiskPpvCse}{2.8}
\newcommand{\VarResCmRiskPpvRaw}{.23}
\newcommand{\VarResCmRiskTn}{857}
\newcommand{\VarResCmRiskTp}{204}
\newcommand{\VarResCmRiskTpr}{44}
\newcommand{\VarResCmRiskTprCse}{4.5}
\newcommand{\VarResCmRiskTprRaw}{.44}
\newcommand{\VarResContentLengthMarkNum}{3}
\newcommand{\VarResContentLengthMarkRatio}{0.05}
\newcommand{\VarResCseDcptDcptFnFilesystemBackup}{10.8}
\newcommand{\VarResCseDcptDcptFnFilesystemCardrz}{9.7}
\newcommand{\VarResCseDcptDcptFnFilesystemConfig}{11.5}
\newcommand{\VarResCseDcptDcptFnFilesystemCustomerList}{8.1}
\newcommand{\VarResCseDcptDcptFnFilesystemKeys}{6.3}
\newcommand{\VarResCseDcptDcptFnFilesystemPasswords}{7.4}
\newcommand{\VarResCseDcptDcptFnFilesystemPrivateKey}{9.2}
\newcommand{\VarResCseDcptDcptFnFilesystemRowe}{13.6}
\newcommand{\VarResCseDcptDcptFnFilesystemSpamList}{9.6}
\newcommand{\VarResCseDcptDcptFnHtaccessAdminRedirect}{5.9}
\newcommand{\VarResCseDcptDcptFnHttpheadersAdminCookie}{10.4}
\newcommand{\VarResCseDcptDcptFnHttpheadersApiserver}{7.7}
\newcommand{\VarResCseDcptDcptFnHttpheadersDevtoken}{5.0}
\newcommand{\VarResCseDcptDcptFnHttpheadersProxyReferer}{7.2}
\newcommand{\VarResCseDcptDcptFnNetworkrequestsAdminFalse}{12.0}
\newcommand{\VarResCseDcptDcptFnNetworkrequestsCleartextPassword}{10.3}
\newcommand{\VarResCseDcptDcptFnNetworkrequestsDevEndpoint}{10.6}
\newcommand{\VarResCseDcptDcptFnNetworkrequestsIdorReadSecrets}{24.4}
\newcommand{\VarResCseDcptDcptFnNetworkrequestsLogEndpoint}{11.2}
\newcommand{\VarResCseDcptDcptFnNetworkrequestsMassAssignment}{14.1}
\newcommand{\VarResCseDcptDcptFnNetworkrequestsPathTraversal}{7.9}
\newcommand{\VarResCseDcptDcptFnNetworkrequestsSessidParameter}{10.0}
\newcommand{\VarResCseDcptDcptFnNetworkrequestsSystemParameter}{17.2}
\newcommand{\VarResCseDcptDcptFnNetworkrequestsUnescapedJavascript}{19.5}
\newcommand{\VarResCseDcptDcptFnNetworkrequestsUnescapedJson}{19.0}
\newcommand{\VarResCseDcptFellTpFilesystemBackup}{14.8}
\newcommand{\VarResCseDcptFellTpFilesystemCardrz}{14.3}
\newcommand{\VarResCseDcptFellTpFilesystemConfig}{13.6}
\newcommand{\VarResCseDcptFellTpFilesystemCustomerList}{12.9}
\newcommand{\VarResCseDcptFellTpFilesystemKeys}{10.3}
\newcommand{\VarResCseDcptFellTpFilesystemPasswords}{14.6}
\newcommand{\VarResCseDcptFellTpFilesystemPrivateKey}{15.6}
\newcommand{\VarResCseDcptFellTpFilesystemRowe}{10.1}
\newcommand{\VarResCseDcptFellTpFilesystemSpamList}{9.6}
\newcommand{\VarResCseDcptFellTpHtaccessAdminRedirect}{8.5}
\newcommand{\VarResCseDcptFellTpHttpheadersAdminCookie}{13.3}
\newcommand{\VarResCseDcptFellTpHttpheadersApiserver}{7.7}
\newcommand{\VarResCseDcptFellTpHttpheadersDevtoken}{8.2}
\newcommand{\VarResCseDcptFellTpHttpheadersProxyReferer}{6.9}
\newcommand{\VarResCseDcptFellTpNetworkrequestsAdminFalse}{13.0}
\newcommand{\VarResCseDcptFellTpNetworkrequestsCleartextPassword}{17.3}
\newcommand{\VarResCseDcptFellTpNetworkrequestsDevEndpoint}{9.2}
\newcommand{\VarResCseDcptFellTpNetworkrequestsIdorReadSecrets}{24.4}
\newcommand{\VarResCseDcptFellTpNetworkrequestsLogEndpoint}{6.3}
\newcommand{\VarResCseDcptFellTpNetworkrequestsMassAssignment}{10.2}
\newcommand{\VarResCseDcptFellTpNetworkrequestsPathTraversal}{8.5}
\newcommand{\VarResCseDcptFellTpNetworkrequestsSessidParameter}{12.2}
\newcommand{\VarResCseDcptFellTpNetworkrequestsSystemParameter}{15.9}
\newcommand{\VarResCseDcptFellTpNetworkrequestsUnescapedJavascript}{18.7}
\newcommand{\VarResCseDcptFellTpNetworkrequestsUnescapedJson}{16.7}
\newcommand{\VarResCseDcptTrapTpFilesystemBackup}{6.5}
\newcommand{\VarResCseDcptTrapTpFilesystemCardrz}{11.4}
\newcommand{\VarResCseDcptTrapTpFilesystemConfig}{5.6}
\newcommand{\VarResCseDcptTrapTpFilesystemCustomerList}{11.0}
\newcommand{\VarResCseDcptTrapTpFilesystemKeys}{9.1}
\newcommand{\VarResCseDcptTrapTpFilesystemPasswords}{14.6}
\newcommand{\VarResCseDcptTrapTpFilesystemPrivateKey}{13.9}
\newcommand{\VarResCseDcptTrapTpFilesystemRowe}{11.1}
\newcommand{\VarResCseDcptTrapTpFilesystemSpamList}{10.3}
\newcommand{\VarResCseDcptTrapTpHtaccessAdminRedirect}{7.0}
\newcommand{\VarResCseDcptTrapTpHttpheadersAdminCookie}{10.0}
\newcommand{\VarResCseDcptTrapTpHttpheadersApiserver}{5.2}
\newcommand{\VarResCseDcptTrapTpHttpheadersDevtoken}{6.6}
\newcommand{\VarResCseDcptTrapTpHttpheadersProxyReferer}{4.3}
\newcommand{\VarResCseDcptTrapTpNetworkrequestsAdminFalse}{11.4}
\newcommand{\VarResCseDcptTrapTpNetworkrequestsCleartextPassword}{16.7}
\newcommand{\VarResCseDcptTrapTpNetworkrequestsDevEndpoint}{6.0}
\newcommand{\VarResCseDcptTrapTpNetworkrequestsIdorReadSecrets}{12.1}
\newcommand{\VarResCseDcptTrapTpNetworkrequestsLogEndpoint}{4.5}
\newcommand{\VarResCseDcptTrapTpNetworkrequestsMassAssignment}{5.7}
\newcommand{\VarResCseDcptTrapTpNetworkrequestsPathTraversal}{6.1}
\newcommand{\VarResCseDcptTrapTpNetworkrequestsSessidParameter}{9.3}
\newcommand{\VarResCseDcptTrapTpNetworkrequestsSystemParameter}{8.5}
\newcommand{\VarResCseDcptTrapTpNetworkrequestsUnescapedJavascript}{7.7}
\newcommand{\VarResCseDcptTrapTpNetworkrequestsUnescapedJson}{10.5}
\newcommand{\VarResCseRiskHackTpFilesystemBackup}{12.3}
\newcommand{\VarResCseRiskHackTpFilesystemDnsUpdateKey}{16.8}
\newcommand{\VarResCseRiskHackTpFilesystemKubernetesManifests}{17.0}
\newcommand{\VarResCseRiskHackTpFilesystemOpenvpnConfig}{17.4}
\newcommand{\VarResCseRiskHackTpFilesystemPrivateKey}{16.8}
\newcommand{\VarResCseRiskHackTpHttpheadersCrossDomainRefererLeakage}{13.3}
\newcommand{\VarResCseRiskHackTpHttpheadersOutdatedApache}{13.0}
\newcommand{\VarResCseRiskHackTpHttpheadersOutdatedPhp}{14.8}
\newcommand{\VarResCseRiskHackTpHttpheadersProxyAuthLeak}{16.7}
\newcommand{\VarResCseRiskHackTpHttpheadersRequestSmugglingClte}{23.8}
\newcommand{\VarResCseRiskHackTpNetworkrequestsBrokenFunctionLevelAuthorization}{20.1}
\newcommand{\VarResCseRiskHackTpNetworkrequestsBrokenObjectLevelAuthorization}{22.2}
\newcommand{\VarResCseRiskHackTpNetworkrequestsDevEndpointAccessible}{20.1}
\newcommand{\VarResCseRiskHackTpNetworkrequestsInsecureHttp}{9.7}
\newcommand{\VarResCseRiskHackTpNetworkrequestsLogSpamEndpoint}{14.3}
\newcommand{\VarResCseRiskHackTpNetworkrequestsMassAssignment}{25.4}
\newcommand{\VarResCseRiskHackTpNetworkrequestsNoRateLimiting}{17.0}
\newcommand{\VarResCseRiskHackTpNetworkrequestsNosqlInjection}{12.1}
\newcommand{\VarResCseRiskHackTpNetworkrequestsPasswordHashesInQueryParameters}{22.2}
\newcommand{\VarResCseRiskNewTpFilesystemBackup}{8.1}
\newcommand{\VarResCseRiskNewTpFilesystemDnsUpdateKey}{13.3}
\newcommand{\VarResCseRiskNewTpFilesystemKubernetesManifests}{8.5}
\newcommand{\VarResCseRiskNewTpFilesystemOpenvpnConfig}{8.8}
\newcommand{\VarResCseRiskNewTpFilesystemPrivateKey}{15.8}
\newcommand{\VarResCseRiskNewTpHttpheadersCrossDomainRefererLeakage}{10.7}
\newcommand{\VarResCseRiskNewTpHttpheadersOutdatedApache}{3.6}
\newcommand{\VarResCseRiskNewTpHttpheadersOutdatedPhp}{9.8}
\newcommand{\VarResCseRiskNewTpHttpheadersProxyAuthLeak}{12.5}
\newcommand{\VarResCseRiskNewTpHttpheadersRequestSmugglingClte}{19.0}
\newcommand{\VarResCseRiskNewTpNetworkrequestsBrokenFunctionLevelAuthorization}{17.0}
\newcommand{\VarResCseRiskNewTpNetworkrequestsBrokenObjectLevelAuthorization}{12.1}
\newcommand{\VarResCseRiskNewTpNetworkrequestsDevEndpointAccessible}{20.1}
\newcommand{\VarResCseRiskNewTpNetworkrequestsInsecureHttp}{8.0}
\newcommand{\VarResCseRiskNewTpNetworkrequestsLogSpamEndpoint}{8.8}
\newcommand{\VarResCseRiskNewTpNetworkrequestsMassAssignment}{18.1}
\newcommand{\VarResCseRiskNewTpNetworkrequestsNoRateLimiting}{12.1}
\newcommand{\VarResCseRiskNewTpNetworkrequestsNosqlInjection}{12.1}
\newcommand{\VarResCseRiskNewTpNetworkrequestsPasswordHashesInQueryParameters}{17.0}
\newcommand{\VarResCseRiskRiskFnFilesystemBackup}{6.3}
\newcommand{\VarResCseRiskRiskFnFilesystemDnsUpdateKey}{10.7}
\newcommand{\VarResCseRiskRiskFnFilesystemKubernetesManifests}{12.9}
\newcommand{\VarResCseRiskRiskFnFilesystemOpenvpnConfig}{10.7}
\newcommand{\VarResCseRiskRiskFnFilesystemPrivateKey}{6.2}
\newcommand{\VarResCseRiskRiskFnHttpheadersCrossDomainRefererLeakage}{17.6}
\newcommand{\VarResCseRiskRiskFnHttpheadersOutdatedApache}{10.5}
\newcommand{\VarResCseRiskRiskFnHttpheadersOutdatedPhp}{12.4}
\newcommand{\VarResCseRiskRiskFnHttpheadersProxyAuthLeak}{14.7}
\newcommand{\VarResCseRiskRiskFnHttpheadersRequestSmugglingClte}{21.0}
\newcommand{\VarResCseRiskRiskFnNetworkrequestsBrokenFunctionLevelAuthorization}{17.0}
\newcommand{\VarResCseRiskRiskFnNetworkrequestsBrokenObjectLevelAuthorization}{24.4}
\newcommand{\VarResCseRiskRiskFnNetworkrequestsDevEndpointAccessible}{20.1}
\newcommand{\VarResCseRiskRiskFnNetworkrequestsInsecureHttp}{16.7}
\newcommand{\VarResCseRiskRiskFnNetworkrequestsLogSpamEndpoint}{17.6}
\newcommand{\VarResCseRiskRiskFnNetworkrequestsMassAssignment}{25.4}
\newcommand{\VarResCseRiskRiskFnNetworkrequestsNoRateLimiting}{17.0}
\newcommand{\VarResCseRiskRiskFnNetworkrequestsNosqlInjection}{20.1}
\newcommand{\VarResCseRiskRiskFnNetworkrequestsPasswordHashesInQueryParameters}{17.0}
\newcommand{\VarResFellTpDcpt}{37}
\newcommand{\VarResFellTpDcptCse}{2.4}
\newcommand{\VarResNumHackMarksOnLineBashHistory}{151}
\newcommand{\VarResNumHackMarksOnLineDataCsv}{32}
\newcommand{\VarResNumHackMarksOnLineModHeader}{32}
\newcommand{\VarResNumHackMarksOnLineSsh}{260}
\newcommand{\VarResNumTrapMarksOnLineBashHistory}{0}
\newcommand{\VarResNumTrapMarksOnLineDataCsv}{3}
\newcommand{\VarResNumTrapMarksOnLineModHeader}{6}
\newcommand{\VarResNumTrapMarksOnLineSsh}{7}
\newcommand{\VarResRatioDcptDcptFnFilesystemBackup}{13}
\newcommand{\VarResRatioDcptDcptFnFilesystemCardrz}{12}
\newcommand{\VarResRatioDcptDcptFnFilesystemConfig}{20}
\newcommand{\VarResRatioDcptDcptFnFilesystemCustomerList}{9}
\newcommand{\VarResRatioDcptDcptFnFilesystemKeys}{9}
\newcommand{\VarResRatioDcptDcptFnFilesystemPasswords}{5}
\newcommand{\VarResRatioDcptDcptFnFilesystemPrivateKey}{6}
\newcommand{\VarResRatioDcptDcptFnFilesystemRowe}{57}
\newcommand{\VarResRatioDcptDcptFnFilesystemSpamList}{17}
\newcommand{\VarResRatioDcptDcptFnHtaccessAdminRedirect}{13}
\newcommand{\VarResRatioDcptDcptFnHttpheadersAdminCookie}{17}
\newcommand{\VarResRatioDcptDcptFnHttpheadersApiserver}{30}
\newcommand{\VarResRatioDcptDcptFnHttpheadersDevtoken}{9}
\newcommand{\VarResRatioDcptDcptFnHttpheadersProxyReferer}{25}
\newcommand{\VarResRatioDcptDcptFnNetworkrequestsAdminFalse}{27}
\newcommand{\VarResRatioDcptDcptFnNetworkrequestsCleartextPassword}{7}
\newcommand{\VarResRatioDcptDcptFnNetworkrequestsDevEndpoint}{42}
\newcommand{\VarResRatioDcptDcptFnNetworkrequestsIdorReadSecrets}{42}
\newcommand{\VarResRatioDcptDcptFnNetworkrequestsLogEndpoint}{59}
\newcommand{\VarResRatioDcptDcptFnNetworkrequestsMassAssignment}{55}
\newcommand{\VarResRatioDcptDcptFnNetworkrequestsPathTraversal}{28}
\newcommand{\VarResRatioDcptDcptFnNetworkrequestsSessidParameter}{19}
\newcommand{\VarResRatioDcptDcptFnNetworkrequestsSystemParameter}{43}
\newcommand{\VarResRatioDcptDcptFnNetworkrequestsUnescapedJavascript}{57}
\newcommand{\VarResRatioDcptDcptFnNetworkrequestsUnescapedJson}{41}
\newcommand{\VarResRatioDcptFellTpFilesystemBackup}{61}
\newcommand{\VarResRatioDcptFellTpFilesystemCardrz}{51}
\newcommand{\VarResRatioDcptFellTpFilesystemConfig}{38}
\newcommand{\VarResRatioDcptFellTpFilesystemCustomerList}{43}
\newcommand{\VarResRatioDcptFellTpFilesystemKeys}{52}
\newcommand{\VarResRatioDcptFellTpFilesystemPasswords}{46}
\newcommand{\VarResRatioDcptFellTpFilesystemPrivateKey}{66}
\newcommand{\VarResRatioDcptFellTpFilesystemRowe}{15}
\newcommand{\VarResRatioDcptFellTpFilesystemSpamList}{17}
\newcommand{\VarResRatioDcptFellTpHtaccessAdminRedirect}{54}
\newcommand{\VarResRatioDcptFellTpHttpheadersAdminCookie}{40}
\newcommand{\VarResRatioDcptFellTpHttpheadersApiserver}{30}
\newcommand{\VarResRatioDcptFellTpHttpheadersDevtoken}{55}
\newcommand{\VarResRatioDcptFellTpHttpheadersProxyReferer}{22}
\newcommand{\VarResRatioDcptFellTpNetworkrequestsAdminFalse}{37}
\newcommand{\VarResRatioDcptFellTpNetworkrequestsCleartextPassword}{54}
\newcommand{\VarResRatioDcptFellTpNetworkrequestsDevEndpoint}{24}
\newcommand{\VarResRatioDcptFellTpNetworkrequestsIdorReadSecrets}{58}
\newcommand{\VarResRatioDcptFellTpNetworkrequestsLogEndpoint}{7}
\newcommand{\VarResRatioDcptFellTpNetworkrequestsMassAssignment}{14}
\newcommand{\VarResRatioDcptFellTpNetworkrequestsPathTraversal}{39}
\newcommand{\VarResRatioDcptFellTpNetworkrequestsSessidParameter}{40}
\newcommand{\VarResRatioDcptFellTpNetworkrequestsSystemParameter}{29}
\newcommand{\VarResRatioDcptFellTpNetworkrequestsUnescapedJavascript}{33}
\newcommand{\VarResRatioDcptFellTpNetworkrequestsUnescapedJson}{23}
\newcommand{\VarResRatioDcptTrapTpFilesystemBackup}{3}
\newcommand{\VarResRatioDcptTrapTpFilesystemCardrz}{19}
\newcommand{\VarResRatioDcptTrapTpFilesystemConfig}{2}
\newcommand{\VarResRatioDcptTrapTpFilesystemCustomerList}{23}
\newcommand{\VarResRatioDcptTrapTpFilesystemKeys}{26}
\newcommand{\VarResRatioDcptTrapTpFilesystemPasswords}{49}
\newcommand{\VarResRatioDcptTrapTpFilesystemPrivateKey}{22}
\newcommand{\VarResRatioDcptTrapTpFilesystemRowe}{19}
\newcommand{\VarResRatioDcptTrapTpFilesystemSpamList}{21}
\newcommand{\VarResRatioDcptTrapTpHtaccessAdminRedirect}{21}
\newcommand{\VarResRatioDcptTrapTpHttpheadersAdminCookie}{15}
\newcommand{\VarResRatioDcptTrapTpHttpheadersApiserver}{10}
\newcommand{\VarResRatioDcptTrapTpHttpheadersDevtoken}{20}
\newcommand{\VarResRatioDcptTrapTpHttpheadersProxyReferer}{7}
\newcommand{\VarResRatioDcptTrapTpNetworkrequestsAdminFalse}{22}
\newcommand{\VarResRatioDcptTrapTpNetworkrequestsCleartextPassword}{36}
\newcommand{\VarResRatioDcptTrapTpNetworkrequestsDevEndpoint}{8}
\newcommand{\VarResRatioDcptTrapTpNetworkrequestsIdorReadSecrets}{0}
\newcommand{\VarResRatioDcptTrapTpNetworkrequestsLogEndpoint}{3}
\newcommand{\VarResRatioDcptTrapTpNetworkrequestsMassAssignment}{2}
\newcommand{\VarResRatioDcptTrapTpNetworkrequestsPathTraversal}{14}
\newcommand{\VarResRatioDcptTrapTpNetworkrequestsSessidParameter}{16}
\newcommand{\VarResRatioDcptTrapTpNetworkrequestsSystemParameter}{4}
\newcommand{\VarResRatioDcptTrapTpNetworkrequestsUnescapedJavascript}{0}
\newcommand{\VarResRatioDcptTrapTpNetworkrequestsUnescapedJson}{5}
\newcommand{\VarResRatioRiskHackTpFilesystemBackup}{61}
\newcommand{\VarResRatioRiskHackTpFilesystemDnsUpdateKey}{33}
\newcommand{\VarResRatioRiskHackTpFilesystemKubernetesManifests}{39}
\newcommand{\VarResRatioRiskHackTpFilesystemOpenvpnConfig}{41}
\newcommand{\VarResRatioRiskHackTpFilesystemPrivateKey}{67}
\newcommand{\VarResRatioRiskHackTpHttpheadersCrossDomainRefererLeakage}{15}
\newcommand{\VarResRatioRiskHackTpHttpheadersOutdatedApache}{62}
\newcommand{\VarResRatioRiskHackTpHttpheadersOutdatedPhp}{59}
\newcommand{\VarResRatioRiskHackTpHttpheadersProxyAuthLeak}{69}
\newcommand{\VarResRatioRiskHackTpHttpheadersRequestSmugglingClte}{54}
\newcommand{\VarResRatioRiskHackTpNetworkrequestsBrokenFunctionLevelAuthorization}{83}
\newcommand{\VarResRatioRiskHackTpNetworkrequestsBrokenObjectLevelAuthorization}{25}
\newcommand{\VarResRatioRiskHackTpNetworkrequestsDevEndpointAccessible}{17}
\newcommand{\VarResRatioRiskHackTpNetworkrequestsInsecureHttp}{7}
\newcommand{\VarResRatioRiskHackTpNetworkrequestsLogSpamEndpoint}{19}
\newcommand{\VarResRatioRiskHackTpNetworkrequestsMassAssignment}{45}
\newcommand{\VarResRatioRiskHackTpNetworkrequestsNoRateLimiting}{8}
\newcommand{\VarResRatioRiskHackTpNetworkrequestsNosqlInjection}{0}
\newcommand{\VarResRatioRiskHackTpNetworkrequestsPasswordHashesInQueryParameters}{75}
\newcommand{\VarResRatioRiskNewTpFilesystemBackup}{11}
\newcommand{\VarResRatioRiskNewTpFilesystemDnsUpdateKey}{15}
\newcommand{\VarResRatioRiskNewTpFilesystemKubernetesManifests}{4}
\newcommand{\VarResRatioRiskNewTpFilesystemOpenvpnConfig}{4}
\newcommand{\VarResRatioRiskNewTpFilesystemPrivateKey}{26}
\newcommand{\VarResRatioRiskNewTpHttpheadersCrossDomainRefererLeakage}{7}
\newcommand{\VarResRatioRiskNewTpHttpheadersOutdatedApache}{0}
\newcommand{\VarResRatioRiskNewTpHttpheadersOutdatedPhp}{10}
\newcommand{\VarResRatioRiskNewTpHttpheadersProxyAuthLeak}{12}
\newcommand{\VarResRatioRiskNewTpHttpheadersRequestSmugglingClte}{15}
\newcommand{\VarResRatioRiskNewTpNetworkrequestsBrokenFunctionLevelAuthorization}{8}
\newcommand{\VarResRatioRiskNewTpNetworkrequestsBrokenObjectLevelAuthorization}{0}
\newcommand{\VarResRatioRiskNewTpNetworkrequestsDevEndpointAccessible}{17}
\newcommand{\VarResRatioRiskNewTpNetworkrequestsInsecureHttp}{3}
\newcommand{\VarResRatioRiskNewTpNetworkrequestsLogSpamEndpoint}{4}
\newcommand{\VarResRatioRiskNewTpNetworkrequestsMassAssignment}{9}
\newcommand{\VarResRatioRiskNewTpNetworkrequestsNoRateLimiting}{0}
\newcommand{\VarResRatioRiskNewTpNetworkrequestsNosqlInjection}{0}
\newcommand{\VarResRatioRiskNewTpNetworkrequestsPasswordHashesInQueryParameters}{8}
\newcommand{\VarResRatioRiskRiskFnFilesystemBackup}{5}
\newcommand{\VarResRatioRiskRiskFnFilesystemDnsUpdateKey}{7}
\newcommand{\VarResRatioRiskRiskFnFilesystemKubernetesManifests}{14}
\newcommand{\VarResRatioRiskRiskFnFilesystemOpenvpnConfig}{7}
\newcommand{\VarResRatioRiskRiskFnFilesystemPrivateKey}{0}
\newcommand{\VarResRatioRiskRiskFnHttpheadersCrossDomainRefererLeakage}{44}
\newcommand{\VarResRatioRiskRiskFnHttpheadersOutdatedApache}{18}
\newcommand{\VarResRatioRiskRiskFnHttpheadersOutdatedPhp}{21}
\newcommand{\VarResRatioRiskRiskFnHttpheadersProxyAuthLeak}{19}
\newcommand{\VarResRatioRiskRiskFnHttpheadersRequestSmugglingClte}{23}
\newcommand{\VarResRatioRiskRiskFnNetworkrequestsBrokenFunctionLevelAuthorization}{8}
\newcommand{\VarResRatioRiskRiskFnNetworkrequestsBrokenObjectLevelAuthorization}{58}
\newcommand{\VarResRatioRiskRiskFnNetworkrequestsDevEndpointAccessible}{17}
\newcommand{\VarResRatioRiskRiskFnNetworkrequestsInsecureHttp}{57}
\newcommand{\VarResRatioRiskRiskFnNetworkrequestsLogSpamEndpoint}{52}
\newcommand{\VarResRatioRiskRiskFnNetworkrequestsMassAssignment}{45}
\newcommand{\VarResRatioRiskRiskFnNetworkrequestsNoRateLimiting}{92}
\newcommand{\VarResRatioRiskRiskFnNetworkrequestsNosqlInjection}{83}
\newcommand{\VarResRatioRiskRiskFnNetworkrequestsPasswordHashesInQueryParameters}{8}
\newcommand{\VarResRisksFirstOverall}{49}
\newcommand{\VarResRoweDeceptiveHack}{23}
\newcommand{\VarResRoweDeceptiveHackCse}{11.8}
\newcommand{\VarResRoweDeceptiveTrap}{26}
\newcommand{\VarResRoweDeceptiveTrapCse}{12.1}
\newcommand{\VarResRoweNeutralHack}{30}
\newcommand{\VarResRoweNeutralHackCse}{12.9}
\newcommand{\VarResRoweNeutralTrap}{11}
\newcommand{\VarResRoweNeutralTrapCse}{9.2}
\newcommand{\VarResTestMarkPreferenceDcptOppositePvalue}{0.00000089}
\newcommand{\VarResTestMarkPreferenceDcptPvalue}{1.0000}
\newcommand{\VarResTestMarkPreferenceRiskOppositePvalue}{0.39855117}
\newcommand{\VarResTestMarkPreferenceRiskPvalue}{0.6659}
\newcommand{\VarResTestRiskDcptMarksNumDeceptive}{1561}
\newcommand{\VarResTestRiskDcptMarksNumRisky}{461}
\newcommand{\VarResTestRiskDcptMarksNumTotal}{2022}
\newcommand{\VarResTestRiskDcptMarksPvalue}{0.0076}
\newcommand{\VarResTrapsFirstOverall}{36}
\newcommand{\VarTableLimitMrkHackLong}{60}
\newcommand{\VarTableLimitMrkHackShort}{40}
\newcommand{\VarTableLimitMrkTrapLong}{60}
\newcommand{\VarTableLimitMrkTrapShort}{15}

%%
%% long and short title
\title{Honeyquest: Rapidly Measuring the Enticingness of
  Cyber~Deception~Techniques~with~Code-based~Questionnaires}

%%
%% author names and affiliations
\author{Mario Kahlhofer}
\email{mario.kahlhofer@dynatrace.com}
\orcid{0000-0002-6820-4953}
\affiliation{%
  \institution{Dynatrace Research}
  \city{Linz}
  \country{Austria}
}
\affiliation{%
  \institution{Johannes Kepler University}
  \city{Linz}
  \country{Austria}
}

\author{Stefan Achleitner}
\email{stefan.achleitner@dynatrace.com}
\orcid{0000-0002-5499-6101}
\affiliation{%
  \institution{Dynatrace Research}
  \city{Linz}
  \country{Austria}
}

\author{Stefan Rass}
\email{stefan.rass@jku.at}
\orcid{0000-0003-2821-2489}
% \affiliation{%
%   \institution{Johannes Kepler University}
%   \city{Linz}
%   \country{Austria}
% }

\author{René Mayrhofer}
\email{rene.mayrhofer@jku.at}
\orcid{0000-0003-1566-4646}
\affiliation{%
  \institution{Johannes Kepler University}
  \city{Linz}
  \country{Austria}
}

%%
%% [ACM] short author names
\renewcommand{\shortauthors}{Kahlhofer et al.}

%%
%% artifact url
\newcommand{\repositoryurl}{https://github.com/dynatrace-oss/honeyquest}

%%
%% abstract
\begin{abstract}
  % State the problem
  Fooling adversaries with traps such as honeytokens
  can slow down cyber attacks and create strong
  indicators of compromise.
  % Say why it’s an interesting problem
  Unfortunately, cyber deception techniques are often poorly specified.
  Also, realistically measuring their effectiveness requires
  a well-exposed software system together with a production-ready
  implementation of these techniques.
  This makes rapid prototyping challenging.
  % Say what your solution achieves
  Our work translates \VarNumCDTsWithLiteratureRef{}~previously researched
  and \VarNumCDTsWithoutLiteratureRef{}~self-defined techniques
  into a high-level, machine-readable specification.
  Our open-source tool, Honeyquest, allows researchers
  to quickly evaluate the enticingness of deception techniques without implementing them.
  % Say what follows from you solution
  We test the enticingness of \VarNumCDTs{}~cyber deception techniques
  and \VarNumRisks{}~true security risks
  in an experiment with \VarNumParticipants{}~humans. % security-aware
  We successfully replicate the goals of previous work with many consistent findings,
  but without a time-consuming implementation of these techniques on real computer systems.
  % making Honeyquest a rapid method to measure the enticingness of cyber deception techniques.
  We provide valuable insights for the design of enticing deception
  and also show that the presence of cyber deception can significantly
  % reduce the risk of true weaknesses being exploited.
  reduce the risk that adversaries will find a true security risk % genuine
  by about \VarResBeforeAfterOverallRiskReduction{}\% on average.
\end{abstract}

%%
%% [ACM] ccs concepts
\begin{CCSXML}
  <ccs2012>
  <concept>
  <concept_id>10002978.10003022.10003026</concept_id>
  <concept_desc>Security and privacy~Web application security</concept_desc>
  <concept_significance>500</concept_significance>
  </concept>
  <concept>
  <concept_id>10002978.10002997.10002999</concept_id>
  <concept_desc>Security and privacy~Intrusion detection systems</concept_desc>
  <concept_significance>300</concept_significance>
  </concept>
  <concept>
  <concept_id>10002978.10003006</concept_id>
  <concept_desc>Security and privacy~Systems security</concept_desc>
  <concept_significance>300</concept_significance>
  </concept>
  <concept>
  <concept_id>10002978.10003014</concept_id>
  <concept_desc>Security and privacy~Network security</concept_desc>
  <concept_significance>300</concept_significance>
  </concept>
  </ccs2012>
\end{CCSXML}

\ccsdesc[500]{Security and privacy~Web application security}
\ccsdesc[300]{Security and privacy~Intrusion detection systems}
\ccsdesc[300]{Security and privacy~Systems security}
\ccsdesc[300]{Security and privacy~Network security}

%%
%% [ACM] keywords
\keywords{cyber deception, effective deception, honeytokens, honeypots}

%%
%% [ACM] show page numbers
\settopmatter{printfolios=true}

%%
%% [ACM] always show ACM citation reference
\settopmatter{printacmref=true}
\settopmatter{printccs=true}

%%
%% title
\maketitle
\enlargethispage{-6pt}

%%
%% sections
\section{Introduction}
\label{sec:introduction}

% =================================================================================================

Cyber deception deceives adversaries about the true appearance of a software
system, tricking them into taking (or not taking) actions that are not in their favor%
~\cite{%
  Yuill2006:UsingDeceptionHide,
  Yuill2007:DefensiveComputerSecurityDeception,
  Wang2018:CyberDeceptionOverview}.
Imagine that an attacker has already broken into a container somewhere
in your infrastructure, completely undetected by any security measures.
At this stage, the goal of such an adversary could be to move laterally
through your infrastructure and take over additional resources. % computer
We can defend against that by placing honeytokens%
~\cite{Spitzner2003:HoneypotsCatchingInsider,Spitzner2003:HoneytokensOtherHoneypot}
in the container: Fake credentials or tokens that trigger an alarm when used.
Such incidents may then be escalated to a human operator for further investigation.
Benefits of honeytokens are:%
~\cite{Ferguson-Walter2021:ExaminingEfficacyDecoybased,Ferguson-Walter2023:CyberExpertFeedback}

\begin{enumerate}
  \item \textbf{Adversaries are slowed down}
        as they waste time with unsuccessful exploit attempts.
  \item \textbf{Defenders get strong indicators of compromise (IoCs)}
        from such alarms for incident resolution.
  \item \textbf{Reduces the risk of adversaries exploiting true weaknesses}
        because they are distracted by honeytokens.
\end{enumerate}

\begin{figure*}[t]
  \centering
  \resizebox{\textwidth}{!}{\begin{tikzpicture}[scale=1.0]

  \def\boxw{2.45}  % width of boxes
  \def\smlw{1.8}   % width of slightly smaller boxes
  \def\smls{0.75}  % scale of text in slightly smaller boxes
  \def\boxh{1.5}   % height of boxes
  \def\smlh{1}     % height of slightly smaller box
  \def\arrw{0.7}   % width of arrow between boxes
  \def\arsw{0.5}   % width of arrow between slightly smaller boxes
  \def\subw{3.5cm} % width of subtext
  \def\subt{0.2}   % margin on the top of the subtext
  \def\subs{0.65}  % scale of text below boxes
  \def\gryh{1.4}   % extra height of grey background

  \newcommand{\starbullet}{\ding{72}}
  \newcommand{\posbullet}{\textcolor{teal}{\ding{51}}}
  \newcommand{\negbullet}{\textcolor{purple}{\ding{56}}}

  \coordinate (B1) at (0,0);
  \coordinate (B1L) at (0,-1.25*\boxh);
  \coordinate (B2) at ($(B1)+(\boxw,0)+(\arrw,0)$);
  \coordinate (B3) at ($(B2)+(\smlw,0)+(\arsw,0)$);
  \coordinate (B4) at ($(B3)+(\smlw,0)+(\arsw,0)$);
  \coordinate (B5) at ($(B4)+(\smlw,0)+(\arsw,0)$);
  \coordinate (B6) at ($(B5)+(\boxw,0)+(\arrw,0)$);
  \coordinate (B7) at ($(B6)+(\boxw,0)+(\arrw,0)$);

  % arrow from honeyaml box to deployment (behind grey background)
  \draw[-,color=ACMDarkBlue,dotted,very thick] ($(B1L)+0.5*(\boxw,-0.2)$) to[out=-22,in=180] ($(B4)+(0,-\gryh-1.5)$) to[out=0,in=-120] ($(B6)+(0,-0.5*\boxh)$);

  % grey background around the small boxes (put first to draw it behind all the other elements)
  \path[dashed,draw=lightgray!80,fill=lightgray!10,thick] ($(B2)+0.7*(-\smlw,1.8*\boxh)$) rectangle ($(B5)+0.7*(\smlw,-1.2*\boxh)-(0,\gryh)$);

  % text above smaller evaluation boxes
  \node[text width=5cm, anchor=west] at ($(B2)+0.7*(-\smlw+0.2,1.2*\boxh+0.3)$)
  {\textbf{Evaluation} of CDTs};

  % design box
  \draw[thick] ($(B1)+0.5*(-\boxw,\boxh)$) rectangle ($(B1)+0.5*(\boxw,-\boxh)$) node[midway,align=center]
    {\textbf{CDT Design}};

  % honeyaml box
  \draw[thick,color=ACMDarkBlue,fill=ACMBlue!10] ($(B1L)+0.5*(-\boxw,\boxh)$) rectangle ($(B1L)+0.5*(\boxw,-\boxh)$) node[midway,align=center]
    {\textcolor{ACMDarkBlue}{\textsc{HoneYAML}}\\\textcolor{black}{\small{(\secname\ref{sec:honeyaml})}}};
  \node[text width=3.8cm,scale=\subs,anchor=north,align=center] at ($(B1L)-0.5*(0,\boxh)-(0,\subt)$)
  {\textcolor{ACMDarkBlue}{Shared CDT Specification}};

  % the four small evaluation boxes
  \draw[thick] ($(B2)+0.5*(-\smlw,\boxh)$) rectangle ($(B2)+0.5*(\smlw,-\boxh)$) node[midway,align=center,scale=\smls]
    {Text-based\\Questionnaires};
  \node[text width=\subw,scale=\subs,anchor=north,align=center] at ($(B2)-0.5*(0,\boxh)-(0,\subt)$)
  {\posbullet~simple, easy, and implementation-free\\~\\\negbullet~requires humans and often lacks connection to technical aspects};

  \draw[thick,color=ACMDarkBlue,fill=ACMBlue!10] ($(B3)+0.5*(-\smlw,\boxh)$) rectangle ($(B3)+0.5*(\smlw,-\boxh)$) node[midway,align=center,scale=\smls]
    {\textcolor{ACMDarkBlue}{\textsc{Honeyquest}}\\\textcolor{black}{\small{(\secname\ref{sec:honeyquest})}}};
  \node[text width=\subw,scale=\subs,anchor=north,align=center] at ($(B3)-0.5*(0,\boxh)-(0,\subt)$)
  {\posbullet~mimics system properties without implementation needs\\~\\\negbullet~still requires humans right now};

  \draw[thick] ($(B4)+0.5*(-\smlw,\boxh)$) rectangle ($(B4)+0.5*(\smlw,-\boxh)$) node[midway,align=center,scale=\smls]
    {CTF\\Experiments};
  \node[text width=\subw,scale=\subs,anchor=north,align=center] at ($(B4)-0.5*(0,\boxh)-(0,\subt)$)
  {\posbullet~close to real-world\\~\\\negbullet~requires humans and often labor-intensive implementations of apps and services};

  \draw[thick] ($(B5)+0.5*(-\smlw,\boxh)$) rectangle ($(B5)+0.5*(\smlw,-\boxh)$) node[midway,align=center,scale=\smls]
    {Honeypot\\Deployments};
  \node[text width=\subw,scale=\subs,anchor=north,align=center] at ($(B5)-0.5*(0,\boxh)-(0,\subt)$)
  {\posbullet~real-world\\~\\\negbullet~dependent upon attracting real attackers and a production-ready implementation of traps};

  % deployment box
  \draw[thick] ($(B6)+0.5*(-\boxw,\boxh)$) rectangle ($(B6)+0.5*(\boxw,-\boxh)$) node[midway,align=center]
    {\textbf{Deployment}\\in Real-World\\Applications};

  % engage box
  \draw[thick] ($(B7)+0.5*(-\boxw,\boxh)$) rectangle ($(B7)+0.5*(\boxw,-\boxh)$) node[midway,align=center]
    {\textbf{Engage}\\Real Attackers};
  \node[text width=\subw,scale=\subs,anchor=north,align=center] at ($(B7)-0.5*(0,\boxh)-(0,\subt)$)
  {In the Reconnaissance\\Phase of their Attacks};

  % arrows between boxes
  \draw[-{Stealth[scale=1.5]},thick] ($(B1)+0.5*(\boxw,0)$) -- ($(B2)-0.7*(\smlw,0)$);
  \draw[-{Stealth[scale=1.5]},thick] ($(B5)+0.7*(\smlw,0)$) -- ($(B6)-0.5*(\boxw,0)$);
  \draw[-{Stealth[scale=1.5]},thick] ($(B6)+0.5*(\boxw,0)$) -- ($(B7)-0.5*(\boxw,0)$);

  % arrow from honeyaml box to honeyquest
  \draw[-,color=ACMDarkBlue,dotted,very thick] ($(B1L)+0.5*(\boxw,0.2)$) to[out=10,in=-90] ($(B1)+(0.5*\boxw+0.3*\arrw,0)$) to[out=90,in=160] ($(B3)+(-0.5*\smlw,0.5*\boxh)$);

  % arrow from honeyquest to ctf experiments and honeypot deployments
  \draw[-{Stealth[scale=1.0]},color=ACMDarkBlue] ($(B3)+(0,0.5*\boxh)$) to[out=30,in=120] ($(B4)+(-0.3*\smlw,0.5*\boxh)$);
  \draw[-{Stealth[scale=1.0]},color=ACMDarkBlue] ($(B3)+(0,0.5*\boxh)$) to[out=30,in=120] ($(B5)+(-0.3*\smlw,0.5*\boxh)$);

  % text right to those two arrows
  \node[text width=6cm,scale=\subs,anchor=north,align=center] at ($(B5)+0.5*(0.1,\boxh+2)$)
  {\textcolor{ACMDarkBlue}{Replicated Findings}\\(\secname\ref{sec:discuss-replicated-work})};

\end{tikzpicture}}
  \caption{%
    The lifecycle of designing, evaluating, and deploying CDTs,
    with the ultimate goal of engaging real adversaries.
  }
  \Description[Deception lifecycle, depicting design, evaluation, deployment, and engagement.]{
    A figure showing the deception lifecycle,
    which is depicted by four large boxes that are connected by arrows:
    (1) CDT Design, (2) Evaluation of CDTS,
    (3) Deployment in Real-World Applications, and (4) Engage with Real Attackers.
    The second box is further divided into four sub-boxes:
    (a) Text-based Questionnaires, which are simple, easy, and implementation-free,
    but require humans and often lack connection to technical aspects.
    (b) Honeyquest, which mimics system properties without implementation needs,
    but still requires humans right now,
    (c) CTF Experiments, which are close to real-world,
    but requre humans often a labor-intensive implementation of apps and services,
    and finally (d) Honeypot Deployments,
    which happen in the real-world,
    but depend upon attracting real attackers and need
    a production-ready implementation of traps.
    Lastly, there is one more box on the side with the text HoneYAML,
    which is a shared specification of CDTs.
    That box is connected with a dashed line to Honeyquest,
    and the third box on deployment in real-world applications.
  }
  \label{fig:deception-lifecycle}
\end{figure*}
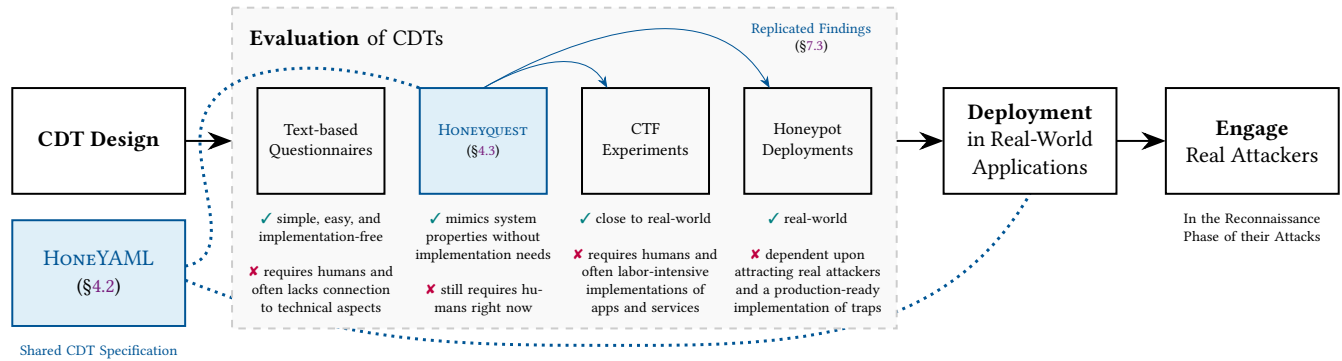

Recent research has come up with great techniques to deceive attackers
(\secname\ref{sec:honeypots-honeytokens}).
But are they effective?
Will attackers fall for such traps, or will they see through them?
After all, hackers are neither lazy nor stupid.
\citeauthor{Bowen2009:BaitingAttackersUsing}~\cite{Bowen2009:BaitingAttackersUsing}
introduced various properties that can guide us in designing effective decoys.
\citeauthor{BenSalem2011:DecoyDocumentDeployment}~\cite{BenSalem2011:DecoyDocumentDeployment}
found six of them to be very important, the first three being
detectability, conspicuousness, and enticingness.
\emph{Detectability} describes the necessary requirement to detect when a trap has been triggered.
\emph{Enticingness} describes how attractive a trap is for an adversary,
how well it lures them and awakens desires and hopes to achieve their mission.
\emph{Conspicuousness} is similar to enticingness, but conspicuous traps are chosen
by adversaries because they are easily found, clearly visible, or obvious,
but not necessarily because they are attractive.
To measure these properties with real humans, researchers typically
use one of three methods (depicted in \figurename~\ref{fig:deception-lifecycle}):
Capture The Flag~(CTF) events, honeypots, or questionnaires. % honeypot deployments

\textbf{CTF~events, red team engagements, or cyber ranges}~\cite{%
  Sahin2022:MeasuringDevelopersWeb,
  Ferguson-Walter2019:TularosaStudyExperimental,
  Araujo2019:ImprovingIntrusionDetectors,
  Aljohani2022:PitfallsEvaluatingCyber,
  Gabrys2023:EmotionalStateClassification,
  Ferguson-Walter2021:ExaminingEfficacyDecoybased,
  Shade2020:MoonrakerStudyExperimental,
  Heckman2013:ActiveCyberDefense,
  BenSalem2011:DecoyDocumentDeployment,
  Voris2015:FoxTrapThwarting,
  Cranford2021:CognitiveTheoryCyber,
  Cranford2020:AdaptiveCyberDeception,
  Sahin2020:LessonsLearnedSunDEW,
  Barron2021:ClickThisNot,
  Han2017:EvaluationDeceptionBasedWeb,
  Aggarwal2022:HumanSubjectExperimentsRiskBased,
  Aggarwal2022:DesigningEffectiveMasking,
  Aggarwal2020:ExploratoryStudyMasking}
are competitions where participants attack and defend software systems.
% often offering prizes for the winners.
Creating such environments for deception experiments is very labor-intensive
because engineers have to setup the infrastructure, mimic a realistic app,
and implement traps.
The latter also presents various technical challenges%
~\cite{
  Sahin2020:LessonsLearnedSunDEW,
  Han2018:DeceptionTechniquesComputer,
  Kahlhofer2024:ApplicationLayerCyber}.

\textbf{Honeypots in the wild}~\cite{%
  Bowen2009:BaitingAttackersUsing,
  Han2017:EvaluationDeceptionBasedWeb,
  Rowe2006:FakeHoneypotsDefensive,
  Rowe2007:DefendingCyberspaceFake,
  Bowen2010:AutomatingInjectionBelievable,
  Fraunholz2018:DemystifyingDeceptionTechnology}
are software systems that want to be attacked.
While deploying such honeypots % in the wild
brings the closest contact to real adversaries,
it typically requires a well-exposed software system that is of
interest to adversaries, along with a production-ready implementation of traps.
In addition, it relies on waiting for attackers to come along and fall for the traps,
resulting in slow feedback loops.

\newpage

\textbf{Questionnaires}~\cite{%
  Ferguson-Walter2023:CyberExpertFeedback,
  Sahin2022:MeasuringDevelopersWeb,
  Aggarwal2021:DecoysCybersecurityExploratory,
  Ferguson-Walter2020:EmpiricalAssessmentEffectiveness,
  Ferguson-Walter2019:TularosaStudyExperimental,
  Rowe2007:DefendingCyberspaceFake,
  Rowe2006:FakeHoneypotsDefensive,
  Aljohani2022:PitfallsEvaluatingCyber,
  Araujo2015:ExperiencesHoneyPatchingActive,
  Gabrys2023:EmotionalStateClassification,
  Ferguson-Walter2021:ExaminingEfficacyDecoybased,
  Bercovitch2011:HoneyGenAutomatedHoneytokens,
  Sahin2020:LessonsLearnedSunDEW}
can rapidly test specific deception hypotheses. % conceptualize and
CTF events and honeypots can hardly measure psychological properties%
~\cite{Ferguson-Walter2019:WorldCTFNot},
which explains why they are often accompanied by questionnaires~\cite{%
  Sahin2020:LessonsLearnedSunDEW,
  Sahin2022:MeasuringDevelopersWeb,
  Ferguson-Walter2019:TularosaStudyExperimental,
  Ferguson-Walter2021:ExaminingEfficacyDecoybased,
  Gabrys2023:EmotionalStateClassification,
  Aljohani2022:PitfallsEvaluatingCyber,
  Rowe2006:FakeHoneypotsDefensive,
  Rowe2007:DefendingCyberspaceFake}.
However, we argue that text-based questionnaires quickly become detached
from the technical ``views'' that adversaries typically gain from a system.

Our work introduces \textbf{Honeyquest} as a method that combines the benefits of
questionnaires with the realism of CTF events and honeypots.
Questions in Honeyquest -- we call them \emph{queries} --
imitate the technical views that adversaries typically
gain of a software system, e.g., by presenting a real file listing with honeytokens in it.
We ask participants to mark what they would try to \expli~exploit
and where they spot potential \trapi~traps.
This allows us to measure the enticingness of % and test
various \textbf{Cyber~Deception~Techniques~(CDTs)}
in a fast and controlled manner.
% We believe that one may also be able to draw conclusions
% on the believability and conspicuousness from the results.
%
To bridge the gap to an actual technical implementation of CDTs
within a software system, we introduce \textbf{\mbox{HoneYAML}}.
We describe traps in our questionnaires with \mbox{HoneYAML},
but also use it to directly configure deception products.
\mbox{HoneYAML} further allows us to clearly define traps
and conduct easily reproducible experiments with them.
We contribute:
% Our contributions are:

\begin{enumerate}
  \item A method to test the enticingness of CDTs (\secname\ref{sec:enticing-deception}).
  \item A translation of \VarNumCDTsWithLiteratureRef{}~previously researched~\citereplicatedworks{}
        and \VarNumCDTsWithoutLiteratureRef{}~self-defined CDTs, into \textbf{HoneYAML}:
        A high-level, machine-readable specification of CDTs (\secname\ref{sec:honeyaml}).
  \item \textbf{Honeyquest}: A flexible open-source%
        \footnote{%
          \anon[Repository URL anonymized. Artifact provided after initial review.]%
          {\url{\repositoryurl}}%
        }
        tool for setting up studies that measure the enticingness of CDTs % rapidly
        % with code-based questionnaires
        (\secname\ref{sec:honeyquest}).
  \item Results of a human subject study using Honeyquest:
        We show \VarNumParticipants{}~humans
        \ntrli~\VarNumQueriesNeutral{}~neutral, \riski~\VarNumQueriesRisky{}~risky,
        and \dcpti~\VarNumQueriesDeceptive{}~deceptive components of a web application
        (\secname\ref{sec:results}).
        Our results validate many previous findings and also unveil new insights
        (\secname\ref{sec:discussion}).
        Raw data from that study is available in our
        \anon[repository]{\href{\repositoryurl}{repository}}.
        % Our \anon[dataset]{\href{\repositoryurl}{dataset}} with \VarNumMarks{}~data points
        % is contributed to the research community.
\end{enumerate}

\section{Problem Statement}

% =================================================================================================

Ultimately, we want to use cyber deception to defend against adversaries.
But first, we highlight the problem of designing reproducible experiments
to measure the enticingness of CDTs.
Then, as a case study, we consider CDTs that can secure web applications. % modern

% ==============================================================================
% ------------------------------------------------------------------------------

\subsection{Lack of Reproducible Experiments}

Experiments on deceiving humans necessarily involve real humans,
which makes conducting and replicating such studies challenging.
\citeauthor{Han2018:DeceptionTechniquesComputer}~\cite{Han2018:DeceptionTechniquesComputer}
point out that ``it is often impossible to test deception techniques offline'' and that
``[properties for achieving effective deception] are difficult to formalize and measure'',
which contributed to a widespread ``lack of reproducible experiments''%
~\cite[Sec. 6-7]{Han2018:DeceptionTechniquesComputer}.

% Sahin2022:MeasuringDevelopersWeb        => replicated (1), but had different (2) + (3)
% Sahin2020:LessonsLearnedSunDEW          => TODO
% Han2017:EvaluationDeceptionBasedWeb     => partially replicated (1), but had different (2) + (3)
% Petrunic2015:HoneytokensActiveDefense   => replicated (1), lacked (2) + (3)
% Nikiforakis2011:ExposingLackPrivacy     => fully replicated (1) + (2) + (3)
% Rowe2006:FakeHoneypotsDefensive         => fully replicated (1) + (2) + (3)
% Rowe2007:DefendingCyberspaceFake        => fully replicated (1) + (2) + (3)

To align with and replicate prior work (\secname\ref{sec:discuss-replicated-work}),
we looked for works that provided at least three ingredients:
(1) A detailed description of the tested CDTs % or implementation
beyond vague terms like ``honeyfiles''.
(2) A quantitative evaluation of the effectiveness of these CDTs,
beyond assumptions about attacker behavior.
(3) A report on the results obtained, beyond aggregate statistics.
We found most of these items in seven works~\citereplicatedworks{},
whose findings we could hence validate at least partially.
Further work often lacked some details for confidentiality reasons.
These items also inspired us to define CDTs with HoneYAML,
have a theoretically-grounded approach to measure enticingness,
and open-source raw results.

% ==============================================================================
% ------------------------------------------------------------------------------

\subsection{Defending Threats with Cyber Deception}

We consider adversaries in cloud environments in the
reconnaissance phase of an attack~\cite{Mandiant2013:APT1ExposingOne}.
They may aim to establish a foothold on a system or are already inside it, % Those adversaries
trying to move laterally to complete their mission.
Our work proposes a novel approach to evaluate what CDTs
are most effective against adversaries at this stage of an attack,
by measuring how well they entice attackers.

To demonstrate feasibility,
we study four components of a web application, where CDTs can be applied.
We chose these four because they are
``mostly invisible to benign users''~\cite{Han2017:EvaluationDeceptionBasedWeb}
and will not interfere~\cite{BenSalem2011:DecoyDocumentDeployment} with legitimate activities:

\begin{itemize}
  \item \textbf{\mfus{\filesystem{}}.}
        Honeyfiles like ``keys.txt'' that appear sensitive, % to carry sensitive data,
        allow us to detect unauthorized access attempts.
  \item \textbf{\htaccess{}s} configure Apache servers.
        These should never be publicly accessible. We deliberately expose these files
        with sensitive paths in them (e.g., to a fake admin site)
        and detect attackers who access these paths.
  \item Attackers might observe \textbf{\header{}} packets by probing endpoints.
        If we add HTTP headers that are indicative of known vulnerabilities,
        we aim to lure attackers into trying unsuccessful exploits for them.
  \item Attackers could monitor all \textbf{\network{}s} of an application.
        By adding fake tokens to those requests, % or session keys
        we aim to lure attackers into using them for subsequent attacks.
\end{itemize}

A deception systems can be structured into decoys and captors~\cite{Fan2018:EnablingAnatomicView}.
\emph{Decoys} are the entities being attacked, e.g., a honeytoken, while \emph{captors} perform the
security-related functions, e.g., logging and alerting on access attempts.
Our work focuses solely on the evaluation of decoys, which we call CDTs.
Decoys and captors can be readily implemented:%
~\cite{Kahlhofer2024:ApplicationLayerCyber}
Creating files is trivial in most operating systems.
Monitoring access attempts to them can be achieved with architectures such as SELinux%
~\cite{McCarty2004:SELinuxNSAOpen}.
Intercepting, modifying, and monitoring HTTP packets is often achieved
with a reverse proxy in front of applications%
~\cite{%
  Han2017:EvaluationDeceptionBasedWeb,
  Araujo2014:PatchesHoneyPatchesLightweight,
  Barron2021:ClickThisNot,
  Fraunholz2018:CloxyContextawareDeceptionasaService,
  Sahin2020:LessonsLearnedSunDEW,
  Pohl2015:HiveZeroConfiguration}.

\section{Measuring the Enticingness of Cyber Deception Techniques}
\label{sec:enticing-deception}

% =================================================================================================

This section presents an approach to quantify the enticingness of CDTs.
% Enticing deception can deliberately divert an adversary's attention away from true weaknesses.
This lays the groundwork for the design of
Honeyquest (\figurename~\ref{fig:queries-and-marks}) in \secname\ref{sec:prototype-design},
and its evaluation in \secname\ref{sec:experiment-design}.

% ==============================================================================
% ------------------------------------------------------------------------------

\subsection{Queries, Labels, Marks, and Annotations}

In the reconnaissance phase, attackers explore their target.
While probing our system, they might find certain properties depending on the technique used,
i.e., they gain different ``views'' of our system.

Assume that an attacker has already managed to break into a container.
They might perform the naive technique of ``listing files''
and observe Listing~\ref{lst:filesystem}.
In Honeyquest, we call this a query.
A query is just plain text, i.e., a collection of lines.

\begin{lstlisting}[
  label=lst:filesystem,
  caption={
    A ``\filesystem{}'' query with
    CDT~\RowRefSingle{Results}{DcptFilesystemKeys} injected in it.
  }
]
    drwxr-xr-x 25 elsa 4.0K Dec 30 08:36 .
    drwxr-xr-x  4 root 4.0K Jun 21  2019 ..
    -rw-------  1 elsa  57K Jan 13 14:48 .bash_history
    -rw-r--r--  1 elsa 3.5K Sep 17  2017 .bashrc
    drwx------  6 elsa 4.0K Sep 25 17:40 .config
    drwxr-xr-x  6 elsa 4.0K Nov 14 09:08 .npm
    drwx------  6 elsa 4.0K Nov 14 09:12 .yarn
    -rwxr-xr-x  1 elsa 3.9K Dec  5 16:02 buildcsv.py
    -rw-r--r--  1 elsa  12K Feb  6  2022 keys.json
\end{lstlisting}

\emph{We are now curious about the next move of an adversary.}
In a file system, possible actions may be reading a file,
visiting a directory, or doing nothing at all.
So we allow our adversary to place either \expli~\textbf{exploit marks}
or \trapi~\textbf{trap marks} on each line in a query.
Not marking anything is also a valid action -- and the default,
since no lines are marked initially.
Placing an exploit mark means that an adversary sees a potential security weakness on that line.
When presented with a file system, this signifies that the adversary would like
to examine the file or directory on that line or attack it somehow.
On the other hand, marking something as a trap means that an adversary
definitely wants to avoid interacting with that line. In a file system, this would
mean that these particular files must not be opened in order to avoid triggering an alarm.

\emph{Adversaries may want to try the most promising attack vector first.}
To let them express this, we number \textbf{answer marks}
in the order in which they are placed on a line.
These numbers are also visible to the user (Figure~\ref{fig:ui}).
This feature allows us to find out which parts of a query attract an adversary's attention first
(\secname\ref{sec:aspect-deception-first}).

\begin{figure}[!t]
  \centering
  \resizebox{\columnwidth}{!}{\input{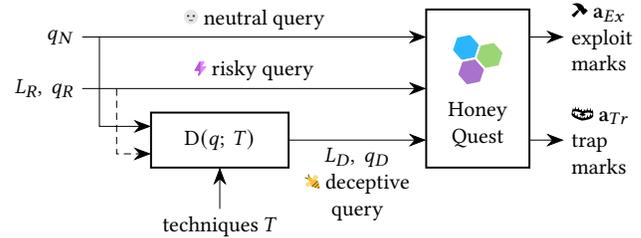}}
  \caption[]{
    In Honeyquest, users are presented with
    \ntrli~neutral, \riski~risky, and \dcpti~deceptive queries.
    A line annotation set~$L$ indicates the risky or deceptive lines
    in the associated query~$q$.
    An answer marks vector~$\mathbf{a}$ holds placed marks in order.
    The probabilistic algorithm~$\mathrm{D}(q;~T)$ makes queries deceptive.
  }
  \Description{
    A block diagram that illustrates the components of Honeyquest,
    exactly as described by the caption already.
  }
  \label{fig:queries-and-marks}
\end{figure}

To summarize, we imitate views on systems, frame them as queries,
and let users mark what they want to exploit or avoid, in order.
Recalling that one defined goal of cyber deception is
``tricking [adversaries] into taking (or not taking) actions that are not in their favor''%
~\cite{%
  Yuill2006:UsingDeceptionHide,
  Yuill2007:DefensiveComputerSecurityDeception,
  Wang2018:CyberDeceptionOverview},
and that ``enticement depends upon the attacker's intent or preference''%
~\cite{Bowen2009:BaitingAttackersUsing},
\emph{we are interested in queries,
  where adversaries mark the deceptive elements to be exploited,
  and not to be avoided, thus falling for a trap.}
This expresses the ``enticement'' property
from \citeauthor{Bowen2009:BaitingAttackersUsing}~\cite{Bowen2009:BaitingAttackersUsing}
with exploit and trap marks.

Each query has a \textbf{label} that indicates
which of these three strategies (\secname\ref{sec:query-design}) was followed in its design:

\begin{itemize}
  \item \ntrli~\textbf{Neutral}
        queries may be harmless, secure, benign,
        well-pro\-tect\-ed, or of neutral appearance.\footnote{%
          We intentionally chose the term ``neutral'' over ``benign'', because
          we do not want to imply the positive connotation that ``benign'' typically expresses.}
  \item \riski~\textbf{Risky}
        queries may be harmful, insecure, malicious,
        lack security measures, or have negative intent. Here, % In this case,
        the system owner bears that risk, not the adversary.\footnote{%
          We have deliberately chosen not to use the common terms ``malicious'' or
          ``vulnerable''. Maliciousness expresses a harmful intent, which can rarely arise
          from a textual query alone. Vulnerabilities are commonly defined as weaknesses
          that might be exploitable. So while our risky queries could be interpreted as
          ``weaknesses'', we do not want to imply that they are exploitable.}
  \item \dcpti~\textbf{Deceptive} queries want to grab the attention of an adversary,
        often by seeming risky. They contain CDTs.
\end{itemize}

Risky and deceptive queries have so-called \textbf{line annotations},
which store the exact line numbers that are risky or deceptive.

Put together, we get \textbf{Honeyquest}~(\figurename~\ref{fig:queries-and-marks}).
Honeyquest shows users queries of different types and labels,
where they can mark every line with \expli~exploit marks,
\trapi~trap marks, or nothing at all. Deceptive queries are created by modifying
a neutral or risky query (\secname\ref{sec:honeyaml}). For example, in a \filesystem{} query,
we might purposefully add a ``keys.json'' entry as a trap.
Later, in a deployed deception system, we would monitor if a potential
adversary accesses this file or tries to use one of the fake passwords
that we deliberately placed inside of it for authentication.
% that we placed inside of it.

Careful readers may find that it is often impossible to tell without context
(e.g., implementation details) whether certain query elements are risky or deceptive.
This is true, and is why we do not evaluate if users accurately identify a query's label,
but what parts of a query users perceive as exploitable or deceptive.
% which ultimately serves as a predictor for the enticingness of CDTs.

% ==============================================================================
% ------------------------------------------------------------------------------

\subsubsection{Matching Answer Marks and Line Annotations}
\label{sec:matching-marks}

After users have placed their marks,
we want to determine if they identified potential traps or risks.
In other words, we just check if answer marks intersect with line annotations.

Table~\ref{tab:terminology} introduces some terminology to express that clearly.
The set~$L$ reflects the ``ground truth'' labels,
and the set~$A$ holds user's answer marks.
We later specialize the notation by talking about
deceptive or risky line annotations, $L_D$ and $L_R$, respectively,
and about \expli~exploit and \trapi~trap marks, $A_{Ex}$ and $A_{Tr}$, respectively.
When we need the order in which the marks have been placed,
we will refer to the vectors $\mathbf{a}_{Ex}$ and $\mathbf{a}_{Tr}$ instead.
%
% Table~\ref{tab:query} shows a query with that terminology.
Table~\ref{tab:query} shows an example of a query that uses this terminology.

Consider a user answering a \dcpti~deceptive or \riski~risky query.
A simple way to express that ``answer marks~$A$ match line annotations~$L$''
is to check whether they intersect:

\noindent
\begin{align} \label{eq:marks-intersect-lines}
  \text{\ymark~match~:}    &  & L \cap A \neq \varnothing &  &
  \text{\nmark~no match~:} &  & L \cap A = \varnothing
\end{align}

This matching criterion is sufficient for our experiment
and is also well suited for expressing results with typical confusion matrices
(Appendix~\ref{sec:appendix:confusion-matrices}).
It has the small drawback that answers that intersect the line annotations only partially
are also ``matching''. Even worse, answers that place marks
on every single line will always intersect -- and therefore ``match'' --
with every possible set of line annotations.
However, in our experiment, only \VarRatioAnswerAllMarked{}\% of answers
had marks on every single line. Further, only \VarRatioVariantSubsetMax{}\% of answers
marked risky or deceptive lines only partially.
This is not surprising, since only \VarRatioQueriesWithGtOneRiskyLine{}\% of risky queries
and only \VarRatioQueriesWithGtOneDeceptiveLine{}\% of deceptive queries
in our current dataset have more than one line annotated anyway.
We therefore conclude that this simple matching criterion will not significantly bias our results.
Appendix~\ref{sec:appendix:matching-criteria} discusses alternative matching criteria
for different experimental conditions.
% that might be necessary under different experimental conditions.

\begin{table}[!b]
  \begin{threeparttable}
    % \small
    \centering
    \renewcommand{\arraystretch}{1.2}
    \caption{Terminology and common specializations.}
    \Description[An overview of the terminology and notation used within the paper.]{
      Queries are denoted with $q$ and a subscript that indicates the query type.
      Subscript N denotes neutral queries, D deceptive queries, and R risky queries.
      All queries are part of a respective set~$Q$.
      Subscript N denotes the set of neutral queries, D deceptive queries, and R risky queries.
      All query sets are pairwise disjoint.
      Cyber deception techniques (CDTs) are denoted with $t$.
      Line annotations are denoted with $L$ and a subscript that indicates the query type.
      Subscript D denotes deceptive lines, and R risky lines.
      Answer marks are denoted with $\mathbf{a}$ and a subscript that indicates the mark type.
      Subscript Ex denotes exploit marks, and Tr trap marks.
      $\mathbf{a}$ is a vector since users place marks in order.
      For equations that do not need ordered marks,
      we define $A$ as the set of unique elements from $\mathbf{a}$.
      The probabilistic algorithm~$\mathrm{D}$ makes a query~$q$ deceptive,
      by selecting a suitable but random element from~$T$ and applying it to~$q$.
      Our experiments chose the particular technique~$t$ manually for consistency with~$q$.
    }
    \label{tab:terminology}
    \begin{tabularx}{\columnwidth}{ l X r }
      \toprule
      Queries\tnote{a}      & \multicolumn{2}{l}{
        $ q_N \in Q_N, ~ q_R \in Q_R, ~ q_D \in Q_D $
      }                                                                                                           \\
      Techniques\tnote{b}   & $ t\in T $                                                                          \\
      Line Annotations      & $ L \, := \{ \ell^1, ~ \ell^2, ~ ... \} $  & $ L_D, ~ L_R $                         \\
      Answer Marks\tnote{c} & $ \mathbf{a} \, := ( a^1, ~ a^2, ~ ... ) $ & $ \mathbf{a}_{Ex}, ~ \mathbf{a}_{Tr} $ \\
                            & $ A := \{ a^1, ~ a^2, ~ ... \} $           & $ A_{Ex}, ~ A_{Tr} $                   \\
      \addlinespace[0.1cm]
      Algorithm\tnote{d}    & \multicolumn{2}{l}{
        $ q_D\gets \mathrm{D}(q; ~ T) \in Q_D ~ \text{with} ~ q \in Q_N \cup Q_R$
      }                                                                                                           \\
      \bottomrule
    \end{tabularx}
    \begin{tablenotes}
      \footnotesize
      \item \textbf{Query and Annotation Types:}
      \ntrli~N = Neutral. \riski~R = Risky. \dcpti~D = Deceptive.
      \item \textbf{Answer Types:} \expli~Ex = Exploit. \trapi~Tr = Trap.
      \vspace{0.1cm}
      \item[a] All query sets are pairwise disjoint.
      \item[b] We use the terms technique and CDT interchangeably throughout the paper.
      \item[c] $\mathbf{a}$ is a vector since users place marks in order.
      For equations that do not need ordered marks,
      we define $A$ as the set of unique elements from $\mathbf{a}$.
      \item[d] The probabilistic algorithm~$\mathrm{D}$ makes a query~$q$ deceptive,
      by selecting a suitable but random element from~$T$ and applying it to~$q$.
      Our experiments chose the particular technique~$t$ manually for consistency with~$q$.
    \end{tablenotes}
  \end{threeparttable}
\end{table}

\begin{table}[!b]
  \begin{threeparttable}
    % \small
    \centering
    \renewcommand{\arraystretch}{1.2}
    \caption{Query with line annotations and answer marks.}
    \Description[An example query with four lines, line annotations, and answer marks.]{
      A deceptive ``\header{}'' query~$q_D$ with one risky
      (\VarIdRiskHttpheadersOutdatedPhp{}, true vulnerability, purple) % shade
      and one deceptive
      (\VarIdDcptHttpheadersApiserver{}, injected weakness, orange) % shade
      line annotation. The resulting query contains a true vulnerability as well as a trap. % now
      A user placed three marks here.
      A trap mark in line~2 that was no trap,
      an exploit mark in line~3 on the true vulnerability,
      and another exploit mark in line~4 on the trap.
      The order indicates that our user would exploit line~4 first,
      and therefore fall for the trap first.
    }
    \label{tab:query}
    \begin{tabularx}{\columnwidth}{ l l @{\hspace{6pt}} X l }
      \toprule
      Line Annot.            & \#         & Query Line                                                          & Ans. Marks                        \\
      \addlinespace[0.11cm]
      \riski~$L_R = \{ 3 \}$ &            &                                                                     & \expli~$\mathbf{a}_{Ex} = (4, 3)$ \\
      \dcpti~$L_D = \{ 4 \}$ &            &                                                                     & \trapi~$\mathbf{a}_{Tr} = (2)$    \\
      \addlinespace[0.11cm]
      \midrule
                             & \texttt{1} & \small\texttt{HTTP/1.1 200 OK}                                      &                                   \\
                             & \texttt{2} & \small\texttt{Server: Apache/2.4.1}                                 & $a^1_{Tr} = 2$                    \\
      $\ell^1_R = 3$         & \texttt{3} & \sethlcolor{Thistle1}\hl{\small\texttt{X-Powered-By: PHP/5.1.6}}    & $a^2_{Ex} = 3$                    \\
      % & \texttt{4} & \texttt{Pragma: no-cache}                                     &                              \\
      $\ell^1_D = 4$         & \texttt{4} & \sethlcolor{NavajoWhite1}\hl{\small\texttt{X-Api-Server: /hko/api}} & $a^1_{Ex} = 4$                    \\
      %  & \texttt{5} & \small\texttt{Content-Type: text/html}                              &                                   \\
      \bottomrule
    \end{tabularx}
    \begin{tablenotes}
      \footnotesize
      \item \textbf{Description:} A deceptive ``\header{}'' query~$q_D$ with one \riski~risky
      (\RowRefSingle{Results}{RiskHttpheadersOutdatedPhp}, true vulnerability, purple) % shade
      and one \dcpti~deceptive
      (\RowRefSingle{Results}{DcptHttpheadersApiserver}, injected weakness, orange) % shade
      line annotation. The resulting query contains a true vulnerability as well as a trap. % now
      \item \textbf{Answer Marks:} A user placed three marks here.
      A \trapi~trap mark in line~2 that was no trap,
      an \expli~exploit mark in line~3 on the true vulnerability,
      and another \expli~exploit mark in line~4 on the trap.
      The order indicates that our user would exploit line~4 first,
      and therefore fall for the trap first.
      % \vspace{0.1cm}
      % \item[a] We still label queries with risky and deceptive elements as being deceptive.
    \end{tablenotes}
  \end{threeparttable}
\end{table}

% ==============================================================================
% ------------------------------------------------------------------------------

\subsection{Research Questions}

We differentiate between the \textbf{enticingness of deception} by itself,
in the sense that humans fall for traps (Aspect~A),
and its \textbf{ability to be defensive},
by deliberately diverting attention (Aspect~B).
Thus, we ask the following research questions:

\begin{bfseries}
  \begin{itshape}
    ``To what degree are humans enticed by deceptive elements,
    true weaknesses and vulnerabilities, and will deceptive elements divert their
    attention away from true risks?''
  \end{itshape}
\end{bfseries}

This section formulates hypotheses on that question.
We then select CDTs and risks for testing (\secname\ref{sec:experiment-design}),
and then report (\secname\ref{sec:results}) and discuss (\secname\ref{sec:discussion})
results that answer this question.

% ==============================================================================
% ------------------------------------------------------------------------------

\subsubsection{\textbf{Aspect~A:}
  To what degree are humans enticed by deceptive and risky elements?}
\label{sec:aspect-enticingness}

Consider that we show humans
\ntrli~neutral ($Q_N$),
\dcpti~deceptive ($Q_D$), and
\riski~risky ($Q_R$) queries.
%
% In our experiments,
We know which technique~$t$ was used to create deceptive queries
and which risk is present in the risky ones (\secname\ref{sec:experiment-design}).
To measure the enticingness of individual \dcpti~CDTs,
we group answers by CDT and count how often participants
fell for traps, detected traps, or did not react to traps.
Likewise, for \riski~risks, % (weaknesses and vulnerabilities)
we group by risks and count how often participants
detected risks, have mistaken risks for traps, or did not react to risks.
%
% All those counts are computed by following the method from \secname\ref{sec:matching-marks}.
Appendix~\ref{sec:appendix:counting-marks} lists the explicit formulation of those counts.

% ==============================================================================
% ------------------------------------------------------------------------------

\subsubsection{\textbf{Aspect~B1:}
  Do humans exploit deceptive elements before non-deceptive elements?}
\label{sec:aspect-deception-first}

Let \dcpti~${q_D \in Q_D}$ be a deceptive query with deceptive lines~$L_D$
where more than one \expli~exploit mark was placed, i.e., ${|A_{Ex}| > 1}$.
As before, we assume that a human ``fell for a trap'' when exploit marks
intersect deceptive lines, i.e., ${L \cap A_{Ex} \neq \varnothing}$.
Remember that participants were instructed to place marks in an order
that indicates what they would like to exploit first.
Let ${a^\prime \in A_{Ex} \cap L_D}$ be the first exploit mark that
marked a deceptive line and let ${a^{\prime\prime} \in A_{Ex} \setminus L_D}$
be the first exploit mark that marked a non-deceptive line.
$a^{\prime\prime}$ may be a risky or neutral line then.
Let $d^\prime_B$ be the number of times that $a^\prime$ is ranked before
$a^{\prime\prime}$ out of $d^{}_B$~samples where all of these aforementioned conditions hold.

We can phrase this null hypothesis~$H_0$:
When users place exploit marks~$A_{Ex}$
that intersect with deceptive lines~$L_D$,
whether $a^\prime$ or $a^{\prime\prime}$ is
ranked first is up to chance.
We chose a Binomial~test
that tries to reject ${t = \nicefrac{1}{2}}$
with the one-sided alternative that $t$ is greater.%
\footnote{%
  We used the \texttt{binomtest} function
  from SciPy~\cite{Virtanen2020:SciPyFundamentalAlgorithms} and
  computed the test's power with the \texttt{binom.power} function
  from the binom package~\cite{Dorai-Raj2022:BinomialConfidenceIntervals}.
}
\begin{equation}
  t = \frac{ d^\prime_B }{ d^{}_B } ~ \sim ~ B( d^{}_B, \nicefrac{1}{2} )
\end{equation}

A greater ratio hints at a greater preference to mark traps first.
This formulation is equivalent to the ``believability'' property for
``a perfect decoy [...] that is completely indistinguishable from one that is not'',
as proposed by%
~\citeauthor{Bowen2009:BaitingAttackersUsing}~\cite{Bowen2009:BaitingAttackersUsing}.

% ==============================================================================
% ------------------------------------------------------------------------------

\subsubsection{\textbf{Aspect~B2:}
  Are deceptive elements diverting an attacker's interest away from risky elements?}
\label{sec:aspect-divert-attention}

Instead of looking at what is marked first,
we test if the presence of deception is so distracting that % alone
an attacker misses weaknesses and vulnerabilities entirely. % marking potential

Consider that we have a set of risky queries~$Q_R$.
Let \riski~$q_R \in Q_R$ be a risky query with risky lines~$L_R$.
From that query, we derive a new deceptive query \dcpti~${q_D ~ = ~ \mathrm{D}(q_R; ~ T)}$
with deceptive lines $L_D$ and risky lines~$L^\prime_R$.
Note that we only introduce $L^\prime_R$ because making a query deceptive
means that we insert new lines, which could also change the line numbers of risky lines.
We now present~$q_R$ and~$q_D$ to each participant (within-subject)
and record the \expli~exploit marks~$A_{Ex}$ that each of the two queries receives.
As in all our experiments, we assume that a human has
``interest in exploiting a line'' when exploit marks intersect risky lines.

We phrase this null hypothesis~$H_0$:
When we show participants a risky query~$q_R$
and a derived deceptive query~$q_D$,
there is no difference in what they mark to exploit, i.e.,
the presence of deceptive lines does not distract them.

Table~\ref{tab:ct-deceptive-risky} formulates this
with a $2 \times 2$~contingency table over two factors:
``Did the participant mark the risky lines~$L_R$ to exploit in the risky query~$q_R$''
(``before'' condition),
and ``did the (same) participant mark the risky lines~$L^\prime_R$ to exploit
in the derived deceptive (and still risky) query~$q_D$''
(``after'' condition).
To draw an analogy, % to a medical setting,
think of deception as being the treatment for risky queries,
where the disease ``breaks out'' when patients detect the risk.
We test if the ``deception treatment'' effects the ``disease break-out''.

\begin{table}[h]
  \begin{threeparttable}
    % \small
    \centering
    \renewcommand{\arraystretch}{1.2}
    \caption{Contingency table on attention diversion.}
    \Description[%
      A 2x2 contingency table, with four cells associating
      $\alpha$, $\beta$, $\gamma$, and $\delta$ to each case.]{%
      The letter $\alpha$ is associated with ``no match'' in $q_R$ and ``no match'' in $q_D$.
      The letter $\beta$ is associated with ``match'' in $q_R$ and ``no match'' in $q_D$.
      The letter $\gamma$ is associated with ``no match'' in $q_R$ and ``match'' in $q_D$.
      The letter $\delta$ is associated with ``match'' in $q_R$ and ``match'' in $q_D$.
    }
    \label{tab:ct-deceptive-risky}
    \begin{tabularx}{\columnwidth}{ll @{\extracolsep{\fill}} cc }
      \toprule
      \multicolumn{2}{c}{Match in $q_D$?} & \multicolumn{2}{c}{Match in $q_R$? \emph{(``before'')}}                                                                                   \\
      \addlinespace[0.15cm]
      \multicolumn{2}{c}{\em (``after'')} & \nmark \hspace{.75em} $L^\prime_R \cap A_{Ex} = \varnothing$ & \ymark \hspace{.75em} $L^\prime_R \cap A_{Ex} \neq \varnothing$            \\
      \addlinespace[0.05cm]
      \midrule % \cmidrule{3-4}
      \nmark                              & $L_R \cap A_{Ex} = \varnothing$                              & $\alpha$                                                        & $\beta$  \\
      \ymark                              & $L_R \cap A_{Ex} \neq \varnothing$                           & $\gamma$                                                        & $\delta$ \\
      \bottomrule
    \end{tabularx}
    \begin{tablenotes}
      \footnotesize
      \vspace{0.1cm}
      \item \textbf{Legend:} A \nmark~cross-mark indicates that answer marks~$A_{Ex}$ do not
      intersect with risky lines. A \ymark~check mark indicates that they do intersect / match.
    \end{tablenotes}
  \end{threeparttable}
\end{table}

Our two factors and the single outcome are nominal,
and our subjects are paired because every participant sees both~$q_R$ and~$q_D$.
This scenario is usually tested with a McNemar's test~\cite{McNemar1947:NoteSamplingError}
and a two-sided alternative hypothesis.
The one-sided alternative hypothesis that the presence of deception reduces the risk of marking
risky lines is a Binomial~test~\cite{Fay2014:ExactMcNemarTest} with the following test statistic:
\begin{equation}
  t = \frac{ \gamma }{ \beta + \gamma } ~ \sim ~ B( \beta + \gamma, \nicefrac{1}{2} )
\end{equation}

To make this more intuitive, we compute the relative risk that describes
how much the risk that humans mark risky lines is reduced (or increased),
when deceptive lines are present:
\begin{displaymath}
  \text{RR} = \frac{\Pr(\text{Match in}~q_D)}{\Pr(\text{Match in}~q_R)}
  = \frac{ \Pr(L^\prime_R \cap A_{Ex} \neq \varnothing) }{ \Pr(L_R \cap A_{Ex} \neq \varnothing) }
  = \frac{ \beta + \delta }{ \gamma + \delta } \\[1em]
\end{displaymath}

%
% OUT OF SCOPE LIST
%
% - advanced deception
%   - multi-step deception,
%   - longest streaks
%   - advanced repetition studies
%   - what makes human more confused
% - scanner noise reduction
% - discarded research questions
%   - "Are the answers of humans consistent?"
%   - "What triggers tools, but no humans, and vice versa?"
%   - "Are humans incapable to tell the difference for some CDTs?"
%   - "Are there differences between different professional backgrounds?"
%   - "What factors influence the time it does take a human to answer a query?"
%   - "How must queries be patched to make them look like a true weakness to humans?"
%   - "Can we have a model learn how to best deceive humans or predict their judgment?"
%   - "How can we design CDTs so that humans always fall for the trap when presented with both?"
%

\section{Prototype Design}
\label{sec:prototype-design}

% =================================================================================================

Honeyquest, which is our tool to run interactive questionnaires, uses the three
\textbf{query labels}~\ntrli~neutral, \riski~risky, and \dcpti~deceptive
and our four \textbf{query types}, as shown in Table~\ref{tab:query-examples}.
We generate deceptive queries with \mbox{HoneYAML} files,
our description language for CDTs.

\subsection{Risky Queries}
\label{sec:query-risk-types}

There are three types of risky queries, inspired by MITRE's three knowledge bases
(CVE for vulnerabilities~\cite{Mann1999:CommonEnumerationVulnerabilities},
CWE for weaknesses~\cite{Martin2006:CaseCommonFlaw},
and CAPEC for attack patterns~\cite{Barnum2007:AttackPatternsKnowledge}):

\begin{itemize}
  \item \textbf{Vulnerability queries} contain at least one indicator in the
        query that points to a known vulnerability.
  \item \textbf{Weakness queries} display an insecure pattern that might lead to
        a vulnerability.
  \item \textbf{Attack queries} showcase a deliberate attempt to do harm, often
        by exploiting a vulnerability or weakness.
\end{itemize}

The risky query in Table~\ref{tab:query-examples} is an example for a vulnerability query
by indicating that the server is running Apache 1.0.3, which is vulnerable to
\href{https://nvd.nist.gov/vuln/detail/CVE-1999-0067}{CVE-1999-0067}.
Participants are not expected to know this, but to be suspicious of the version text.

Weaknesses on the other hand may lead to vulnerabilities.
Listing~\ref{lst:http-risky-weakness} shows an example
that is indicative for a potential path traversal weakness
(\href{{https://cwe.mitre.org/data/definitions/22.html}}{CWE-22}) in a web application.
In this example, it appears that a user can potentially control the
``file'' parameter to request arbitrary files from the remote server file system.

\noindent\begin{minipage}{\columnwidth}
  \begin{lstlisting}[%
  label=lst:http-risky-weakness,
  caption={%
    A risky ``\header{}'' query of risk type ``weakness'' which inspired us to
    derive CDT \RowRefSingle{Results}{DcptNetworkrequestsPathTraversal} from that risk.
  }
]
GET /view?file=../Overview.php
Host: github.io
User-Agent: curl/7.68.0
Accept: */*
\end{lstlisting}
\end{minipage}

To obtain an attack query, we can change the query parameter in
Listing~\ref{lst:http-risky-weakness} to something like \texttt{file=../../etc/passwd}
and make it look like a concrete path traversal attack
(\href{https://capec.mitre.org/data/definitions/126.html}{CAPEC-126}).%
\enlargethispage{-6pt}

% ==============================================================================
% ------------------------------------------------------------------------------

\begin{table*}[t]
  \begin{threeparttable}
    % \small
    \centering
    \caption{One representative example for each of the four query types in Honeyquest.}
    \Description[%
      An example for a risky \header{} query, a deceptive \filesystem{} query,
      a deceptive \htaccess{} query, and a neutral \network{}s query.]{%
      The example on the risky \header{} query
      shows a vulnerability \VarIdRiskHttpheadersOutdatedApache{},
      which is the outdated Apache version 1.0.3.
      The example on the deceptive \filesystem{} query
      shows CDT \VarIdDcptFilesystemKeys{}, which is a ``keys.json'' file.
      The example on the deceptive \htaccess{} query
      shows CDT \VarIdDcptHtaccessAdminRedirect{}, which is a suspicious redirect to an admin site.
      The example on the neutral \network{}s
      shows five \network{}s to different endpoints of a web application.
    }
    \label{tab:query-examples}
    \begin{tabular*}{\textwidth}{l @{\extracolsep{\fill}} l}
      \toprule
      \addlinespace[0.2cm]
      \riski~\textbf{Risky \header{}}\,\tnote{\VarAbbrvHttpheaders{}} query (with vulnerability~\RowRefSingle{Results}{RiskHttpheadersOutdatedApache})                                   & \dcpti~\textbf{Deceptive \filesystem{}}\,\tnote{\VarAbbrvFilesystem{}} query (with technique~\RowRefSingle{Results}{DcptFilesystemKeys} injected) \\
      \begin{lstlisting}[frame=none]
            HTTP/1.1 200 OK
            Date: Tue, 02 May 2018 04:32:14 GMT
            (*@\sethlcolor{Thistle1}\hl{Server: Apache/1.0.3 (Debian)}@*)
            Vary: Accept-Encoding
            Content-Type: text/html
        \end{lstlisting}              &
      \begin{lstlisting}[frame=none]
            drwxr-xr-x 25 elsa 4.0K Dec 30 08:36 .
            drwxr-xr-x  4 root 4.0K Jun 21  2019 ..
            -rw-------  1 elsa  57K Jan 13 14:48 .bash_history
            drwx------  6 elsa 4.0K Sep 25 17:40 .config
            (*@\sethlcolor{NavajoWhite1}\hl{-rw-r\texttt{-{}-}r\texttt{-{}-}~~1 elsa~~12K Feb~~6~~2022 keys.json}@*)
        \end{lstlisting}                  \\
      \addlinespace[0.3cm]
      \dcpti~\textbf{Deceptive \htaccess{}}\,\tnote{\VarAbbrvHtaccess{}} query (with CDT~\RowRefSingle{Results}{DcptHtaccessAdminRedirect} injected)                                         & \ntrli~\textbf{Neutral \network{}s}\,\tnote{\VarAbbrvNetworkrequests{}} query         \\
      \begin{lstlisting}[frame=none]
            <IfModule mod_rewrite.c>
              RewriteEngine on
            (*@\sethlcolor{NavajoWhite1}\hl{~~Redirect 301 "/admin" \textbackslash}@*)
            (*@\sethlcolor{NavajoWhite1}\hl{~~~~"/plugins/kul/panel?role=view"}@*)
            </IfModule>
        \end{lstlisting} &
      \begin{lstlisting}[frame=none]
            0.120 POST https://shop.com/rest/user/export 200 OK (0.4 kB)
            0.215 GET https://shop.com/rest/image-captcha/ 200 OK (4.1 kB)
            0.381 GET https://shop.com/rest/user/whoami 200 OK (0.1 kB)
            2.031 GET https://shop.com/rest/history 200 OK (30 bytes)
            2.876 GET https://shop.com/api/Quantitys/ 200 OK (0.6 kB)
        \end{lstlisting}                                                            \\
      \addlinespace[0.2cm]
      \bottomrule
    \end{tabular*}
    \begin{tablenotes}
      \footnotesize
      \item \textbf{Legend:} Purple shades indicate \riski~risky lines
      and orange shades indicate \dcpti~deceptive lines.
      \vspace{0.1cm}
      \item[\VarAbbrvHttpheaders{}] \mfus{\header} queries
      show HTTP response headers, but always without any payload.
      \item[\VarAbbrvFilesystem{}] \mfus{\filesystem} queries
      show the output of the UNIX command \texttt{ls -lah},
      which lists all files in the current working directory and their metadata.
      % For each file, the following metadata is included in order:
      % type and permissions, the number of links to the file, the owner,
      % the group, the size, the last modification date, and the filename.
      \item[\VarAbbrvHtaccess{}] \htaccess queries
      show the configuration directives in an .htaccess file,
      which is used to configure Apache web servers.
      \item[\VarAbbrvNetworkrequests{}] \mfus{\network} queries
      show requests made by a web application:
      Seconds since load, method, URL, response status code,
      and response size, unless empty.
    \end{tablenotes}
  \end{threeparttable}
\end{table*}

\subsection{Deceptive Queries and HoneYAML}
\label{sec:honeyaml}

While dry-running deception experiments is valuable,
ultimately, we want to deploy them into real systems. To build a bridge to future work
we designed a specification for CDTs that we use to make queries deceptive
and which also serves as a configuration for tools that can deploy them%
~\cite{Nawrocki2016:SurveyHoneypotSoftware,Kahlhofer2024:ApplicationLayerCyber}.
We envision HoneYAML to become an enumeration of CDTs some day, much like we have an
enumeration of Common Vulnerabilities and Exposures (CVEs)%
~\cite{Mann1999:CommonEnumerationVulnerabilities}.

Listing~\ref{lst:honeyaml} shows how to define a CDT that adds a deceptive HTTP header.
Within Honeyquest, the implementation of the ``decoy-apiserver'' CDT%
~\RowRef{Results}{DcptHttpheadersApiserver}
that we show here is simply inserting a new line in the query payload.
The resulting query will be labeled as deceptive, regardless of its original type.
Our \anon[open-source repository]{\href{\repositoryurl}{open-source repository}}
contains all HoneYAML specifications that we created.
Real-world systems that add deceptive elements to the HTTP protocol
often use reverse proxies to do so~\cite{
  Han2017:EvaluationDeceptionBasedWeb,
  Araujo2014:PatchesHoneyPatchesLightweight,
  Barron2021:ClickThisNot,
  Fraunholz2018:CloxyContextawareDeceptionasaService,
  Sahin2020:LessonsLearnedSunDEW,
  Pohl2015:HiveZeroConfiguration}.
The same HoneYAML specification can be used to first evaluate CDTs
with Honeyquest and later configure proxies.

\begin{lstlisting}[%
  label=lst:honeyaml,
  caption={%
    A HoneYAML specification for CDT~\RowRefSingle{Results}{DcptHttpheadersApiserver}.
  }
]
kind: httpheader
name: decoy-apiserver
description: Header that points to an API endpoint
operations:
  - op: add
    key: X-Kube-ApiServer
    value: /hko/api
\end{lstlisting}

Because it is smart to imitate risks in deceptive queries,
we might generate deceptive queries that look similar to risky ones.
This is fine, as labels only say something about design strategies anyway.
% This is okay, because labels only tell which design strategies were used.

% ==============================================================================
% ------------------------------------------------------------------------------

\subsection{Honeyquest}
\label{sec:honeyquest}

Honeyquest is a web-based application. % The app consists of a Python backend and a React frontend.
Queries are read from a pre-computed query store that we have prepared from different sources
(\secname\ref{sec:query-design}).
New users experience the following:

\begin{enumerate}
  \item We ask for consent to collect anonymized data. % for research.
  \item We show them eight tutorial queries to teach them about
        queries, labels, and marks~(Appendix~\ref{sec:appendix:tutorial}).
  \item We collect profile information%
        ~(\secname\ref{sec:ethics} and Appendix~\ref{sec:appendix:profiling}).
  \item We sample one random query after another until we run out of queries.
        Queries are never shown twice to the same user.
        % The random sampling strategy was slightly biased:
        We also made sure that the first 100~queries included an equal
        number of neutral queries, deceptive queries with all possible CDTs,
        and risky queries with all possible risks.
\end{enumerate}

\section{Experiment Design}
\label{sec:experiment-design}

% =================================================================================================

This section summarizes the queries that we tested,
the CDTs and risks we injected, and how we recruited participants.

We pre-tested the tutorial queries (Appendix~\ref{sec:appendix:tutorial})
with two colleagues who did not participate in the actual experiment.
All of our \VarNumRisks{}~risks and \VarNumCDTsWithLiteratureRef{} of our \VarNumCDTs{}~CDTs
were indirectly pre-tested because they had been used successfully in previous work.
Nevertheless, we consulted domain experts who did not participate in the actual experiment
to pre-test all the risks and CDTs that we developed.

% ==============================================================================
% ------------------------------------------------------------------------------

\subsection{Query Design}
\label{sec:query-design}

\begin{figure}[hbt]
  \centering
  \includegraphics[width=\columnwidth,page=1]{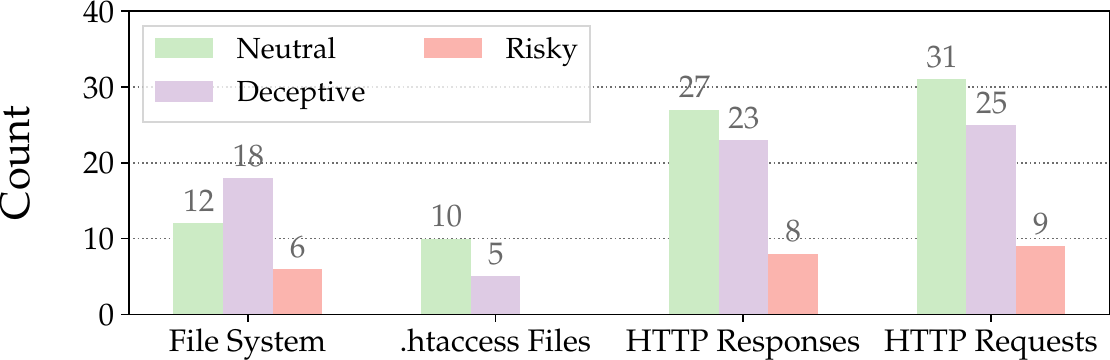}
  \caption{Distribution of neutral, deceptive, and risky labels.}
  \Description[Grouped bar plot showing the distribution of neutral, deceptive, and risky labels.]{
    A grouped bar with one group for each query type.
    The distribution is slightly imbalanced, with more neutral queries
    than deceptive queries and even fewer risky queries.
    In total, there are
    \VarNumQueriesFilesystem{}~\filesystem{},
    \VarNumQueriesHtaccess{}~\htaccess{},
    \VarNumQueriesHttpHeaders{}~\header{}, and
    \VarNumQueriesNetworkRequests{}~\network{} queries.
    The \filesystem{} group has
    \VarNumQueriesFilesystemNeut{}~neutral,
    \VarNumQueriesFilesystemDcpt{}~deceptive, and
    \VarNumQueriesFilesystemRisk{}~risky queries.
    The \htaccess{} group has
    \VarNumQueriesHtaccessNeut{}~neutral,
    \VarNumQueriesHtaccessDcpt{}~deceptive, and
    \VarNumQueriesHtaccessRisk{}~risky queries.
    The \header{} group has
    \VarNumQueriesHttpHeadersNeut{}~neutral,
    \VarNumQueriesHttpHeadersDcpt{}~deceptive, and
    \VarNumQueriesHttpHeadersRisk{}~risky queries.
    The \network{} group has
    \VarNumQueriesNetworkRequestsNeut{}~neutral,
    \VarNumQueriesNetworkRequestsDcpt{}~deceptive, and
    \VarNumQueriesNetworkRequestsRisk{}~risky queries.
  }
  \label{fig:query-types}
\end{figure}

Our dataset consists of a total of \VarNumQueries{}~queries (\figurename~\ref{fig:query-types}).
We first collected \VarNumQueriesNeutral{}~neutral queries.
We then browsed through well-known vulnerabilities and weaknesses in web applications
and manually derived \VarNumQueriesRisky{}~risky queries. Many risky queries were built
by taking a neutral one and adding indicators of risks.
A total of \VarNumQueriesDeceptive{}~deceptive queries were generated with the
method explained in \secname\ref{sec:honeyaml}.

Since we wanted to test many interesting risks and CDTs,
we did not aim for a perfectly balanced dataset.
However, this is not a problem because we do not evaluate metrics such as accuracy
that would be sensitive to imbalanced datasets.

% ==============================================================================
% ------------------------------------------------------------------------------

\subsubsection{Design of Neutral Queries}

Every single query in our \anon[open-source dataset]{\href{\repositoryurl}{open-source dataset}}
carries a reference to its original source.
Most were collected from the following real-world environments:

\begin{itemize}
  \item \textbf{\VarNumQueriesFilesystemNeut{}~\filesystem{}}
        payloads capture the output of the \texttt{ls~-lah} command
        in the home directories of a few servers, containers, and personal computers in our lab.
        Sensitive content was manually anonymized or removed.
  \item \textbf{\VarNumQueriesHtaccessNeut{}~\htaccess{}s}
        were randomly picked by searching for ``.htaccess'' % files
        in open-source projects with Sourcegraph.\footnote{\url{https://sourcegraph.com/search}}
  \item \textbf{\VarNumQueriesHttpHeadersNeut{}~\header{}s}
        were randomly sampled from the ``500K HTTP Headers'' dataset%
        ~\cite{HackerTargetPtyLtd2014:500KHTTPHeaders}
        that crawled the HTTP responses from the 500,000 most-visited websites in
        \citeyear{HackerTargetPtyLtd2014:500KHTTPHeaders},
        as ranked by the now discontinued company Alexa Internet.
  \item \textbf{\VarNumQueriesNetworkRequestsNeut{}~sets of \network{}s}
        were gathered by manually using the websites
        of popular web services and recording all HTTP requests that happened.
        We recorded traces for the websites of Amazon, Dropbox, \anon[]{Dynatrace,} GitHub, Gmail,
        Google, Jira, the OWASP Juice Shop~\cite{TheOWASPFoundationInc.2014:OWASPJuiceShop},
        TikTok, Wikipedia, and YouTube.
        Sensitive fields, names, and identifiers were manually anonymized or removed.
\end{itemize}

% ==============================================================================
% ------------------------------------------------------------------------------

\subsubsection{Design of Deceptive Queries}

We manually picked and designed \VarNumCDTs{}~CDTs:
\VarNumCDTsWithoutLiteratureRef{}~were self-defined
% and \VarNumCDTsWithLiteratureRef{}~were from previous work.
and \VarNumCDTsWithLiteratureRef{}~have been mentioned or evaluated in previous work.
Table~\ref{tab:results-deceptive-overview} lists all of them.

% ==============================================================================
% ------------------------------------------------------------------------------

\subsubsection{Design of Risky Queries}

Most of our \VarNumRisks{}~risks are designed to resemble
the categories of the OWASP Top~10~\cite{TheOWASPFoundationInc.2021:OWASPTop10}
and OWASP API Security Top~10~\cite{TheOWASPFoundationInc.2019:OWASPAPITop}.
Some were inspired by OWASP~ZAP security scanner rules~\cite{TheOWASPFoundationInc.2023:OWASPZAP}.
Table~\ref{tab:results-risky-overview} explains each risk with an example.

% ==============================================================================
% ------------------------------------------------------------------------------

\subsection{User Study Details and Ethics}
\label{sec:ethics}

We carefully reviewed our experiment to conform to ethical standards,
protect the privacy of all participants,
and follow best practices in user research~\cite{Sauro2012:QuantifyingUserExperience}.
Our institution has no IRB, so we instead conducted
an ethics self-assessment~\cite{EuropeanCommission2021:EUGrantsHow}
% thoroughly discussed it within our research group,
and obtained approval from our legal and privacy counsel.
More details of the user study and ethical considerations
are described in Appendix~\ref{sec:appendix:ethics}.

Participants were recruited by posting messages to
Slack channels of security professionals
% that develop enterprise security products,
and to a Mattermost server of a local CTF team.
\VarNumParticipantsAndDroppedUsers{}~volunteers
responded to that message and started the experiment.
We had to discard all answers from \VarNumDroppedUsers{}~of them because they answered
fewer than \VarNumMinimumResponsesWithoutTutorial{}~warm-up queries
(consisting of two pre-selected queries for each of the four query types;
same for each participant),
which left us with \textbf{\VarNumParticipants{}~participants}:
\VarNumParticipantsCtf{}~CTF players,
where most of them are graduate students, and
\VarNumParticipantsRes{}~security professionals,
where most of them build enterprise security products.
The skills of this target audience are very similar to those of real attackers.
Unlike typical user studies on cyber security~\cite{%
  Han2018:DeceptionTechniquesComputer,
  Aljohani2022:PitfallsEvaluatingCyber},
our study only had \VarNumParticipantsStudentsRatio{}\% students.
% Note that, whether this has a significant influence on the results is still an open question%
% ~\cite{Aljohani2022:PitfallsEvaluatingCyber}.
%
Demographics, consent collection, timeline, and the preceding tutorial queries are described
in Appendix~\ref{sec:appendix:ethics}.

During the experiment, we collected self-reported profile information,
and recorded how long it took users to answer queries.
No personally-identifying information was collected.
Participants were informed about the purpose of the experiment,
about the presence of neutral, deceptive, and risky queries, and about their
option to stop answering queries at any time, without negative consequences and
without giving reasons. Participants were allowed to continue where they left
off by visiting our web application again.
In a second run of the experiment with \VarNumParticipantsResPhaseB{}
security professionals, participants could win a 50€ Amazon gift card in a lottery,
if they answered at least 50\% of the queries.
% \VarNumParticipantsResLotteryPlayers{} of them joined that lottery.
All others received no incentives to participate.
Security professionals were allowed to do this during their work time,
CTF players participated in their free time without any compensation.
Participants did not receive a performance report % upon completion
and where never informed if their answer marks were ``correct''. % or not.

Participants could comment on any query during the experiment
and those that were freely disclosing their identity in the comments
were invited to discuss them with us (\secname\ref{sec:challenges}).

\section{Results}
\label{sec:results}
\label{sec:with-reference-ids}

% =================================================================================================

% We summarize the results from the experiment
% that we described in \secname\ref{sec:experiment-design}.

\begin{figure}[!htb]
  \centering
  \includegraphics[width=\columnwidth,page=1]{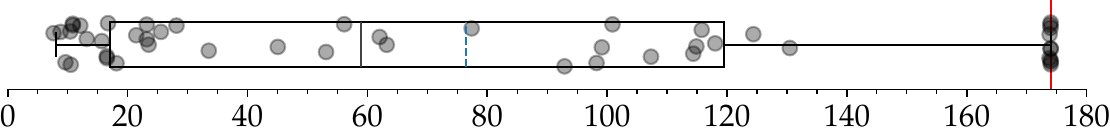}
  \caption{
    Boxplot on the number of answered queries per participant. % Tukey boxplot
    Total Queries~=~\VarNumQueries{}.
    Mean~=~\VarResAnswerQueriesMean{}.
    Median~=~\VarResAnswerQueriesMedian{}.
  }
  \label{fig:query-counts}
  \Description{
    A boxplot that shows the distribution of the number of answered queries,
    with the numbers as already described in the caption.
  }
\end{figure}

Our \VarNumParticipants{}~participants answered \VarNumResponses{}~queries in total%
~(\figurename~\ref{fig:query-counts}).
All of them answered at least \VarNumMinimumResponsesWithoutTutorial{} queries.
The median answer time per query was \VarResAnswerTimeMedian{}~seconds.
Participants needed 45~-~60 minutes on average to answer all \VarNumQueries{}~queries.

The subscript on the following percentage numbers is the 95\%~CI of the mean,
calculated using Wilson's method~\cite{Wilson1927:ProbableInferenceLaw}.

% On a very high level, we observed the following results:

\begin{itemize}
  \item \textbf{Aspect A.} (Table~\ref{tab:results-neutral-overview},
        \ref{tab:results-deceptive-overview}, \ref{tab:results-risky-overview}) \quad
        Participants fell for \dcpti~traps in
        ${\VarResFellTpDcpt{}_{\pm\VarResFellTpDcptCse{}}\%}$
        of their answers. They recognized traps in
        ${\VarResCmDcptTpr{}_{\pm\VarResCmDcptTprCse{}}\%}$
        of their answers. In
        ${\VarResCmDcptFpr{}_{\pm\VarResCmDcptFprCse{}}\%}$
        of answers to \ntrli~neutral queries, they mistakenly classified something as a trap.
        Participants correctly identified \riski~risks in
        ${\VarResCmRiskTpr{}_{\pm\VarResCmRiskTprCse{}}\%}$
        of their answers and mistakenly classified something as a risk in
        ${\VarResCmRiskFpr{}_{\pm\VarResCmRiskFprCse{}}\%}$
        of their answers to \ntrli~neutral queries.
  \item \textbf{Aspect B1.} (Table~\ref{tab:results-deceptive-overview}) \quad
        Deceptive lines were marked first in only \VarResTrapsFirstOverall{}\% of answers.
        We cannot reject the null from \secname\ref{sec:aspect-deception-first}, i.e., % hypothesis
        participants did not prefer CDTs over other elements. % to mark
        % that this preference is random or higher than 50\%.
        % Our participants did not prefer to mark CDTs before other elements.
        % Unfortunately, participants still preferred true risks over CDTs
        % (Binomial test, $p \ll 0.001$, $n=\VarAspectThreeDcptOverallSampleSize{}$).
        % While this sounds like a negative result, remember that we
        % now have data on what participants actually marked first.
        % This allows us to design better CDTs next time.
  \item \textbf{Aspect B2.} (Table~\ref{tab:results-contingency-table}) \quad
        The presence of deception reduced the risk of marking a weakness or vulnerability
        by \VarResBeforeAfterOverallRiskReduction{}\% on average.
        We tested this on a small set of
        \VarNumCDTsWithRiskyAndDeceptiveQueries{}~CDTs:
        % (injected in \VarNumPairedQueries{}~paired queries).
        The null hypothesis from \secname\ref{sec:aspect-divert-attention}
        can be rejected (${ \alpha = 0.05 }$)
        for all techniques combined ${ (p = \VarResBeforeAfterOverallPvalue{}) }$,
        and for techniques~\RowRefSingle{Results}{DcptHttpheadersApiserver}
        and~\RowRefSingle{Results}{DcptHttpheadersDevtoken} alone.
        We cannot reject the null for the other CDTs.
\end{itemize}

% Results for neutral queries are shown in Table~\ref{tab:results-neutral-overview},
% grouped by \dcpti~CDTs in Table~\ref{tab:results-deceptive-overview},
% and grouped by \riski~risks in Table~\ref{tab:results-risky-overview}.

\begin{table}[!htb]
  \newcommand{\fileResultsNeutralOverview}{\bfseries All Neutral Queries & 80 & 1647 & \detectionbar{41.0}{8.0}{7.0}{44.0}[0.0238][0.0133] \\
\addlinespace[0.05cm]
File System & 12 & 302 & \detectionbar{53.0}{5.0}{14.0}{28.0}[0.0559][0.0250] \\
.htaccess Files & 10 & 195 & \detectionbar{31.0}{3.0}{3.0}{63.0}[0.0645][0.0238] \\
HTTP Responses & 27 & 559 & \detectionbar{43.0}{10.0}{5.0}{42.0}[0.0410][0.0253] \\
HTTP Requests & 31 & 591 & \detectionbar{37.0}{10.0}{6.0}{47.0}[0.0387][0.0240] \\
}
  \begin{threeparttable}
    \renewcommand{\defaultothercolor}{Sienna}
    \renewcommand{\defaultotherdrawopacity}{0.5}
    \renewcommand{\defaultrectwidth}{3.25cm}
    \renewcommand{\defaultlabelthreshold}{15}
    \small
    \centering
    \caption{Results on the enticingness of neutral queries.}
    \label{tab:results-neutral-overview}
    \begin{tabularx}{\columnwidth}{l @{\hspace{10pt}} @{\extracolsep{\fill}} rr l}
      \toprule
      \addlinespace[0.1cm]
       & $k$ & $n$ & Mark Distribution \\
      \midrule
      \bfseries All Neutral Queries & 80 & 1647 & \detectionbar{41.0}{8.0}{7.0}{44.0}[0.0238][0.0133] \\
\addlinespace[0.05cm]
File System & 12 & 302 & \detectionbar{53.0}{5.0}{14.0}{28.0}[0.0559][0.0250] \\
.htaccess Files & 10 & 195 & \detectionbar{31.0}{3.0}{3.0}{63.0}[0.0645][0.0238] \\
HTTP Responses & 27 & 559 & \detectionbar{43.0}{10.0}{5.0}{42.0}[0.0410][0.0253] \\
HTTP Requests & 31 & 591 & \detectionbar{37.0}{10.0}{6.0}{47.0}[0.0387][0.0240] \\

      \bottomrule
    \end{tabularx}
    \begin{tablenotes}
      \footnotesize
      \item \textbf{Description:}
      Bars show the \%~of answers~$n$ to neutral queries with
      \dhackbox \enspace $\nicefrac{n_{Ex}}{n}$ \expli~exploit marks,~
      \dtrapbox \enspace $\nicefrac{n_{Tr}}{n}$ \trapi~trap marks,~
      \dbothbox \enspace $\nicefrac{n_\land}{n}$ \expli~exploit and \trapi~trap marks,~
      \dnonebox \enspace $\nicefrac{n_\varnothing}{n}$ no marks at all.
      $k$~are the number of neutral queries.
      $n$~are the number of answers (not marks) to them.
      Bars without percentage numbers account for less than \defaultlabelthreshold\%.
      The tiny bars denote the 95\%~CI of that mean.
      % $n_{Ex}$~are the num. of ans. with \expli~exploit marks on neutral lines.
      % $n_{Tr}$~are the num. of ans. with \trapi~trap marks on neutral lines.
      % $n_\land$~are the num. of ans. with \expli~exploit and \trapi~trap marks on neutral lines.
      % $n_\varnothing$~are the num. of ans. where nothing was marked.
    \end{tablenotes}
  \end{threeparttable}
\end{table}

\begin{table}[!t]
  \newcommand{\fileResultsRiskyOverview}{\addlinespace[0.05cm]
\multicolumn{2}{l}{\bfseries All Risky Queries} & 461 & \detectionbar{44.0}{8.0}{23.0}{25.0}[0.0452][0.0249] \\
\midrule
\addlinespace[0.1cm]
\multicolumn{2}{l}{\bfseries File System} & 166 & \detectionbar{51.0}{11.0}{31.0}{7.0}[0.0752][0.0487] \\
\addlinespace[0.05cm]
\rhypertarget{row:Results:RiskFilesystemPrivateKey}\scriptsize\VarIdRiskFilesystemPrivateKey & \RowRiskyFilesystemPrivateKey{} & 27 & \detectionbar{67.0}{26.0}{7.0}{0.0}[0.1677][0.1575] \\
\rhypertarget{row:Results:RiskFilesystemBackup}\scriptsize\VarIdRiskFilesystemBackup & \RowRiskyFilesystemBackup{} & 57 & \detectionbar{61.0}{11.0}{23.0}{5.0}[0.1225][0.0810] \\
\rhypertarget{row:Results:RiskFilesystemOpenvpnConfig}\scriptsize\VarIdRiskFilesystemOpenvpnConfig & \RowRiskyFilesystemOpenvpnConfig{} & 27 & \detectionbar{41.0}{4.0}{48.0}{7.0}[0.1738][0.0881] \\
\rhypertarget{row:Results:RiskFilesystemKubernetesManifests}\scriptsize\VarIdRiskFilesystemKubernetesManifests & \RowRiskyFilesystemKubernetesManifests{} & 28 & \detectionbar{39.0}{4.0}{43.0}{14.0}[0.1701][0.0854] \\
\rhypertarget{row:Results:RiskFilesystemDnsUpdateKey}\scriptsize\VarIdRiskFilesystemDnsUpdateKey & \RowRiskyFilesystemDnsUpdateKey{} & 27 & \detectionbar{33.0}{15.0}{44.0}{8.0}[0.1677][0.1328] \\
\midrule
\addlinespace[0.1cm]
\multicolumn{2}{l}{\bfseries HTTP Responses} & 155 & \detectionbar{54.0}{7.0}{15.0}{24.0}[0.0776][0.0413] \\
\addlinespace[0.05cm]
\rhypertarget{row:Results:RiskHttpheadersProxyAuthLeak}\scriptsize\VarIdRiskHttpheadersProxyAuthLeak & \RowRiskyHttpheadersProxyAuthLeak{} & 26 & \detectionbar{69.0}{12.0}{0.0}{19.0}[0.1674][0.1249] \\
\rhypertarget{row:Results:RiskHttpheadersOutdatedApache}\scriptsize\VarIdRiskHttpheadersOutdatedApache & \RowRiskyHttpheadersOutdatedApache{} & 50 & \detectionbar{62.0}{0.0}{20.0}{18.0}[0.1299][0.0357] \\
\rhypertarget{row:Results:RiskHttpheadersOutdatedPhp}\scriptsize\VarIdRiskHttpheadersOutdatedPhp & \RowRiskyHttpheadersOutdatedPhp{} & 39 & \detectionbar{59.0}{10.0}{10.0}{21.0}[0.1475][0.0976] \\
\rhypertarget{row:Results:RiskHttpheadersRequestSmugglingClte}\scriptsize\VarIdRiskHttpheadersRequestSmugglingClte & \RowRiskyHttpheadersRequestSmugglingClte{} & 13 & \detectionbar{54.0}{15.0}{8.0}{23.0}[0.2383][0.1895] \\
\rhypertarget{row:Results:RiskHttpheadersCrossDomainRefererLeakage}\scriptsize\VarIdRiskHttpheadersCrossDomainRefererLeakage & \RowRiskyHttpheadersCrossDomainRefererLeakage{} & 27 & \detectionbar{15.0}{7.0}{33.0}{45.0}[0.1328][0.1066] \\
\midrule
\addlinespace[0.1cm]
\multicolumn{2}{l}{\bfseries HTTP Requests} & 140 & \detectionbar{26.0}{5.0}{20.0}{49.0}[0.0723][0.0376] \\
\addlinespace[0.05cm]
\rhypertarget{row:Results:RiskNetworkrequestsBrokenFunctionLevelAuthorization}\scriptsize\VarIdRiskNetworkrequestsBrokenFunctionLevelAuthorization & \RowRiskyNetworkrequestsBrokenFunctionLevelAuthorization{} & 12 & \detectionbar{83.0}{8.0}{0.0}{9.0}[0.2005][0.1695] \\
\rhypertarget{row:Results:RiskNetworkrequestsPasswordHashesInQueryParameters}\scriptsize\VarIdRiskNetworkrequestsPasswordHashesInQueryParameters & \RowRiskyNetworkrequestsPasswordHashesInQueryParameters{} & 12 & \detectionbar{75.0}{8.0}{8.0}{9.0}[0.2217][0.1695] \\
\rhypertarget{row:Results:RiskNetworkrequestsMassAssignment}\scriptsize\VarIdRiskNetworkrequestsMassAssignment & \RowRiskyNetworkrequestsMassAssignment{} & 11 & \detectionbar{45.0}{9.0}{0.0}{46.0}[0.2536][0.1806] \\
\rhypertarget{row:Results:RiskNetworkrequestsBrokenObjectLevelAuthorization}\scriptsize\VarIdRiskNetworkrequestsBrokenObjectLevelAuthorization & \RowRiskyNetworkrequestsBrokenObjectLevelAuthorization{} & 12 & \detectionbar{25.0}{0.0}{17.0}{58.0}[0.2217][0.1212] \\
\rhypertarget{row:Results:RiskNetworkrequestsLogSpamEndpoint}\scriptsize\VarIdRiskNetworkrequestsLogSpamEndpoint & \RowRiskyNetworkrequestsLogSpamEndpoint{} & 27 & \detectionbar{19.0}{4.0}{26.0}{51.0}[0.1426][0.0881] \\
\rhypertarget{row:Results:RiskNetworkrequestsDevEndpointAccessible}\scriptsize\VarIdRiskNetworkrequestsDevEndpointAccessible & \RowRiskyNetworkrequestsDevEndpointAccessible{} & 12 & \detectionbar{17.0}{17.0}{50.0}{16.0}[0.2005][0.2005] \\
\rhypertarget{row:Results:RiskNetworkrequestsNoRateLimiting}\scriptsize\VarIdRiskNetworkrequestsNoRateLimiting & \RowRiskyNetworkrequestsNoRateLimiting{} & 12 & \detectionbar{8.0}{0.0}{0.0}{92.0}[0.1695][0.1212] \\
\rhypertarget{row:Results:RiskNetworkrequestsInsecureHttp}\scriptsize\VarIdRiskNetworkrequestsInsecureHttp & \RowRiskyNetworkrequestsInsecureHttp{} & 30 & \detectionbar{7.0}{3.0}{33.0}{57.0}[0.0974][0.0804] \\
\rhypertarget{row:Results:RiskNetworkrequestsNosqlInjection}\scriptsize\VarIdRiskNetworkrequestsNosqlInjection & \RowRiskyNetworkrequestsNosqlInjection{} & 12 & \detectionbar{0.0}{0.0}{17.0}{83.0}[0.1212][0.1212] \\
}
  \begin{threeparttable}
    \renewcommand{\defaultrectwidth}{3cm}
    \renewcommand{\defaultlabelthreshold}{15}
    \setlength{\tabcolsep}{3pt}
    \small
    \centering
    \caption{Results on the enticingness of risks.}
    \label{tab:results-risky-overview}
    \begin{tabularx}{\columnwidth}{l @{\hspace{3pt}} @{\extracolsep{\fill}} X r l}
      \toprule
      \addlinespace[0.1cm]
      \multicolumn{2}{l}{\riski~Risk} & $r$ & Mark Distribution \\
      \midrule
      \addlinespace[0.05cm]
\multicolumn{2}{l}{\bfseries All Risky Queries} & 461 & \detectionbar{44.0}{8.0}{23.0}{25.0}[0.0452][0.0249] \\
\midrule
\addlinespace[0.1cm]
\multicolumn{2}{l}{\bfseries File System} & 166 & \detectionbar{51.0}{11.0}{31.0}{7.0}[0.0752][0.0487] \\
\addlinespace[0.05cm]
\rhypertarget{row:Results:RiskFilesystemPrivateKey}\scriptsize\VarIdRiskFilesystemPrivateKey & \RowRiskyFilesystemPrivateKey{} & 27 & \detectionbar{67.0}{26.0}{7.0}{0.0}[0.1677][0.1575] \\
\rhypertarget{row:Results:RiskFilesystemBackup}\scriptsize\VarIdRiskFilesystemBackup & \RowRiskyFilesystemBackup{} & 57 & \detectionbar{61.0}{11.0}{23.0}{5.0}[0.1225][0.0810] \\
\rhypertarget{row:Results:RiskFilesystemOpenvpnConfig}\scriptsize\VarIdRiskFilesystemOpenvpnConfig & \RowRiskyFilesystemOpenvpnConfig{} & 27 & \detectionbar{41.0}{4.0}{48.0}{7.0}[0.1738][0.0881] \\
\rhypertarget{row:Results:RiskFilesystemKubernetesManifests}\scriptsize\VarIdRiskFilesystemKubernetesManifests & \RowRiskyFilesystemKubernetesManifests{} & 28 & \detectionbar{39.0}{4.0}{43.0}{14.0}[0.1701][0.0854] \\
\rhypertarget{row:Results:RiskFilesystemDnsUpdateKey}\scriptsize\VarIdRiskFilesystemDnsUpdateKey & \RowRiskyFilesystemDnsUpdateKey{} & 27 & \detectionbar{33.0}{15.0}{44.0}{8.0}[0.1677][0.1328] \\
\midrule
\addlinespace[0.1cm]
\multicolumn{2}{l}{\bfseries HTTP Responses} & 155 & \detectionbar{54.0}{7.0}{15.0}{24.0}[0.0776][0.0413] \\
\addlinespace[0.05cm]
\rhypertarget{row:Results:RiskHttpheadersProxyAuthLeak}\scriptsize\VarIdRiskHttpheadersProxyAuthLeak & \RowRiskyHttpheadersProxyAuthLeak{} & 26 & \detectionbar{69.0}{12.0}{0.0}{19.0}[0.1674][0.1249] \\
\rhypertarget{row:Results:RiskHttpheadersOutdatedApache}\scriptsize\VarIdRiskHttpheadersOutdatedApache & \RowRiskyHttpheadersOutdatedApache{} & 50 & \detectionbar{62.0}{0.0}{20.0}{18.0}[0.1299][0.0357] \\
\rhypertarget{row:Results:RiskHttpheadersOutdatedPhp}\scriptsize\VarIdRiskHttpheadersOutdatedPhp & \RowRiskyHttpheadersOutdatedPhp{} & 39 & \detectionbar{59.0}{10.0}{10.0}{21.0}[0.1475][0.0976] \\
\rhypertarget{row:Results:RiskHttpheadersRequestSmugglingClte}\scriptsize\VarIdRiskHttpheadersRequestSmugglingClte & \RowRiskyHttpheadersRequestSmugglingClte{} & 13 & \detectionbar{54.0}{15.0}{8.0}{23.0}[0.2383][0.1895] \\
\rhypertarget{row:Results:RiskHttpheadersCrossDomainRefererLeakage}\scriptsize\VarIdRiskHttpheadersCrossDomainRefererLeakage & \RowRiskyHttpheadersCrossDomainRefererLeakage{} & 27 & \detectionbar{15.0}{7.0}{33.0}{45.0}[0.1328][0.1066] \\
\midrule
\addlinespace[0.1cm]
\multicolumn{2}{l}{\bfseries HTTP Requests} & 140 & \detectionbar{26.0}{5.0}{20.0}{49.0}[0.0723][0.0376] \\
\addlinespace[0.05cm]
\rhypertarget{row:Results:RiskNetworkrequestsBrokenFunctionLevelAuthorization}\scriptsize\VarIdRiskNetworkrequestsBrokenFunctionLevelAuthorization & \RowRiskyNetworkrequestsBrokenFunctionLevelAuthorization{} & 12 & \detectionbar{83.0}{8.0}{0.0}{9.0}[0.2005][0.1695] \\
\rhypertarget{row:Results:RiskNetworkrequestsPasswordHashesInQueryParameters}\scriptsize\VarIdRiskNetworkrequestsPasswordHashesInQueryParameters & \RowRiskyNetworkrequestsPasswordHashesInQueryParameters{} & 12 & \detectionbar{75.0}{8.0}{8.0}{9.0}[0.2217][0.1695] \\
\rhypertarget{row:Results:RiskNetworkrequestsMassAssignment}\scriptsize\VarIdRiskNetworkrequestsMassAssignment & \RowRiskyNetworkrequestsMassAssignment{} & 11 & \detectionbar{45.0}{9.0}{0.0}{46.0}[0.2536][0.1806] \\
\rhypertarget{row:Results:RiskNetworkrequestsBrokenObjectLevelAuthorization}\scriptsize\VarIdRiskNetworkrequestsBrokenObjectLevelAuthorization & \RowRiskyNetworkrequestsBrokenObjectLevelAuthorization{} & 12 & \detectionbar{25.0}{0.0}{17.0}{58.0}[0.2217][0.1212] \\
\rhypertarget{row:Results:RiskNetworkrequestsLogSpamEndpoint}\scriptsize\VarIdRiskNetworkrequestsLogSpamEndpoint & \RowRiskyNetworkrequestsLogSpamEndpoint{} & 27 & \detectionbar{19.0}{4.0}{26.0}{51.0}[0.1426][0.0881] \\
\rhypertarget{row:Results:RiskNetworkrequestsDevEndpointAccessible}\scriptsize\VarIdRiskNetworkrequestsDevEndpointAccessible & \RowRiskyNetworkrequestsDevEndpointAccessible{} & 12 & \detectionbar{17.0}{17.0}{50.0}{16.0}[0.2005][0.2005] \\
\rhypertarget{row:Results:RiskNetworkrequestsNoRateLimiting}\scriptsize\VarIdRiskNetworkrequestsNoRateLimiting & \RowRiskyNetworkrequestsNoRateLimiting{} & 12 & \detectionbar{8.0}{0.0}{0.0}{92.0}[0.1695][0.1212] \\
\rhypertarget{row:Results:RiskNetworkrequestsInsecureHttp}\scriptsize\VarIdRiskNetworkrequestsInsecureHttp & \RowRiskyNetworkrequestsInsecureHttp{} & 30 & \detectionbar{7.0}{3.0}{33.0}{57.0}[0.0974][0.0804] \\
\rhypertarget{row:Results:RiskNetworkrequestsNosqlInjection}\scriptsize\VarIdRiskNetworkrequestsNosqlInjection & \RowRiskyNetworkrequestsNosqlInjection{} & 12 & \detectionbar{0.0}{0.0}{17.0}{83.0}[0.1212][0.1212] \\

      \bottomrule
    \end{tabularx}
    \begin{tablenotes}
      \footnotesize
      \item \textbf{Aspect A:}
      Bars show the \%~of answers~$r$ that~
      \dhackbox \enspace $\nicefrac{r_{Ex}}{r}$
      match \expli~exploit marks (=~``risk detected''),~
      \dtrapbox \enspace $\nicefrac{r_{Tr}}{r}$
      match \trapi~trap marks (=~``risk mistaken for trap''),~
      \dotherbox \enspace $\nicefrac{r_\triangle}{r}$ placed marks elsewhere,~
      \dnonebox \enspace $\nicefrac{r_\varnothing}{r}$ had no marks at all.
      $r$~are the number of answers (not marks) to queries with this risk.
      Bars without percentage numbers account for less than \defaultlabelthreshold\%.
      The tiny bars denote the 95\%~CI of that mean.
      % $r_{Ex}$~are the num. of ans. where an \expli~exploit mark intersected the risky lines.
      % $r_{Tr}$~are the num. of ans. where a \trapi~trap mark intersected the risky lines.
      % $r_\triangle$~are the num. of ans. where marks do not intersect risky lines.
      % $r_\varnothing$~are the num. of ans. where nothing was marked.
      \item \textbf{Risk Presence:}
      In our dataset, every risk was present in exactly one query,
      with the exception of \RowRefSingle{Results}{RiskFilesystemBackup} (2x),
      \RowRefSingle{Results}{RiskHttpheadersOutdatedPhp} (2x), and
      \RowRefSingle{Results}{RiskHttpheadersOutdatedApache} (3x).
      This explains the relatively higher number of answers~$r$ on those three risks.
      \vspace{0.1cm}
      \item[a]
      HTTP request smuggling exploits how web servers handle HTTP requests,
      such that they initiate illegitimate requests.
    \end{tablenotes}
  \end{threeparttable}
\end{table}

\section{Discussion}
\label{sec:discussion}

% =================================================================================================

This section discusses new insights about enticing and defensive deception,
compares our results to previous work, and points out possible improvements
for future experiments.

% ==============================================================================
% ------------------------------------------------------------------------------

\subsection{Aspect A: Enticing Deception}
\label{sec:discuss-enticing-deception}

To discuss enticingness, % To discuss what query elements are enticing,
we will begin to blur the distinction between traps and risks in this section.
Ultimately, we do not want attackers to be able to distinguish between them,
but rather want to learn how they react to certain query elements.

% ==============================================================================
% ------------------------------------------------------------------------------

\textbf{Enticing deception should be neither too obvious nor too camouflaged.}
Our participants were most tempted to exploit
authentication (passwords, tokens, hashes, cookies) and configuration elements.
But, obvious elements like a ``passwords.txt'' file%
~\RowRef{Results}{DcptFilesystemPasswords}
or parameters with clear-text passwords%
~\RowRef{Results}{DcptNetworkrequestsCleartextPassword},
while still being comparably enticing, were often recognized as traps
(all~$\geq~\VarResRatioDcptTrapTpNetworkrequestsCleartextPassword{}\%$).
On the other end, harder-to-find traps or risks like a logging endpoint
or mass assignment weaknesses%
~\RowRef{Results}{DcptNetworkrequestsLogEndpoint,DcptNetworkrequestsMassAssignment}
were rarely discovered, neither as something to exploit
(all~$\leq~\VarResRatioDcptFellTpNetworkrequestsMassAssignment{}\%$)
nor as a trap
(all~$\leq~\VarResRatioDcptTrapTpNetworkrequestsLogEndpoint{}\%$).
\citeauthor{Sahin2022:MeasuringDevelopersWeb} also observed that more complex risks
were tried less often in their CTF experiment~\cite{Sahin2022:MeasuringDevelopersWeb}.

% Specifically,
In our \filesystem{} queries,
we saw that filenames containing the terms
``backup'', ``config'', ``ovpn'', or ``k8s-manifests''
received less \trapi~trap marks than more obvious terms like ``key'' or ``password''.
This makes us believe that CDTs should be neither too obvious nor too camouflaged.
\citeauthor{Han2017:EvaluationDeceptionBasedWeb} also
speculated that the placement of CDTs should be neither too sparse nor too aggressive%
~\cite{Han2017:EvaluationDeceptionBasedWeb}.

% ==============================================================================
% ------------------------------------------------------------------------------

\textbf{Imitating true risks is a promising method for designing deceptive elements.}
Our participants placed significantly more \expli~exploit marks % put
on \riski~true risks than on \dcpti~traps
(${\VarResCmRiskTpr{}_{\pm\VarResCmRiskTprCse{}}\%}$ vs.
${\VarResFellTpDcpt{}_{\pm\VarResFellTpDcptCse{}}\%}$;
% (\VarResCmRiskTpr{}\% vs. \VarResFellTpDcpt{}\%;
$\chi^2$-test, $p=\VarResTestRiskDcptMarksPvalue{}$, $n=\VarResTestRiskDcptMarksNumTotal{}$).
This is reasonable, since true risks should be more enticing than traps,
but, it shows that the best traps may need to only imitate true risks.
This strengthens the idea proposed by % of
\citeauthor{Araujo2014:PatchesHoneyPatchesLightweight}%
~\cite{Araujo2014:PatchesHoneyPatchesLightweight}
on ``honeypatching'' true vulnerabilities
such that they are technically fixed but still respond as if they were vulnerable when attacked.

Specifically, in our \header{} and \network{} queries,
true risks like outdated Apache or PHP versions%
~\RowRef{Results}{RiskHttpheadersOutdatedApache,RiskHttpheadersOutdatedPhp}
and password hashes in a parameter
~\RowRef{Results}{RiskNetworkrequestsPasswordHashesInQueryParameters}
were more often exploited
(all~$\geq~\VarResRatioRiskHackTpHttpheadersOutdatedPhp{}\%$)
than deceptive tokens or cookies
\RowRef{Results}{DcptHttpheadersDevtoken,DcptHttpheadersAdminCookie}
in header fields
(all~$\geq~\VarResRatioDcptFellTpHttpheadersAdminCookie{}\%$).
Please note that findings for \RowRefSingle{Results}{RiskHttpheadersOutdatedPhp}
and \RowRefSingle{Results}{RiskHttpheadersOutdatedApache}
might include a bias since we showcased similar risks in the tutorial,
but without disclosing whether they are risky or deceptive.
Also, the path traversal trap%
~\RowRef{Results}{DcptNetworkrequestsPathTraversal}
was part of the tutorial but was surprisingly rarely identified as a trap
(${\VarResRatioDcptTrapTpNetworkrequestsPathTraversal{}_{\pm\VarResCseDcptTrapTpNetworkrequestsPathTraversal{}}\%}$)
in the actual experiment.

% ==============================================================================
% ------------------------------------------------------------------------------

\textbf{Letting participants place marks on individual lines proves valuable for inventing new CDTs.}
Instead of only imitating known risks that received many \expli~exploit marks,
we can also devise new traps by looking at what marks
individual lines received (Table~\ref{tab:results-mark-hack}, \ref{tab:results-mark-trap}).
For example, filenames ``.ssh'', ``.bash\_history'', and ``data.csv'' received
\VarResNumHackMarksOnLineSsh{}, \VarResNumHackMarksOnLineBashHistory{},
and \VarResNumHackMarksOnLineDataCsv{}~exploit marks, respectively.
All of them had $\leq~\VarResNumTrapMarksOnLineSsh{}$~trap marks,
the ``.bash\_history'' even had zero.
HTTP headers where version strings leaked Apache modules, e.g. ``mod\_ssl/2.2.17'',
received \VarResNumHackMarksOnLineModHeader{}~exploit marks
and only \VarResNumTrapMarksOnLineModHeader{}~trap marks.

Gaining such insights from otherwise neutral queries is possible
because Honeyquest lets participants place marks on individual lines.
We believe that this allows for more fine-grained analysis
without increasing participants' cognitive load.
Previous work often presented participants with pairs of questions
(one knowingly genuine, one knowingly deceptive),
and asked them to find the deceptive one%
~\cite{
  Sahin2022:ApproachGenerateRealistic,
  Pohl2015:HiveZeroConfiguration,
  Bercovitch2011:HoneyGenAutomatedHoneytokens,
  Rowe2006:FakeHoneypotsDefensive,
  Rowe2007:DefendingCyberspaceFake}.
\citeauthor{Sahin2022:ApproachGenerateRealistic}~\cite{Sahin2022:ApproachGenerateRealistic}
let participants choose between placing genuine and deceptive marks on HTTP parameter names.
It should be noted that Honeyquest's \trapi~trap marks are most similar to their deceptive marks,
but Honeyquest distinguishes adversaries' intentions further by letting
participants place either \expli~exploit marks or no marks at all.
In Honeyquest, not placing any mark
could be interpreted as placing a genuine mark. % in other works

% ==============================================================================
% ------------------------------------------------------------------------------

\textbf{Multiple iterations of Honeyquest can inform the design of more enticing CDTs.}
Our experiment is just the beginning of a feedback cycle
that can inform the design of future, more enticing CDTs.
Table~\ref{tab:decision-matrix} shows a possible ranking that
would reward ``enticing traps'' and punish ``ineffective traps''.
We see this as an early attempt to rank enticement,
useful for continuous experiments with humans
or for training autonomous agents. % reinforcement learning
%
% The highest reward shall go to CDTs,
% where many adversaries mark the deceptive elements to be exploited,
% and not to be avoided, thus falling for the trap.

% ==============================================================================
% ------------------------------------------------------------------------------

\begin{table}[!hb]
  \begin{threeparttable}
    % \small
    \centering
    \caption{Reward matrix to maximize CDT enticement.}
    \Description{
      There are three columns: No marks, trap marks, and exploit marks.
      There are three rows: Neutral, deceptive, and risky queries.
      A positive reward is given for all exploit marks, with the strongest reward for risky queries.
      A negative reward is given for all trap marks, with the strongest penalty for risky queries.
      A slightly negative reward is given for no marks.
    }
    \label{tab:decision-matrix}
    \begin{tabularx}{\columnwidth}{ c @{\hspace{6pt}} @{\extracolsep{\fill}} ccc }
      \toprule
      Qry.   & No Marks              & \trapi~Trap Marks      & \expli~Exploit Marks  \\
      \midrule
      \ntrli &                       & \rewardfigure{-0.1cm}  & \rewardfigure{0.1cm}  \\
      \riski & \rewardfigure{-0.1cm} & \rewardfigure{-0.2cm}  & \rewardfigure{0.2cm}  \\
      \dcpti & \rewardfigure{-0.1cm} & \rewardfigure{-0.35cm} & \rewardfigure{0.35cm} \\
      \bottomrule
    \end{tabularx}
    \begin{tablenotes}
      \footnotesize
      \item \textbf{Legend:} Arrow direction and strength indicate our
      subjective judgment on how to rank enticement.
      $\textcolor{BoxColorRed}{\blacktriangleleft}$~=~Less enticing.
      $\textcolor{BoxColorBlue}{\blacktriangleright}$~=~More enticing.
    \end{tablenotes}
  \end{threeparttable}
\end{table}

% ==============================================================================
% ------------------------------------------------------------------------------

\subsection{Aspect B: Defensive Deception}
\label{sec:discuss-defensive-deception}

Cyber deception is seen as an active form of cyber defense~\cite{
  Heckman2013:ActiveCyberDefense,
  Petrunic2015:HoneytokensActiveDefense,
  Zhang2021:ThreeDecadesDeception,
  Sahin2022:ApproachGenerateRealistic}.
Our results support and enrich this claim.

% ==============================================================================
% ------------------------------------------------------------------------------

\textbf{Some CDTs significantly reduce the risk of true weaknesses being exploited.}
Fascinatingly, we see that our participants were in fact
distracted by the presence of deception (Table~\ref{tab:results-contingency-table}).
Adding \mbox{``X-ApiServer''} and \mbox{``X-DevToken''} headers%
~\RowRef{Results}{DcptHttpheadersApiserver,DcptHttpheadersDevtoken}
reduced the risk that participants marked the true vulnerability
by \VarResBeforeAfterHttpheadersApiserverRiskReduction{}\%
and \VarResBeforeAfterHttpheadersDevtokenRiskReduction{}\%, respectively.
The true vulnerabilities that were missed due to the presence of the deceptive headers
were vulnerable versions of Apache and PHP%
~\RowRef{Results}{RiskHttpheadersOutdatedApache,RiskHttpheadersOutdatedPhp}.
In all cases where we have obtained enough statistical power,
we can measure a significant reduction in risk.
Our findings demonstrate that cyber deception can be an
active form of cyber defense, reducing the risk of exploitation of true system weaknesses.
We see more tests on hypothesis like this one as a promising direction for future work.

% ==============================================================================
% ------------------------------------------------------------------------------

\textbf{When participants fell for traps, the trap was not the first thing they marked
  --- at least not in Honeyquest.}
Surprisingly, traps were clearly not what participants marked first for exploitation
(Binomial test, ${p \ll 0.001}$, ${n=\VarAspectThreeDcptOverallSampleSize{}}$).
% \VarResTestMarkPreferenceDcptOppositePvalue{} = 0.000
We suspect that comparably enticing neutral query elements or
the exact location of a CDT in the query
(e.g., participants possibly read queries from top to bottom) might explain this finding.
Thus, we do not recommend drawing conclusions from our participants' order of actions.
Instead, we believe that CTF experiments are more appropriate for studying this aspect.
However, to our knowledge, no CTF experiment has investigated this question yet.
Also, little can be said if participants showed a preference to mark risks first,
whenever they placed their \expli~exploit marks.
Risks were marked first in \VarResRisksFirstOverall{}\% of cases
(${n=\VarAspectThreeRiskOverallSampleSize{}}$),
which is not significantly different from random.
Investigating which factors influence participants' mark preferences
may be an interesting direction for future work.

% ==============================================================================
% ------------------------------------------------------------------------------

\subsection{Replication of Prior Findings}
\label{sec:discuss-replicated-work}

Compared to existing work, namely
\cite{%
  Rowe2006:FakeHoneypotsDefensive,
  Rowe2007:DefendingCyberspaceFake,
  Nikiforakis2011:ExposingLackPrivacy,
  Sahin2020:LessonsLearnedSunDEW},
our results seem consistent.
In all cases, we can enrich previous results.
Appendix~\ref{sec:appendix:prior-work-mapping}
details how we aligned our results to previous work.

% ==============================================================================
% ------------------------------------------------------------------------------

\textbf{What rarely received marks in Honeyquest, was also rarely exploited in a real-world CTF game.}
In most CTF games~\cite{Han2017:EvaluationDeceptionBasedWeb,
  Sahin2020:LessonsLearnedSunDEW,
  Sahin2022:MeasuringDevelopersWeb},
all exploitable elements had to be discovered by CTF players,
while ours were clearly presented to participants, which makes a comparison unfair.
\citeauthor{Han2017:EvaluationDeceptionBasedWeb}~\cite{Han2017:EvaluationDeceptionBasedWeb}
primarily evaluated placement strategies rather than specific techniques,
thus, we refrain from making a direct comparison to their results.

However, we can still partially align our results with previous work.
The SunDEW experiment~\cite{Sahin2020:LessonsLearnedSunDEW}
compared the elements that participants considered deceptive in a questionnaire
with the elements with which the CTF players interacted (``considered deceptive'' ratio in parentheses):
``username''~(53\%) and ``role''~(61\%) cookies,
and deceptive GET parameters~(7\%).
The ranking remained the same in our experiment.
${\VarResRatioDcptTrapTpHttpheadersAdminCookie{}_{\pm\VarResCseDcptTrapTpHttpheadersAdminCookie{}}\%}$
thought our cookie was deceptive~%
\RowRef{Results}{DcptHttpheadersAdminCookie}
and
${\VarResRatioDcptTrapTpNetworkrequestsIdorReadSecrets{}_{\pm\VarResCseDcptTrapTpNetworkrequestsIdorReadSecrets{}}\%}$
thought the IDOR trap was deceptive~%
\RowRef{Results}{DcptNetworkrequestsIdorReadSecrets}.
In a different CTF game by \citeauthor{Sahin2022:MeasuringDevelopersWeb}%
~\cite{Sahin2022:MeasuringDevelopersWeb}
developers rarely tried to modify the ``Content-Type'' header field
(between 5\% and 13\% of players).
This is consistent with our results, where these headers were only marked
\VarResContentLengthMarkNum{}~times (\VarResContentLengthMarkRatio{}\% of all marks).

% ==============================================================================
% ------------------------------------------------------------------------------

\textbf{Our most enticing filenames were also most enticing in a real-world honeypot.}
\citeauthor{Nikiforakis2011:ExposingLackPrivacy}~\cite{Nikiforakis2011:ExposingLackPrivacy}
placed six files on public file hosting services
and recorded how often attackers attempted to downloaded them.
We showed three of them to our participants (reported download attempts in parentheses):
``card3rz\_reg\_details.html'' (22\%), % 21.81%
``customer\_list\_2010.html'' (9\%), % 9.09%
and ``SPAM\_list.pdf'' (5\%). % 5.09%
The ranking remained the same in our experiment, with
${\VarResRatioDcptFellTpFilesystemCardrz{}_{\pm\VarResCseDcptFellTpFilesystemCardrz{}}\%}$
\RowRef{Results}{DcptFilesystemCardrz},
${\VarResRatioDcptFellTpFilesystemCustomerList{}_{\pm\VarResCseDcptFellTpFilesystemCustomerList{}}\%}$
\RowRef{Results}{DcptFilesystemCustomerList},
and ${\VarResRatioDcptFellTpFilesystemSpamList{}_{\pm\VarResCseDcptFellTpFilesystemSpamList{}}\%}$
\RowRef{Results}{DcptFilesystemSpamList}
of answers exploiting these names, respectively.

% ==============================================================================
% ------------------------------------------------------------------------------

\textbf{Our participants were also good at recognizing obvious traps.}
\citeauthor{Rowe2006:FakeHoneypotsDefensive}%
~\cite{Rowe2006:FakeHoneypotsDefensive,Rowe2007:DefendingCyberspaceFake}
showed participants a genuine and a deceptive file listing.
Participants had to chose the one that ``most likely came from a real system''.
53.6\% chose the real one.
In our experiment,
${\VarResRoweDeceptiveTrap{}_{\pm\VarResRoweDeceptiveTrapCse{}}\%}$
of answers marked traps in the deceptive query%
~\RowRef{Results}{DcptFilesystemRowe}
and ${\VarResRoweNeutralTrap{}_{\pm\VarResRoweNeutralTrapCse{}}\%}$
marked traps in the neutral one.
This reinforces the hypothesis that humans show a tendency
to detect traps for these specific queries.

\citeauthor{Petrunic2015:HoneytokensActiveDefense}~\cite{Petrunic2015:HoneytokensActiveDefense}
proposed adding an ``Admin=false'' GET parameter to network requests
(\RowRefSingle{Results}{DcptNetworkrequestsAdminFalse}),
but did not report measures on attacker enticement.
Our results showed that many answers wanted to exploit that
(${\VarResRatioDcptFellTpNetworkrequestsAdminFalse{}_{\pm\VarResCseDcptFellTpNetworkrequestsAdminFalse{}}\%}$).
But, this trap might be too obvious with
${\VarResRatioDcptTrapTpNetworkrequestsAdminFalse{}_{\pm\VarResCseDcptTrapTpNetworkrequestsAdminFalse{}}\%}$
of answers also seeing a trap in it.

% ==============================================================================
% ------------------------------------------------------------------------------

\begin{table}[!t]
  \newcommand{\fileResultsAspectTwo}{\multicolumn{2}{l}{\bfseries All Techniques} & \bfseries 120 & \bfseries 9 & \bfseries 28 & \bfseries 59 & \bfseries 0.0013 & \bfseries 0.9090 & \bfseries -22\% \\
\addlinespace[0.05cm]
\scriptsize \RowRefSingle{Results}{DcptHttpheadersApiserver} & API Server & 19 & 1 & 9 & 16 & 0.0107 & 0.8415 & -32\% \\
\scriptsize \RowRefSingle{Results}{DcptHttpheadersDevtoken} & Dev. Token & 18 & 1 & 8 & 18 & 0.0195 & 0.7999 & -27\% \\
\scriptsize \RowRefSingle{Results}{DcptHttpheadersProxyReferer} & Proxy Referer & 17 & 3 & 7 & 20 & 0.1719 & 0.3497 & -15\% \\
\scriptsize \RowRefSingle{Results}{DcptNetworkrequestsPathTraversal} & Path Traversal & 43 & 2 & 3 & 4 & \st{0.5000} & \st{0.1038} & -14\% \\
\scriptsize \RowRefSingle{Results}{DcptNetworkrequestsSessidParameter} & SESSID Param. & 23 & 2 & 1 & 1 & \st{0.8750} & \st{0.0598} & +50\% \\
}
  \begin{threeparttable}
    \setlength{\tabcolsep}{4pt}
    \small
    \centering
    \caption{Resulting contingency table on defensive deception.}
    \Description{
      The overall risk reduction for all techniques combined is
      -\VarResBeforeAfterOverallRiskReduction{}\%
      with a $p$-value of \VarResBeforeAfterOverallPvalue{}.
      The risk reduction for the API server technique (\VarIdDcptHttpheadersApiserver{}) is
      -\VarResBeforeAfterHttpheadersApiserverRiskReduction{}\%
      with a $p$-value of \VarResBeforeAfterHttpheadersApiserverPvalue{}.
      The risk reduction for the developer token technique (\VarIdDcptHttpheadersDevtoken{}) is
      -\VarResBeforeAfterHttpheadersDevtokenRiskReduction{}\%
      with a $p$-value of \VarResBeforeAfterHttpheadersDevtokenPvalue{}.
      The risk reduction for the proxy referer technique (\VarIdDcptHttpheadersProxyReferer{}) is
      -\VarResBeforeAfterHttpheadersProxyRefererRiskReduction{}\%
      with a $p$-value of \VarResBeforeAfterHttpheadersProxyRefererPvalue{}.
      All other techniques did not show a significant risk reduction.
    }
    \label{tab:results-contingency-table}
    \begin{tabularx}{\columnwidth}{r @{\hspace{3pt}} X rrrr rrr}
      \toprule
      \multicolumn{2}{l}{exploited before} & \nmark   & \nmark  & \ymark   & \ymark   &                                                                                 \\
      \multicolumn{2}{l}{exploited after}  & \nmark   & \ymark  & \nmark   & \ymark   &                                                                                 \\
      \addlinespace[0.1cm]
      \multicolumn{2}{l}{\dcpti~CDT}       & $\alpha$ & $\beta$ & $\gamma$ & $\delta$ & $\blacktriangledown$~$p$-val. & $\blacktriangle$~pwr. & $\blacktriangledown$~RR \\
      \midrule
      \multicolumn{2}{l}{\bfseries All Techniques} & \bfseries 120 & \bfseries 9 & \bfseries 28 & \bfseries 59 & \bfseries 0.0013 & \bfseries 0.9090 & \bfseries -22\% \\
\addlinespace[0.05cm]
\scriptsize \RowRefSingle{Results}{DcptHttpheadersApiserver} & API Server & 19 & 1 & 9 & 16 & 0.0107 & 0.8415 & -32\% \\
\scriptsize \RowRefSingle{Results}{DcptHttpheadersDevtoken} & Dev. Token & 18 & 1 & 8 & 18 & 0.0195 & 0.7999 & -27\% \\
\scriptsize \RowRefSingle{Results}{DcptHttpheadersProxyReferer} & Proxy Referer & 17 & 3 & 7 & 20 & 0.1719 & 0.3497 & -15\% \\
\scriptsize \RowRefSingle{Results}{DcptNetworkrequestsPathTraversal} & Path Traversal & 43 & 2 & 3 & 4 & \st{0.5000} & \st{0.1038} & -14\% \\
\scriptsize \RowRefSingle{Results}{DcptNetworkrequestsSessidParameter} & SESSID Param. & 23 & 2 & 1 & 1 & \st{0.8750} & \st{0.0598} & +50\% \\

      \bottomrule
    \end{tabularx}
    \begin{tablenotes}
      \footnotesize
      \item \textbf{Legend:}
      Striked out tests violate the rule of thumb that all of the expected values are greater than~5,
      making the test less informative~\cite{Yates1934:ContingencyTablesInvolving}.
    \end{tablenotes}
  \end{threeparttable}
\end{table}

% ==============================================================================
% ------------------------------------------------------------------------------

\textbf{Our participants were also subjectively aggressive when placing marks.}
The work from \citeauthor{Ferguson-Walter2021:ExaminingEfficacyDecoybased}
provided experimental evidence that adversaries who know about the presence of deception
tend to act more aggressively than unaware adversaries%
~\cite{Ferguson-Walter2021:ExaminingEfficacyDecoybased}.
In the past, the opposite was believed to be true, i.e.,
that deception is only effective when it is well hidden and attackers are unaware of it%
~\cite{Fraunholz2018:DemystifyingDeceptionTechnology,Rowe2006:FakeHoneypotsDefensive}.
Results by \citeauthor{Sahin2020:LessonsLearnedSunDEW}~\cite{Sahin2020:LessonsLearnedSunDEW}
also suggest that informing attackers about deceptive measures deters them,
which ultimately benefits defenders.
Our experiment did not intend to provide evidence for or against this conjecture.
What can be said is that our participants placed
an average of \VarNumAvgMarksPerResponse{}~marks per answer.
But, only \VarNumTotalRiskyLinesPercent{}\% of all query lines were risky
and only \VarNumTotalDeceptiveLinesPercent{}\% were deceptive.
Some participants told us afterwards that they
``thought [that] every single query has a lot of traps in it''
and that they ``better mark too much than too little''.
This fits in with a challenge mentioned later in \secname\ref{sec:challenges},
which is that some participants also felt the urge to answer queries correctly.
In a similar survey~\cite{Sahin2022:ApproachGenerateRealistic},
participants mislabeled at least 10\% of genuine parameters as deceptive.
This seems consistent with our results, where
${\VarResCmDcptFpr{}_{\pm\VarResCmDcptFprCse{}}\%}$
of answers to neutral queries saw traps in them.

% ==============================================================================
% ------------------------------------------------------------------------------

\begin{table*}[!t]
  \newcommand{\fileResultsDeceptiveOverview}{\addlinespace[0.05cm]
\multicolumn{2}{l}{\bfseries All Deceptive Queries} &  & \RowDeceptiveDeceptive{} & 71 & 1561 & \detectionbar{37.0}{15.0}{23.0}{25.0}[0.0240][0.0177] &  & 313 &  36\% \\
\midrule
\addlinespace[0.1cm]
\multicolumn{2}{l}{\bfseries File System} &  & \RowDeceptiveFilesystemDeceptive{} & 18 & 443 & \detectionbar{42.0}{21.0}{21.0}{16.0}[0.0458][0.0377] &  & 139 &  39\% \\
\addlinespace[0.05cm]
\rhypertarget{row:Results:DcptFilesystemPrivateKey}\scriptsize\VarIdDcptFilesystemPrivateKey & \texttt{private-key.pem} &  & \RowDeceptiveFilesystemPrivateKey{} & 2 & 32 & \detectionbar{66.0}{22.0}{6.0}{6.0}[0.1564][0.1387] &  & 17 &  35\% \\
\rhypertarget{row:Results:DcptFilesystemBackup}\scriptsize\VarIdDcptFilesystemBackup & \texttt{backup.tar.gz} &  & \RowDeceptiveFilesystemBackup{} & 2 & 38 & \detectionbar{61.0}{3.0}{24.0}{12.0}[0.1484][0.0651] &  & 19 &  42\% \\
\rhypertarget{row:Results:DcptFilesystemKeys}\scriptsize\VarIdDcptFilesystemKeys & \texttt{keys.json} &  & \RowDeceptiveFilesystemKeys{} & 3 & 86 & \detectionbar{52.0}{26.0}{13.0}{9.0}[0.1033][0.0908] &  & 37 &  51\% \\
\rhypertarget{row:Results:DcptFilesystemCardrz}\scriptsize\VarIdDcptFilesystemCardrz & \makebox[0pt][l]{\texttt{card3rz\_reg\_details.html}~\cite{Nikiforakis2011:ExposingLackPrivacy}} &  & \RowDeceptiveFilesystemCardrz{} & 2 & 43 & \detectionbar{51.0}{19.0}{19.0}{11.0}[0.1432][0.1144] &  & 15 &  47\% \\
\rhypertarget{row:Results:DcptFilesystemPasswords}\scriptsize\VarIdDcptFilesystemPasswords & \texttt{passwords.txt} &  & \RowDeceptiveFilesystemPasswords{} & 2 & 41 & \detectionbar{46.0}{49.0}{0.0}{5.0}[0.1460][0.1463] &  & 12 &  25\% \\
\rhypertarget{row:Results:DcptFilesystemCustomerList}\scriptsize\VarIdDcptFilesystemCustomerList & \makebox[0pt][l]{\texttt{customer\_list\_2010.html}~\cite{Nikiforakis2011:ExposingLackPrivacy}} &  & \RowDeceptiveFilesystemCustomerList{} & 2 & 53 & \detectionbar{43.0}{23.0}{25.0}{9.0}[0.1289][0.1104] &  & 17 &  29\% \\
\rhypertarget{row:Results:DcptFilesystemConfig}\scriptsize\VarIdDcptFilesystemConfig & \texttt{config.ini} &  & \RowDeceptiveFilesystemConfig{} & 2 & 45 & \detectionbar{38.0}{2.0}{40.0}{20.0}[0.1363][0.0559] &  & 15 &  33\% \\
\rhypertarget{row:Results:DcptFilesystemSpamList}\scriptsize\VarIdDcptFilesystemSpamList & \texttt{SPAM\_list.pdf}~\cite{Nikiforakis2011:ExposingLackPrivacy} &  & \RowDeceptiveFilesystemSpamList{} & 2 & 58 & \detectionbar{17.0}{21.0}{45.0}{17.0}[0.0963][0.1026] &  & 7 &  14\% \\
\rhypertarget{row:Results:DcptFilesystemRowe}\scriptsize\VarIdDcptFilesystemRowe & Rowe et al.~\cite{Rowe2006:FakeHoneypotsDefensive,Rowe2007:DefendingCyberspaceFake} &  & \RowDeceptiveFilesystemRowe{} & 1 & 47 & \detectionbar{15.0}{19.0}{9.0}{57.0}[0.1014][0.1106] &  & 0 & \textminus \\
\midrule
\addlinespace[0.1cm]
\multicolumn{2}{l}{\bfseries .htaccess Files} &  & \RowDeceptiveHtaccessDeceptive{} & 5 & 128 & \detectionbar{54.0}{21.0}{12.0}{13.0}[0.0851][0.0701] &  & 14 &  36\% \\
\addlinespace[0.05cm]
\rhypertarget{row:Results:DcptHtaccessAdminRedirect}\scriptsize\VarIdDcptHtaccessAdminRedirect & Admin Redirect &  & \RowDeceptiveHtaccessAdminRedirect{} & 5 & 128 & \detectionbar{54.0}{21.0}{12.0}{13.0}[0.0851][0.0701] &  & 14 &  36\% \\
\midrule
\addlinespace[0.1cm]
\multicolumn{2}{l}{\bfseries HTTP Responses} &  & \RowDeceptiveHttpheadersDeceptive{} & 23 & 456 & \detectionbar{36.0}{12.0}{30.0}{22.0}[0.0439][0.0304] &  & 103 &  24\% \\
\addlinespace[0.05cm]
\rhypertarget{row:Results:DcptHttpheadersDevtoken}\scriptsize\VarIdDcptHttpheadersDevtoken & Developer Token & \riski & \RowDeceptiveHttpheadersDevtoken{} & 7 & 137 & \detectionbar{55.0}{20.0}{15.0}{10.0}[0.0821][0.0662] &  & 46 &  33\% \\
\rhypertarget{row:Results:DcptHttpheadersAdminCookie}\scriptsize\VarIdDcptHttpheadersAdminCookie & Cookie~\cite{Han2017:EvaluationDeceptionBasedWeb,Sahin2020:LessonsLearnedSunDEW,Sahin2022:MeasuringDevelopersWeb} &  & \RowDeceptiveHttpheadersAdminCookie{} & 2 & 48 & \detectionbar{40.0}{15.0}{29.0}{16.0}[0.1333][0.0996] &  & 13 &  38\% \\
\rhypertarget{row:Results:DcptHttpheadersApiserver}\scriptsize\VarIdDcptHttpheadersApiserver & API Server & \riski & \RowDeceptiveHttpheadersApiserver{} & 7 & 134 & \detectionbar{30.0}{10.0}{30.0}{30.0}[0.0766][0.0522] &  & 21 &  5\% \\
\rhypertarget{row:Results:DcptHttpheadersProxyReferer}\scriptsize\VarIdDcptHttpheadersProxyReferer & Proxy Referer & \riski & \RowDeceptiveHttpheadersProxyReferer{} & 7 & 137 & \detectionbar{22.0}{7.0}{47.0}{24.0}[0.0687][0.0426] &  & 23 &  17\% \\
\midrule
\addlinespace[0.1cm]
\multicolumn{2}{l}{\bfseries HTTP Requests} &  & \RowDeceptiveNetworkrequestsDeceptive{} & 25 & 534 & \detectionbar{30.0}{11.0}{22.0}{37.0}[0.0387][0.0264] &  & 57 &  53\% \\
\addlinespace[0.05cm]
\rhypertarget{row:Results:DcptNetworkrequestsIdorReadSecrets}\scriptsize\VarIdDcptNetworkrequestsIdorReadSecrets & IDOR Secrets~\cite{Sahin2020:LessonsLearnedSunDEW} &  & \RowDeceptiveNetworkrequestsIdorReadSecrets{} & 1 & 12 & \detectionbar{58.0}{0.0}{0.0}{42.0}[0.2436][0.1212] &  & 2 & \textminus \\
\rhypertarget{row:Results:DcptNetworkrequestsCleartextPassword}\scriptsize\VarIdDcptNetworkrequestsCleartextPassword & Clear-Text Pass.~\cite{Sahin2022:MeasuringDevelopersWeb} &  & \RowDeceptiveNetworkrequestsCleartextPassword{} & 1 & 28 & \detectionbar{54.0}{36.0}{4.0}{6.0}[0.1733][0.1673] &  & 12 &  67\% \\
\rhypertarget{row:Results:DcptNetworkrequestsSessidParameter}\scriptsize\VarIdDcptNetworkrequestsSessidParameter & SESSID Param.~\cite{Han2017:EvaluationDeceptionBasedWeb} & \riski & \RowDeceptiveNetworkrequestsSessidParameter{} & 2 & 58 & \detectionbar{40.0}{16.0}{26.0}{18.0}[0.1221][0.0927] &  & 10 &  70\% \\
\rhypertarget{row:Results:DcptNetworkrequestsPathTraversal}\scriptsize\VarIdDcptNetworkrequestsPathTraversal & Path Traversal~\cite{Sahin2022:MeasuringDevelopersWeb} & \riski & \RowDeceptiveNetworkrequestsPathTraversal{} & 5 & 122 & \detectionbar{39.0}{14.0}{20.0}{27.0}[0.0851][0.0615] &  & 10 &  40\% \\
\rhypertarget{row:Results:DcptNetworkrequestsAdminFalse}\scriptsize\VarIdDcptNetworkrequestsAdminFalse & Admin Param.~\cite{Han2017:EvaluationDeceptionBasedWeb,Petrunic2015:HoneytokensActiveDefense} &  & \RowDeceptiveNetworkrequestsAdminFalse{} & 2 & 49 & \detectionbar{37.0}{22.0}{14.0}{27.0}[0.1303][0.1143] &  & 6 &  50\% \\
\rhypertarget{row:Results:DcptNetworkrequestsUnescapedJavascript}\scriptsize\VarIdDcptNetworkrequestsUnescapedJavascript & Unescaped JS~\cite{Sahin2022:MeasuringDevelopersWeb} &  & \RowDeceptiveNetworkrequestsUnescapedJavascript{} & 2 & 21 & \detectionbar{33.0}{0.0}{10.0}{57.0}[0.1872][0.0773] &  & 0 & \textminus \\
\rhypertarget{row:Results:DcptNetworkrequestsSystemParameter}\scriptsize\VarIdDcptNetworkrequestsSystemParameter & System Param.~\cite{Han2017:EvaluationDeceptionBasedWeb,Sahin2022:MeasuringDevelopersWeb} &  & \RowDeceptiveNetworkrequestsSystemParameter{} & 1 & 28 & \detectionbar{29.0}{4.0}{25.0}{42.0}[0.1590][0.0854] &  & 4 & \textminus \\
\rhypertarget{row:Results:DcptNetworkrequestsDevEndpoint}\scriptsize\VarIdDcptNetworkrequestsDevEndpoint & Developer Endpoint &  & \RowDeceptiveNetworkrequestsDevEndpoint{} & 4 & 80 & \detectionbar{24.0}{8.0}{26.0}{42.0}[0.0919][0.0596] &  & 6 &  33\% \\
\rhypertarget{row:Results:DcptNetworkrequestsUnescapedJson}\scriptsize\VarIdDcptNetworkrequestsUnescapedJson & Unespaced JSON~\cite{Sahin2022:MeasuringDevelopersWeb} &  & \RowDeceptiveNetworkrequestsUnescapedJson{} & 2 & 22 & \detectionbar{23.0}{5.0}{32.0}{40.0}[0.1666][0.1050] &  & 3 & \textminus \\
\rhypertarget{row:Results:DcptNetworkrequestsMassAssignment}\scriptsize\VarIdDcptNetworkrequestsMassAssignment & Mass Assignment &  & \RowDeceptiveNetworkrequestsMassAssignment{} & 2 & 44 & \detectionbar{14.0}{2.0}{30.0}{54.0}[0.1015][0.0570] &  & 2 & \textminus \\
\rhypertarget{row:Results:DcptNetworkrequestsLogEndpoint}\scriptsize\VarIdDcptNetworkrequestsLogEndpoint & Log Endpoint &  & \RowDeceptiveNetworkrequestsLogEndpoint{} & 3 & 70 & \detectionbar{7.0}{3.0}{31.0}{59.0}[0.0628][0.0452] &  & 2 & \textminus \\
}
  \begin{threeparttable}
    \renewcommand{\defaultrectwidth}{3.25cm}
    \renewcommand{\defaultlabelthreshold}{15}
    \small
    \centering
    \caption{Results on the enticingness of Cyber Deception Techniques~(CDTs).}
    \label{tab:results-deceptive-overview}
    \begin{tabularx}{\textwidth}{l @{\hspace{3pt}} X @{\hspace{2pt}} c @{\hspace{2pt}} l rr l p{2pt} @{\extracolsep{\fill}} rr} % added two extra r's
      \toprule
                                     &  &                            &     & \multicolumn{3}{c}{{\bfseries Aspect A} (\secname\ref{sec:aspect-enticingness})} &                   & \multicolumn{2}{c}{{\bfseries B1} (\secname\ref{sec:aspect-deception-first})}                                              \\
      \cmidrule{5-7} \cmidrule{9-10}
      \multicolumn{2}{l}{\dcpti~CDT} &  & Representative Description & $k$ & $d$                                                                              & Mark Distribution &                                                                               & $d^{}_B$ & $\nicefrac{d^\prime_B}{d^{}_B}$ \\
      \midrule
      \addlinespace[0.05cm]
\multicolumn{2}{l}{\bfseries All Deceptive Queries} &  & \RowDeceptiveDeceptive{} & 71 & 1561 & \detectionbar{37.0}{15.0}{23.0}{25.0}[0.0240][0.0177] &  & 313 &  36\% \\
\midrule
\addlinespace[0.1cm]
\multicolumn{2}{l}{\bfseries File System} &  & \RowDeceptiveFilesystemDeceptive{} & 18 & 443 & \detectionbar{42.0}{21.0}{21.0}{16.0}[0.0458][0.0377] &  & 139 &  39\% \\
\addlinespace[0.05cm]
\rhypertarget{row:Results:DcptFilesystemPrivateKey}\scriptsize\VarIdDcptFilesystemPrivateKey & \texttt{private-key.pem} &  & \RowDeceptiveFilesystemPrivateKey{} & 2 & 32 & \detectionbar{66.0}{22.0}{6.0}{6.0}[0.1564][0.1387] &  & 17 &  35\% \\
\rhypertarget{row:Results:DcptFilesystemBackup}\scriptsize\VarIdDcptFilesystemBackup & \texttt{backup.tar.gz} &  & \RowDeceptiveFilesystemBackup{} & 2 & 38 & \detectionbar{61.0}{3.0}{24.0}{12.0}[0.1484][0.0651] &  & 19 &  42\% \\
\rhypertarget{row:Results:DcptFilesystemKeys}\scriptsize\VarIdDcptFilesystemKeys & \texttt{keys.json} &  & \RowDeceptiveFilesystemKeys{} & 3 & 86 & \detectionbar{52.0}{26.0}{13.0}{9.0}[0.1033][0.0908] &  & 37 &  51\% \\
\rhypertarget{row:Results:DcptFilesystemCardrz}\scriptsize\VarIdDcptFilesystemCardrz & \makebox[0pt][l]{\texttt{card3rz\_reg\_details.html}~\cite{Nikiforakis2011:ExposingLackPrivacy}} &  & \RowDeceptiveFilesystemCardrz{} & 2 & 43 & \detectionbar{51.0}{19.0}{19.0}{11.0}[0.1432][0.1144] &  & 15 &  47\% \\
\rhypertarget{row:Results:DcptFilesystemPasswords}\scriptsize\VarIdDcptFilesystemPasswords & \texttt{passwords.txt} &  & \RowDeceptiveFilesystemPasswords{} & 2 & 41 & \detectionbar{46.0}{49.0}{0.0}{5.0}[0.1460][0.1463] &  & 12 &  25\% \\
\rhypertarget{row:Results:DcptFilesystemCustomerList}\scriptsize\VarIdDcptFilesystemCustomerList & \makebox[0pt][l]{\texttt{customer\_list\_2010.html}~\cite{Nikiforakis2011:ExposingLackPrivacy}} &  & \RowDeceptiveFilesystemCustomerList{} & 2 & 53 & \detectionbar{43.0}{23.0}{25.0}{9.0}[0.1289][0.1104] &  & 17 &  29\% \\
\rhypertarget{row:Results:DcptFilesystemConfig}\scriptsize\VarIdDcptFilesystemConfig & \texttt{config.ini} &  & \RowDeceptiveFilesystemConfig{} & 2 & 45 & \detectionbar{38.0}{2.0}{40.0}{20.0}[0.1363][0.0559] &  & 15 &  33\% \\
\rhypertarget{row:Results:DcptFilesystemSpamList}\scriptsize\VarIdDcptFilesystemSpamList & \texttt{SPAM\_list.pdf}~\cite{Nikiforakis2011:ExposingLackPrivacy} &  & \RowDeceptiveFilesystemSpamList{} & 2 & 58 & \detectionbar{17.0}{21.0}{45.0}{17.0}[0.0963][0.1026] &  & 7 &  14\% \\
\rhypertarget{row:Results:DcptFilesystemRowe}\scriptsize\VarIdDcptFilesystemRowe & Rowe et al.~\cite{Rowe2006:FakeHoneypotsDefensive,Rowe2007:DefendingCyberspaceFake} &  & \RowDeceptiveFilesystemRowe{} & 1 & 47 & \detectionbar{15.0}{19.0}{9.0}{57.0}[0.1014][0.1106] &  & 0 & \textminus \\
\midrule
\addlinespace[0.1cm]
\multicolumn{2}{l}{\bfseries .htaccess Files} &  & \RowDeceptiveHtaccessDeceptive{} & 5 & 128 & \detectionbar{54.0}{21.0}{12.0}{13.0}[0.0851][0.0701] &  & 14 &  36\% \\
\addlinespace[0.05cm]
\rhypertarget{row:Results:DcptHtaccessAdminRedirect}\scriptsize\VarIdDcptHtaccessAdminRedirect & Admin Redirect &  & \RowDeceptiveHtaccessAdminRedirect{} & 5 & 128 & \detectionbar{54.0}{21.0}{12.0}{13.0}[0.0851][0.0701] &  & 14 &  36\% \\
\midrule
\addlinespace[0.1cm]
\multicolumn{2}{l}{\bfseries HTTP Responses} &  & \RowDeceptiveHttpheadersDeceptive{} & 23 & 456 & \detectionbar{36.0}{12.0}{30.0}{22.0}[0.0439][0.0304] &  & 103 &  24\% \\
\addlinespace[0.05cm]
\rhypertarget{row:Results:DcptHttpheadersDevtoken}\scriptsize\VarIdDcptHttpheadersDevtoken & Developer Token & \riski & \RowDeceptiveHttpheadersDevtoken{} & 7 & 137 & \detectionbar{55.0}{20.0}{15.0}{10.0}[0.0821][0.0662] &  & 46 &  33\% \\
\rhypertarget{row:Results:DcptHttpheadersAdminCookie}\scriptsize\VarIdDcptHttpheadersAdminCookie & Cookie~\cite{Han2017:EvaluationDeceptionBasedWeb,Sahin2020:LessonsLearnedSunDEW,Sahin2022:MeasuringDevelopersWeb} &  & \RowDeceptiveHttpheadersAdminCookie{} & 2 & 48 & \detectionbar{40.0}{15.0}{29.0}{16.0}[0.1333][0.0996] &  & 13 &  38\% \\
\rhypertarget{row:Results:DcptHttpheadersApiserver}\scriptsize\VarIdDcptHttpheadersApiserver & API Server & \riski & \RowDeceptiveHttpheadersApiserver{} & 7 & 134 & \detectionbar{30.0}{10.0}{30.0}{30.0}[0.0766][0.0522] &  & 21 &  5\% \\
\rhypertarget{row:Results:DcptHttpheadersProxyReferer}\scriptsize\VarIdDcptHttpheadersProxyReferer & Proxy Referer & \riski & \RowDeceptiveHttpheadersProxyReferer{} & 7 & 137 & \detectionbar{22.0}{7.0}{47.0}{24.0}[0.0687][0.0426] &  & 23 &  17\% \\
\midrule
\addlinespace[0.1cm]
\multicolumn{2}{l}{\bfseries HTTP Requests} &  & \RowDeceptiveNetworkrequestsDeceptive{} & 25 & 534 & \detectionbar{30.0}{11.0}{22.0}{37.0}[0.0387][0.0264] &  & 57 &  53\% \\
\addlinespace[0.05cm]
\rhypertarget{row:Results:DcptNetworkrequestsIdorReadSecrets}\scriptsize\VarIdDcptNetworkrequestsIdorReadSecrets & IDOR Secrets~\cite{Sahin2020:LessonsLearnedSunDEW} &  & \RowDeceptiveNetworkrequestsIdorReadSecrets{} & 1 & 12 & \detectionbar{58.0}{0.0}{0.0}{42.0}[0.2436][0.1212] &  & 2 & \textminus \\
\rhypertarget{row:Results:DcptNetworkrequestsCleartextPassword}\scriptsize\VarIdDcptNetworkrequestsCleartextPassword & Clear-Text Pass.~\cite{Sahin2022:MeasuringDevelopersWeb} &  & \RowDeceptiveNetworkrequestsCleartextPassword{} & 1 & 28 & \detectionbar{54.0}{36.0}{4.0}{6.0}[0.1733][0.1673] &  & 12 &  67\% \\
\rhypertarget{row:Results:DcptNetworkrequestsSessidParameter}\scriptsize\VarIdDcptNetworkrequestsSessidParameter & SESSID Param.~\cite{Han2017:EvaluationDeceptionBasedWeb} & \riski & \RowDeceptiveNetworkrequestsSessidParameter{} & 2 & 58 & \detectionbar{40.0}{16.0}{26.0}{18.0}[0.1221][0.0927] &  & 10 &  70\% \\
\rhypertarget{row:Results:DcptNetworkrequestsPathTraversal}\scriptsize\VarIdDcptNetworkrequestsPathTraversal & Path Traversal~\cite{Sahin2022:MeasuringDevelopersWeb} & \riski & \RowDeceptiveNetworkrequestsPathTraversal{} & 5 & 122 & \detectionbar{39.0}{14.0}{20.0}{27.0}[0.0851][0.0615] &  & 10 &  40\% \\
\rhypertarget{row:Results:DcptNetworkrequestsAdminFalse}\scriptsize\VarIdDcptNetworkrequestsAdminFalse & Admin Param.~\cite{Han2017:EvaluationDeceptionBasedWeb,Petrunic2015:HoneytokensActiveDefense} &  & \RowDeceptiveNetworkrequestsAdminFalse{} & 2 & 49 & \detectionbar{37.0}{22.0}{14.0}{27.0}[0.1303][0.1143] &  & 6 &  50\% \\
\rhypertarget{row:Results:DcptNetworkrequestsUnescapedJavascript}\scriptsize\VarIdDcptNetworkrequestsUnescapedJavascript & Unescaped JS~\cite{Sahin2022:MeasuringDevelopersWeb} &  & \RowDeceptiveNetworkrequestsUnescapedJavascript{} & 2 & 21 & \detectionbar{33.0}{0.0}{10.0}{57.0}[0.1872][0.0773] &  & 0 & \textminus \\
\rhypertarget{row:Results:DcptNetworkrequestsSystemParameter}\scriptsize\VarIdDcptNetworkrequestsSystemParameter & System Param.~\cite{Han2017:EvaluationDeceptionBasedWeb,Sahin2022:MeasuringDevelopersWeb} &  & \RowDeceptiveNetworkrequestsSystemParameter{} & 1 & 28 & \detectionbar{29.0}{4.0}{25.0}{42.0}[0.1590][0.0854] &  & 4 & \textminus \\
\rhypertarget{row:Results:DcptNetworkrequestsDevEndpoint}\scriptsize\VarIdDcptNetworkrequestsDevEndpoint & Developer Endpoint &  & \RowDeceptiveNetworkrequestsDevEndpoint{} & 4 & 80 & \detectionbar{24.0}{8.0}{26.0}{42.0}[0.0919][0.0596] &  & 6 &  33\% \\
\rhypertarget{row:Results:DcptNetworkrequestsUnescapedJson}\scriptsize\VarIdDcptNetworkrequestsUnescapedJson & Unespaced JSON~\cite{Sahin2022:MeasuringDevelopersWeb} &  & \RowDeceptiveNetworkrequestsUnescapedJson{} & 2 & 22 & \detectionbar{23.0}{5.0}{32.0}{40.0}[0.1666][0.1050] &  & 3 & \textminus \\
\rhypertarget{row:Results:DcptNetworkrequestsMassAssignment}\scriptsize\VarIdDcptNetworkrequestsMassAssignment & Mass Assignment &  & \RowDeceptiveNetworkrequestsMassAssignment{} & 2 & 44 & \detectionbar{14.0}{2.0}{30.0}{54.0}[0.1015][0.0570] &  & 2 & \textminus \\
\rhypertarget{row:Results:DcptNetworkrequestsLogEndpoint}\scriptsize\VarIdDcptNetworkrequestsLogEndpoint & Log Endpoint &  & \RowDeceptiveNetworkrequestsLogEndpoint{} & 3 & 70 & \detectionbar{7.0}{3.0}{31.0}{59.0}[0.0628][0.0452] &  & 2 & \textminus \\

      \bottomrule
    \end{tabularx}
    \begin{tablenotes}
      \footnotesize
      \item \textbf{Aspect A:} (\secname\ref{sec:aspect-enticingness})
      The best CDTs are the ones that participants often fall for
      (higher \dhackbox \enspace $\nicefrac{d_{Ex}}{d}$ ratio is better)
      and rarely avoid (lower \dtrapbox \enspace $\nicefrac{d_{Tr}}{d}$ ratio is better).
      Bars show the \%~of answers~$d$ that~
      \dhackbox \enspace $\nicefrac{d_{Ex}}{d}$
      match \expli~exploit marks (=~``fallen for trap''),~
      \dtrapbox \enspace $\nicefrac{d_{Tr}}{d}$
      match \trapi~trap marks (=~``trap detected''),~
      \dotherbox \enspace $\nicefrac{d_\triangle}{d}$ placed marks elsewhere,~
      \dnonebox \enspace $\nicefrac{d_\varnothing}{d}$ had no marks at all.
      $k$~are the number of queries in the dataset that had the associated CTD
      injected. $d$~are the number of answers (not marks) that these queries received.
      Bars without percentage numbers account for less than \defaultlabelthreshold\%.
      The tiny bars denote the 95\%~CI of that mean.
      % $d_{Ex}$~are the num. of ans. where an \expli~exploit mark intersected the deceptive lines.
      % $d_{Tr}$~are the num. of ans. where a \trapi~trap mark intersected the deceptive lines.
      % $d_\triangle$~are the num. of ans. where marks do not intersect deceptive lines.
      % $d_\varnothing$~are the num. of ans. where nothing was marked.
      %
      \item \textbf{Aspect B1:} (\secname\ref{sec:aspect-deception-first})
      In the $d_B$~answers where participants placed multiple \expli~exploit marks, $d^\prime_B$~is
      the number of times that a deceptive line was marked before a non-deceptive one.
      If there are at least ${d^{}_B \geq \VarAspectTwoMinimumSampleSize{}}$~such cases,
      we perform a Binomial test on the null hypothesis that this ratio is random, i.e.,
      ${\nicefrac{d^\prime_B}{d^{}_B}=\nicefrac{1}{2}}$, with the one-sided alternative
      that deceptive lines were marked first more often than random.
      No test was significant (all $p \geq \VarAspectRankedFirstMinimumPvalue{}$)
      with $\alpha = 0.05$.
      \item \textbf{Aspect B2:} (\secname\ref{sec:aspect-divert-attention})
      The \riski~lightning bolt symbol indicates that this CDT was also evaluated
      on Aspect B2, but in Table~\ref{tab:results-contingency-table}.
      \vspace{0.1cm}
      \item[a]
      Insecure direct object references (IDOR) are vulnerabilities where
      potentially sensitive content can be retrieved by guessing
      (predictable) identifiers (IDs).
      \item[b]
      Applications may automatically bind HTTP parameter names to fields
      in the underlying object, potentially enabling attackers to manipulate restricted fields.
    \end{tablenotes}
  \end{threeparttable}
\end{table*}

% ==============================================================================
% ------------------------------------------------------------------------------

\textbf{Honeyquest is a useful tool for repeating cyber deception experiments
  in a sample-efficient and cost-effective way,
  while still yielding qualitatively similar results to CTF experiments.}
Compared to real-world deployments of CDTs,
Honeyquest is more cost-effective since it does not require
the time and effort to set up and maintain CTF experiments or honeypots.
Experiments are also more sample-efficient than others
because participants can answer a query within seconds
% (median: \VarResAnswerTimeMedian{}~seconds)
instead of spending time on coding actual attacks.
The decision-making processes of attackers and humans in general
also include fast -- but, not necessarily correct -- heuristics%
~\cite{Gutzwiller2019:AreCyberAttackers,Kahneman2011:ThinkingFastSlow},
which probably explains why participants
can respond quickly to most queries.
%
% We encourage researchers to replicate our results and conduct similar experiments.
% We encourage researchers to replicate our results and conduct similar experiments,
% by following our approach or by using our
% \anon[open-source tool]{\href{\repositoryurl}{open-source tool}} Honeyquest.
% Investigating how slow or fast attackers tend to be enticed by CDTs

% ==============================================================================
% ------------------------------------------------------------------------------

\subsection{Challenges and Limitations}
\label{sec:challenges}

While our results are insightful and seem consistent with prior research,
we also faced several design challenges % and limitations of our approach
that we would like to share with future researchers.

% ==============================================================================
% ------------------------------------------------------------------------------

\begin{table*}[!p]
  \ifdefined\UseLongTablesInAppendix%
    \newcommand{\fileResultsAspectFourHack}{\dcpti & \footnotesize \RowRefSingle{Results}{DcptHttpheadersDevtoken} & \texttt{P} & \raggedright\arraybackslash\texttt{X{-{\allowbreak}}DevToken: {\allowbreak}eyJhbGciOiJIUzI1NiIsInR5cCI6IkpXVCJ9.{\allowbreak}eyJ0b2tlbiI6IjMwZDk4NGI4Iiwicm9sZSI6ImNsdXN0ZXIifQ.{\allowbreak}J2c1XH2RsXDjaWJhCHSWR4bBNxEm7l2Q7WxrlG2vph4} & 76 & 27 \\
\hdashline
 &  & \texttt{F} & \raggedright\arraybackslash\texttt{drwx{-{\allowbreak}}{-{\allowbreak}}{-{\allowbreak}}{-{\allowbreak}}{-{\allowbreak}}{-{\allowbreak}}~~2 {\allowbreak}lpc~~lpc~~4.{\allowbreak}0K {\allowbreak}Nov~~2 {\allowbreak}09:21 {\allowbreak}.{\allowbreak}ssh} & 58 & 1 \\
\hdashline
\riski & \footnotesize \RowRefSingle{Results}{RiskHttpheadersOutdatedApache} & \texttt{P} & \raggedright\arraybackslash\texttt{Server: {\allowbreak}Apache/{\allowbreak}1.{\allowbreak}0.{\allowbreak}3 {\allowbreak}(Debian)} & 55 & 1 \\
\hdashline
 &  & \texttt{F} & \raggedright\arraybackslash\texttt{drwx{-{\allowbreak}}{-{\allowbreak}}{-{\allowbreak}}{-{\allowbreak}}{-{\allowbreak}}{-{\allowbreak}}~~4 {\allowbreak}leonard {\allowbreak}leonard {\allowbreak}4.{\allowbreak}0K {\allowbreak}Jan {\allowbreak}14~~2022 {\allowbreak}.{\allowbreak}gnupg} & 48 & 1 \\
\hdashline
 &  & \texttt{F} & \raggedright\arraybackslash\texttt{drwx{-{\allowbreak}}{-{\allowbreak}}{-{\allowbreak}}{-{\allowbreak}}{-{\allowbreak}}{-{\allowbreak}}.{\allowbreak}~~2 {\allowbreak}root {\allowbreak}root {\allowbreak}4.{\allowbreak}0K {\allowbreak}Sep {\allowbreak}27 {\allowbreak}19:30 {\allowbreak}.{\allowbreak}ssh} & 46 & 2 \\
\hdashline
\riski & \footnotesize \RowRefSingle{Results}{RiskHttpheadersOutdatedPhp} & \texttt{P} & \raggedright\arraybackslash\texttt{X{-{\allowbreak}}Powered{-{\allowbreak}}By: {\allowbreak}PHP/{\allowbreak}5.{\allowbreak}1.{\allowbreak}6} & 41 & 9 \\
\hdashline
 &  & \texttt{P} & \raggedright\arraybackslash\texttt{Set{-{\allowbreak}}Cookie: {\allowbreak}PHPSESSID=hLAGcA9qClz36kOr71sSgw; {\allowbreak}path=/{\allowbreak}} & 41 & 10 \\
\hdashline
 &  & \texttt{F} & \raggedright\arraybackslash\texttt{drwx{-{\allowbreak}}{-{\allowbreak}}{-{\allowbreak}}{-{\allowbreak}}{-{\allowbreak}}{-{\allowbreak}}~~2 {\allowbreak}furi0zuc {\allowbreak}furi0zuc {\allowbreak}4.{\allowbreak}0K {\allowbreak}Jun {\allowbreak}21~~2019 {\allowbreak}.{\allowbreak}ssh} & 40 & 1 \\
\hdashline
 &  & \texttt{F} & \raggedright\arraybackslash\texttt{{-{\allowbreak}}rw{-{\allowbreak}}{-{\allowbreak}}{-{\allowbreak}}{-{\allowbreak}}{-{\allowbreak}}{-{\allowbreak}}{-{\allowbreak}}~~1 {\allowbreak}lpc~~lpc~~ {\allowbreak}40K {\allowbreak}Dec {\allowbreak}29 {\allowbreak}15:58 {\allowbreak}.{\allowbreak}bash\_history} & 39 & 0 \\
\hdashline
\dcpti & \footnotesize \RowRefSingle{Results}{DcptHttpheadersApiserver} & \texttt{P} & \raggedright\arraybackslash\texttt{X{-{\allowbreak}}Kube{-{\allowbreak}}ApiServer: {\allowbreak}/{\allowbreak}hko/{\allowbreak}api} & 36 & 13 \\
\hdashline
 &  & \texttt{P} & \raggedright\arraybackslash\texttt{X{-{\allowbreak}}AspNet{-{\allowbreak}}Version: {\allowbreak}4.{\allowbreak}0.{\allowbreak}30319} & 34 & 4 \\
\hdashline
 &  & \texttt{P} & \raggedright\arraybackslash\texttt{Server: {\allowbreak}Microsoft{-{\allowbreak}}IIS/{\allowbreak}7.{\allowbreak}5} & 33 & 0 \\
\hdashline
 &  & \texttt{P} & \raggedright\arraybackslash\texttt{Server: {\allowbreak}Apache/{\allowbreak}2.{\allowbreak}2.{\allowbreak}17 {\allowbreak}(Unix) {\allowbreak}mod\_ssl/{\allowbreak}2.{\allowbreak}2.{\allowbreak}17 {\allowbreak}OpenSSL/{\allowbreak}0.{\allowbreak}9.{\allowbreak}8e{-{\allowbreak}}fips{-{\allowbreak}}rhel5 {\allowbreak}mod\_auth\_passthrough/{\allowbreak}2.{\allowbreak}1 {\allowbreak}mod\_bwlimited/{\allowbreak}1.{\allowbreak}4 {\allowbreak}FrontPage/{\allowbreak}5.{\allowbreak}0.{\allowbreak}2.{\allowbreak}2635} & 31 & 6 \\
\hdashline
 &  & \texttt{F} & \raggedright\arraybackslash\texttt{{-{\allowbreak}}rw{-{\allowbreak}}{-{\allowbreak}}{-{\allowbreak}}{-{\allowbreak}}{-{\allowbreak}}{-{\allowbreak}}{-{\allowbreak}}.{\allowbreak}~~1 {\allowbreak}root {\allowbreak}root~~21K {\allowbreak}Oct {\allowbreak}25 {\allowbreak}19:26 {\allowbreak}.{\allowbreak}bash\_history} & 30 & 0 \\
\hdashline
 &  & \texttt{S} & \raggedright\arraybackslash\texttt{6.{\allowbreak}588 {\allowbreak}GET {\allowbreak}https:/{\allowbreak}/{\allowbreak}juice{-{\allowbreak}}shop.{\allowbreak}herokuapp.{\allowbreak}com/{\allowbreak}rest/{\allowbreak}admin/{\allowbreak}application{-{\allowbreak}}configuration {\allowbreak}200 {\allowbreak}OK {\allowbreak}(7.{\allowbreak}2 {\allowbreak}kB)} & 29 & 7 \\
\hdashline
 &  & \texttt{P} & \raggedright\arraybackslash\texttt{Transfer{-{\allowbreak}}Encoding: {\allowbreak}chunked} & 28 & 0 \\
\hdashline
\dcpti & \footnotesize \RowRefSingle{Results}{DcptFilesystemKeys} & \texttt{F} & \raggedright\arraybackslash\texttt{{-{\allowbreak}}rw{-{\allowbreak}}r{-{\allowbreak}}{-{\allowbreak}}r{-{\allowbreak}}{-{\allowbreak}}~~1 {\allowbreak}elsa {\allowbreak}elsa~~12K {\allowbreak}Feb~~6~~2022 {\allowbreak}keys.{\allowbreak}json} & 28 & 13 \\
\hdashline
 &  & \texttt{P} & \raggedright\arraybackslash\texttt{Server: {\allowbreak}nginx/{\allowbreak}1.{\allowbreak}2.{\allowbreak}4} & 27 & 0 \\
\hdashline
\dcpti & \footnotesize \RowRefSingle{Results}{DcptHtaccessAdminRedirect} & \texttt{H} & \raggedright\arraybackslash\texttt{Redirect {\allowbreak}301 {\allowbreak}"/{\allowbreak}admin" {\allowbreak}"/{\allowbreak}plugins/{\allowbreak}kul/{\allowbreak}pages/{\allowbreak}admin/{\allowbreak}index.{\allowbreak}php?role=view"} & 26 & 11 \\
\hdashline
 &  & \texttt{P} & \raggedright\arraybackslash\texttt{Server: {\allowbreak}Microsoft{-{\allowbreak}}IIS/{\allowbreak}6.{\allowbreak}0} & 25 & 0 \\
\hdashline
 &  & \texttt{F} & \raggedright\arraybackslash\texttt{drwx{-{\allowbreak}}{-{\allowbreak}}{-{\allowbreak}}{-{\allowbreak}}{-{\allowbreak}}{-{\allowbreak}}~~2 {\allowbreak}cathy {\allowbreak}cathy {\allowbreak}4.{\allowbreak}0K {\allowbreak}Dec~~7 {\allowbreak}14:04 {\allowbreak}.{\allowbreak}ssh} & 25 & 0 \\
\hdashline
 &  & \texttt{F} & \raggedright\arraybackslash\texttt{{-{\allowbreak}}rw{-{\allowbreak}}{-{\allowbreak}}{-{\allowbreak}}{-{\allowbreak}}{-{\allowbreak}}{-{\allowbreak}}{-{\allowbreak}}~~1 {\allowbreak}IconThor {\allowbreak}IconThor {\allowbreak}231K {\allowbreak}Dec {\allowbreak}30 {\allowbreak}08:36 {\allowbreak}.{\allowbreak}zsh\_history} & 25 & 1 \\
\hdashline
 &  & \texttt{F} & \raggedright\arraybackslash\texttt{{-{\allowbreak}}rwxr{-{\allowbreak}}{-{\allowbreak}}r{-{\allowbreak}}{-{\allowbreak}}~~1 {\allowbreak}lpc~~lpc~~ {\allowbreak}919 {\allowbreak}Dec {\allowbreak}28 {\allowbreak}14:31 {\allowbreak}query.{\allowbreak}sh} & 25 & 1 \\
\hdashline
 &  & \texttt{P} & \raggedright\arraybackslash\texttt{X{-{\allowbreak}}Powered{-{\allowbreak}}By: {\allowbreak}PHP/{\allowbreak}5.{\allowbreak}3.{\allowbreak}18} & 25 & 2 \\
\hdashline
 &  & \texttt{P} & \raggedright\arraybackslash\texttt{X{-{\allowbreak}}Powered{-{\allowbreak}}By: {\allowbreak}PHP/{\allowbreak}5.{\allowbreak}2.{\allowbreak}16} & 23 & 5 \\
\hdashline
 &  & \texttt{F} & \raggedright\arraybackslash\texttt{drwx{-{\allowbreak}}{-{\allowbreak}}{-{\allowbreak}}{-{\allowbreak}}{-{\allowbreak}}{-{\allowbreak}}~~2 {\allowbreak}donald {\allowbreak}donald {\allowbreak}4.{\allowbreak}0K {\allowbreak}May {\allowbreak}16~~2022 {\allowbreak}.{\allowbreak}ssh} & 22 & 0 \\
\hdashline
 &  & \texttt{P} & \raggedright\arraybackslash\texttt{Set{-{\allowbreak}}Cookie: {\allowbreak}ASP.{\allowbreak}NET\_SessionId=fyi3sylqfunbwtdy03s4fdqv; {\allowbreak}path=/{\allowbreak}; {\allowbreak}HttpOnly} & 22 & 2 \\
\hdashline
 &  & \texttt{P} & \raggedright\arraybackslash\texttt{X{-{\allowbreak}}Powered{-{\allowbreak}}By: {\allowbreak}PHP/{\allowbreak}5.{\allowbreak}4.{\allowbreak}5} & 22 & 2 \\
\hdashline
 &  & \texttt{F} & \raggedright\arraybackslash\texttt{{-{\allowbreak}}rw{-{\allowbreak}}r{-{\allowbreak}}{-{\allowbreak}}r{-{\allowbreak}}{-{\allowbreak}}~~1 {\allowbreak}lpc~~lpc~~8.{\allowbreak}1K {\allowbreak}Nov {\allowbreak}24 {\allowbreak}10:18 {\allowbreak}data.{\allowbreak}csv} & 22 & 3 \\
\hdashline
 &  & \texttt{S} & \raggedright\arraybackslash\texttt{9.{\allowbreak}912 {\allowbreak}GET {\allowbreak}https:/{\allowbreak}/{\allowbreak}juice{-{\allowbreak}}shop.{\allowbreak}herokuapp.{\allowbreak}com/{\allowbreak}rest/{\allowbreak}admin/{\allowbreak}application{-{\allowbreak}}configuration {\allowbreak}200 {\allowbreak}OK {\allowbreak}(7.{\allowbreak}2 {\allowbreak}kB)} & 22 & 3 \\
\hdashline
 &  & \texttt{H} & \raggedright\arraybackslash\texttt{RewriteCond {\allowbreak}\%\{THE\_REQUEST\} {\allowbreak}\^{}{[}A{-{\allowbreak}}Z{]}\{3,\}\textbackslash{}s({[}\^{}.{\allowbreak}{]}+)\textbackslash{}.{\allowbreak}php {\allowbreak}{[}NC{]}} & 22 & 3 \\
\hdashline
 &  & \texttt{F} & \raggedright\arraybackslash\texttt{{-{\allowbreak}}rw{-{\allowbreak}}{-{\allowbreak}}{-{\allowbreak}}{-{\allowbreak}}{-{\allowbreak}}{-{\allowbreak}}{-{\allowbreak}}~~1 {\allowbreak}elsa {\allowbreak}elsa {\allowbreak}231K {\allowbreak}Dec {\allowbreak}30 {\allowbreak}08:36 {\allowbreak}.{\allowbreak}zsh\_history} & 21 & 0 \\
\hdashline
 &  & \texttt{P} & \raggedright\arraybackslash\texttt{Server: {\allowbreak}nginx/{\allowbreak}1.{\allowbreak}6.{\allowbreak}0} & 21 & 0 \\
\hdashline
 &  & \texttt{S} & \raggedright\arraybackslash\texttt{2.{\allowbreak}460 {\allowbreak}POST {\allowbreak}https:/{\allowbreak}/{\allowbreak}api.{\allowbreak}site.{\allowbreak}com/{\allowbreak}\_private/{\allowbreak}browser/{\allowbreak}stats} & 21 & 3 \\
\hdashline
 &  & \texttt{F} & \raggedright\arraybackslash\texttt{{-{\allowbreak}}rw{-{\allowbreak}}r{-{\allowbreak}}{-{\allowbreak}}r{-{\allowbreak}}{-{\allowbreak}}~~1 {\allowbreak}IconThor {\allowbreak}IconThor~~ {\allowbreak}88 {\allowbreak}Jul {\allowbreak}11~~2021 {\allowbreak}test.{\allowbreak}gpg} & 21 & 8 \\
\hdashline
 &  & \texttt{F} & \raggedright\arraybackslash\texttt{{-{\allowbreak}}rw{-{\allowbreak}}{-{\allowbreak}}{-{\allowbreak}}{-{\allowbreak}}{-{\allowbreak}}{-{\allowbreak}}{-{\allowbreak}}~~1 {\allowbreak}furi0zuc {\allowbreak}furi0zuc {\allowbreak}8.{\allowbreak}9K {\allowbreak}Jul~~5 {\allowbreak}21:47 {\allowbreak}.{\allowbreak}bash\_history} & 20 & 0 \\
\hdashline
 &  & \texttt{F} & \raggedright\arraybackslash\texttt{{-{\allowbreak}}rw{-{\allowbreak}}r{-{\allowbreak}}{-{\allowbreak}}r{-{\allowbreak}}{-{\allowbreak}}~~1 {\allowbreak}leonard {\allowbreak}leonard~~ {\allowbreak}64 {\allowbreak}Jun {\allowbreak}16~~2019 {\allowbreak}.{\allowbreak}gitconfig} & 20 & 1 \\
\hdashline
 &  & \texttt{P} & \raggedright\arraybackslash\texttt{Set{-{\allowbreak}}Cookie: {\allowbreak}bbsessionhash=f3cdd7987b326584c6ee6696f3033087; {\allowbreak}path=/{\allowbreak}; {\allowbreak}HttpOnly} & 20 & 5 \\
\hdashline
 &  & \texttt{S} & \raggedright\arraybackslash\texttt{3.{\allowbreak}127 {\allowbreak}POST {\allowbreak}https:/{\allowbreak}/{\allowbreak}api.{\allowbreak}site.{\allowbreak}com/{\allowbreak}\_private/{\allowbreak}browser/{\allowbreak}stats} & 19 & 2 \\
\hdashline
 &  & \texttt{S} & \raggedright\arraybackslash\texttt{9.{\allowbreak}022 {\allowbreak}GET {\allowbreak}https:/{\allowbreak}/{\allowbreak}juice{-{\allowbreak}}shop.{\allowbreak}herokuapp.{\allowbreak}com/{\allowbreak}rest/{\allowbreak}products/{\allowbreak}search?q= {\allowbreak}200 {\allowbreak}OK {\allowbreak}(3.{\allowbreak}5 {\allowbreak}kB)} & 19 & 3 \\
\hdashline
 &  & \texttt{P} & \raggedright\arraybackslash\texttt{Server: {\allowbreak}nginx/{\allowbreak}1.{\allowbreak}2.{\allowbreak}2} & 18 & 0 \\
\hdashline
 &  & \texttt{F} & \raggedright\arraybackslash\texttt{drwx{-{\allowbreak}}{-{\allowbreak}}{-{\allowbreak}}{-{\allowbreak}}{-{\allowbreak}}{-{\allowbreak}}~~6 {\allowbreak}leonard {\allowbreak}leonard {\allowbreak}4.{\allowbreak}0K {\allowbreak}Sep {\allowbreak}25 {\allowbreak}17:40 {\allowbreak}.{\allowbreak}config} & 18 & 1 \\
\hdashline
 &  & \texttt{F} & \raggedright\arraybackslash\texttt{{-{\allowbreak}}rw{-{\allowbreak}}r{-{\allowbreak}}{-{\allowbreak}}r{-{\allowbreak}}{-{\allowbreak}}~~1 {\allowbreak}goofy {\allowbreak}goofy {\allowbreak}816M {\allowbreak}Nov {\allowbreak}18~~2020 {\allowbreak}nginx{-{\allowbreak}}proxy{-{\allowbreak}}logs.{\allowbreak}txt} & 18 & 2 \\
\hdashline
\riski & \footnotesize \RowRefSingle{Results}{RiskHttpheadersProxyAuthLeak} & \texttt{P} & \raggedright\arraybackslash\texttt{Proxy{-{\allowbreak}}Authorization: {\allowbreak}Basic {\allowbreak}dXNlcj1wcm94eTtwYXNzPTdMOUFDQ0FKS3BLU3A0} & 18 & 3 \\
\hdashline
\riski & \footnotesize \RowRefSingle{Results}{RiskFilesystemBackup} & \texttt{F} & \raggedright\arraybackslash\texttt{{-{\allowbreak}}rw{-{\allowbreak}}r{-{\allowbreak}}{-{\allowbreak}}r{-{\allowbreak}}{-{\allowbreak}}.{\allowbreak}~~1 {\allowbreak}root {\allowbreak}root~~13M {\allowbreak}Jun {\allowbreak}16~~2022 {\allowbreak}mariadb.{\allowbreak}tar.{\allowbreak}gz} & 18 & 5 \\
\hdashline
 &  & \texttt{S} & \raggedright\arraybackslash\texttt{8.{\allowbreak}281 {\allowbreak}GET {\allowbreak}https:/{\allowbreak}/{\allowbreak}juice{-{\allowbreak}}shop.{\allowbreak}herokuapp.{\allowbreak}com/{\allowbreak}rest/{\allowbreak}user/{\allowbreak}whoami {\allowbreak}200 {\allowbreak}OK {\allowbreak}(0.{\allowbreak}1 {\allowbreak}kB)} & 18 & 5 \\
\hdashline
 &  & \texttt{P} & \raggedright\arraybackslash\texttt{X{-{\allowbreak}}Pingback: {\allowbreak}http:/{\allowbreak}/{\allowbreak}www.{\allowbreak}loksatta.{\allowbreak}com/{\allowbreak}xmlrpc.{\allowbreak}php} & 18 & 6 \\
\hdashline
 &  & \texttt{P} & \raggedright\arraybackslash\texttt{Server: {\allowbreak}nginx/{\allowbreak}1.{\allowbreak}4.{\allowbreak}7} & 17 & 0 \\
\hdashline
\riski & \footnotesize \RowRefSingle{Results}{RiskFilesystemBackup} & \texttt{F} & \raggedright\arraybackslash\texttt{{-{\allowbreak}}rw{-{\allowbreak}}r{-{\allowbreak}}{-{\allowbreak}}r{-{\allowbreak}}{-{\allowbreak}}.{\allowbreak}~~1 {\allowbreak}Plut0 {\allowbreak}Plut0~~13M {\allowbreak}Jun {\allowbreak}16~~2022 {\allowbreak}backup.{\allowbreak}tar.{\allowbreak}gz} & 17 & 1 \\
\hdashline
 &  & \texttt{S} & \raggedright\arraybackslash\texttt{9.{\allowbreak}743 {\allowbreak}POST {\allowbreak}https:/{\allowbreak}/{\allowbreak}juice{-{\allowbreak}}shop.{\allowbreak}herokuapp.{\allowbreak}com/{\allowbreak}api/{\allowbreak}SecurityAnswers/{\allowbreak} {\allowbreak}201 {\allowbreak}Created {\allowbreak}(0.{\allowbreak}2 {\allowbreak}kB)} & 17 & 1 \\
\hdashline
 &  & \texttt{F} & \raggedright\arraybackslash\texttt{drwx{-{\allowbreak}}{-{\allowbreak}}{-{\allowbreak}}{-{\allowbreak}}{-{\allowbreak}}{-{\allowbreak}}~~2 {\allowbreak}salphard {\allowbreak}salphard {\allowbreak}4.{\allowbreak}0K {\allowbreak}Jun {\allowbreak}21~~2019 {\allowbreak}.{\allowbreak}ssh} & 17 & 1 \\
\hdashline
 &  & \texttt{F} & \raggedright\arraybackslash\texttt{{-{\allowbreak}}rw{-{\allowbreak}}r{-{\allowbreak}}{-{\allowbreak}}r{-{\allowbreak}}{-{\allowbreak}}~~1 {\allowbreak}lpc~~lpc~~7.{\allowbreak}1K {\allowbreak}Nov {\allowbreak}30 {\allowbreak}16:34 {\allowbreak}data.{\allowbreak}json} & 17 & 5 \\
\hdashline
\dcpti & \footnotesize \RowRefSingle{Results}{DcptNetworkrequestsPathTraversal} & \texttt{S} & \raggedright\arraybackslash\texttt{0.{\allowbreak}415 {\allowbreak}GET {\allowbreak}https:/{\allowbreak}/{\allowbreak}assets.{\allowbreak}siteusercontent.{\allowbreak}com/{\allowbreak}u/{\allowbreak}418982Sccs?file=.{\allowbreak}.{\allowbreak}/{\allowbreak}dist/{\allowbreak}img/{\allowbreak}4FqAq.{\allowbreak}svg\&w=53px {\allowbreak}200 {\allowbreak}OK {\allowbreak}(1.{\allowbreak}1 {\allowbreak}kB)} & 17 & 5 \\
\hdashline
\dcpti & \footnotesize \RowRefSingle{Results}{DcptNetworkrequestsPathTraversal} & \texttt{S} & \raggedright\arraybackslash\texttt{2.{\allowbreak}140 {\allowbreak}GET {\allowbreak}https:/{\allowbreak}/{\allowbreak}en.{\allowbreak}site.{\allowbreak}org/{\allowbreak}cdn/{\allowbreak}commons/{\allowbreak}df?file=.{\allowbreak}.{\allowbreak}/{\allowbreak}dist/{\allowbreak}img/{\allowbreak}4FqAq.{\allowbreak}svg\&w=53px\&h=53px {\allowbreak}200 {\allowbreak}OK {\allowbreak}(5.{\allowbreak}3 {\allowbreak}kB)} & 17 & 6 \\
\hdashline
\dcpti & \footnotesize \RowRefSingle{Results}{DcptFilesystemBackup} & \texttt{F} & \raggedright\arraybackslash\texttt{{-{\allowbreak}}rw{-{\allowbreak}}r{-{\allowbreak}}{-{\allowbreak}}r{-{\allowbreak}}{-{\allowbreak}}~~1 {\allowbreak}IconThor {\allowbreak}IconThor {\allowbreak}129M {\allowbreak}Feb {\allowbreak}21 {\allowbreak}12:40 {\allowbreak}backup.{\allowbreak}tar.{\allowbreak}gz} & 16 & 1 \\
\hdashline
 &  & \texttt{F} & \raggedright\arraybackslash\texttt{drwx{-{\allowbreak}}{-{\allowbreak}}{-{\allowbreak}}{-{\allowbreak}}{-{\allowbreak}}{-{\allowbreak}}~~2 {\allowbreak}root~~~~ {\allowbreak}root~~~~ {\allowbreak}4.{\allowbreak}0K {\allowbreak}Jul {\allowbreak}10~~2021 {\allowbreak}.{\allowbreak}gnupg} & 16 & 3 \\
\hdashline
 &  & \texttt{F} & \raggedright\arraybackslash\texttt{drwxr{-{\allowbreak}}xr{-{\allowbreak}}x~~5 {\allowbreak}goofy {\allowbreak}goofy {\allowbreak}4.{\allowbreak}0K {\allowbreak}Mar {\allowbreak}21~~2021 {\allowbreak}terraform{-{\allowbreak}}saas} & 16 & 3 \\
\hdashline
 &  & \texttt{F} & \raggedright\arraybackslash\texttt{lrwxrwxrwx~~1 {\allowbreak}leonard {\allowbreak}leonard~~ {\allowbreak}19 {\allowbreak}Jan {\allowbreak}23~~2018 {\allowbreak}html {\allowbreak}{-{\allowbreak}}> {\allowbreak}/{\allowbreak}srv/{\allowbreak}nginx{-{\allowbreak}}www/{\allowbreak}html} & 16 & 5 \\
\hdashline
 &  & \texttt{S} & \raggedright\arraybackslash\texttt{0.{\allowbreak}000 {\allowbreak}GET {\allowbreak}https:/{\allowbreak}/{\allowbreak}juice{-{\allowbreak}}shop.{\allowbreak}herokuapp.{\allowbreak}com/{\allowbreak}rest/{\allowbreak}user/{\allowbreak}whoami {\allowbreak}200 {\allowbreak}OK {\allowbreak}(11 {\allowbreak}bytes)} & 16 & 8 \\
\hdashline
 &  & \texttt{F} & \raggedright\arraybackslash\texttt{{-{\allowbreak}}rw{-{\allowbreak}}{-{\allowbreak}}{-{\allowbreak}}{-{\allowbreak}}{-{\allowbreak}}{-{\allowbreak}}{-{\allowbreak}}~~1 {\allowbreak}donald {\allowbreak}donald~~57K {\allowbreak}Dec {\allowbreak}29 {\allowbreak}17:18 {\allowbreak}.{\allowbreak}bash\_history} & 15 & 0 \\
}%
  \else\newcommand{\fileResultsAspectFourHack}{\dcpti & \footnotesize \RowRefSingle{Results}{DcptHttpheadersDevtoken} & \texttt{P} & \raggedright\arraybackslash\texttt{X{-{\allowbreak}}DevToken: {\allowbreak}eyJhbGciOiJIUzI1NiIsInR5cCI6IkpXVCJ9.{\allowbreak}eyJ0b2tlbiI6IjMwZDk4NGI4Iiwicm9sZSI6ImNsdXN0ZXIifQ.{\allowbreak}J2c1XH2RsXDjaWJhCHSWR4bBNxEm7l2Q7WxrlG2vph4} & 76 & 27 \\
\hdashline
 &  & \texttt{F} & \raggedright\arraybackslash\texttt{drwx{-{\allowbreak}}{-{\allowbreak}}{-{\allowbreak}}{-{\allowbreak}}{-{\allowbreak}}{-{\allowbreak}}~~2 {\allowbreak}lpc~~lpc~~4.{\allowbreak}0K {\allowbreak}Nov~~2 {\allowbreak}09:21 {\allowbreak}.{\allowbreak}ssh} & 58 & 1 \\
\hdashline
\riski & \footnotesize \RowRefSingle{Results}{RiskHttpheadersOutdatedApache} & \texttt{P} & \raggedright\arraybackslash\texttt{Server: {\allowbreak}Apache/{\allowbreak}1.{\allowbreak}0.{\allowbreak}3 {\allowbreak}(Debian)} & 55 & 1 \\
\hdashline
 &  & \texttt{F} & \raggedright\arraybackslash\texttt{drwx{-{\allowbreak}}{-{\allowbreak}}{-{\allowbreak}}{-{\allowbreak}}{-{\allowbreak}}{-{\allowbreak}}~~4 {\allowbreak}leonard {\allowbreak}leonard {\allowbreak}4.{\allowbreak}0K {\allowbreak}Jan {\allowbreak}14~~2022 {\allowbreak}.{\allowbreak}gnupg} & 48 & 1 \\
\hdashline
 &  & \texttt{F} & \raggedright\arraybackslash\texttt{drwx{-{\allowbreak}}{-{\allowbreak}}{-{\allowbreak}}{-{\allowbreak}}{-{\allowbreak}}{-{\allowbreak}}.{\allowbreak}~~2 {\allowbreak}root {\allowbreak}root {\allowbreak}4.{\allowbreak}0K {\allowbreak}Sep {\allowbreak}27 {\allowbreak}19:30 {\allowbreak}.{\allowbreak}ssh} & 46 & 2 \\
\hdashline
\riski & \footnotesize \RowRefSingle{Results}{RiskHttpheadersOutdatedPhp} & \texttt{P} & \raggedright\arraybackslash\texttt{X{-{\allowbreak}}Powered{-{\allowbreak}}By: {\allowbreak}PHP/{\allowbreak}5.{\allowbreak}1.{\allowbreak}6} & 41 & 9 \\
\hdashline
 &  & \texttt{P} & \raggedright\arraybackslash\texttt{Set{-{\allowbreak}}Cookie: {\allowbreak}PHPSESSID=hLAGcA9qClz36kOr71sSgw; {\allowbreak}path=/{\allowbreak}} & 41 & 10 \\
\hdashline
 &  & \texttt{F} & \raggedright\arraybackslash\texttt{drwx{-{\allowbreak}}{-{\allowbreak}}{-{\allowbreak}}{-{\allowbreak}}{-{\allowbreak}}{-{\allowbreak}}~~2 {\allowbreak}furi0zuc {\allowbreak}furi0zuc {\allowbreak}4.{\allowbreak}0K {\allowbreak}Jun {\allowbreak}21~~2019 {\allowbreak}.{\allowbreak}ssh} & 40 & 1 \\
\hdashline
 &  & \texttt{F} & \raggedright\arraybackslash\texttt{{-{\allowbreak}}rw{-{\allowbreak}}{-{\allowbreak}}{-{\allowbreak}}{-{\allowbreak}}{-{\allowbreak}}{-{\allowbreak}}{-{\allowbreak}}~~1 {\allowbreak}lpc~~lpc~~ {\allowbreak}40K {\allowbreak}Dec {\allowbreak}29 {\allowbreak}15:58 {\allowbreak}.{\allowbreak}bash\_history} & 39 & 0 \\
\hdashline
\dcpti & \footnotesize \RowRefSingle{Results}{DcptHttpheadersApiserver} & \texttt{P} & \raggedright\arraybackslash\texttt{X{-{\allowbreak}}Kube{-{\allowbreak}}ApiServer: {\allowbreak}/{\allowbreak}hko/{\allowbreak}api} & 36 & 13 \\
\hdashline
 &  & \texttt{P} & \raggedright\arraybackslash\texttt{X{-{\allowbreak}}AspNet{-{\allowbreak}}Version: {\allowbreak}4.{\allowbreak}0.{\allowbreak}30319} & 34 & 4 \\
\hdashline
 &  & \texttt{P} & \raggedright\arraybackslash\texttt{Server: {\allowbreak}Microsoft{-{\allowbreak}}IIS/{\allowbreak}7.{\allowbreak}5} & 33 & 0 \\
\hdashline
 &  & \texttt{P} & \raggedright\arraybackslash\texttt{Server: {\allowbreak}Apache/{\allowbreak}2.{\allowbreak}2.{\allowbreak}17 {\allowbreak}(Unix) {\allowbreak}mod\_ssl/{\allowbreak}2.{\allowbreak}2.{\allowbreak}17 {\allowbreak}OpenSSL/{\allowbreak}0.{\allowbreak}9.{\allowbreak}8e{-{\allowbreak}}fips{-{\allowbreak}}rhel5 {\allowbreak}mod\_auth\_passthrough/{\allowbreak}2.{\allowbreak}1 {\allowbreak}mod\_bwlimited/{\allowbreak}1.{\allowbreak}4 {\allowbreak}FrontPage/{\allowbreak}5.{\allowbreak}0.{\allowbreak}2.{\allowbreak}2635} & 31 & 6 \\
\hdashline
 &  & \texttt{F} & \raggedright\arraybackslash\texttt{{-{\allowbreak}}rw{-{\allowbreak}}{-{\allowbreak}}{-{\allowbreak}}{-{\allowbreak}}{-{\allowbreak}}{-{\allowbreak}}{-{\allowbreak}}.{\allowbreak}~~1 {\allowbreak}root {\allowbreak}root~~21K {\allowbreak}Oct {\allowbreak}25 {\allowbreak}19:26 {\allowbreak}.{\allowbreak}bash\_history} & 30 & 0 \\
\hdashline
 &  & \texttt{S} & \raggedright\arraybackslash\texttt{6.{\allowbreak}588 {\allowbreak}GET {\allowbreak}https:/{\allowbreak}/{\allowbreak}juice{-{\allowbreak}}shop.{\allowbreak}herokuapp.{\allowbreak}com/{\allowbreak}rest/{\allowbreak}admin/{\allowbreak}application{-{\allowbreak}}configuration {\allowbreak}200 {\allowbreak}OK {\allowbreak}(7.{\allowbreak}2 {\allowbreak}kB)} & 29 & 7 \\
\hdashline
 &  & \texttt{P} & \raggedright\arraybackslash\texttt{Transfer{-{\allowbreak}}Encoding: {\allowbreak}chunked} & 28 & 0 \\
\hdashline
\dcpti & \footnotesize \RowRefSingle{Results}{DcptFilesystemKeys} & \texttt{F} & \raggedright\arraybackslash\texttt{{-{\allowbreak}}rw{-{\allowbreak}}r{-{\allowbreak}}{-{\allowbreak}}r{-{\allowbreak}}{-{\allowbreak}}~~1 {\allowbreak}elsa {\allowbreak}elsa~~12K {\allowbreak}Feb~~6~~2022 {\allowbreak}keys.{\allowbreak}json} & 28 & 13 \\
\hdashline
 &  & \texttt{P} & \raggedright\arraybackslash\texttt{Server: {\allowbreak}nginx/{\allowbreak}1.{\allowbreak}2.{\allowbreak}4} & 27 & 0 \\
\hdashline
\dcpti & \footnotesize \RowRefSingle{Results}{DcptHtaccessAdminRedirect} & \texttt{H} & \raggedright\arraybackslash\texttt{Redirect {\allowbreak}301 {\allowbreak}"/{\allowbreak}admin" {\allowbreak}"/{\allowbreak}plugins/{\allowbreak}kul/{\allowbreak}pages/{\allowbreak}admin/{\allowbreak}index.{\allowbreak}php?role=view"} & 26 & 11 \\
\hdashline
 &  & \texttt{P} & \raggedright\arraybackslash\texttt{Server: {\allowbreak}Microsoft{-{\allowbreak}}IIS/{\allowbreak}6.{\allowbreak}0} & 25 & 0 \\
\hdashline
 &  & \texttt{F} & \raggedright\arraybackslash\texttt{drwx{-{\allowbreak}}{-{\allowbreak}}{-{\allowbreak}}{-{\allowbreak}}{-{\allowbreak}}{-{\allowbreak}}~~2 {\allowbreak}cathy {\allowbreak}cathy {\allowbreak}4.{\allowbreak}0K {\allowbreak}Dec~~7 {\allowbreak}14:04 {\allowbreak}.{\allowbreak}ssh} & 25 & 0 \\
\hdashline
 &  & \texttt{F} & \raggedright\arraybackslash\texttt{{-{\allowbreak}}rw{-{\allowbreak}}{-{\allowbreak}}{-{\allowbreak}}{-{\allowbreak}}{-{\allowbreak}}{-{\allowbreak}}{-{\allowbreak}}~~1 {\allowbreak}IconThor {\allowbreak}IconThor {\allowbreak}231K {\allowbreak}Dec {\allowbreak}30 {\allowbreak}08:36 {\allowbreak}.{\allowbreak}zsh\_history} & 25 & 1 \\
\hdashline
 &  & \texttt{F} & \raggedright\arraybackslash\texttt{{-{\allowbreak}}rwxr{-{\allowbreak}}{-{\allowbreak}}r{-{\allowbreak}}{-{\allowbreak}}~~1 {\allowbreak}lpc~~lpc~~ {\allowbreak}919 {\allowbreak}Dec {\allowbreak}28 {\allowbreak}14:31 {\allowbreak}query.{\allowbreak}sh} & 25 & 1 \\
\hdashline
 &  & \texttt{P} & \raggedright\arraybackslash\texttt{X{-{\allowbreak}}Powered{-{\allowbreak}}By: {\allowbreak}PHP/{\allowbreak}5.{\allowbreak}3.{\allowbreak}18} & 25 & 2 \\
\hdashline
 &  & \texttt{P} & \raggedright\arraybackslash\texttt{X{-{\allowbreak}}Powered{-{\allowbreak}}By: {\allowbreak}PHP/{\allowbreak}5.{\allowbreak}2.{\allowbreak}16} & 23 & 5 \\
\hdashline
 &  & \texttt{F} & \raggedright\arraybackslash\texttt{drwx{-{\allowbreak}}{-{\allowbreak}}{-{\allowbreak}}{-{\allowbreak}}{-{\allowbreak}}{-{\allowbreak}}~~2 {\allowbreak}donald {\allowbreak}donald {\allowbreak}4.{\allowbreak}0K {\allowbreak}May {\allowbreak}16~~2022 {\allowbreak}.{\allowbreak}ssh} & 22 & 0 \\
\hdashline
 &  & \texttt{P} & \raggedright\arraybackslash\texttt{Set{-{\allowbreak}}Cookie: {\allowbreak}ASP.{\allowbreak}NET\_SessionId=fyi3sylqfunbwtdy03s4fdqv; {\allowbreak}path=/{\allowbreak}; {\allowbreak}HttpOnly} & 22 & 2 \\
\hdashline
 &  & \texttt{P} & \raggedright\arraybackslash\texttt{X{-{\allowbreak}}Powered{-{\allowbreak}}By: {\allowbreak}PHP/{\allowbreak}5.{\allowbreak}4.{\allowbreak}5} & 22 & 2 \\
\hdashline
 &  & \texttt{F} & \raggedright\arraybackslash\texttt{{-{\allowbreak}}rw{-{\allowbreak}}r{-{\allowbreak}}{-{\allowbreak}}r{-{\allowbreak}}{-{\allowbreak}}~~1 {\allowbreak}lpc~~lpc~~8.{\allowbreak}1K {\allowbreak}Nov {\allowbreak}24 {\allowbreak}10:18 {\allowbreak}data.{\allowbreak}csv} & 22 & 3 \\
\hdashline
 &  & \texttt{S} & \raggedright\arraybackslash\texttt{9.{\allowbreak}912 {\allowbreak}GET {\allowbreak}https:/{\allowbreak}/{\allowbreak}juice{-{\allowbreak}}shop.{\allowbreak}herokuapp.{\allowbreak}com/{\allowbreak}rest/{\allowbreak}admin/{\allowbreak}application{-{\allowbreak}}configuration {\allowbreak}200 {\allowbreak}OK {\allowbreak}(7.{\allowbreak}2 {\allowbreak}kB)} & 22 & 3 \\
\hdashline
 &  & \texttt{H} & \raggedright\arraybackslash\texttt{RewriteCond {\allowbreak}\%\{THE\_REQUEST\} {\allowbreak}\^{}{[}A{-{\allowbreak}}Z{]}\{3,\}\textbackslash{}s({[}\^{}.{\allowbreak}{]}+)\textbackslash{}.{\allowbreak}php {\allowbreak}{[}NC{]}} & 22 & 3 \\
\hdashline
 &  & \texttt{F} & \raggedright\arraybackslash\texttt{{-{\allowbreak}}rw{-{\allowbreak}}{-{\allowbreak}}{-{\allowbreak}}{-{\allowbreak}}{-{\allowbreak}}{-{\allowbreak}}{-{\allowbreak}}~~1 {\allowbreak}elsa {\allowbreak}elsa {\allowbreak}231K {\allowbreak}Dec {\allowbreak}30 {\allowbreak}08:36 {\allowbreak}.{\allowbreak}zsh\_history} & 21 & 0 \\
\hdashline
 &  & \texttt{P} & \raggedright\arraybackslash\texttt{Server: {\allowbreak}nginx/{\allowbreak}1.{\allowbreak}6.{\allowbreak}0} & 21 & 0 \\
\hdashline
 &  & \texttt{S} & \raggedright\arraybackslash\texttt{2.{\allowbreak}460 {\allowbreak}POST {\allowbreak}https:/{\allowbreak}/{\allowbreak}api.{\allowbreak}site.{\allowbreak}com/{\allowbreak}\_private/{\allowbreak}browser/{\allowbreak}stats} & 21 & 3 \\
\hdashline
 &  & \texttt{F} & \raggedright\arraybackslash\texttt{{-{\allowbreak}}rw{-{\allowbreak}}r{-{\allowbreak}}{-{\allowbreak}}r{-{\allowbreak}}{-{\allowbreak}}~~1 {\allowbreak}IconThor {\allowbreak}IconThor~~ {\allowbreak}88 {\allowbreak}Jul {\allowbreak}11~~2021 {\allowbreak}test.{\allowbreak}gpg} & 21 & 8 \\
\hdashline
 &  & \texttt{F} & \raggedright\arraybackslash\texttt{{-{\allowbreak}}rw{-{\allowbreak}}{-{\allowbreak}}{-{\allowbreak}}{-{\allowbreak}}{-{\allowbreak}}{-{\allowbreak}}{-{\allowbreak}}~~1 {\allowbreak}furi0zuc {\allowbreak}furi0zuc {\allowbreak}8.{\allowbreak}9K {\allowbreak}Jul~~5 {\allowbreak}21:47 {\allowbreak}.{\allowbreak}bash\_history} & 20 & 0 \\
\hdashline
 &  & \texttt{F} & \raggedright\arraybackslash\texttt{{-{\allowbreak}}rw{-{\allowbreak}}r{-{\allowbreak}}{-{\allowbreak}}r{-{\allowbreak}}{-{\allowbreak}}~~1 {\allowbreak}leonard {\allowbreak}leonard~~ {\allowbreak}64 {\allowbreak}Jun {\allowbreak}16~~2019 {\allowbreak}.{\allowbreak}gitconfig} & 20 & 1 \\
\hdashline
 &  & \texttt{P} & \raggedright\arraybackslash\texttt{Set{-{\allowbreak}}Cookie: {\allowbreak}bbsessionhash=f3cdd7987b326584c6ee6696f3033087; {\allowbreak}path=/{\allowbreak}; {\allowbreak}HttpOnly} & 20 & 5 \\
\hdashline
 &  & \texttt{S} & \raggedright\arraybackslash\texttt{3.{\allowbreak}127 {\allowbreak}POST {\allowbreak}https:/{\allowbreak}/{\allowbreak}api.{\allowbreak}site.{\allowbreak}com/{\allowbreak}\_private/{\allowbreak}browser/{\allowbreak}stats} & 19 & 2 \\
\hdashline
 &  & \texttt{S} & \raggedright\arraybackslash\texttt{9.{\allowbreak}022 {\allowbreak}GET {\allowbreak}https:/{\allowbreak}/{\allowbreak}juice{-{\allowbreak}}shop.{\allowbreak}herokuapp.{\allowbreak}com/{\allowbreak}rest/{\allowbreak}products/{\allowbreak}search?q= {\allowbreak}200 {\allowbreak}OK {\allowbreak}(3.{\allowbreak}5 {\allowbreak}kB)} & 19 & 3 \\
}\fi
  \begin{threeparttable}
    \setlength\dashlinedash{0.2pt}
    \setlength\dashlinegap{1.5pt}
    \setlength\arrayrulewidth{0.3pt}
    \centering
    \footnotesize
    \caption[]{
      Top~\ifdefined\UseLongTablesInAppendix%
        \VarTableLimitMrkHackLong{}\else\VarTableLimitMrkHackShort{}\fi~individual query lines,
      ranked by the total number of \expli~exploit marks that the line received.
    }
    \label{tab:results-mark-hack}
    \begin{tabularx}{\textwidth}{c @{\hspace{3pt}} lc X rr}
      \toprule
                             &      &            &          & \expli~exploit & \trapi~trap \\
      \multicolumn{2}{c}{ID} & Type & Query Line & $n_{Ex}$ & $n_{Tr}$                     \\
      \midrule
      \dcpti & \footnotesize \RowRefSingle{Results}{DcptHttpheadersDevtoken} & \texttt{P} & \raggedright\arraybackslash\texttt{X{-{\allowbreak}}DevToken: {\allowbreak}eyJhbGciOiJIUzI1NiIsInR5cCI6IkpXVCJ9.{\allowbreak}eyJ0b2tlbiI6IjMwZDk4NGI4Iiwicm9sZSI6ImNsdXN0ZXIifQ.{\allowbreak}J2c1XH2RsXDjaWJhCHSWR4bBNxEm7l2Q7WxrlG2vph4} & 76 & 27 \\
\hdashline
 &  & \texttt{F} & \raggedright\arraybackslash\texttt{drwx{-{\allowbreak}}{-{\allowbreak}}{-{\allowbreak}}{-{\allowbreak}}{-{\allowbreak}}{-{\allowbreak}}~~2 {\allowbreak}lpc~~lpc~~4.{\allowbreak}0K {\allowbreak}Nov~~2 {\allowbreak}09:21 {\allowbreak}.{\allowbreak}ssh} & 58 & 1 \\
\hdashline
\riski & \footnotesize \RowRefSingle{Results}{RiskHttpheadersOutdatedApache} & \texttt{P} & \raggedright\arraybackslash\texttt{Server: {\allowbreak}Apache/{\allowbreak}1.{\allowbreak}0.{\allowbreak}3 {\allowbreak}(Debian)} & 55 & 1 \\
\hdashline
 &  & \texttt{F} & \raggedright\arraybackslash\texttt{drwx{-{\allowbreak}}{-{\allowbreak}}{-{\allowbreak}}{-{\allowbreak}}{-{\allowbreak}}{-{\allowbreak}}~~4 {\allowbreak}leonard {\allowbreak}leonard {\allowbreak}4.{\allowbreak}0K {\allowbreak}Jan {\allowbreak}14~~2022 {\allowbreak}.{\allowbreak}gnupg} & 48 & 1 \\
\hdashline
 &  & \texttt{F} & \raggedright\arraybackslash\texttt{drwx{-{\allowbreak}}{-{\allowbreak}}{-{\allowbreak}}{-{\allowbreak}}{-{\allowbreak}}{-{\allowbreak}}.{\allowbreak}~~2 {\allowbreak}root {\allowbreak}root {\allowbreak}4.{\allowbreak}0K {\allowbreak}Sep {\allowbreak}27 {\allowbreak}19:30 {\allowbreak}.{\allowbreak}ssh} & 46 & 2 \\
\hdashline
\riski & \footnotesize \RowRefSingle{Results}{RiskHttpheadersOutdatedPhp} & \texttt{P} & \raggedright\arraybackslash\texttt{X{-{\allowbreak}}Powered{-{\allowbreak}}By: {\allowbreak}PHP/{\allowbreak}5.{\allowbreak}1.{\allowbreak}6} & 41 & 9 \\
\hdashline
 &  & \texttt{P} & \raggedright\arraybackslash\texttt{Set{-{\allowbreak}}Cookie: {\allowbreak}PHPSESSID=hLAGcA9qClz36kOr71sSgw; {\allowbreak}path=/{\allowbreak}} & 41 & 10 \\
\hdashline
 &  & \texttt{F} & \raggedright\arraybackslash\texttt{drwx{-{\allowbreak}}{-{\allowbreak}}{-{\allowbreak}}{-{\allowbreak}}{-{\allowbreak}}{-{\allowbreak}}~~2 {\allowbreak}furi0zuc {\allowbreak}furi0zuc {\allowbreak}4.{\allowbreak}0K {\allowbreak}Jun {\allowbreak}21~~2019 {\allowbreak}.{\allowbreak}ssh} & 40 & 1 \\
\hdashline
 &  & \texttt{F} & \raggedright\arraybackslash\texttt{{-{\allowbreak}}rw{-{\allowbreak}}{-{\allowbreak}}{-{\allowbreak}}{-{\allowbreak}}{-{\allowbreak}}{-{\allowbreak}}{-{\allowbreak}}~~1 {\allowbreak}lpc~~lpc~~ {\allowbreak}40K {\allowbreak}Dec {\allowbreak}29 {\allowbreak}15:58 {\allowbreak}.{\allowbreak}bash\_history} & 39 & 0 \\
\hdashline
\dcpti & \footnotesize \RowRefSingle{Results}{DcptHttpheadersApiserver} & \texttt{P} & \raggedright\arraybackslash\texttt{X{-{\allowbreak}}Kube{-{\allowbreak}}ApiServer: {\allowbreak}/{\allowbreak}hko/{\allowbreak}api} & 36 & 13 \\
\hdashline
 &  & \texttt{P} & \raggedright\arraybackslash\texttt{X{-{\allowbreak}}AspNet{-{\allowbreak}}Version: {\allowbreak}4.{\allowbreak}0.{\allowbreak}30319} & 34 & 4 \\
\hdashline
 &  & \texttt{P} & \raggedright\arraybackslash\texttt{Server: {\allowbreak}Microsoft{-{\allowbreak}}IIS/{\allowbreak}7.{\allowbreak}5} & 33 & 0 \\
\hdashline
 &  & \texttt{P} & \raggedright\arraybackslash\texttt{Server: {\allowbreak}Apache/{\allowbreak}2.{\allowbreak}2.{\allowbreak}17 {\allowbreak}(Unix) {\allowbreak}mod\_ssl/{\allowbreak}2.{\allowbreak}2.{\allowbreak}17 {\allowbreak}OpenSSL/{\allowbreak}0.{\allowbreak}9.{\allowbreak}8e{-{\allowbreak}}fips{-{\allowbreak}}rhel5 {\allowbreak}mod\_auth\_passthrough/{\allowbreak}2.{\allowbreak}1 {\allowbreak}mod\_bwlimited/{\allowbreak}1.{\allowbreak}4 {\allowbreak}FrontPage/{\allowbreak}5.{\allowbreak}0.{\allowbreak}2.{\allowbreak}2635} & 31 & 6 \\
\hdashline
 &  & \texttt{F} & \raggedright\arraybackslash\texttt{{-{\allowbreak}}rw{-{\allowbreak}}{-{\allowbreak}}{-{\allowbreak}}{-{\allowbreak}}{-{\allowbreak}}{-{\allowbreak}}{-{\allowbreak}}.{\allowbreak}~~1 {\allowbreak}root {\allowbreak}root~~21K {\allowbreak}Oct {\allowbreak}25 {\allowbreak}19:26 {\allowbreak}.{\allowbreak}bash\_history} & 30 & 0 \\
\hdashline
 &  & \texttt{S} & \raggedright\arraybackslash\texttt{6.{\allowbreak}588 {\allowbreak}GET {\allowbreak}https:/{\allowbreak}/{\allowbreak}juice{-{\allowbreak}}shop.{\allowbreak}herokuapp.{\allowbreak}com/{\allowbreak}rest/{\allowbreak}admin/{\allowbreak}application{-{\allowbreak}}configuration {\allowbreak}200 {\allowbreak}OK {\allowbreak}(7.{\allowbreak}2 {\allowbreak}kB)} & 29 & 7 \\
\hdashline
 &  & \texttt{P} & \raggedright\arraybackslash\texttt{Transfer{-{\allowbreak}}Encoding: {\allowbreak}chunked} & 28 & 0 \\
\hdashline
\dcpti & \footnotesize \RowRefSingle{Results}{DcptFilesystemKeys} & \texttt{F} & \raggedright\arraybackslash\texttt{{-{\allowbreak}}rw{-{\allowbreak}}r{-{\allowbreak}}{-{\allowbreak}}r{-{\allowbreak}}{-{\allowbreak}}~~1 {\allowbreak}elsa {\allowbreak}elsa~~12K {\allowbreak}Feb~~6~~2022 {\allowbreak}keys.{\allowbreak}json} & 28 & 13 \\
\hdashline
 &  & \texttt{P} & \raggedright\arraybackslash\texttt{Server: {\allowbreak}nginx/{\allowbreak}1.{\allowbreak}2.{\allowbreak}4} & 27 & 0 \\
\hdashline
\dcpti & \footnotesize \RowRefSingle{Results}{DcptHtaccessAdminRedirect} & \texttt{H} & \raggedright\arraybackslash\texttt{Redirect {\allowbreak}301 {\allowbreak}"/{\allowbreak}admin" {\allowbreak}"/{\allowbreak}plugins/{\allowbreak}kul/{\allowbreak}pages/{\allowbreak}admin/{\allowbreak}index.{\allowbreak}php?role=view"} & 26 & 11 \\
\hdashline
 &  & \texttt{P} & \raggedright\arraybackslash\texttt{Server: {\allowbreak}Microsoft{-{\allowbreak}}IIS/{\allowbreak}6.{\allowbreak}0} & 25 & 0 \\
\hdashline
 &  & \texttt{F} & \raggedright\arraybackslash\texttt{drwx{-{\allowbreak}}{-{\allowbreak}}{-{\allowbreak}}{-{\allowbreak}}{-{\allowbreak}}{-{\allowbreak}}~~2 {\allowbreak}cathy {\allowbreak}cathy {\allowbreak}4.{\allowbreak}0K {\allowbreak}Dec~~7 {\allowbreak}14:04 {\allowbreak}.{\allowbreak}ssh} & 25 & 0 \\
\hdashline
 &  & \texttt{F} & \raggedright\arraybackslash\texttt{{-{\allowbreak}}rw{-{\allowbreak}}{-{\allowbreak}}{-{\allowbreak}}{-{\allowbreak}}{-{\allowbreak}}{-{\allowbreak}}{-{\allowbreak}}~~1 {\allowbreak}IconThor {\allowbreak}IconThor {\allowbreak}231K {\allowbreak}Dec {\allowbreak}30 {\allowbreak}08:36 {\allowbreak}.{\allowbreak}zsh\_history} & 25 & 1 \\
\hdashline
 &  & \texttt{F} & \raggedright\arraybackslash\texttt{{-{\allowbreak}}rwxr{-{\allowbreak}}{-{\allowbreak}}r{-{\allowbreak}}{-{\allowbreak}}~~1 {\allowbreak}lpc~~lpc~~ {\allowbreak}919 {\allowbreak}Dec {\allowbreak}28 {\allowbreak}14:31 {\allowbreak}query.{\allowbreak}sh} & 25 & 1 \\
\hdashline
 &  & \texttt{P} & \raggedright\arraybackslash\texttt{X{-{\allowbreak}}Powered{-{\allowbreak}}By: {\allowbreak}PHP/{\allowbreak}5.{\allowbreak}3.{\allowbreak}18} & 25 & 2 \\
\hdashline
 &  & \texttt{P} & \raggedright\arraybackslash\texttt{X{-{\allowbreak}}Powered{-{\allowbreak}}By: {\allowbreak}PHP/{\allowbreak}5.{\allowbreak}2.{\allowbreak}16} & 23 & 5 \\
\hdashline
 &  & \texttt{F} & \raggedright\arraybackslash\texttt{drwx{-{\allowbreak}}{-{\allowbreak}}{-{\allowbreak}}{-{\allowbreak}}{-{\allowbreak}}{-{\allowbreak}}~~2 {\allowbreak}donald {\allowbreak}donald {\allowbreak}4.{\allowbreak}0K {\allowbreak}May {\allowbreak}16~~2022 {\allowbreak}.{\allowbreak}ssh} & 22 & 0 \\
\hdashline
 &  & \texttt{P} & \raggedright\arraybackslash\texttt{Set{-{\allowbreak}}Cookie: {\allowbreak}ASP.{\allowbreak}NET\_SessionId=fyi3sylqfunbwtdy03s4fdqv; {\allowbreak}path=/{\allowbreak}; {\allowbreak}HttpOnly} & 22 & 2 \\
\hdashline
 &  & \texttt{P} & \raggedright\arraybackslash\texttt{X{-{\allowbreak}}Powered{-{\allowbreak}}By: {\allowbreak}PHP/{\allowbreak}5.{\allowbreak}4.{\allowbreak}5} & 22 & 2 \\
\hdashline
 &  & \texttt{F} & \raggedright\arraybackslash\texttt{{-{\allowbreak}}rw{-{\allowbreak}}r{-{\allowbreak}}{-{\allowbreak}}r{-{\allowbreak}}{-{\allowbreak}}~~1 {\allowbreak}lpc~~lpc~~8.{\allowbreak}1K {\allowbreak}Nov {\allowbreak}24 {\allowbreak}10:18 {\allowbreak}data.{\allowbreak}csv} & 22 & 3 \\
\hdashline
 &  & \texttt{S} & \raggedright\arraybackslash\texttt{9.{\allowbreak}912 {\allowbreak}GET {\allowbreak}https:/{\allowbreak}/{\allowbreak}juice{-{\allowbreak}}shop.{\allowbreak}herokuapp.{\allowbreak}com/{\allowbreak}rest/{\allowbreak}admin/{\allowbreak}application{-{\allowbreak}}configuration {\allowbreak}200 {\allowbreak}OK {\allowbreak}(7.{\allowbreak}2 {\allowbreak}kB)} & 22 & 3 \\
\hdashline
 &  & \texttt{H} & \raggedright\arraybackslash\texttt{RewriteCond {\allowbreak}\%\{THE\_REQUEST\} {\allowbreak}\^{}{[}A{-{\allowbreak}}Z{]}\{3,\}\textbackslash{}s({[}\^{}.{\allowbreak}{]}+)\textbackslash{}.{\allowbreak}php {\allowbreak}{[}NC{]}} & 22 & 3 \\
\hdashline
 &  & \texttt{F} & \raggedright\arraybackslash\texttt{{-{\allowbreak}}rw{-{\allowbreak}}{-{\allowbreak}}{-{\allowbreak}}{-{\allowbreak}}{-{\allowbreak}}{-{\allowbreak}}{-{\allowbreak}}~~1 {\allowbreak}elsa {\allowbreak}elsa {\allowbreak}231K {\allowbreak}Dec {\allowbreak}30 {\allowbreak}08:36 {\allowbreak}.{\allowbreak}zsh\_history} & 21 & 0 \\
\hdashline
 &  & \texttt{P} & \raggedright\arraybackslash\texttt{Server: {\allowbreak}nginx/{\allowbreak}1.{\allowbreak}6.{\allowbreak}0} & 21 & 0 \\
\hdashline
 &  & \texttt{S} & \raggedright\arraybackslash\texttt{2.{\allowbreak}460 {\allowbreak}POST {\allowbreak}https:/{\allowbreak}/{\allowbreak}api.{\allowbreak}site.{\allowbreak}com/{\allowbreak}\_private/{\allowbreak}browser/{\allowbreak}stats} & 21 & 3 \\
\hdashline
 &  & \texttt{F} & \raggedright\arraybackslash\texttt{{-{\allowbreak}}rw{-{\allowbreak}}r{-{\allowbreak}}{-{\allowbreak}}r{-{\allowbreak}}{-{\allowbreak}}~~1 {\allowbreak}IconThor {\allowbreak}IconThor~~ {\allowbreak}88 {\allowbreak}Jul {\allowbreak}11~~2021 {\allowbreak}test.{\allowbreak}gpg} & 21 & 8 \\
\hdashline
 &  & \texttt{F} & \raggedright\arraybackslash\texttt{{-{\allowbreak}}rw{-{\allowbreak}}{-{\allowbreak}}{-{\allowbreak}}{-{\allowbreak}}{-{\allowbreak}}{-{\allowbreak}}{-{\allowbreak}}~~1 {\allowbreak}furi0zuc {\allowbreak}furi0zuc {\allowbreak}8.{\allowbreak}9K {\allowbreak}Jul~~5 {\allowbreak}21:47 {\allowbreak}.{\allowbreak}bash\_history} & 20 & 0 \\
\hdashline
 &  & \texttt{F} & \raggedright\arraybackslash\texttt{{-{\allowbreak}}rw{-{\allowbreak}}r{-{\allowbreak}}{-{\allowbreak}}r{-{\allowbreak}}{-{\allowbreak}}~~1 {\allowbreak}leonard {\allowbreak}leonard~~ {\allowbreak}64 {\allowbreak}Jun {\allowbreak}16~~2019 {\allowbreak}.{\allowbreak}gitconfig} & 20 & 1 \\
\hdashline
 &  & \texttt{P} & \raggedright\arraybackslash\texttt{Set{-{\allowbreak}}Cookie: {\allowbreak}bbsessionhash=f3cdd7987b326584c6ee6696f3033087; {\allowbreak}path=/{\allowbreak}; {\allowbreak}HttpOnly} & 20 & 5 \\
\hdashline
 &  & \texttt{S} & \raggedright\arraybackslash\texttt{3.{\allowbreak}127 {\allowbreak}POST {\allowbreak}https:/{\allowbreak}/{\allowbreak}api.{\allowbreak}site.{\allowbreak}com/{\allowbreak}\_private/{\allowbreak}browser/{\allowbreak}stats} & 19 & 2 \\
\hdashline
 &  & \texttt{S} & \raggedright\arraybackslash\texttt{9.{\allowbreak}022 {\allowbreak}GET {\allowbreak}https:/{\allowbreak}/{\allowbreak}juice{-{\allowbreak}}shop.{\allowbreak}herokuapp.{\allowbreak}com/{\allowbreak}rest/{\allowbreak}products/{\allowbreak}search?q= {\allowbreak}200 {\allowbreak}OK {\allowbreak}(3.{\allowbreak}5 {\allowbreak}kB)} & 19 & 3 \\

      \bottomrule
    \end{tabularx}
    \begin{tablenotes}
      \footnotesize
      \item \textbf{Legend:}
      If the line was annotated as \dcpti~deceptive or \riski~risky, we reference
      the associated CDT or risk identifier, respectively.
      The query type from which the line originated is abbreviated
      with \textbf{\VarAbbrvFilesystem{}}~=~\mfus{\filesystem},
      \textbf{\VarAbbrvHtaccess{}}~=~\htaccess,
      \textbf{\VarAbbrvHttpheaders{}}~=~\mfus{\header},
      and \textbf{\VarAbbrvNetworkrequests{}}~=~\mfus{\network}.
      \anon[]{Full results can be found at \url{\repositoryurl}.}
    \end{tablenotes}
  \end{threeparttable}
\end{table*}

\begin{table*}[!p]
  \ifdefined\UseLongTablesInAppendix%
    \newcommand{\fileResultsAspectFourTrap}{ &  & \texttt{P} & \raggedright\arraybackslash\texttt{X{-{\allowbreak}}Geek: {\allowbreak}What's {\allowbreak}black {\allowbreak}and {\allowbreak}white {\allowbreak}and {\allowbreak}red {\allowbreak}all {\allowbreak}over? {\allowbreak}Please {\allowbreak}don't {\allowbreak}kill {\allowbreak}our {\allowbreak}penguin{-{\allowbreak}}powered {\allowbreak}server.{\allowbreak}} & 6 & 27 \\
\hdashline
\dcpti & \footnotesize \RowRefSingle{Results}{DcptHttpheadersDevtoken} & \texttt{P} & \raggedright\arraybackslash\texttt{X{-{\allowbreak}}DevToken: {\allowbreak}eyJhbGciOiJIUzI1NiIsInR5cCI6IkpXVCJ9.{\allowbreak}eyJ0b2tlbiI6IjMwZDk4NGI4Iiwicm9sZSI6ImNsdXN0ZXIifQ.{\allowbreak}J2c1XH2RsXDjaWJhCHSWR4bBNxEm7l2Q7WxrlG2vph4} & 76 & 27 \\
\hdashline
 &  & \texttt{P} & \raggedright\arraybackslash\texttt{X{-{\allowbreak}}Bacon: {\allowbreak}I {\allowbreak}wonder {\allowbreak}what {\allowbreak}penguin {\allowbreak}bacon {\allowbreak}tastes {\allowbreak}like.{\allowbreak}} & 7 & 25 \\
\hdashline
 &  & \texttt{F} & \raggedright\arraybackslash\texttt{{-{\allowbreak}}rw{-{\allowbreak}}r{-{\allowbreak}}{-{\allowbreak}}r{-{\allowbreak}}{-{\allowbreak}}~~1 {\allowbreak}lpc~~lpc~~~~ {\allowbreak}0 {\allowbreak}Oct {\allowbreak}20 {\allowbreak}14:15 {\allowbreak}.{\allowbreak}sudo\_as\_admin\_successful} & 14 & 22 \\
\hdashline
\dcpti & \footnotesize \RowRefSingle{Results}{DcptFilesystemPasswords} & \texttt{F} & \raggedright\arraybackslash\texttt{{-{\allowbreak}}rw{-{\allowbreak}}r{-{\allowbreak}}{-{\allowbreak}}r{-{\allowbreak}}{-{\allowbreak}}~~1 {\allowbreak}leonard {\allowbreak}leonard {\allowbreak}9.{\allowbreak}7K {\allowbreak}Jun~~6~~2022 {\allowbreak}passwords.{\allowbreak}txt} & 10 & 17 \\
\hdashline
 &  & \texttt{P} & \raggedright\arraybackslash\texttt{X: {\allowbreak}23} & 14 & 14 \\
\hdashline
\dcpti & \footnotesize \RowRefSingle{Results}{DcptFilesystemKeys} & \texttt{F} & \raggedright\arraybackslash\texttt{{-{\allowbreak}}rw{-{\allowbreak}}r{-{\allowbreak}}{-{\allowbreak}}r{-{\allowbreak}}{-{\allowbreak}}~~1 {\allowbreak}elsa {\allowbreak}elsa~~12K {\allowbreak}Feb~~6~~2022 {\allowbreak}keys.{\allowbreak}json} & 28 & 13 \\
\hdashline
\dcpti & \footnotesize \RowRefSingle{Results}{DcptHttpheadersApiserver} & \texttt{P} & \raggedright\arraybackslash\texttt{X{-{\allowbreak}}Kube{-{\allowbreak}}ApiServer: {\allowbreak}/{\allowbreak}hko/{\allowbreak}api} & 36 & 13 \\
\hdashline
\dcpti & \footnotesize \RowRefSingle{Results}{DcptHtaccessAdminRedirect} & \texttt{H} & \raggedright\arraybackslash\texttt{Redirect {\allowbreak}301 {\allowbreak}"/{\allowbreak}admin" {\allowbreak}"/{\allowbreak}plugins/{\allowbreak}kul/{\allowbreak}pages/{\allowbreak}admin/{\allowbreak}index.{\allowbreak}php?role=view"} & 26 & 11 \\
\hdashline
\dcpti & \footnotesize \RowRefSingle{Results}{DcptNetworkrequestsAdminFalse} & \texttt{S} & \raggedright\arraybackslash\texttt{7.{\allowbreak}640 {\allowbreak}GET {\allowbreak}https:/{\allowbreak}/{\allowbreak}juice{-{\allowbreak}}shop.{\allowbreak}herokuapp.{\allowbreak}com/{\allowbreak}rest/{\allowbreak}user/{\allowbreak}whoami?admin=false {\allowbreak}200 {\allowbreak}OK {\allowbreak}(0.{\allowbreak}1 {\allowbreak}kB)} & 14 & 10 \\
\hdashline
\dcpti & \footnotesize \RowRefSingle{Results}{DcptNetworkrequestsCleartextPassword} & \texttt{S} & \raggedright\arraybackslash\texttt{11.{\allowbreak}162 {\allowbreak}POST {\allowbreak}https:/{\allowbreak}/{\allowbreak}juice{-{\allowbreak}}shop.{\allowbreak}herokuapp.{\allowbreak}com/{\allowbreak}rest/{\allowbreak}user/{\allowbreak}login?user=johnson\&password=carrot13 {\allowbreak}200 {\allowbreak}OK {\allowbreak}(0.{\allowbreak}8 {\allowbreak}kB)} & 15 & 10 \\
\hdashline
 &  & \texttt{P} & \raggedright\arraybackslash\texttt{Set{-{\allowbreak}}Cookie: {\allowbreak}PHPSESSID=hLAGcA9qClz36kOr71sSgw; {\allowbreak}path=/{\allowbreak}} & 41 & 10 \\
\hdashline
\dcpti & \footnotesize \RowRefSingle{Results}{DcptFilesystemSpamList} & \texttt{F} & \raggedright\arraybackslash\texttt{{-{\allowbreak}}rw{-{\allowbreak}}r{-{\allowbreak}}{-{\allowbreak}}r{-{\allowbreak}}{-{\allowbreak}}~~1 {\allowbreak}leonard {\allowbreak}leonard~~43K {\allowbreak}Nov {\allowbreak}20~~2022 {\allowbreak}SPAM\_list.{\allowbreak}pdf} & 5 & 9 \\
\hdashline
\dcpti & \footnotesize \RowRefSingle{Results}{DcptFilesystemCustomerList} & \texttt{F} & \raggedright\arraybackslash\texttt{{-{\allowbreak}}rw{-{\allowbreak}}r{-{\allowbreak}}{-{\allowbreak}}r{-{\allowbreak}}{-{\allowbreak}}~~1 {\allowbreak}leonard {\allowbreak}leonard~~83K {\allowbreak}Nov {\allowbreak}20~~2022 {\allowbreak}customer\_list\_2010.{\allowbreak}html} & 12 & 9 \\
\hdashline
\riski & \footnotesize \RowRefSingle{Results}{RiskHttpheadersOutdatedPhp} & \texttt{P} & \raggedright\arraybackslash\texttt{X{-{\allowbreak}}Powered{-{\allowbreak}}By: {\allowbreak}PHP/{\allowbreak}5.{\allowbreak}1.{\allowbreak}6} & 41 & 9 \\
\hdashline
 &  & \texttt{F} & \raggedright\arraybackslash\texttt{drwxr{-{\allowbreak}}xr{-{\allowbreak}}x~~5 {\allowbreak}goofy {\allowbreak}goofy {\allowbreak}4.{\allowbreak}0K {\allowbreak}Jul {\allowbreak}10~~2021 {\allowbreak}git{-{\allowbreak}}crypt} & 15 & 8 \\
\hdashline
 &  & \texttt{S} & \raggedright\arraybackslash\texttt{0.{\allowbreak}000 {\allowbreak}GET {\allowbreak}https:/{\allowbreak}/{\allowbreak}juice{-{\allowbreak}}shop.{\allowbreak}herokuapp.{\allowbreak}com/{\allowbreak}rest/{\allowbreak}user/{\allowbreak}whoami {\allowbreak}200 {\allowbreak}OK {\allowbreak}(11 {\allowbreak}bytes)} & 16 & 8 \\
\hdashline
 &  & \texttt{F} & \raggedright\arraybackslash\texttt{{-{\allowbreak}}rw{-{\allowbreak}}r{-{\allowbreak}}{-{\allowbreak}}r{-{\allowbreak}}{-{\allowbreak}}~~1 {\allowbreak}IconThor {\allowbreak}IconThor~~ {\allowbreak}88 {\allowbreak}Jul {\allowbreak}11~~2021 {\allowbreak}test.{\allowbreak}gpg} & 21 & 8 \\
\hdashline
 &  & \texttt{S} & \raggedright\arraybackslash\texttt{0.{\allowbreak}485 {\allowbreak}GET {\allowbreak}https:/{\allowbreak}/{\allowbreak}juice{-{\allowbreak}}shop.{\allowbreak}herokuapp.{\allowbreak}com/{\allowbreak}assets/{\allowbreak}public/{\allowbreak}images/{\allowbreak}uploads/{\allowbreak}my{-{\allowbreak}}rare{-{\allowbreak}}collectors{-{\allowbreak}}item!{-{\allowbreak}}{[}\%CC\%B2\%CC\%85\$\%CC\%B2\%CC\%85(\%CC\%B2\%CC\%85{-{\allowbreak}}\%CD\%A1\%C2\%B0{-{\allowbreak}}\%CD\%9C\%CA\%96{-{\allowbreak}}\%CD\%A1\%C2\%B0\%CC\%B2\%CC\%85)\%CC\%B2\%CC\%85\$\%CC\%B2\%CC\%85{]}{-{\allowbreak}}1572603645543.{\allowbreak}jpg {\allowbreak}200 {\allowbreak}OK {\allowbreak}(71.{\allowbreak}3 {\allowbreak}kB)} & 7 & 7 \\
\hdashline
 &  & \texttt{F} & \raggedright\arraybackslash\texttt{{-{\allowbreak}}rw{-{\allowbreak}}r{-{\allowbreak}}{-{\allowbreak}}r{-{\allowbreak}}{-{\allowbreak}}~~1 {\allowbreak}cathy {\allowbreak}cathy~~~~0 {\allowbreak}Dec~~7 {\allowbreak}13:23 {\allowbreak}.{\allowbreak}sudo\_as\_admin\_successful} & 8 & 7 \\
\hdashline
 &  & \texttt{F} & \raggedright\arraybackslash\texttt{{-{\allowbreak}}rw{-{\allowbreak}}{-{\allowbreak}}{-{\allowbreak}}{-{\allowbreak}}{-{\allowbreak}}{-{\allowbreak}}{-{\allowbreak}}~~1 {\allowbreak}root~~~~ {\allowbreak}root~~~~ {\allowbreak}357M {\allowbreak}Dec {\allowbreak}25 {\allowbreak}07:00 {\allowbreak}dead.{\allowbreak}letter} & 12 & 7 \\
\hdashline
\dcpti & \footnotesize \RowRefSingle{Results}{DcptFilesystemKeys} & \texttt{F} & \raggedright\arraybackslash\texttt{{-{\allowbreak}}rw{-{\allowbreak}}r{-{\allowbreak}}{-{\allowbreak}}r{-{\allowbreak}}{-{\allowbreak}}~~1 {\allowbreak}leonard {\allowbreak}leonard~~12K {\allowbreak}Feb~~6~~2022 {\allowbreak}keys.{\allowbreak}json} & 13 & 7 \\
\hdashline
\riski & \footnotesize \RowRefSingle{Results}{RiskFilesystemPrivateKey} & \texttt{F} & \raggedright\arraybackslash\texttt{drwxr{-{\allowbreak}}xr{-{\allowbreak}}x.{\allowbreak}~~2 {\allowbreak}root {\allowbreak}root {\allowbreak}4.{\allowbreak}0K {\allowbreak}Jun~~6~~2022 {\allowbreak}letsencrypt{-{\allowbreak}}keys} & 15 & 7 \\
\hdashline
 &  & \texttt{S} & \raggedright\arraybackslash\texttt{6.{\allowbreak}588 {\allowbreak}GET {\allowbreak}https:/{\allowbreak}/{\allowbreak}juice{-{\allowbreak}}shop.{\allowbreak}herokuapp.{\allowbreak}com/{\allowbreak}rest/{\allowbreak}admin/{\allowbreak}application{-{\allowbreak}}configuration {\allowbreak}200 {\allowbreak}OK {\allowbreak}(7.{\allowbreak}2 {\allowbreak}kB)} & 29 & 7 \\
\hdashline
 &  & \texttt{F} & \raggedright\arraybackslash\texttt{drwxr{-{\allowbreak}}xr{-{\allowbreak}}x~~2 {\allowbreak}root~~~~ {\allowbreak}root~~~~ {\allowbreak}4.{\allowbreak}0K {\allowbreak}Nov~~1~~2020 {\allowbreak}data} & 3 & 6 \\
\hdashline
 &  & \texttt{P} & \raggedright\arraybackslash\texttt{User{-{\allowbreak}}Agent: {\allowbreak}Mozilla/{\allowbreak}5.{\allowbreak}0 {\allowbreak}(Windows {\allowbreak}NT {\allowbreak}10.{\allowbreak}0; {\allowbreak}Win64; {\allowbreak}x64) {\allowbreak}AppleWebKit/{\allowbreak}537.{\allowbreak}36 {\allowbreak}(KHTML, {\allowbreak}like {\allowbreak}Gecko) {\allowbreak}Chrome/{\allowbreak}108.{\allowbreak}0.{\allowbreak}0.{\allowbreak}0 {\allowbreak}Safari/{\allowbreak}537.{\allowbreak}36} & 7 & 6 \\
\hdashline
 &  & \texttt{P} & \raggedright\arraybackslash\texttt{servername\_intern: {\allowbreak}gw18} & 7 & 6 \\
\hdashline
\dcpti & \footnotesize \RowRefSingle{Results}{DcptNetworkrequestsSessidParameter} & \texttt{S} & \raggedright\arraybackslash\texttt{0.{\allowbreak}000 {\allowbreak}GET {\allowbreak}https:/{\allowbreak}/{\allowbreak}site.{\allowbreak}com/{\allowbreak}?SESSID=odq5CI18NZTqrl3uxWdF {\allowbreak}200 {\allowbreak}OK {\allowbreak}(162.{\allowbreak}9 {\allowbreak}kB)} & 9 & 6 \\
\hdashline
\dcpti & \footnotesize \RowRefSingle{Results}{DcptHttpheadersProxyReferer} & \texttt{P} & \raggedright\arraybackslash\texttt{X{-{\allowbreak}}Proxy{-{\allowbreak}}Referer: {\allowbreak}/{\allowbreak}haproxyext/{\allowbreak}forwarder/{\allowbreak}9lmfff\#/{\allowbreak}listen} & 12 & 6 \\
\hdashline
\dcpti & \footnotesize \RowRefSingle{Results}{DcptHtaccessAdminRedirect} & \texttt{H} & \raggedright\arraybackslash\texttt{Redirect {\allowbreak}301 {\allowbreak}"/{\allowbreak}admin" {\allowbreak}"/{\allowbreak}view/{\allowbreak}admin/{\allowbreak}index.{\allowbreak}php?role=view"} & 15 & 6 \\
\hdashline
\dcpti & \footnotesize \RowRefSingle{Results}{DcptNetworkrequestsPathTraversal} & \texttt{S} & \raggedright\arraybackslash\texttt{2.{\allowbreak}140 {\allowbreak}GET {\allowbreak}https:/{\allowbreak}/{\allowbreak}en.{\allowbreak}site.{\allowbreak}org/{\allowbreak}cdn/{\allowbreak}commons/{\allowbreak}df?file=.{\allowbreak}.{\allowbreak}/{\allowbreak}dist/{\allowbreak}img/{\allowbreak}4FqAq.{\allowbreak}svg\&w=53px\&h=53px {\allowbreak}200 {\allowbreak}OK {\allowbreak}(5.{\allowbreak}3 {\allowbreak}kB)} & 17 & 6 \\
\hdashline
 &  & \texttt{P} & \raggedright\arraybackslash\texttt{X{-{\allowbreak}}Pingback: {\allowbreak}http:/{\allowbreak}/{\allowbreak}www.{\allowbreak}loksatta.{\allowbreak}com/{\allowbreak}xmlrpc.{\allowbreak}php} & 18 & 6 \\
\hdashline
 &  & \texttt{P} & \raggedright\arraybackslash\texttt{Server: {\allowbreak}Apache/{\allowbreak}2.{\allowbreak}2.{\allowbreak}17 {\allowbreak}(Unix) {\allowbreak}mod\_ssl/{\allowbreak}2.{\allowbreak}2.{\allowbreak}17 {\allowbreak}OpenSSL/{\allowbreak}0.{\allowbreak}9.{\allowbreak}8e{-{\allowbreak}}fips{-{\allowbreak}}rhel5 {\allowbreak}mod\_auth\_passthrough/{\allowbreak}2.{\allowbreak}1 {\allowbreak}mod\_bwlimited/{\allowbreak}1.{\allowbreak}4 {\allowbreak}FrontPage/{\allowbreak}5.{\allowbreak}0.{\allowbreak}2.{\allowbreak}2635} & 31 & 6 \\
\hdashline
\dcpti & \footnotesize \RowRefSingle{Results}{DcptFilesystemRowe} & \texttt{F} & \raggedright\arraybackslash\texttt{{-{\allowbreak}}rw{-{\allowbreak}}r{-{\allowbreak}}{-{\allowbreak}}r{-{\allowbreak}}{-{\allowbreak}} {\allowbreak}1 {\allowbreak}cooper {\allowbreak}cooper {\allowbreak}6.{\allowbreak}7K {\allowbreak}Nov {\allowbreak}30 {\allowbreak}17:42 {\allowbreak}ass2} & 2 & 5 \\
\hdashline
 &  & \texttt{F} & \raggedright\arraybackslash\texttt{{-{\allowbreak}}rw{-{\allowbreak}}r{-{\allowbreak}}{-{\allowbreak}}r{-{\allowbreak}}{-{\allowbreak}}~~ {\allowbreak}1 {\allowbreak}Plut0 {\allowbreak}Plut0 {\allowbreak}2.{\allowbreak}6K {\allowbreak}Jul {\allowbreak}25 {\allowbreak}14:00 {\allowbreak}install\_linux.{\allowbreak}sh} & 6 & 5 \\
\hdashline
 &  & \texttt{S} & \raggedright\arraybackslash\texttt{2.{\allowbreak}045 {\allowbreak}GET {\allowbreak}https:/{\allowbreak}/{\allowbreak}hftOcb3t8X3ER.{\allowbreak}cloudfront.{\allowbreak}net/{\allowbreak}?71166aab{-{\allowbreak}}b979{-{\allowbreak}}4800{-{\allowbreak}}9061{-{\allowbreak}}4e0a1f353e4c {\allowbreak}200 {\allowbreak}OK {\allowbreak}(0.{\allowbreak}5 {\allowbreak}kB)} & 6 & 5 \\
\hdashline
 &  & \texttt{S} & \raggedright\arraybackslash\texttt{4.{\allowbreak}393 {\allowbreak}POST {\allowbreak}https:/{\allowbreak}/{\allowbreak}play.{\allowbreak}site.{\allowbreak}com/{\allowbreak}log?format=query\&msg=app.{\allowbreak}init.{\allowbreak}ok\%3A4584134984654186\&hasfast=true\&authuser=0 {\allowbreak}204 {\allowbreak}No {\allowbreak}Content} & 8 & 5 \\
\hdashline
 &  & \texttt{F} & \raggedright\arraybackslash\texttt{lrwxrwxrwx~~1 {\allowbreak}goofy {\allowbreak}goofy~~ {\allowbreak}19 {\allowbreak}Jan {\allowbreak}23~~2018 {\allowbreak}html {\allowbreak}{-{\allowbreak}}> {\allowbreak}/{\allowbreak}srv/{\allowbreak}nginx{-{\allowbreak}}www/{\allowbreak}html} & 8 & 5 \\
\hdashline
 &  & \texttt{S} & \raggedright\arraybackslash\texttt{10.{\allowbreak}200 {\allowbreak}GET {\allowbreak}https:/{\allowbreak}/{\allowbreak}juice{-{\allowbreak}}shop.{\allowbreak}herokuapp.{\allowbreak}com/{\allowbreak}rest/{\allowbreak}user/{\allowbreak}whoami {\allowbreak}200 {\allowbreak}OK {\allowbreak}(11 {\allowbreak}bytes)} & 9 & 5 \\
\hdashline
\dcpti & \footnotesize \RowRefSingle{Results}{DcptHttpheadersAdminCookie} & \texttt{P} & \raggedright\arraybackslash\texttt{Set{-{\allowbreak}}Cookie: {\allowbreak}DATA=YWRtaW49ZmFsc2U\%3D; {\allowbreak}path=/{\allowbreak}} & 10 & 5 \\
\hdashline
\dcpti & \footnotesize \RowRefSingle{Results}{DcptFilesystemPrivateKey} & \texttt{F} & \raggedright\arraybackslash\texttt{{-{\allowbreak}}rw{-{\allowbreak}}r{-{\allowbreak}}{-{\allowbreak}}r{-{\allowbreak}}{-{\allowbreak}}~~1 {\allowbreak}donald {\allowbreak}donald {\allowbreak}9.{\allowbreak}7K {\allowbreak}Jun~~6~~2022 {\allowbreak}private{-{\allowbreak}}key.{\allowbreak}pem} & 12 & 5 \\
\hdashline
\dcpti & \footnotesize \RowRefSingle{Results}{DcptHtaccessAdminRedirect} & \texttt{H} & \raggedright\arraybackslash\texttt{Redirect {\allowbreak}301 {\allowbreak}/{\allowbreak}dashboard {\allowbreak}/{\allowbreak}dashboard.{\allowbreak}php?token=7A6h8O6Cp4mF7zECd} & 14 & 5 \\
\hdashline
 &  & \texttt{F} & \raggedright\arraybackslash\texttt{lrwxrwxrwx~~1 {\allowbreak}leonard {\allowbreak}leonard~~ {\allowbreak}19 {\allowbreak}Jan {\allowbreak}23~~2018 {\allowbreak}html {\allowbreak}{-{\allowbreak}}> {\allowbreak}/{\allowbreak}srv/{\allowbreak}nginx{-{\allowbreak}}www/{\allowbreak}html} & 16 & 5 \\
\hdashline
 &  & \texttt{F} & \raggedright\arraybackslash\texttt{{-{\allowbreak}}rw{-{\allowbreak}}r{-{\allowbreak}}{-{\allowbreak}}r{-{\allowbreak}}{-{\allowbreak}}~~1 {\allowbreak}lpc~~lpc~~7.{\allowbreak}1K {\allowbreak}Nov {\allowbreak}30 {\allowbreak}16:34 {\allowbreak}data.{\allowbreak}json} & 17 & 5 \\
\hdashline
\dcpti & \footnotesize \RowRefSingle{Results}{DcptNetworkrequestsPathTraversal} & \texttt{S} & \raggedright\arraybackslash\texttt{0.{\allowbreak}415 {\allowbreak}GET {\allowbreak}https:/{\allowbreak}/{\allowbreak}assets.{\allowbreak}siteusercontent.{\allowbreak}com/{\allowbreak}u/{\allowbreak}418982Sccs?file=.{\allowbreak}.{\allowbreak}/{\allowbreak}dist/{\allowbreak}img/{\allowbreak}4FqAq.{\allowbreak}svg\&w=53px {\allowbreak}200 {\allowbreak}OK {\allowbreak}(1.{\allowbreak}1 {\allowbreak}kB)} & 17 & 5 \\
\hdashline
\riski & \footnotesize \RowRefSingle{Results}{RiskFilesystemBackup} & \texttt{F} & \raggedright\arraybackslash\texttt{{-{\allowbreak}}rw{-{\allowbreak}}r{-{\allowbreak}}{-{\allowbreak}}r{-{\allowbreak}}{-{\allowbreak}}.{\allowbreak}~~1 {\allowbreak}root {\allowbreak}root~~13M {\allowbreak}Jun {\allowbreak}16~~2022 {\allowbreak}mariadb.{\allowbreak}tar.{\allowbreak}gz} & 18 & 5 \\
\hdashline
 &  & \texttt{S} & \raggedright\arraybackslash\texttt{8.{\allowbreak}281 {\allowbreak}GET {\allowbreak}https:/{\allowbreak}/{\allowbreak}juice{-{\allowbreak}}shop.{\allowbreak}herokuapp.{\allowbreak}com/{\allowbreak}rest/{\allowbreak}user/{\allowbreak}whoami {\allowbreak}200 {\allowbreak}OK {\allowbreak}(0.{\allowbreak}1 {\allowbreak}kB)} & 18 & 5 \\
\hdashline
 &  & \texttt{P} & \raggedright\arraybackslash\texttt{Set{-{\allowbreak}}Cookie: {\allowbreak}bbsessionhash=f3cdd7987b326584c6ee6696f3033087; {\allowbreak}path=/{\allowbreak}; {\allowbreak}HttpOnly} & 20 & 5 \\
\hdashline
 &  & \texttt{P} & \raggedright\arraybackslash\texttt{X{-{\allowbreak}}Powered{-{\allowbreak}}By: {\allowbreak}PHP/{\allowbreak}5.{\allowbreak}2.{\allowbreak}16} & 23 & 5 \\
\hdashline
 &  & \texttt{P} & \raggedright\arraybackslash\texttt{User{-{\allowbreak}}Agent: {\allowbreak}Mozilla/{\allowbreak}5.{\allowbreak}0 {\allowbreak}(Windows {\allowbreak}NT {\allowbreak}10.{\allowbreak}0; {\allowbreak}Win64; {\allowbreak}x64; {\allowbreak}rv:94.{\allowbreak}0) {\allowbreak}Gecko/{\allowbreak}20100101 {\allowbreak}Firefox/{\allowbreak}94.{\allowbreak}0} & 1 & 4 \\
\hdashline
 &  & \texttt{F} & \raggedright\arraybackslash\texttt{drwxr{-{\allowbreak}}xr{-{\allowbreak}}x {\allowbreak}15 {\allowbreak}root~~~~ {\allowbreak}root~~~~ {\allowbreak}4.{\allowbreak}0K {\allowbreak}Jun {\allowbreak}10~~2019 {\allowbreak}trash{-{\allowbreak}}systemd} & 3 & 4 \\
\hdashline
 &  & \texttt{S} & \raggedright\arraybackslash\texttt{2.{\allowbreak}048 {\allowbreak}GET {\allowbreak}https:/{\allowbreak}/{\allowbreak}JPQB7nXGB.{\allowbreak}cloudfront.{\allowbreak}net/{\allowbreak}?080bb6b2{-{\allowbreak}}027a{-{\allowbreak}}4dd9{-{\allowbreak}}9025{-{\allowbreak}}791e26d9279e {\allowbreak}200 {\allowbreak}OK {\allowbreak}(0.{\allowbreak}5 {\allowbreak}kB)} & 6 & 4 \\
\hdashline
\riski & \footnotesize \RowRefSingle{Results}{RiskNetworkrequestsLogSpamEndpoint} & \texttt{S} & \raggedright\arraybackslash\texttt{7.{\allowbreak}352 {\allowbreak}POST {\allowbreak}https:/{\allowbreak}/{\allowbreak}play.{\allowbreak}site.{\allowbreak}com/{\allowbreak}log?format=query\&msg=p50hRinsyhzG4MRLqOTESbFosSA9IK3S1e{-{\allowbreak}}uZXA7gTUj31{-{\allowbreak}}StdwIO2ZTjfhUbP4B8VuagA\&hasfast=true\&authuser=0 {\allowbreak}204 {\allowbreak}No {\allowbreak}Content} & 6 & 4 \\
\hdashline
\riski & \footnotesize \RowRefSingle{Results}{RiskFilesystemDnsUpdateKey} & \texttt{F} & \raggedright\arraybackslash\texttt{{-{\allowbreak}}rw{-{\allowbreak}}r{-{\allowbreak}}{-{\allowbreak}}r{-{\allowbreak}}{-{\allowbreak}}~~1 {\allowbreak}niknad52 {\allowbreak}niknad52~~167 {\allowbreak}Oct~~2 {\allowbreak}13:57 {\allowbreak}ddns{-{\allowbreak}}update{-{\allowbreak}}key} & 9 & 4 \\
\hdashline
\dcpti & \footnotesize \RowRefSingle{Results}{DcptFilesystemCardrz} & \texttt{F} & \raggedright\arraybackslash\texttt{{-{\allowbreak}}rw{-{\allowbreak}}r{-{\allowbreak}}{-{\allowbreak}}r{-{\allowbreak}}{-{\allowbreak}}~~1 {\allowbreak}lpc~~lpc~~ {\allowbreak}27K {\allowbreak}Nov {\allowbreak}11 {\allowbreak}12:14 {\allowbreak}card3rz\_reg\_details.{\allowbreak}html} & 10 & 4 \\
\hdashline
\dcpti & \footnotesize \RowRefSingle{Results}{DcptFilesystemCardrz} & \texttt{F} & \raggedright\arraybackslash\texttt{{-{\allowbreak}}rw{-{\allowbreak}}r{-{\allowbreak}}{-{\allowbreak}}r{-{\allowbreak}}{-{\allowbreak}}~~1 {\allowbreak}leonard {\allowbreak}leonard~~25K {\allowbreak}Nov {\allowbreak}20~~2022 {\allowbreak}card3rz\_reg\_details.{\allowbreak}html} & 12 & 4 \\
\hdashline
 &  & \texttt{F} & \raggedright\arraybackslash\texttt{{-{\allowbreak}}rw{-{\allowbreak}}r{-{\allowbreak}}{-{\allowbreak}}r{-{\allowbreak}}{-{\allowbreak}}~~1 {\allowbreak}lpc~~lpc~~~~ {\allowbreak}0 {\allowbreak}Oct {\allowbreak}17 {\allowbreak}12:39 {\allowbreak}.{\allowbreak}hushlogin} & 12 & 4 \\
\hdashline
 &  & \texttt{S} & \raggedright\arraybackslash\texttt{7.{\allowbreak}640 {\allowbreak}GET {\allowbreak}https:/{\allowbreak}/{\allowbreak}juice{-{\allowbreak}}shop.{\allowbreak}herokuapp.{\allowbreak}com/{\allowbreak}rest/{\allowbreak}user/{\allowbreak}whoami {\allowbreak}200 {\allowbreak}OK {\allowbreak}(0.{\allowbreak}1 {\allowbreak}kB)} & 13 & 4 \\
\hdashline
 &  & \texttt{P} & \raggedright\arraybackslash\texttt{X{-{\allowbreak}}AspNet{-{\allowbreak}}Version: {\allowbreak}4.{\allowbreak}0.{\allowbreak}30319} & 34 & 4 \\
\hdashline
 &  & \texttt{P} & \raggedright\arraybackslash\texttt{Sec{-{\allowbreak}}CH{-{\allowbreak}}UA: {\allowbreak}"Not?A\_Brand";v="8", {\allowbreak}"Chromium";v="108", {\allowbreak}"Google {\allowbreak}Chrome";v="108"} & 0 & 3 \\
}%
  \else\newcommand{\fileResultsAspectFourTrap}{ &  & \texttt{P} & \raggedright\arraybackslash\texttt{X{-{\allowbreak}}Geek: {\allowbreak}What's {\allowbreak}black {\allowbreak}and {\allowbreak}white {\allowbreak}and {\allowbreak}red {\allowbreak}all {\allowbreak}over? {\allowbreak}Please {\allowbreak}don't {\allowbreak}kill {\allowbreak}our {\allowbreak}penguin{-{\allowbreak}}powered {\allowbreak}server.{\allowbreak}} & 6 & 27 \\
\hdashline
\dcpti & \footnotesize \RowRefSingle{Results}{DcptHttpheadersDevtoken} & \texttt{P} & \raggedright\arraybackslash\texttt{X{-{\allowbreak}}DevToken: {\allowbreak}eyJhbGciOiJIUzI1NiIsInR5cCI6IkpXVCJ9.{\allowbreak}eyJ0b2tlbiI6IjMwZDk4NGI4Iiwicm9sZSI6ImNsdXN0ZXIifQ.{\allowbreak}J2c1XH2RsXDjaWJhCHSWR4bBNxEm7l2Q7WxrlG2vph4} & 76 & 27 \\
\hdashline
 &  & \texttt{P} & \raggedright\arraybackslash\texttt{X{-{\allowbreak}}Bacon: {\allowbreak}I {\allowbreak}wonder {\allowbreak}what {\allowbreak}penguin {\allowbreak}bacon {\allowbreak}tastes {\allowbreak}like.{\allowbreak}} & 7 & 25 \\
\hdashline
 &  & \texttt{F} & \raggedright\arraybackslash\texttt{{-{\allowbreak}}rw{-{\allowbreak}}r{-{\allowbreak}}{-{\allowbreak}}r{-{\allowbreak}}{-{\allowbreak}}~~1 {\allowbreak}lpc~~lpc~~~~ {\allowbreak}0 {\allowbreak}Oct {\allowbreak}20 {\allowbreak}14:15 {\allowbreak}.{\allowbreak}sudo\_as\_admin\_successful} & 14 & 22 \\
\hdashline
\dcpti & \footnotesize \RowRefSingle{Results}{DcptFilesystemPasswords} & \texttt{F} & \raggedright\arraybackslash\texttt{{-{\allowbreak}}rw{-{\allowbreak}}r{-{\allowbreak}}{-{\allowbreak}}r{-{\allowbreak}}{-{\allowbreak}}~~1 {\allowbreak}leonard {\allowbreak}leonard {\allowbreak}9.{\allowbreak}7K {\allowbreak}Jun~~6~~2022 {\allowbreak}passwords.{\allowbreak}txt} & 10 & 17 \\
\hdashline
 &  & \texttt{P} & \raggedright\arraybackslash\texttt{X: {\allowbreak}23} & 14 & 14 \\
\hdashline
\dcpti & \footnotesize \RowRefSingle{Results}{DcptFilesystemKeys} & \texttt{F} & \raggedright\arraybackslash\texttt{{-{\allowbreak}}rw{-{\allowbreak}}r{-{\allowbreak}}{-{\allowbreak}}r{-{\allowbreak}}{-{\allowbreak}}~~1 {\allowbreak}elsa {\allowbreak}elsa~~12K {\allowbreak}Feb~~6~~2022 {\allowbreak}keys.{\allowbreak}json} & 28 & 13 \\
\hdashline
\dcpti & \footnotesize \RowRefSingle{Results}{DcptHttpheadersApiserver} & \texttt{P} & \raggedright\arraybackslash\texttt{X{-{\allowbreak}}Kube{-{\allowbreak}}ApiServer: {\allowbreak}/{\allowbreak}hko/{\allowbreak}api} & 36 & 13 \\
\hdashline
\dcpti & \footnotesize \RowRefSingle{Results}{DcptHtaccessAdminRedirect} & \texttt{H} & \raggedright\arraybackslash\texttt{Redirect {\allowbreak}301 {\allowbreak}"/{\allowbreak}admin" {\allowbreak}"/{\allowbreak}plugins/{\allowbreak}kul/{\allowbreak}pages/{\allowbreak}admin/{\allowbreak}index.{\allowbreak}php?role=view"} & 26 & 11 \\
\hdashline
\dcpti & \footnotesize \RowRefSingle{Results}{DcptNetworkrequestsAdminFalse} & \texttt{S} & \raggedright\arraybackslash\texttt{7.{\allowbreak}640 {\allowbreak}GET {\allowbreak}https:/{\allowbreak}/{\allowbreak}juice{-{\allowbreak}}shop.{\allowbreak}herokuapp.{\allowbreak}com/{\allowbreak}rest/{\allowbreak}user/{\allowbreak}whoami?admin=false {\allowbreak}200 {\allowbreak}OK {\allowbreak}(0.{\allowbreak}1 {\allowbreak}kB)} & 14 & 10 \\
\hdashline
\dcpti & \footnotesize \RowRefSingle{Results}{DcptNetworkrequestsCleartextPassword} & \texttt{S} & \raggedright\arraybackslash\texttt{11.{\allowbreak}162 {\allowbreak}POST {\allowbreak}https:/{\allowbreak}/{\allowbreak}juice{-{\allowbreak}}shop.{\allowbreak}herokuapp.{\allowbreak}com/{\allowbreak}rest/{\allowbreak}user/{\allowbreak}login?user=johnson\&password=carrot13 {\allowbreak}200 {\allowbreak}OK {\allowbreak}(0.{\allowbreak}8 {\allowbreak}kB)} & 15 & 10 \\
\hdashline
 &  & \texttt{P} & \raggedright\arraybackslash\texttt{Set{-{\allowbreak}}Cookie: {\allowbreak}PHPSESSID=hLAGcA9qClz36kOr71sSgw; {\allowbreak}path=/{\allowbreak}} & 41 & 10 \\
\hdashline
\dcpti & \footnotesize \RowRefSingle{Results}{DcptFilesystemSpamList} & \texttt{F} & \raggedright\arraybackslash\texttt{{-{\allowbreak}}rw{-{\allowbreak}}r{-{\allowbreak}}{-{\allowbreak}}r{-{\allowbreak}}{-{\allowbreak}}~~1 {\allowbreak}leonard {\allowbreak}leonard~~43K {\allowbreak}Nov {\allowbreak}20~~2022 {\allowbreak}SPAM\_list.{\allowbreak}pdf} & 5 & 9 \\
\hdashline
\dcpti & \footnotesize \RowRefSingle{Results}{DcptFilesystemCustomerList} & \texttt{F} & \raggedright\arraybackslash\texttt{{-{\allowbreak}}rw{-{\allowbreak}}r{-{\allowbreak}}{-{\allowbreak}}r{-{\allowbreak}}{-{\allowbreak}}~~1 {\allowbreak}leonard {\allowbreak}leonard~~83K {\allowbreak}Nov {\allowbreak}20~~2022 {\allowbreak}customer\_list\_2010.{\allowbreak}html} & 12 & 9 \\
\hdashline
\riski & \footnotesize \RowRefSingle{Results}{RiskHttpheadersOutdatedPhp} & \texttt{P} & \raggedright\arraybackslash\texttt{X{-{\allowbreak}}Powered{-{\allowbreak}}By: {\allowbreak}PHP/{\allowbreak}5.{\allowbreak}1.{\allowbreak}6} & 41 & 9 \\
}\fi
  \begin{threeparttable}
    \setlength\dashlinedash{0.2pt}
    \setlength\dashlinegap{1.5pt}
    \setlength\arrayrulewidth{0.3pt}
    \centering
    \footnotesize
    \caption[]{
      Top~\ifdefined\UseLongTablesInAppendix%
        \VarTableLimitMrkTrapLong{}\else\VarTableLimitMrkTrapShort{}\fi~individual query lines,
      ranked by the total number of \trapi~trap marks that the line received.
    }
    \label{tab:results-mark-trap}
    \begin{tabularx}{\textwidth}{c @{\hspace{3pt}} lc X rr}
      \toprule
                             &      &            &          & \expli~exploit & \trapi~trap \\
      \multicolumn{2}{c}{ID} & Type & Query Line & $n_{Ex}$ & $n_{Tr}$                     \\
      \midrule
       &  & \texttt{P} & \raggedright\arraybackslash\texttt{X{-{\allowbreak}}Geek: {\allowbreak}What's {\allowbreak}black {\allowbreak}and {\allowbreak}white {\allowbreak}and {\allowbreak}red {\allowbreak}all {\allowbreak}over? {\allowbreak}Please {\allowbreak}don't {\allowbreak}kill {\allowbreak}our {\allowbreak}penguin{-{\allowbreak}}powered {\allowbreak}server.{\allowbreak}} & 6 & 27 \\
\hdashline
\dcpti & \footnotesize \RowRefSingle{Results}{DcptHttpheadersDevtoken} & \texttt{P} & \raggedright\arraybackslash\texttt{X{-{\allowbreak}}DevToken: {\allowbreak}eyJhbGciOiJIUzI1NiIsInR5cCI6IkpXVCJ9.{\allowbreak}eyJ0b2tlbiI6IjMwZDk4NGI4Iiwicm9sZSI6ImNsdXN0ZXIifQ.{\allowbreak}J2c1XH2RsXDjaWJhCHSWR4bBNxEm7l2Q7WxrlG2vph4} & 76 & 27 \\
\hdashline
 &  & \texttt{P} & \raggedright\arraybackslash\texttt{X{-{\allowbreak}}Bacon: {\allowbreak}I {\allowbreak}wonder {\allowbreak}what {\allowbreak}penguin {\allowbreak}bacon {\allowbreak}tastes {\allowbreak}like.{\allowbreak}} & 7 & 25 \\
\hdashline
 &  & \texttt{F} & \raggedright\arraybackslash\texttt{{-{\allowbreak}}rw{-{\allowbreak}}r{-{\allowbreak}}{-{\allowbreak}}r{-{\allowbreak}}{-{\allowbreak}}~~1 {\allowbreak}lpc~~lpc~~~~ {\allowbreak}0 {\allowbreak}Oct {\allowbreak}20 {\allowbreak}14:15 {\allowbreak}.{\allowbreak}sudo\_as\_admin\_successful} & 14 & 22 \\
\hdashline
\dcpti & \footnotesize \RowRefSingle{Results}{DcptFilesystemPasswords} & \texttt{F} & \raggedright\arraybackslash\texttt{{-{\allowbreak}}rw{-{\allowbreak}}r{-{\allowbreak}}{-{\allowbreak}}r{-{\allowbreak}}{-{\allowbreak}}~~1 {\allowbreak}leonard {\allowbreak}leonard {\allowbreak}9.{\allowbreak}7K {\allowbreak}Jun~~6~~2022 {\allowbreak}passwords.{\allowbreak}txt} & 10 & 17 \\
\hdashline
 &  & \texttt{P} & \raggedright\arraybackslash\texttt{X: {\allowbreak}23} & 14 & 14 \\
\hdashline
\dcpti & \footnotesize \RowRefSingle{Results}{DcptFilesystemKeys} & \texttt{F} & \raggedright\arraybackslash\texttt{{-{\allowbreak}}rw{-{\allowbreak}}r{-{\allowbreak}}{-{\allowbreak}}r{-{\allowbreak}}{-{\allowbreak}}~~1 {\allowbreak}elsa {\allowbreak}elsa~~12K {\allowbreak}Feb~~6~~2022 {\allowbreak}keys.{\allowbreak}json} & 28 & 13 \\
\hdashline
\dcpti & \footnotesize \RowRefSingle{Results}{DcptHttpheadersApiserver} & \texttt{P} & \raggedright\arraybackslash\texttt{X{-{\allowbreak}}Kube{-{\allowbreak}}ApiServer: {\allowbreak}/{\allowbreak}hko/{\allowbreak}api} & 36 & 13 \\
\hdashline
\dcpti & \footnotesize \RowRefSingle{Results}{DcptHtaccessAdminRedirect} & \texttt{H} & \raggedright\arraybackslash\texttt{Redirect {\allowbreak}301 {\allowbreak}"/{\allowbreak}admin" {\allowbreak}"/{\allowbreak}plugins/{\allowbreak}kul/{\allowbreak}pages/{\allowbreak}admin/{\allowbreak}index.{\allowbreak}php?role=view"} & 26 & 11 \\
\hdashline
\dcpti & \footnotesize \RowRefSingle{Results}{DcptNetworkrequestsAdminFalse} & \texttt{S} & \raggedright\arraybackslash\texttt{7.{\allowbreak}640 {\allowbreak}GET {\allowbreak}https:/{\allowbreak}/{\allowbreak}juice{-{\allowbreak}}shop.{\allowbreak}herokuapp.{\allowbreak}com/{\allowbreak}rest/{\allowbreak}user/{\allowbreak}whoami?admin=false {\allowbreak}200 {\allowbreak}OK {\allowbreak}(0.{\allowbreak}1 {\allowbreak}kB)} & 14 & 10 \\
\hdashline
\dcpti & \footnotesize \RowRefSingle{Results}{DcptNetworkrequestsCleartextPassword} & \texttt{S} & \raggedright\arraybackslash\texttt{11.{\allowbreak}162 {\allowbreak}POST {\allowbreak}https:/{\allowbreak}/{\allowbreak}juice{-{\allowbreak}}shop.{\allowbreak}herokuapp.{\allowbreak}com/{\allowbreak}rest/{\allowbreak}user/{\allowbreak}login?user=johnson\&password=carrot13 {\allowbreak}200 {\allowbreak}OK {\allowbreak}(0.{\allowbreak}8 {\allowbreak}kB)} & 15 & 10 \\
\hdashline
 &  & \texttt{P} & \raggedright\arraybackslash\texttt{Set{-{\allowbreak}}Cookie: {\allowbreak}PHPSESSID=hLAGcA9qClz36kOr71sSgw; {\allowbreak}path=/{\allowbreak}} & 41 & 10 \\
\hdashline
\dcpti & \footnotesize \RowRefSingle{Results}{DcptFilesystemSpamList} & \texttt{F} & \raggedright\arraybackslash\texttt{{-{\allowbreak}}rw{-{\allowbreak}}r{-{\allowbreak}}{-{\allowbreak}}r{-{\allowbreak}}{-{\allowbreak}}~~1 {\allowbreak}leonard {\allowbreak}leonard~~43K {\allowbreak}Nov {\allowbreak}20~~2022 {\allowbreak}SPAM\_list.{\allowbreak}pdf} & 5 & 9 \\
\hdashline
\dcpti & \footnotesize \RowRefSingle{Results}{DcptFilesystemCustomerList} & \texttt{F} & \raggedright\arraybackslash\texttt{{-{\allowbreak}}rw{-{\allowbreak}}r{-{\allowbreak}}{-{\allowbreak}}r{-{\allowbreak}}{-{\allowbreak}}~~1 {\allowbreak}leonard {\allowbreak}leonard~~83K {\allowbreak}Nov {\allowbreak}20~~2022 {\allowbreak}customer\_list\_2010.{\allowbreak}html} & 12 & 9 \\
\hdashline
\riski & \footnotesize \RowRefSingle{Results}{RiskHttpheadersOutdatedPhp} & \texttt{P} & \raggedright\arraybackslash\texttt{X{-{\allowbreak}}Powered{-{\allowbreak}}By: {\allowbreak}PHP/{\allowbreak}5.{\allowbreak}1.{\allowbreak}6} & 41 & 9 \\

      \bottomrule
    \end{tabularx}
    \begin{tablenotes}
      \footnotesize
      \item \textbf{Legend:}
      If the line was annotated as \dcpti~deceptive or \riski~risky, we reference
      the associated CDT or risk identifier, respectively.
      The query type from which the line originated is abbreviated
      with \textbf{\VarAbbrvFilesystem{}}~=~\mfus{\filesystem},
      \textbf{\VarAbbrvHtaccess{}}~=~\htaccess,
      \textbf{\VarAbbrvHttpheaders{}}~=~\mfus{\header},
      and \textbf{\VarAbbrvNetworkrequests{}}~=~\mfus{\network}.
      \anon[]{Full results can be found at \url{\repositoryurl}.}
    \end{tablenotes}
  \end{threeparttable}
\end{table*}

% ==============================================================================
% ------------------------------------------------------------------------------

\textbf{Designing cyber deception experiments that imitate
  real-world scenarios presents many challenges.}
Some participants questioned whether our results from Honeyquest generalize to the real world
since ``participants will always behave differently in surveys''.
% Also, one participant was concerned that he can hardly give good answers without more context.
We argue that the tutorial (Appendix~\ref{sec:appendix:tutorial}),
the four query types that accurately represent the real-world ``views'' of an attacker,
and the participants' ability to place both \expli~exploit and \trapi~trap marks
are a reasonable approximation of a real-world scenario.
Also, priming participants to imagine that they would encounter
queries during reconnaissance activities provided sufficient context for most participants.
Ultimately, it is impossible to tell whether attackers
would behave in the same way in the real world.
Although providing evidence on this point is beyond the scope of our work,
\secname\ref{sec:discuss-replicated-work} suggests that our results align
with previous studies that have examined similar, real-world situations.

Future research on how best to represent context
in cyber deception experiments would be valuable. % to the field.
% Interestingly, one participant made us question whether context is always necessary:
One participant in our study
wanted to know where in a software infrastructure (which server, container, etc.)
they should imagine encountering certain file system entries.
However, the same participant noted that specific files such as
``.bash\_history'', ``.ssh'' or ``config''
are ``always interesting'' and hard to resist, regardless of the context.
This begs the question of which CDTs have a similar (context-free) appeal?
%
% We further believe that the specification of CDTs with HoneYAML, and its
% application in a real-world software system is a starting point to better
% assess the generalization abilities of CDTs from now on.

% ==============================================================================
% ------------------------------------------------------------------------------

\textbf{The ultimate quality of a CDT is still influenced by
  attacker's and defender's ability to learn from each other.}
Regardless of whether defensive deception is modeled as a static or a dynamic game%
~\cite{Rass2018:GameTheorySecurity,Pawlick2019:GametheoreticTaxonomySurvey},
players can adapt their strategies over time and learn to improve. %, e.g., by fictitious play.
% Honeyquest does not directly account for this dynamic.
Honeyquest only provides snapshots of the enticingness of CDTs for one round of such a game.
Repeated experiments, variations of CDTs, identification of contextual factors
and skill levels are necessary to account for these dynamics. % the influence of
% and control over participants' expertise

% ==============================================================================
% ------------------------------------------------------------------------------

\textbf{The trade-off between template-like queries for controllable experiments and
  the need for diverse, neutral queries.}
Three participants noted that queries often looked similar,
encouraging them to remember the differences between them,
which is not what we want to measure.
Ideally, template-like queries are preferred,
providing control over specific elements to isolate the effect of a CDT.
Nevertheless, a greater variety of pre-tested neutral queries are essential for future experiments.
Our \anon[open-source dataset]{\href{\repositoryurl}{open-source dataset}}
can serve as a starting point to build such a collection.

% ==============================================================================
% ------------------------------------------------------------------------------

\textbf{Participants enjoyed that Honeyquest felt like a game,
  but also felt an urge to answer queries correctly.}
% We received a diverse set of feedback on how Honeyquest felt to participants.
% Four participants reported that they found our queries quite challenging
% and that they tried to ``avoid making mistakes'' when answering queries,
% which led them to spend a lot of time on each query.
Although participants knew that we did not score accuracy,
four of them reported that they still felt the urge to ``get it right''
and ``avoid making mistakes'', which led them to spend a lot of time on each query.
This phenomenon was reported by participants of all skill levels.
A time limit within which a query must be answered % strict
might be beneficial in future experiments to prevent this behavior.

Overall, participants enjoyed the ``different'' and ``game-like'' experience of Honeyquest.
We believe that gamification aspects, as also explored in studies on cyber security education%
~\cite{Luh2020:PenQuestGamifiedAttacker,Luh2022:PenQuestReloadedDigital},
are a promising avenue to explore further in studies on cyber deception.

% ==============================================================================
% ------------------------------------------------------------------------------

\textbf{The tutorial and our risky elements proved beneficial as an integrated skill check.}
Results of surveys like ours can easily be distorted by responses from incompetent participants;
hence, we chose two populations with proven expertise to address this.
However, this is more difficult to control in anonymous populations%
~\cite{Aljohani2022:PitfallsEvaluatingCyber}.
Honeyquest can mitigate such problems by removing answers
from participants who barely recognized risks or who failed the tutorial.

\textbf{Some CDTs and true risks can be difficult to distinguish from each other.}
The tutorial (Listing~\ref{lst:tutorial-06}) taught participants
that a final distinction may only be possible by knowing their actual implementation.
% Still, we have reached a limit of what Honeyquest can represent.
%
% We believe that dissecting this problem is a valuable avenue to explore further.
We wonder what properties of CDTs can be represented by % qualitative
questionnaires, and what can only be
represented by CTF experiments or honeypots.

\section{Related Work}
\label{sec:related-work}

% =================================================================================================

\subsection{Honeypots and Honeytokens}
\label{sec:honeypots-honeytokens}

\citeauthor{Spitzner2003:HoneypotsCatchingInsider}
was among the first to introduce honeypots as a measure against insider threats%
~\cite{Spitzner2003:HoneypotsCatchingInsider,Spitzner2003:HoneytokensOtherHoneypot}.
He describes honeypots as ``an information system resource whose value lies in
unauthorized or illicit use of that resource''%
~\cite{Spitzner2003:HoneypotsCatchingInsider}.
Most of the honeypot software that has been researched in the last decades%
~\cite{Nawrocki2016:SurveyHoneypotSoftware,Franco2021:SurveyHoneypotsHoneynets}
focuses on emulating protocols, processes, machines, or entire networks%
~\cite{Provos2004:VirtualHoneypotFramework}.
But, the term ``information system resource'' is broad enough to also cover
honeytokens, which are no computers but rather digital entities.
Their most common forms are:
% A honeytoken, which is not a new concept after all~\cite{Stoll1989:CuckooEggTracking},
% can take on many forms, where one of the most common are:
%
Honeytokens~\cite{Spitzner2003:HoneytokensOtherHoneypot}
and Canarytokens\footnote{\url{https://canarytokens.org}}, % , Honeyaccounts, Honeyprofiles
honeyfiles, -pages, and -urls~\cite{%
  Yuill2004:HoneyfilesDeceptiveFiles,
  Voris2015:FoxTrapThwarting,
  BenSalem2011:DecoyDocumentDeployment,
  Lazarov2016:HoneySheetsWhat,
  Petrunic2015:HoneytokensActiveDefense},
honeypatches%
~\cite{%
  Araujo2014:PatchesHoneyPatchesLightweight,
  Araujo2015:ExperiencesHoneyPatchingActive,
  Araujo2016:EmbeddedHoneypotting}
(silently-patched vulnerabilities that still seem exploitable at the surface),
honeywords% , and -encryption%
~\cite{Juels2013:HoneywordsMakingPasswordCracking}
(can be decrypted with wrong keys and still yield plausible yet incorrect data), and
honeypots, and -ports%
~\cite{Provos2004:VirtualHoneypotFramework}, e.g., classic SSH honeypots.

% ==============================================================================
% ------------------------------------------------------------------------------

\subsection{Taxonomies and Classifications}

Numerous taxonomies and classifications of deception techniques have been % (cyber)
introduced, adapted, and surveyed in the past decades%
~\cite{%
  Han2018:DeceptionTechniquesComputer,
  Zhang2021:ThreeDecadesDeception,
  Fraunholz2018:DemystifyingDeceptionTechnology}.
\citeauthor{Whaley1982:GeneralTheoryDeception}~\cite{Whaley1982:GeneralTheoryDeception}
proposed one of the first military-focused theories on (non-cyber) deception
back in~\citeyear{Whaley1982:GeneralTheoryDeception}, which still influenced
cyber deception taxonomies in \citeyear{Rowe2004:TwoTaxonomiesDeception}, as
introduced by \citeauthor{Rowe2004:TwoTaxonomiesDeception}%
~\cite{Rowe2004:TwoTaxonomiesDeception,Rowe2016:IntroductionCyberdeception}.
\citeauthor{Yuill2006:UsingDeceptionHide}~\cite{Yuill2006:UsingDeceptionHide}
described processes, principles, and techniques to hide things from adversaries.
Later work by \citeauthor{Mokube2007:HoneypotsConceptsApproaches}%
~\cite{Mokube2007:HoneypotsConceptsApproaches}
in~\citeyear{Mokube2007:HoneypotsConceptsApproaches} and
\citeauthor{Almeshekah2014:PlanningIntegratingDeception}%
~\cite{Almeshekah2014:PlanningIntegratingDeception}
in~\citeyear{Almeshekah2014:PlanningIntegratingDeception}
focused more closely on the technical aspects and human biases of cyber deception.
Recent work like the one by \citeauthor{Zhang2021:ThreeDecadesDeception}%
~\cite{Zhang2021:ThreeDecadesDeception} in \citeyear{Zhang2021:ThreeDecadesDeception}
aligned taxonomies closer to the cyber kill chain model
and illustrated proposals on deception lifecycles.

Many surveys on cyber deception have been conducted%
~\cite{%
  Han2018:DeceptionTechniquesComputer,
  Fraunholz2018:DemystifyingDeceptionTechnology,
  Fan2018:EnablingAnatomicView,
  Lu2020:CyberDeceptionComputer,
  Bringer2012:SurveyRecentAdvances,
  Mohan2022:LeveragingComputationalIntelligence,
  Qin2023:HybridCyberDefense,
  Javadpour2024:ComprehensiveSurveyCyber},
specifically, ones with a focus on honeypot software%
~\cite{Nawrocki2016:SurveyHoneypotSoftware},
on securing web applications~\cite{Efendi2019:SurveyDeceptionTechniques},
on application layer deception~\cite{Kahlhofer2024:ApplicationLayerCyber},
on IoT~honeypots~\cite{Franco2021:SurveyHoneypotsHoneynets},
or, on approaches using game theory and machine learning%
~\cite{Zhu2021:SurveyDefensiveDeception,Pawlick2019:GametheoreticTaxonomySurvey}.

Closely related,
\citeauthor{Han2018:DeceptionTechniquesComputer}%
~\cite{Han2018:DeceptionTechniquesComputer}
examined how the efficiency and effectiveness of a wide range of CDTs have been
evaluated in the past~(\secname\ref{sec:evaluation}).
\citeauthor{Zhu2021:SurveyDefensiveDeception}~\cite{Zhu2021:SurveyDefensiveDeception}
summarized how approaches that use game theory and machine learning have been evaluated.

The OWASP AppSensor project~\cite{Watson2015:AppSensorGuideApplicationSpecific},
despite only partially addressing deception, serves as a hallmark for
how to describe and taxonomize runtime application defense techniques
and also inspired us to propose HoneYAML.

\subsection{Evaluating Cyber Deception}
\label{sec:evaluation}

\citeauthor{Han2018:DeceptionTechniquesComputer} proposed four aspects
for evaluating CDTs:~\cite{Han2018:DeceptionTechniquesComputer}
\begin{enumerate}
  \item \textbf{Deception placement strategies} have been evaluated by simulation%
        ~\cite{Gavrilis2007:FlashCrowdDetection,Alford2021:CausalModelsAdversary},
        or by studies with students%
        ~\cite{Voris2015:FoxTrapThwarting,BenSalem2011:DecoyDocumentDeployment}.
  \item \textbf{Plausibility and realism of deception}, i.e., measuring how well
        deceptive assets are discernible from genuine assets.
        \citeauthor{Zhu2021:SurveyDefensiveDeception}~\cite{Zhu2021:SurveyDefensiveDeception}
        structured these evaluation testbeds into real testbeds,
        and ones based on probability models, simulation models~\cite{%
          Garg2007:DeceptionHoneynetsGameTheoretic,
          Underbrink2016:EffectiveCyberDeception,
          Schlenker2018:DeceivingCyberAdversaries,
          Wu2020:EffectivenessEvaluationMethod,
          Niakanlahiji2020:HoneyBugPersonalizedCyber},
        and emulation models~\cite{%
          Achleitner2016:CyberDeceptionVirtual,
          Achleitner2017:DeceivingNetworkReconnaissance,
          Acosta2020:CybersecurityDeceptionExperimentation}.
  \item \textbf{Effectiveness of deception}, i.e., measuring if it achieves its
        desired functionality. While ``desired functionality'' is open
        to interpretation, such experiments are generally conducted in either
        confined or natural environments.
  \item \textbf{Accuracy and false-positive rate~(FPR) of deception.} While there are a few studies%
        ~\cite{BenSalem2011:DecoyDocumentDeployment,Han2017:EvaluationDeceptionBasedWeb}
        that tried to evaluate this, it is often impossible in most contexts
        because the base-rate of attacks~\cite{Axelsson2000:BaseRateFallacyDifficulty}
        cannot be measured reliably.
\end{enumerate}

% ==============================================================================
% ------------------------------------------------------------------------------

\citeauthor{Fraunholz2018:DemystifyingDeceptionTechnology}%
~\cite{Fraunholz2018:DemystifyingDeceptionTechnology}
summarized 14 studies that evaluated CDTs in natural environments with field studies.
The rest of this section covers studies in confined environments.
References in parentheses name the CDT that we replicated in our query design
(Table~\ref{tab:results-deceptive-overview}).

% Closely related,
The work from
\citeauthor{Sahin2022:ApproachGenerateRealistic}%
~\cite{Sahin2022:ApproachGenerateRealistic}
is closest to ours.
They also evaluate the enticingness of CDTs~(\secname\ref{sec:discuss-enticing-deception}),
but not their ability to be defensive~(\secname\ref{sec:discuss-defensive-deception}).
They automatically generate realistic HTTP parameters
for web application layer deception, and evaluated it with a survey where developers
were given the link to an actual Swagger UI of a web application. % directly
The Swagger UI showed the available API endpoints, but without the possibility
to interact with the application. Their
``automatically generated parameters names were as realistic as manually selected ones''.
Similar work from the same authors provides an extensive list of CDTs for web applications,
which they evaluated with a CTF challenge and questionnaires%
~\cite{Sahin2020:LessonsLearnedSunDEW}.

The HackIT tool%
~\cite{Aggarwal2019:HackITHumanintheLoopSimulation,
  Aggarwal2020:HackITRealTimeSimulation,
  Aggarwal2021:DecoysCybersecurityExploratory}
by \citeauthor{Aggarwal2019:HackITHumanintheLoopSimulation}
is conceptually similar to Honeyquest.
It enables researchers to map real-world cyberattack scenarios into game-like environments.
Unlike our queries, their scenarios require manual design.
However, this makes experiments more flexible.
% However, this allows for more fine-grained experiments over contextual factors.
%
\citeauthor{Chadha2016:CyberVANCyberSecurity} proposed the related CyberVAN tool~\cite{%
  Chadha2016:CyberVANCyberSecurity,
  Aggarwal2022:HumanSubjectExperimentsRiskBased,
  Aggarwal2022:DesigningEffectiveMasking,
  Aggarwal2020:ExploratoryStudyMasking}
which allows for a speedy and flexible setup of network-based deception experiments.
Game-like environments are also used for cyber security education, e.g., with PenQuest%
~\cite{Luh2020:PenQuestGamifiedAttacker,Luh2022:PenQuestReloadedDigital}.

\citeauthor{Rowe2006:FakeHoneypotsDefensive}%
~\cite{Rowe2006:FakeHoneypotsDefensive,Rowe2007:DefendingCyberspaceFake}
\RowRef{Results}{DcptFilesystemRowe}
conducted an experiment where 14 humans were shown pairs of ``real'' and ``fake'' file listings.

\citeauthor{Ferguson-Walter2019:TularosaStudyExperimental}%
~\cite{Ferguson-Walter2019:TularosaStudyExperimental}
carried out a large study on network deception with
130 professional red teamers in a two-day exercise and
\citeauthor{Shade2020:MoonrakerStudyExperimental}%
~\cite{Shade2020:MoonrakerStudyExperimental}
focused on host-based deception with 59 computer specialists.
Unlike our work, they focused on honeypots.
They are also one of the few authors who divided participants into four cohorts,
based on knowledge about deception (informed, uninformed)
and presence of deception (present, not present)%
~\cite{Ferguson-Walter2021:ExaminingEfficacyDecoybased}.

\citeauthor{Nikiforakis2011:ExposingLackPrivacy}%
~\cite{Nikiforakis2011:ExposingLackPrivacy}
\RowRef{Results}{DcptFilesystemCardrz,DcptFilesystemCustomerList,DcptFilesystemSpamList}
demonstrated that attackers are actively searching
for sensitive files on public file hosting services.
They placed honeyfiles on them and recorded downloads
from 80 unique IP addresses within one month.

\citeauthor{Petrunic2015:HoneytokensActiveDefense}%
~\cite{Petrunic2015:HoneytokensActiveDefense}
\RowRef{Results}{DcptNetworkrequestsAdminFalse}
suggested adding a seemingly deceptive \mbox{``Admin=false''} GET parameter to URLs,
which would presumably only ever be changed by an attacker.

\citeauthor{Han2017:EvaluationDeceptionBasedWeb}%
~\cite{Han2017:EvaluationDeceptionBasedWeb}
\RowRef{Results}{%
  DcptHttpheadersAdminCookie,
  DcptNetworkrequestsSessidParameter,
  DcptNetworkrequestsAdminFalse,
  DcptNetworkrequestsSystemParameter}
held a CTF game with 258 participants on a CMS system, % publicly exposed
where a transparent reverse proxy injected deceptive elements.
They primarily evaluated placement strategies rather than specific techniques.

\citeauthor{Sahin2020:LessonsLearnedSunDEW}%
~\cite{Sahin2020:LessonsLearnedSunDEW}
\RowRef{Results}{%
  DcptHttpheadersAdminCookie,
  DcptNetworkrequestsIdorReadSecrets}
used questionnaires and a CTF game (98 players)
to evaluate their deception framework SunDEW.
% a deception framework that provides self-defense capabilities to web applications.
Other work by \citeauthor{Sahin2022:MeasuringDevelopersWeb}%
~\cite{Sahin2022:MeasuringDevelopersWeb}
\RowRef{Results}{%
  DcptHttpheadersAdminCookie,
  DcptNetworkrequestsCleartextPassword,
  DcptNetworkrequestsPathTraversal,
  DcptNetworkrequestsUnescapedJavascript,
  DcptNetworkrequestsSystemParameter,
  DcptNetworkrequestsUnescapedJson}
used questionnaires (21 participants), and a CTF game (82 players)
to evaluate developers' familiarity with web attack and defense mechanisms.

% ==============================================================================
% ------------------------------------------------------------------------------

\subsection{Effective Cyber Deception}

In \citeyear{Tirenin1999:ConceptStrategicCyber},
\citeauthor{Tirenin1999:ConceptStrategicCyber}%
~\cite{Tirenin1999:ConceptStrategicCyber}
were one of the first authors to suggest that deception % effective
must be dynamic in order to be effective, i.e.,
``it must present a continually-changing situational picture to the enemy''%
~\cite{Tirenin1999:ConceptStrategicCyber}.
\citeauthor{Cohen2006:UseDeceptionTechniques}%
~\cite{Cohen2006:UseDeceptionTechniques}
motivated the need for a link between social sciences and technological development.
In \citeyear{Bowen2010:AutomatingInjectionBelievable},
\citeauthor{Bowen2010:AutomatingInjectionBelievable}%
~\cite{Bowen2010:AutomatingInjectionBelievable}
proposed a ``Decoy Turing Test'' that tasks humans to discern real from decoy network traffic.

\citeauthor{Bercovitch2011:HoneyGenAutomatedHoneytokens}%
~\cite{Bercovitch2011:HoneyGenAutomatedHoneytokens}
developed HoneyGen in \citeyear{Bercovitch2011:HoneyGenAutomatedHoneytokens}
to generate honeytokens by mining characteristics from real data.
Recent research has shifted towards creating dynamic and personalized CDTs%
~\cite{Gonzalez2020:DesignDynamicPersonalized}, e.g., by profiling attacker behavior%
~\cite{Niakanlahiji2020:HoneyBugPersonalizedCyber}.

Many works that followed examined psychological aspects
and decision-making processes of attackers%
~\cite{%
  Ferguson-Walter2023:CyberExpertFeedback,
  Ferguson-Walter2019:TularosaStudyExperimental,
  Ferguson-Walter2021:ExaminingEfficacyDecoybased,
  Ferguson-Walter2020:EmpiricalAssessmentEffectiveness,
  Ferguson-Walter2019:WorldCTFNot,
  Huang2022:ADVERTAdaptiveDataDriven,
  Cranford2018:LearningCyberDeception,
  Cranford2021:CognitiveTheoryCyber,
  Cranford2020:AdaptiveCyberDeception,
  Gutzwiller2024:ExploratoryAnalysisDecisionMaking,
  Gutzwiller2018:OhLookButterfly,
  Gabrys2023:EmotionalStateClassification,
  Gonzalez2020:DesignDynamicPersonalized,
  BenSalem2011:DecoyDocumentDeployment}.
\citeauthor{Ferguson-Walter2020:EmpiricalAssessmentEffectiveness}%
~\cite{%
  Ferguson-Walter2023:CyberExpertFeedback,
  Ferguson-Walter2021:ExaminingEfficacyDecoybased}
showed that cyber deception affects an attacker's cognitive and emotional state,
and that CDTs are effective even if attackers are aware of their use
or merely believe it may be in use.
\citeauthor{Gonzalez2020:DesignDynamicPersonalized}
~\cite{Gonzalez2020:DesignDynamicPersonalized}
found that attackers exhibit irrational behavior that leads to cognitive biases.
Similarly,
\citeauthor{Gabrys2023:EmotionalStateClassification}%
~\cite{Gabrys2023:EmotionalStateClassification}
observed a strong correlation between the emotional state of an attacker
% (e.g., confusion and self-doubt)
(confusion, self-doubt, confidence, frustration, and surprise)
and the frequency of their reconnaissance activity.

% IGNORED WORKS
%
% static vs. adaptive cyber deception       => to much in-depth, maybe something for future work
% moving-target defense                     => same as above
% deception placement                       => same as above
% cyber deception lifecycle                 => too much theory that we aren't interested in
% https://engage.mitre.org/                 => didn't look so valuable after all at first glance, maybe revisit later

\section{Future Work}
\label{sec:future-work}

Future work may include enriching Honeyquest with more CTDs
and a way to evaluate deception placement strategies%
~\cite{%
  Voris2015:FoxTrapThwarting,
  BenSalem2011:DecoyDocumentDeployment},
teaching ML models to design CDTs%
~\cite{%
  Araujo2019:ImprovingIntrusionDetectors,
  Ayoade2020:AutomatingCyberdeceptionEvaluation},
mining our \anon[open-source dataset]{\href{\repositoryurl}{open-source dataset}}
for interesting patterns%
~\cite{%
  Sahin2022:MeasuringDevelopersWeb,
  Niakanlahiji2020:HoneyBugPersonalizedCyber,
  Aljohani2022:PitfallsEvaluatingCyber},
incorporating cognitive models into the experiment design%
~\cite{Gutzwiller2018:OhLookButterfly},
embedding educational aspects into Honeyquest%
~\cite{Luh2020:PenQuestGamifiedAttacker,Luh2022:PenQuestReloadedDigital},
implementing more query types, e.g., ``robots.txt'' files%
~\cite{%
  Fraunholz2018:DefendingWebServers,
  Fraunholz2018:CloxyContextawareDeceptionasaService,
  Han2017:EvaluationDeceptionBasedWeb},
evaluating CDTs that secure non-web applications,
adapting Honeyquest to deceive vulnerability scanners%
~\cite{Angeli2024:FalseFlavorHoneypot},
and, of course, replicating our results in more experiments and real-world deployments.

\section{Conclusion}
\label{sec:conclusion}

% =================================================================================================

This work proposes a method to measure the enticingness of CDTs.
We demonstrate its feasibility for four aspects of a web application,
where we designed \VarNumCDTs{}~CDTs and \VarNumRisks{}~risks,
for an experiment with a high-quality sample of \VarNumParticipants{} humans % security-aware
(\VarNumParticipantsCtf{}~CTF players, \VarNumParticipantsRes{}~professionals).
Our results provide a detailed overview of the enticingness of CDTs
(Table~\ref{tab:results-deceptive-overview})
and show that deception can reduce the risk of finding a true risk by about % genuine
\VarResBeforeAfterOverallRiskReduction{}\% on average.
Knowing such statistics, e.g., that humans fall for traps about \VarResFellTpDcpt{}\% of the time,
enables researchers to back up their theoretical models with our empirical numbers.
Notably, we were able to replicate the goals of previous work with many consistent findings
(\secname\ref{sec:discuss-replicated-work}),
but without a time-consuming implementation on real computer systems.
This strengthens the generalizability of our method to the real world.

%%
%% acknowledgments
%% author names and affiliations
\begin{acks}
  We thank Markus, Olivier, Patrick, Simon, Carlo, Alex, and Chris
  for pre-testing our prototype, Alex for proofreading the paper,
  the many anonymous reviewers for their very constructive and valuable feedback, % on our work,
  and all the volunteers who participated in our experiment.
\end{acks}

\clearpage

%%
%% [IEEEtran & usenix] bibliography
% \bibliographystyle{plain}
% {\small\hfuzz=2pt\bibliography{abbrev,references}}

%%
%% [ACM] bibliography
\bibliographystyle{ACM-Reference-Format}
{\hfuzz=2pt\bibliography{abbrev,references}}

% \clearpage

%%
%% appendix
\appendix
% =================================================================================================

\section{Expressing Results with Typical Confusion Matrices}
\label{sec:appendix:confusion-matrices}

% The task of Honeyquest's participants is partly a binary classification problem.
% Thus, it is helpful to express the obtained results using typical confusion matrices.

\textbf{\ntrli~Neutral queries.} \quad
Consider that a user anwers a \ntrli~neutral query ${q_N \in Q_N}$.
Neutral queries never have line annotations.
If the user places no marks on $q_N$, we consider this as a ``true negative''~(TN) classification.
If the user places some, we say this is a ``false positive''~(FP) classification.

\textbf{\dcpti~Deceptive and \riski~risky queries.} \quad
As defined in Equation~\eqref{eq:marks-intersect-lines}, we say that
``answer marks~$A$ match line annotations~$L$'' if they intersect each other.
Given a deceptive query \dcpti~$q_D \in Q_D$, where \trapi~trap marks~$A_{Tr}$
match deceptive lines~$L_D$, we say this as a ``true positive''~(TP) classification.
If they do not match, we count a ``false negative''~(FN) instead.
Given a risky query \riski~$q_R \in Q_R$, the same rule applies,
but \expli~exploit marks~$A_{Ex}$ are matched against risky lines~$L_R$ instead.
There are at least four ways to match one of
the two kinds of answer marks against one of the two kinds of line annotations.
If we want to assess if participants fell for a trap, we would adapt the previous formulation
to match \expli~exploit marks~$A_{Ex}$ against deceptive line annotations~$L_D$ instead.
In all cases, we can arrange the counts in a confusion matrix.
Table~\ref{tab:cm-aspect-deceptive} shows this for the aforementioned formulation.
Then, we can derive metrics such as accuracy, precision, or recall.
Table~\ref{tab:results-aspect-a} shows the confusion matrices of
our experiment, whose results we also presented in \secname\ref{sec:results}.

\begin{table}[htb]
  \centering
  \caption{Matching answer marks and line annotations.}
  \Description{
    A table with two rows and two columns.
    The first cell (true negative) holds the formula $A = \varnothing$.
    The second cell (false positive) holds the formula $A \neq \varnothing$.
    The third cell (false negative) holds the formula $L \cap A = \varnothing$.
    The fourth cell (true positive) holds the formula $L \cap A \neq \varnothing$.
  }
  \label{tab:cm-aspect-deceptive}
  \begin{tabularx}{\columnwidth}{l @{\extracolsep{\fill}} r @{\hskip 1pt} r p{15pt} @{\extracolsep{\fill}} r @{\hskip 1pt} r}
    \addlinespace[0.1cm] % [usenix] add some extra margin
    \toprule
    \addlinespace[0.15cm]
    $q \in Q_N$    & \framebox{\small TN} & $A = \varnothing$        &  & \framebox{\small FP} & $A \neq \varnothing$        \\
    \addlinespace[0.1cm]
    $q \notin Q_N$ & \framebox{\small FN} & $L \cap A = \varnothing$ &  & \framebox{\small TP} & $L \cap A \neq \varnothing$ \\
    \addlinespace[0.1cm]
    \bottomrule
  \end{tabularx}
\end{table}

\begin{table}[htb]
  \begin{threeparttable}
    \setlength{\tabcolsep}{4pt}
    % \small
    \centering
    \caption{Results on the enticingness of traps and risks.}
    \Description{
      For deceptive queries, there are
      \VarResCmDcptTn{} true negatives, \VarResCmDcptFp{} false positives,
      \VarResCmDcptFn{} false negatives, and \VarResCmDcptTp{} true positives.
      For risky queries, there are
      \VarResCmRiskTn{} true negatives, \VarResCmRiskFp{} false positives,
      \VarResCmRiskFn{} false negatives, and \VarResCmRiskTp{} true positives.
    }
    \label{tab:results-aspect-a}
    \begin{tabularx}{\columnwidth}{c @{\hspace{8pt}} @{\extracolsep{\fill}} rrrr rrrr}
      \toprule
      Qry.   & TN                & FP                & FN                & TP                & ACC                  & PPV                  & TPR                  & FPR                  \\
      \midrule
      \dcpti & \VarResCmDcptTn{} & \VarResCmDcptFp{} & \VarResCmDcptFn{} & \VarResCmDcptTp{} & \VarResCmDcptAcc{}\% & \VarResCmDcptPpv{}\% & \VarResCmDcptTpr{}\% & \VarResCmDcptFpr{}\% \\
      \riski & \VarResCmRiskTn{} & \VarResCmRiskFp{} & \VarResCmRiskFn{} & \VarResCmRiskTp{} & \VarResCmRiskAcc{}\% & \VarResCmRiskPpv{}\% & \VarResCmRiskTpr{}\% & \VarResCmRiskFpr{}\% \\
      \bottomrule
    \end{tabularx}
    \begin{tablenotes}
      \footnotesize
      \item \textbf{Description:} Confusion matrix and metrics on how well % derived metrics
      users are enticed by traps and risks. The opposing class was always a neutral query.
    \end{tablenotes}
  \end{threeparttable}
\end{table}

% ==============================================================================
% ------------------------------------------------------------------------------

\section{Alternative Matching of Answer Marks and Line Annotations}
\label{sec:appendix:matching-criteria}

There are five mutually exclusive variations on how answer marks~$A$
can possibly intersect with (non-empty) line annotations ${L \neq \varnothing}$:
\begin{align}
  \tag{A1} A                                 & = L              &  & \text{marked $L$ exactly}             \\
  \tag{A2} A \neq \varnothing \land A        & \subset L        &  & \text{marked $only$ some in $L$}      \\
  \tag{A3} A \not\subseteq L \land L \cap A  & \neq \varnothing &  & \text{marks $only$ overlap with $L$}  \\
  \tag{A4} A \neq \varnothing \land L \cap A & = \varnothing    &  & \text{marked $only$ lines not in $L$} \\
  \tag{A5} A                                 & = \varnothing    &  & \text{no marks, but non-empty $L$}
\end{align}

The criterion in \secname\ref{sec:matching-marks}
assumes that answer marks match line annotations when lines are marked
exactly~(A1) or partially~(A2), while also allowing overlaps~(A3)
with other (not-annotated) lines. % (Table~\ref{tab:set-logic}).
In all three cases, it is valid to imply that
a user at least partially identified a risk or a trap.

% \begin{table}[htb]
%   \begin{threeparttable}
%     % \small
%     \centering
%     \caption{Line annotation and answer mark intersection.}
%     \label{tab:set-logic}
%     \begin{tabularx}{\columnwidth}{ cc @{\extracolsep{\fill}} ccccc }
%       \toprule
%       \# & $L_D~/~L_R$     & (A1)       & (A2)       & (A3)       & (A4)       & (A5)       \\
%       %  &                 & $\cap$     & $\cap$     & $\cap$     &            &            \\
%       \midrule
%       1  & \dcpti~/~\riski & \xmark     & \textminus & \textminus & \textminus & \textminus \\
%       2  & \dcpti~/~\riski & \xmark     & \xmark     & \xmark     & \textminus & \textminus \\
%       3  & \ntrli~neutral  & \textminus & \textminus & \xmark     & \xmark     & \textminus \\
%       \bottomrule
%     \end{tabularx}
%     \begin{tablenotes}
%       \footnotesize
%       \item \textbf{Description:}
%       This example illustrates a query with three lines where the
%       first two lines are annotated (as risky or deceptive).
%     \end{tablenotes}
%   \end{threeparttable}
% \end{table}

% ==============================================================================
% ------------------------------------------------------------------------------

\section{Details on Counting Answer Marks}
\label{sec:appendix:counting-marks}

Aspect~A (\secname\ref{sec:aspect-enticingness})
required a more concrete formalization of the matching criteria
of \secname\ref{sec:matching-marks}.
We grouped answers to our queries by query type, and computed the following counts:

% \ntrli~\textbf{Neutral queries}~(Table~\ref{tab:results-neutral-overview}).
% test if humans are not enticed to place any marks at all.

% consistent width for first words
\newlength\acmiwidth
\settowidth{\acmiwidth}{$1234$}

\begin{itemize}
  \item % [ntrl_fp_hack]
        \makebox[\acmiwidth][l]{$n_{Ex}$} \enspace
        \emph{How often were neutral lines mistaken for traps?}
        Number of answers that \emph{only} received \expli~exploit marks~$A_{Ex}$.
  \item % [ntrl_fp_trap]
        \makebox[\acmiwidth][l]{$n_{Tr}$} \enspace
        \emph{How often were neutral lines mistaken for risks?}
        Number of answers that \emph{only} received \trapi~trap marks~$A_{Tr}$.
  \item % [ntrl_fp_mark]
        \makebox[\acmiwidth][l]{$n_\land$} \enspace
        \emph{How often were neutral lines mistaken for risks and traps in the same answer?}
        Number of answers that received both \expli~exploit marks~$A_{Ex}$ and \trapi~trap marks~$A_{Tr}$
  \item % [ntrl_tn]
        \makebox[\acmiwidth][l]{$n_\varnothing$} \enspace
        \emph{How often have humans not reacted to neutral lines?} % at all
        Number of answers to neutral queries without any marks.
        \vskip12pt
        % \end{itemize}

        % \dcpti~\textbf{Deceptive queries}~(Table~\ref{tab:results-deceptive-overview}).
        % test if humans are enticed to place \expli~exploit marks but no \trapi~trap marks on~$L_D$.

        % \begin{itemize}
  \item % [fell_tp]
        \makebox[\acmiwidth][l]{$d_{Ex}$} \enspace
        \emph{How often fell humans for traps?}
        Number of answers, where \expli~exploit marks~$A_{Ex}$ match deceptive lines~$L_D$.
  \item % [trap_tp]
        \makebox[\acmiwidth][l]{$d_{Tr}$} \enspace
        \emph{How often were traps detected?}
        Number of answers, where \trapi~trap marks~$A_{Tr}$ match deceptive lines~$L_D$.
  \item % [dcpt_fn_rest]
        \makebox[\acmiwidth][l]{$d_\triangle$} \enspace
        \emph{How often have humans reacted to other lines?} % at all
        Number of answers with marks but no match on decept. lines~$L_D$.
  \item % [dcpt_fn_skip]
        \makebox[\acmiwidth][l]{$d_\varnothing$} \enspace
        \emph{How often have humans not reacted to traps?} % at all
        Number of answers to deceptive queries without any marks.
        \vskip12pt
        % \end{itemize}

        % \riski~\textbf{Risky queries}~(Table~\ref{tab:results-risky-overview}).
        % test if humans are enticed to place \expli~exploit marks on~$L_R$.

        % \begin{itemize}
  \item % [hack_tp]
        \makebox[\acmiwidth][l]{$r_{Ex}$} \enspace
        \emph{How often were risks detected?}
        Number of answers, where \expli~exploit marks~$A_{Ex}$ match risky lines~$L_R$.
  \item % [new_tp]
        \makebox[\acmiwidth][l]{$r_{Tr}$} \enspace
        \emph{How often were risks mistaken for traps?}
        Number of answers, where \trapi~trap marks~$A_{Tr}$ match risky lines~$L_R$.
  \item % [risk_fn_rest]
        \makebox[\acmiwidth][l]{$r_\triangle$} \enspace
        \emph{How often have humans reacted to other lines?} % at all
        Number of answers with marks but no match on risky lines~$L_R$.
  \item % [risk_fn_skip]
        \makebox[\acmiwidth][l]{$r_\varnothing$} \enspace
        \emph{How often have humans not reacted to risks?} % at all
        Number of answers to risky queries without any marks.
\end{itemize}

% ==============================================================================
% ------------------------------------------------------------------------------

\section{Aligning Prior Work to Our Cyber Deception Techniques}
\label{sec:appendix:prior-work-mapping}

This section describes how we mapped our CDTs (Table~\ref{tab:results-deceptive-overview})
to techniques from prior work~\citereplicatedworks{}. A comparison and discussion of the
results can be found in \secname\ref{sec:discuss-replicated-work}.

% ------------------------------------------------------------------------------

\textbf{\citeauthor{Nikiforakis2011:ExposingLackPrivacy}}%
~\cite{Nikiforakis2011:ExposingLackPrivacy}
\RowRef{Results}{%
  DcptFilesystemCardrz,
  DcptFilesystemCustomerList,
  DcptFilesystemSpamList}
\enspace
Their experiment placed six files on public file hosting services.
We randomly injected three of those, whose names seem most likely to represent a real weakness,
in some of our \filesystem{}~queries:
``SPAM\_list.pdf''~\RowRef{Results}{DcptFilesystemSpamList},
``customer\_list\_2010.html''~\RowRef{Results}{DcptFilesystemCustomerList},
``card3rz\_reg\_details.html''~\RowRef{Results}{DcptFilesystemCardrz}.
We omitted the other names
(``phished\_paypal\_details.html'',
``Paypal\_account\_gen.exe'',
``Sniffed\_email1.doc'')
because these names sound more like they would only be found on an
adversary's computer and not on a real server.

% ------------------------------------------------------------------------------

\textbf{\citeauthor{Petrunic2015:HoneytokensActiveDefense}}%
~\cite{Petrunic2015:HoneytokensActiveDefense}
\RowRef{Results}{%
  DcptNetworkrequestsAdminFalse}
\enspace
% Their work suggested to add an additional \mbox{``Admin=false''} parameter to URLs.
We randomly appended the suggested \mbox{``admin=false''} parameter in our \network{}~queries.

% ------------------------------------------------------------------------------

\textbf{\citeauthor{Han2017:EvaluationDeceptionBasedWeb}}%
~\cite{Han2017:EvaluationDeceptionBasedWeb}
\RowRef{Results}{%
  DcptHttpheadersAdminCookie,
  DcptNetworkrequestsSessidParameter,
  DcptNetworkrequestsAdminFalse,
  DcptNetworkrequestsSystemParameter}
\enspace
Their experiment primarily evaluated placement strategies rather than specific techniques.
We mapped their mention of an ``additional cookie''
to our CDT that injects a deceptive cookie into HTTP headers%
~\RowRef{Results}{DcptHttpheadersAdminCookie}.
We assumed that their mention of ``honey GET parameters''
is similar to our three CDTs that inject parameters into URLs:
\mbox{``admin=false''}~\RowRef{Results}{DcptNetworkrequestsAdminFalse},
\mbox{``SESSID=odq...''}~\RowRef{Results}{DcptNetworkrequestsSessidParameter}, and
\mbox{``system=prod''}~\RowRef{Results}{DcptNetworkrequestsSystemParameter}.

% ------------------------------------------------------------------------------

\textbf{\citeauthor{Rowe2006:FakeHoneypotsDefensive}}%
~\cite{Rowe2006:FakeHoneypotsDefensive,Rowe2007:DefendingCyberspaceFake}
\RowRef{Results}{%
  DcptFilesystemRowe}
\enspace
Their experiment showed humans pairs of ``real'' and ``fake'' file listings.
We used the pair that they illustrated in the paper to create two \filesystem{}~queries:
Listing~\ref{lst:rowe-gaitan} shows the \ntrli~``real'' file listing
and Listing~\ref{lst:rowe-cooper} shows the \dcpti~``fake'' file listing
% (with every line except for the current directory ``.''
% and the parent directoy ``..'' annotated as deceptive).
(with every line except for ``.'' and ``..'' annotated as deceptive).

\begin{lstlisting}[
  label=lst:rowe-gaitan,
  belowskip=0pt,
  caption={
    Representation of ``real'' file listing%
    ~\cite{Rowe2006:FakeHoneypotsDefensive,Rowe2007:DefendingCyberspaceFake}.
  }
]
  drwxr-xr-x 2 gaitan gaitan 4.0K Nov 30 17:42 .
  drwxr-xr-x 4 gaitan gaitan 4.0K Nov 30 17:42 ..
  -rw-r--r-- 1 gaitan gaitan 1.4K Nov 30 17:42 specv
  -rw-r--r-- 1 gaitan gaitan 3.2K Nov 30 17:42 other
  -rw-r--r-- 1 gaitan gaitan 1.1K Nov 30 17:42 rp
  -rw-r--r-- 1 gaitan gaitan 6.7K Nov 30 17:42 gilmore
  -rw-r--r-- 1 gaitan gaitan 1.2K Nov 30 17:42 int
  -rw-r--r-- 1 gaitan gaitan 5.2K Nov 30 17:42 trash
  -rw-r--r-- 1 gaitan gaitan 2.0K Nov 30 17:42 old.imsl
  -rw-r--r-- 1 gaitan gaitan 2.7K Nov 30 17:42 flynn
\end{lstlisting}

\begin{lstlisting}[
  label=lst:rowe-cooper,
  aboveskip=3pt,
  caption={
    Representation of ``fake'' file listing%
    ~\cite{Rowe2006:FakeHoneypotsDefensive,Rowe2007:DefendingCyberspaceFake}.
  }
]
  drwxr-xr-x 2 cooper cooper 4.0K Nov 30 17:42 .
  drwxr-xr-x 4 cooper cooper 4.0K Nov 30 17:42 ..
  -rw-r--r-- 1 cooper cooper 1.2K Nov 30 17:42 examples
  -rw-r--r-- 1 cooper cooper 2.0K Nov 30 17:42 gif_files
  -rw-r--r-- 1 cooper cooper 3.2K Nov 30 17:42 idlold
  -rw-r--r-- 1 cooper cooper 5.2K Nov 30 17:42 wizard
  -rw-r--r-- 1 cooper cooper 2.7K Nov 30 17:42 114564-01
  -rw-r--r-- 1 cooper cooper 1.4K Nov 30 17:42 template
  -rw-r--r-- 1 cooper cooper 1.1K Nov 30 17:42 target_n_horiz
  -rw-r--r-- 1 cooper cooper 6.7K Nov 30 17:42 ass2
\end{lstlisting}

% ------------------------------------------------------------------------------

\textbf{\citeauthor{Sahin2020:LessonsLearnedSunDEW}}%
~\cite{Sahin2020:LessonsLearnedSunDEW}
\RowRef{Results}{%
  DcptHttpheadersAdminCookie,
  DcptNetworkrequestsIdorReadSecrets}
\enspace
Their experiment tested seven CDTs in a CTF experiment.
They had a \mbox{``Username''} and \mbox{``Role''} cookie with similar detection rates
that we mapped to our CDT that injects a deceptive cookie into HTTP headers%
~\RowRef{Results}{DcptHttpheadersAdminCookie}.
They also had a deceptive GET parameter on a \mbox{``/view\_patient/\$id''} endpoint
that we mapped to our CDT with a \mbox{``/secrets/\$id''} endpoint%
~\RowRef{Results}{DcptNetworkrequestsIdorReadSecrets}.

% ------------------------------------------------------------------------------

\textbf{\citeauthor{Sahin2022:MeasuringDevelopersWeb}}%
~\cite{Sahin2022:MeasuringDevelopersWeb}
\RowRef{Results}{%
  DcptHttpheadersAdminCookie,
  DcptNetworkrequestsCleartextPassword,
  DcptNetworkrequestsPathTraversal,
  DcptNetworkrequestsUnescapedJavascript,
  DcptNetworkrequestsSystemParameter,
  DcptNetworkrequestsUnescapedJson}
\enspace
Their experiment recorded 17 attack vectors that participants tried in their CTF experiment.
We designed CDTs for six of them:
``Cross-site scripting'' (found in payloads with \mbox{``<script>''} tags)
as a CDT that adds unescaped JavaScript%
~\RowRef{Results}{DcptNetworkrequestsUnescapedJavascript}.
``Credential guessing'' (found in payloads with clear-text credentials)
as a CDT that adds clear-text passwords%
~\RowRef{Results}{DcptNetworkrequestsCleartextPassword}.
``SQL injection'' (found in payloads with unescaped quotes)
as a CDT that adds unescaped JSON%
~\RowRef{Results}{DcptNetworkrequestsUnescapedJson}.
``Cookie tampering''
as a CDT that adds a deceptive cookie into HTTP headers%
~\RowRef{Results}{DcptHttpheadersAdminCookie}.
``Client-side bypass'' (found by tampering with a ``system'' parameter)
as a CDT that adds a ``system=prod'' parameter into URLs%
~\RowRef{Results}{DcptNetworkrequestsSystemParameter}.
``Path traversal'' (found in payloads with ``..'' strings)
as a CDT that imitates a path traversal vulnerability%
~\RowRef{Results}{DcptNetworkrequestsPathTraversal}.
Lastly, their ``Content-Type header attack'' (found by header tampering)
was not mapped to any of our CDTs, but we counted how many participants
marked lines containing ``Content-Type'' in our queries.

% ==============================================================================
% ------------------------------------------------------------------------------

\section{User Study Details}
\label{sec:appendix:ethics}

\subsection{Experiment Website and Tutorial}

\begin{figure*}[!bt]
  \centering
  \includegraphics[width=\textwidth,page=1,clip,trim= 9cm 12.3cm 9cm 6.25cm]{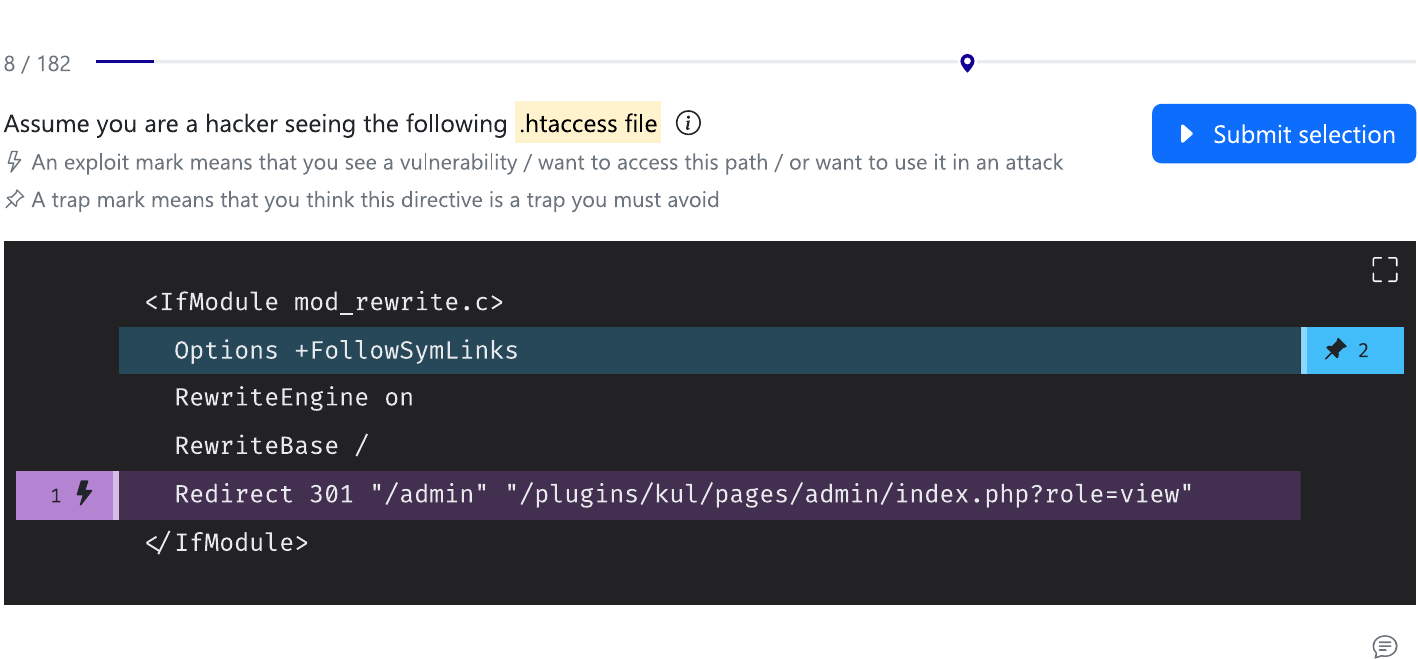}
  \caption[]{
    A screenshot of the web-based user interface of Honeyquest that shows a query of the type
    ``.htaccess file''. In this example, the user placed an \expli~exploit mark on line~5 and a
    \trapi~trap mark on line~2. When hovering over the info icon right next to the query type,
    a tooltip with an extensive description of the syntax in the query is shown.
    The progress bar at the top shows how many of the queries have already been answered.
    The little pin on the progress bar indicates how many queries an average player has answered.
    Users can submit feedback or report mistakes
    by clicking on the speech bubble in the lower right corner.
  }
  \Description{A screenshot of the web-based user interface of Honeyquest.}
  \label{fig:ui}
\end{figure*}

Participants who have agreed to share their data with us (Appendix~\ref{sec:appendix:privacy-form}),
were then directed to a tutorial (Appendix~\ref{sec:appendix:tutorial})
to familiarize them with the experiment.
After answering the profiling questions (Appendix~\ref{sec:appendix:profiling}),
the actual experiment began.
\figurename~\ref{fig:ui} shows the user interface for all subsequent \VarNumQueries{}~queries.

A manual investigation of the answers to the tutorial questions revealed that
all participants understood the interface and the experiment.
This is not surprising, as the tutorial was also pre-tested to ensure that it is understandable.
Two colleagues who did not participate in the actual experiment pre-tested the tutorial.

\subsection{Recruitment Message}
\label{sec:appendix:recruitment-message}

\begin{framed}%
  \itshape\raggedright
  % \textbf{Fellow security-aware people!}
  % \vskip2pt
  Would you lend me some of your valuable time to advance research on cyber deception and prove
  your secure coding skills? We created an interactive game, where you have to think and act
  like you were a hacker: \textbf{LINK} % It should be fun!
  \vskip2pt

  % You can get to the experiment with this link: \textbf{X}
  % \vskip2pt

  If you can participate, please do so by \textbf{DATE}.
  Answering all questions will take you between 30 - 60 minutes.
  But, you can stop any time. You can also continue later. Progress saves automatically. If
  you would like to discuss some queries with us afterwards, leave us a comment with your name.
\end{framed}

\subsection{Data Privacy Consent and Intent Form}
\label{sec:appendix:privacy-form}

\begin{framed}%
  \itshape\raggedright
  Honeyquest is a game where you have to identify security vulnerabilities in
  web applications. Be careful, some of the vulnerabilities are traps trying
  to trick you into thinking something is vulnerable. During the game, we will
  collect some data to help us advance research on cyber deception:
  \vskip2pt

  We store a cookie on your computer to identify you.\textbf{Why?} So that we
  know which answers belong to the same person, even when you continue the game later.
  \vskip2pt

  We store your profile information, like your job, years of experience, and
  skill level. \textbf{Why?} So that we can research, if there are differences
  among professions. % and skill levels.
  \vskip2pt

  We store your answers and the time of your answers. \textbf{Why?} So that we
  can research what kinds of questions humans are good at and what % answering
  kinds of questions are hard to get right. %, or even impossible to get right.
  \vskip2pt

  We do not store your IP address, location, name, email address, or any other PII.
  \textbf{Why?} Because we don't need it. % and think data privacy is important.
\end{framed}

\subsection{Participant Profiles}
\label{sec:appendix:profiling}

\begin{figure}[htb]
  \centering
  \includegraphics[width=\columnwidth,page=1]{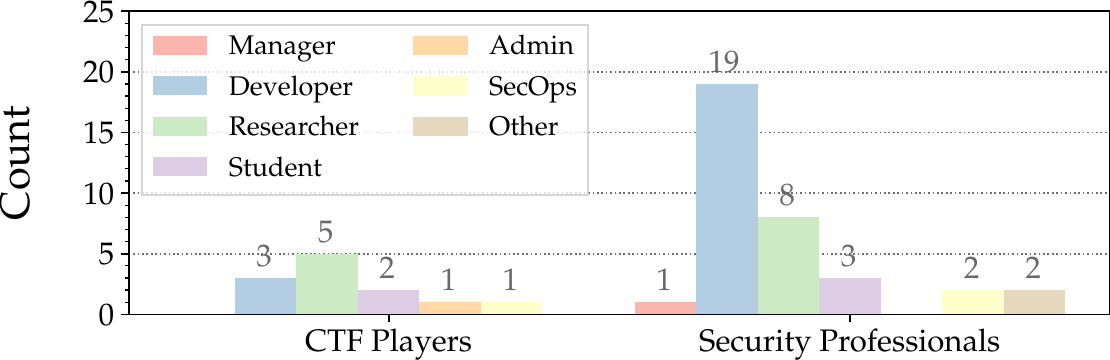} \\
  \vspace{0.75em}
  \includegraphics[width=\columnwidth,page=1]{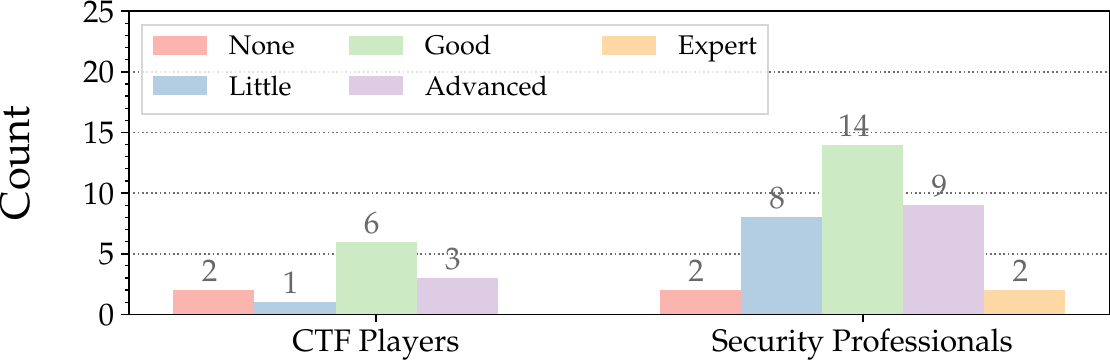} \\
  \vspace{0.75em}
  \includegraphics[width=\columnwidth,page=1]{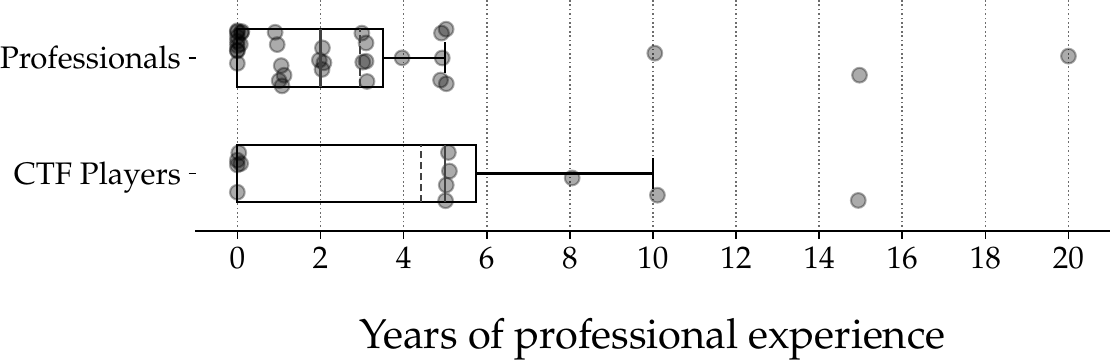}
  \caption{
    Our participants' self-reported profiles.
    % Our participants' self-reported role description, self-assessed secure coding skills,
    % and years of professional experience in the field of cyber security.
    % A small jitter was added to each dot in the boxplot for visual clarity.
  }
  \Description{
    The grouped bar chart on the role description shows two groups.
    Among the CTF players are
    \VarNumParticipantsCtfManagers{}~managers,
    \VarNumParticipantsCtfDevelopers{}~developers,
    \VarNumParticipantsCtfResearcher{}~researchers,
    \VarNumParticipantsCtfStudents{}~students,
    \VarNumParticipantsCtfOps{}~administrators, and
    \VarNumParticipantsCtfSecOps{}~security operations professionals.
    Among the security professionals are
    \VarNumParticipantsResManagers{}~managers,
    \VarNumParticipantsResDevelopers{}~developers,
    \VarNumParticipantsResResearcher{}~researchers,
    \VarNumParticipantsResStudents{}~students,
    \VarNumParticipantsResOps{}~administrators, and
    \VarNumParticipantsResSecOps{}~security operations professionals.
    In both groups, a bell curve with ``good'' skills in the middle can be observed,
    although, the one for the CTF players is more skewed towards lower skills.
    CTF Players reported a mean of \VarNumParticipantsCtfYearsMean{}~and median of
    \VarNumParticipantsCtfYearsMedian{}~years of professional experience.
    Security professionals reported a mean of \VarNumParticipantsResYearsMean{}~and median of
    \VarNumParticipantsResYearsMedian{}~years of professional experience.
  }
  \label{fig:profiles}
\end{figure}

\figurename~\ref{fig:profiles} shows our participant's answers to these questions: % three

\begin{itemize}
  \item What describes your current profession best?
        \begin{itemize}
          \item \textbf{Development:}
                Developer, Engineer, Architect
          \item \textbf{Operations:}
                System Administrator, SRE
          \item \textbf{Security Operations:}
                Penetration Tester, Incident Detection and Response, Product Security
          \item \textbf{Business:}
                Manager, Leader, Sales, Marketing
          \item \textbf{Research:}
                Researcher, Scientist, Innovator
        \end{itemize}
  \item How would you describe your secure coding skills?
        \begin{itemize}
          \item \textbf{None:}
                What do you mean by secure coding?
          \item \textbf{Little:}
                I only heard about a few concepts.
          \item \textbf{Good:}
                I get the basics but still need guidance.
          \item \textbf{Advanced:}
                I apply secure coding concepts regularly.
                % and know where to go to learn more.
          \item \textbf{Expert:}
                I educate others about secure coding.
        \end{itemize}
  \item Roughly, how many years have you been professionally
        involved in the field of cyber security?
\end{itemize}

% Answers are shown in \figurename~\ref{fig:profiles}.
% Participants also had to choose a favorite color,
% which we illustrate in \figurename~\ref{fig:colors}.

% \begin{figure}[htb]
%   \centering
%   \includegraphics[width=0.66\columnwidth,page=1]{figures/fig2d.pdf}
%   \caption{Favorite colors of participants in Honeyquest.}
%   \Description{
%     A colormap with a couple of dots in red, green, black, purple, and yellow shades.
%   }
%   \label{fig:colors}
% \end{figure}

\subsection{Participant Demographics}

We asked participants about their professional role, secure coding skills,
and years of professional experience in the field of cyber security
(\secname\ref{sec:appendix:profiling}).
We did not collect demographic information, but we can describe
the target audience (all of which are located in \anon[ANONYMIZED]{Europe})
to which we posted our recruitment message
(\secname\ref{sec:appendix:recruitment-message}):

\begin{itemize}
  \item \textbf{Security Professionals} were predominantly male and between 20 and 45 years old.
        The majority of them had a university degree in Computer Science.
  \item \textbf{CTF Players} were predominantly male and between 18 and 35 years old.
        The majority of them were graduate Computer Science students.
\end{itemize}

\subsection{Study Timeline and Incentives}

The user study was conducted in two phases, each with exactly the same experimental setup.
The first phase with \VarNumParticipantsPhaseA{}~participants
(\VarNumParticipantsResPhaseA{}~professionals, \VarNumParticipantsCtfPhaseA{}~CTF players)
was held in February 2023.
The second phase with \VarNumParticipantsPhaseB{}~new participants
(\VarNumParticipantsResPhaseB{}~professionals, \VarNumParticipantsCtfPhaseB{}~CTF players)
was held in January 2024.

To increase the number of participants in the second phase,
we introduced an incentive to win a 50€ Amazon gift card,
if they answered at least 50\% of queries.
We promoted this incentive a few days after the second phase started.
\VarNumParticipantsResIncentivized{} of the
\VarNumParticipantsResPhaseB{}~security professionals (in that phase) joined after that promotion.
In the end, \VarNumParticipantsResEligible{}~of them answered
enough queries to qualify for the incentive.

% ==============================================================================
% ------------------------------------------------------------------------------

\subsection{Tutorial Queries}
\label{sec:appendix:tutorial}

Every participant had to answer these 8 tutorial queries prior to the actual experiment.
% The risky and deceptive lines are only highlighted here in the paper, not in Honeyquest.
Lines are only highlighted in the paper.

\begin{lstlisting}[%
  label=lst:tutorial-01,
  frame=single,
  breaklines=true,
  caption={Tutorial query 1~/~8.},
  columns=fullflexible,
  basicstyle=\footnotesize\ttfamily,
  basewidth=0.5em,
  aboveskip=8pt,
  belowskip=2pt
]
  You are reading a tutorial QUERY.
  A query is simply a text of a certain TYPE.
  Honeyquest shows you queries of different types.

  We want to understand how you would respond to them,
  if you act like a hacker or cyber security researcher.
\end{lstlisting}

\begin{lstlisting}[%
  label=lst:tutorial-02,
  frame=single,
  breaklines=true,
  caption={Tutorial query 2~/~8 with neutral HTTP headers.},
  columns=fullflexible,
  basicstyle=\footnotesize\ttfamily,
  basewidth=0.5em,
  aboveskip=0pt,
  belowskip=2pt
]
  The following is a QUERY of type HTTPHEADERS, meaning,
  you observe that an application is making this HTTP request:

  > GET /wiki/Cat HTTP/1.1
  > Host: en.wikipedia.org
  > User-Agent: curl/7.68.0
  > Accept: */*

  Behind the scenes, Honeyquest classified this query as NEUTRAL.
  This means, there is nothing RISKY or DECEPTIVE about it.

  If you agree that this query is NEUTRAL, click the button above.
\end{lstlisting}

\begin{lstlisting}[%
  label=lst:tutorial-03,
  frame=single,
  breaklines=true,
  caption={Tutorial query 3~/~8 with risky ``Server'' header.},
  columns=fullflexible,
  basicstyle=\footnotesize\ttfamily,
  basewidth=0.5em,
  aboveskip=0pt,
  belowskip=2pt
]
  Correct! This query was indeed NEUTRAL.

  Let's look at another query of the same type:

  > HTTP/1.1 200 OK
  > Date: Wed, 04 Jan 2016 23:18:20 GMT
  > Server: Apache/1.0.3 (Debian)
  > Content-Type: text/html
  > Transfer-Encoding: chunked

  If you think this query is NEUTRAL, click the button, as before.
  But, if you see a VULNERABILITY please mark the indicative line.
  You can mark and unmark lines by clicking to the LEFT of a line.

  Hint: There is exactly one vulnerability to be found here.
\end{lstlisting}

\begin{lstlisting}[%
  label=lst:tutorial-04,
  frame=single,
  breaklines=true,
  caption={Tutorial query 4~/~8.},
  columns=fullflexible,
  basicstyle=\footnotesize\ttfamily,
  basewidth=0.5em,
  aboveskip=0pt,
  belowskip=2pt
]
  Well done! You found the vulnerability:

  > Server: Apache/1.0.3 (Debian)

  When we say VULNERABILITY, we mean an indicator for it.
  The vulnerability here is CVE-1999-0067.
  The old version of Apache indicated that.

  We don't expect you to look that up.
  Think like an attacker looking for vulnerabilities to EXPLOIT.
\end{lstlisting}

\begin{lstlisting}[%
  label=lst:tutorial-05,
  frame=single,
  breaklines=true,
  caption={Tutorial query 5~/~8 with deceptive path traversal.},
  columns=fullflexible,
  basicstyle=\footnotesize\ttfamily,
  basewidth=0.5em,
  aboveskip=0pt,
  belowskip=2pt
]
  Let's look at another query.

  > GET /wiki/view?file=../articles/Cat.php 200 OK
  > Host: en.wikipedia.org
  > User-Agent: curl/7.68.0
  > Accept: */*

  This time, you may also mark lines as TRAPS instead.
  A trap wants you to believe that there is something to exploit.
  You want to avoid them so you don't waste time or set off alarms.
  You can mark and unmark traps by clicking to the RIGHT of a line.

  So, you now have three options:

  - Mark lines that you would like to EXPLOIT
  - Mark lines that you think are TRAPS (and therefore, will avoid)
  - Continue without marking, if you think this query is NEUTRAL
\end{lstlisting}

\begin{lstlisting}[%
  label=lst:tutorial-06,
  frame=single,
  breaklines=true,
  caption={Tutorial query 6~/~8.},
  columns=fullflexible,
  basicstyle=\footnotesize\ttfamily,
  basewidth=0.5em,
  aboveskip=0pt,
  belowskip=2pt
]
  This was a bit harder, wasn't it? You probably marked this line:

  > GET /wiki/view?file=../articles/Cat.php 200 OK

  This looks like a path traversal vulnerability, doesn't it?

  Honeyquest classified this query as either RISKY or DECEPTIVE.

  - RISKY queries contain vulnerabilities or weaknesses
  - DECEPTIVE queries contain traps you must avoid

  You may now wonder if this was truly a vulnerability or a trap.
  We don't know either. That depends on the actual implementation.

  What matters is only what you - the hacker - THINK it was.
  Honeyquest wants to understand how you PERCEIVE such queries.
\end{lstlisting}

\balance

\begin{lstlisting}[%
  label=lst:tutorial-07,
  frame=single,
  breaklines=true,
  caption={Tutorial query 7~/~8 with neutral file system.},
  columns=fullflexible,
  basicstyle=\footnotesize\ttfamily,
  basewidth=0.5em,
  aboveskip=0pt,
  belowskip=2pt
]
  Let's look at one last query. This time, it is of type FILESYSTEM.

  > -rw-r--r--  1 goofy goofy 3.5K Sep 17  2017 .bashrc
  > drwx------  7 goofy goofy 4.0K Jan 12  2022 .cache
  > drwx------  6 goofy goofy 4.0K Sep 25 17:40 .config
  > -rw-r--r--  1 goofy goofy   64 Jun 16  2019 .gitconfig
  > drwxr-xr-x  5 goofy goofy 4.0K Jul 10  2021 git-crypt
  > drwx------  4 goofy goofy 4.0K Jan 14  2022 .gnupg
  > lrwxrwxrwx  1 goofy goofy   19 Jan 23  2018 html -> /srv/html
  > drwxr-xr-x  4 goofy goofy 4.0K Aug  1  2021 app-browser
  > -rw-r--r--  1 goofy goofy 816M Nov 18  2020 nginx-proxy-logs.txt
  > drwxr-xr-x  5 goofy goofy 4.0K Mar 21  2021 terraform-saas

  Here, marking something to EXPLOIT means opening the file or folder.
  Marking something as a TRAP means you want to AVOID opening it.

  You might also have noticed that we put numbers next to your marks.
  With them, indicate the ORDER in which you like to EXPLOIT things.

  A few notes:

  - Marking multiple lines is OPTIONAL. Marking one or none is fine.
  - There might be queries where indicating an order makes no sense.
    Ignore the numbers then.
  - We don't expect you to go over every single line and mark it.
    Remember, you are a hacker, rather tell us your next move,
    not an exhaustive list of all possible moves.
\end{lstlisting}

\begin{lstlisting}[%
  label=lst:tutorial-08,
  frame=single,
  breaklines=true,
  caption={Tutorial query 8~/~8.},
  columns=fullflexible,
  basicstyle=\footnotesize\ttfamily,
  basewidth=0.5em,
  aboveskip=0pt,
  belowskip=2pt
]
  You are all set, here is a brief summary.

  Honeyquest shows you NEUTRAL or RISKY or DECEPTIVE queries.
  You can answer as many questions as you like and come back later.
  Honeyquest saves your progress automatically.

  Think like a hacker and tell us your NEXT MOVE.

  - You can CONTINUE without marking anything
  - You can mark lines to EXPLOIT or mark them as a TRAP to avoid
  - You can indicate the ORDER in which you would like to exploit

  That's it. Enjoy the game!
\end{lstlisting}

% ==============================================================================
% ------------------------------------------------------------------------------

\subsection{Image Attribution}
\label{sec:appendix:credits}

\riski~\href{https://www.svgrepo.com/svg/200115/lightning-thunder}{Lightning Thunder Icon} by \textit{svgrepo.com} licensed under CC0
\newline
\dcpti~\href{https://iconduck.com/icons/176847/bee}{Bee Icon} by \textit{bypeople.com} licensed under CC BY 4.0
\newline
\ntrli~\href{https://www.svgrepo.com/svg/402205/neutral-face}{Neutral Face Icon} by \textit{joypixels.com} licensed under CC BY 4.0
\newline
\expli~\href{https://thenounproject.com/icon/hammer-6543915/}{Hammer Icon} by \textit{Muh Zakaria} licensed under CC BY 3.0
\newline
\trapi~\href{https://www.svgrepo.com/svg/499005/bear-trap}{Bear Trap Icon} by \textit{Daniela Howe} licensed under SIL OFL 1.1

\end{document}